\documentclass[12pt,graphics] {article}
\usepackage[dvips]{graphicx,color}
\newcommand{\bea}{\begin{eqnarray}}
\newcommand{\eea}{\end{eqnarray}}
\newcommand{\be}{\begin{equation}}
\newcommand{\ee}{\end{equation}}
\newcommand{\unit}{\mathbf{1}}

\def\lp{l^\prime}
\def\lpp{l^{\prime\prime}}
\def\lppp{l^{\prime\prime\prime}}

\def\llb{{\Bigg {\lbrack}}\!\!{\Bigg {\lbrack}}}
\def\rrb{{\Bigg {\rbrack}}\!\!{\Bigg {\rbrack}}}

\def\rhoAE{\rho_{\!{\atop AE}}}
\def\rhoEB{\rho_{\!{\atop EB}}}
\def\alphaAE{\alpha_{\!{\atop AE}}}
\def\alphaEB{\alpha_{\!{\atop EB}}}
\def\lsim{\mathrel{\lower2.5pt\vbox{\lineskip=0pt\baselineskip=0pt
          \hbox{$<$}\hbox{$\sim$}}}}
\def\gsim{\mathrel{\lower2.5pt\vbox{\lineskip=0pt\baselineskip=0pt
          \hbox{$>$}\hbox{$\sim$}}}}

\begin{document}

\begin{titlepage}
\begin{trivlist}\sffamily\mdseries\large
\item
MTR00W0000052\\[-0.8ex]
\hrule ~\\[-1.2ex]
{\mdseries MITRE TECHNICAL REPORT}\\[1cm]
\LARGE
\begin{center}
\bfseries
Practical Quantum Cryptography:\\
A Comprehensive Analysis\\
(Part One)\\[2.5cm]
\end{center}
\mdseries
\large
G. Gilbert\\[0.8ex]
M. Hamrick\\[0.4ex]
~\\
\large
\textbf{September 2000}\\[3.1cm]
\begingroup\footnotesize
\begin{tabbing}
\textbf{Sponsor:} \phantom{spo} \= The MITRE Corporation \phantom{phantomphantomprospo} \= 
\textbf{Contract No.:} \phantom{pro}\= DAAB07-00-C-C201 \\
\textbf{Dept. No.:} \>W072 \>\textbf{Project No.:} \>51MSR837\\[0.6cm]
The views, opinions and/or findings contained in this report \> \>
   Approved for public release; \\
are those of The MITRE Corporation and should not be \> \>
   distribution unlimited.  \\
construed as an official Government position, policy, or \\
decision, unless designated by other documentation. \\[0.3cm]
\copyright 2000 The MITRE Corporation
\end{tabbing}
\endgroup
~\\
\bfseries
\Large 
MITRE\\
\normalsize
Washington ${\mathbf C^3}$ Center\\
McLean, Virginia\\
\clearpage
\end{trivlist}
\end{titlepage}

\Large
\begin{center}

{\bf Practical Quantum Cryptography:\\A Comprehensive Analysis\\ (Part One)$^{\ast}$}
\normalsize 

\vspace*{15pt}
Gerald Gilbert$^{\dag}$ and Michael Hamrick$^{\ddag}$\\
The MITRE Corporation\\ McLean, Virginia 22102
\end{center}

\begin{abstract}
We perform a comprehensive analysis of practical quantum cryptography (QC) systems
implemented in actual physical environments {\it via} either
free-space or fiber-optic cable quantum channels for ground-ground, ground-satellite,
air-satellite and satellite-satellite links.

(1) We obtain universal
expressions for the effective secrecy capacity and rate for QC systems taking
direct and ancillary processes into account. The analysis in Part
One treats three important individual quantum bit attacks, comprising generic admixtures
of indirect attacks, direct attacks and previously unconsidered simultaneous
combinations of the two types. In all these cases we obtain for the first time
{\it necessary and sufficient} exact closed form expressions for privacy amplification.
Our analysis also includes for the first time
the explicit calculation in detail of the total cost in bits of {\it continuous}
authentication,
thereby obtaining new results for actual ciphers of finite length, as well as
previously obtained limits
for idealized ciphers of infinite length.

(2) We perform for
the first time a detailed, explicit analysis of all systems losses
due to and errors and noises,
including turbulent and static atmospheric propagation losses, optics package losses,
intrinsic channel losses, {\it etc}., as appropriate to both optical fiber cable- and
satellite communications-based implementations of QC.

(3) We calculate for the first time all system load costs
associated to classical communication and computational constraints that are ancillary
to, but essential for carrying out, the pure QC protocol itself, including
the full classical communications bandwidth requirements
and the full computer machine instruction
requirements needed to support actual QC implementations.

(4) We introduce an extended family of generalizations of the Bennett-Brassard
(BB84) QC
protocol that equally provide unconditional secrecy but allow for the possibility of
optimizing throughput rates against specific cryptanalytic attacks. 

(5) We obtain universal predictions for
maximal rates that can be achieved with practical system designs under realistic
environmental conditions, taking into account our results for total system losses and
loads.\\[0.2in]
\hrule width 2.5 in
\vspace*{.175in}
{\footnotesize
$\!\!\!\!\!\!\!\!\!\!\ast$ This research was supported by MITRE under MITRE Sponsored
Research Grant
51MSR837.\\
\dag ~ggilbert@mitre.org\\
\ddag ~mhamrick@mitre.org
}
\clearpage

(6) We propose a specific QC system design that includes
the use of a novel method of high-speed photon detection that may be able to achieve
very high throughput rates for actual implementations in realistic environments.

(7) We deduce the dependence of the effective throughput on
processing block size for actual ciphers of finite length and
derive thereby an upper bound on practical
processing block sizes dictated by current available computing machinery. We use this
to show
how a system employing an array of parallel transmitter and receiver devices can
be multiplexed to substantially increase the throughput of shared secret cipher.
\end{abstract}

\normalsize 

\newpage
\vspace*{.1 in}
{\underbar{PRACTICAL QUANTUM CRYPTOGRAPHY: A COMPREHENSIVE ANALYSIS}}

\vspace{.35 in}

PART ONE - Quantum Cryptography without Entangled States (this volume)

PART TWO - Quantum Cryptography with Entangled States (to appear)
\newpage

\newpage
\tableofcontents
\newpage

\newpage
\listoffigures
\newpage

\newpage
\listoftables
\newpage

\section{Prologue}

What is the reason for the excitement about quantum cryptography? Quantum cryptography
is special because it provides a means for encrypting information that
no amount of analysis can break. This is referred to as ``unconditional secrecy."
The Great Thing is that this property of quantum cryptography
is not a consequence of some ``hard" mathematics problem
that might be solved one day,
nor of some devilishly clever algorithm or fiendishly intricate hardware design
that might be reverse-engineered one day,
but instead is due to what are believed to be inviolable
principles of physical law: the physics of Quantum Mechanics. If our understanding
of quantum mechanics is correct, and after three-quarters of a century of research we
know of no reason to believe it to be incorrect, quantum cryptography is and {\it always
will be} unconditionally secret, irrespective of whatever advances are made
in mathematics or computer science, and probably in any other sphere of human activity.
If the question is ``What is the strongest
{\it cryptographic} protection possible, as constrained directly by physical law?"
the answer is ``Quantum Cryptography."

Quantum cryptography specifically
provides a method of distributing the secret keys required to
provide unconditionally secret communications - these are the famous ``one-time
pads" - and its use is guaranteed to reveal the presence of an enemy attempting to
compromise the transfer.
All quantum communications, such as quantum cryptography, requires the use of a quantum
channel, which is a means of transporting physical objects called quantum bits (or
``qubits") in such a way that the quantum mechanical states of the qubits remain
preserved
from one end of the channel to the other. Two forms of quantum channel for
quantum cryptography have thus far been shown to provide
viable options, namely optical fiber
cable, and (perhaps surprisingly) the atmosphere around us. The demonstrations
conducted thus far have proved that it is possible to carry out quantum
cryptography at low throughput rates, thus far not exceeding a few thousand
bits per second. Our interest is in analyzing the possibilities for increasing
the effective throughput rate for practical quantum cryptography systems to a range
that is high enough to allow for the real-time encryption of useful volumes of
data.

Recent progress in high-speed photon detection, high-speed laser optoelectronics, 
wavelength and time division multiplexing and lasercomm terminal miniaturization has
occurred
which makes it for the first time possible to contemplate the design of high-speed
quantum cryptography systems. In addition to determining how to optimize quantum
cryptography systems built out of currently available technology (and ensure that
such systems will be perfectly secret in the presence of system imperfections),
our analysis identifies the problems, provides
corresponding solutions and demarcates the various constraints that will govern the
development of high-speed quantum cryptography as new technology appears.
The actual implementation of
high-speed quantum cryptography systems would be invaluable, allowing for the
first time the the practical possibility of one-time-pad-encrypted, undecipherable
high-speed communications
in bulk. If this can be achieved it will
offer an essentially new degree of security in future high-bandwidth communications.

\clearpage

\section{Introduction}

\subsection{Overview}

\noindent{Quantum key distribution (QKD) is a promising approach to the ancient problem
of protecting sensitive communications from the enemy.\footnote{
We will sometimes employ the word ``enemy,"
following the usage of Shannon \cite{Shannon}, to denote anyone who may intercept an
enciphered message.}
QKD is not in itself a method of enciphering information: it is instead a means
of arranging that separated parties may share a completely secret, random sequence
of symbols to be used as a {\it key} for the purpose of enciphering
a message. Our objective is to elevate the use of quantum key distribution to the status
of supporting full end-to-end real--time Vernam encryption \cite{Shannon,vernam}.
The Vernam cipher system (or systems based on it) provides
the only known cryptographic method of achieving
unconditionally secret communications.\footnote{
In the Vernam cipher system the message is
referred, via the ``exclusive or" (XOR) logic operation, to a  random
string of symbols, the Vernam cipher (one-time pad), resulting in another random string
of symbols comprising the ciphertext. As a truly random string, the
ciphertext is literally informationless, and cannot be decrypted by anyone not in
possession of the random string used for the encryption. This is true
irrespective of how much computing power they possess or which
algorithms they utilize.}
We thus envisage an end-to-end cryptosystem
that includes an initial phase of quantum key distribution and a subsequent phase of
encryption with the method of the Vernam cipher, continuously and at
useful,
high data throughput rates. There have been a number of experimental demonstrations of
QKD reported recently
\cite{demorefs1,demorefs2,demorefs3,demorefs4,demorefs5,demorefs6,demorefs7,
demorefs8,demorefs9} that have been important in indicating the viability
of the concept and in suggesting that it might be possible to incorporate QKD in
practical
systems applications. The initial demonstrations of QKD have been at low data throughput
rates, none exceeding a few kilobits per second. However, the application of quantum key
distribution at low data throughput rates does not support unconditionally secret Vernam
encryption of modern communications data volumes, although implementation of quantum key
distribution at low rates can indeed be useful as a means of distributing
the {\it cryptovariables}
that are used in traditional, classical symmetrical cryptography.\footnote{
For a review of classical cryptography, see \cite{crypto1,crypto2}.}
For this purpose QKD
systems operating at low data throughput rates of order a few thousand bits per second
are perhaps adequate to play a supporting role for classical enciphering
systems. However, classical cryptography (symmetrical or not) is not {\it
unconditionally} secret, with the sole exception being the Vernam system (or
systems directly based on it). There are situations and circumstances for which
it is desirable, and in some cases absolutely essential, to increase the secrecy to
the level possible only with the use of the Vernam cipher method. However, since the
Vernam method requires a shared cipher at least as long, bit for bit, as the plaintext
message to be enciphered, and moreover may under no circumstances be used more than one
time, it is clear that slow data rates for key distribution\footnote{
We will use the words {\it key} and {\it cipher} to mean the same thing, since we always
have in mind the Vernam system in particular, for which the two are synonymous.}
will not work. Only a high speed QKD system can suffice to distribute, and distribute
again and again as required, sufficiently large amounts of cipher material to support
real-time Vernam encryption. We propose to reserve the phrase {\it quantum key
distribution}
solely to describe the distribution of cipher material, and suggest that the phrase {\it
quantum cryptography} be used to denote the combination in one complete end-to-end protocol
of both QKD {\it and} subsequent Vernam encryption. Since, as discussed above, practical
use in modern communications of the Vernam method requires high speed data  throughput
rates, quantum cryptography defined in this way for such an application
is implicitly {\it high speed} quantum
cryptography. In this sense practical quantum cryptography offers something never before
technologically possible: the use of the Vernam cipher for those applications when
unconditional secrecy is required or desirable, and indeed the fact that this can be
done may lead to the suggestion of its use in circumstances for which it has previously
only been thought of as an abstract idealization.}

\subsection{Summary of Results}

The principal new contributions of this paper are as follows. We obtain:

(1) A universal expression for the effective secrecy capacity that is valid in
the general case of an actual cipher of finite length, that can be specialized to the
case of an abstract cipher of infinite length, in eqs.(\ref{162})
and (\ref{164}). In the course of the
derivation we obtain three categories of new results: (a) exact, closed form
expressions for the {\it necessary and sufficient} amount of privacy amplification
required
to ensure a secret shared key associated with {\it direct}, {\it indirect}
and newly identified {\it combined}
cryptanalytic attacks on the
transmission (eqs.(\ref{138}) through (\ref{140})), (b) a practical,
universal bound on the complete privacy amplification function
that provides for useful data throughput values,
while accounting for direct, indirect and combined individual attacks on the
transmission,\footnote{
In Part One of this work we will consider those attacks that the
enemy can conduct using classical computing machines. In Part Two we will extend the
analysis to include the potential attacks that could be performed in the future if
and when quantum computing machines are available.}
that is {\it always}
at least as large as the minimum number of required subtraction bits
required to ensure a secret shared key,
in eqs.(\ref{142}) through (\ref{151}), and (c) a
complete closed-form expression for the necessary and sufficient
number of bits required to effect continuous authentication, in eqs.(\ref{152})
through (\ref{158}).

(2) Complete characterizations of the total line attenuation losses, for
free-space quantum channels, in eqs.(\ref{179}), (\ref{183}),
(\ref{185}), (\ref{186}), (\ref{196}), (\ref{199}), (\ref{203}) and (\ref{204}).

(3) A closed form relationship between intrinsic fractional
quantum channel error and satellite-ground platform (or satellite-airborne platform
or satellite-satellite) misalignment, in eqs.(\ref{206}) and (\ref{332}).

(4) A closed-form expression relating the necessary amount of classical
communications throughput to the parameters of the system, in the transmitter-receiver
direction as well as in the receiver-transmitter direction (the two are not the same),
and we obtain practical
working values for particular systems, in eqs.(\ref{270}), (\ref{271}), (\ref{273})
and (\ref{274}).

(5) A closed-form expression relating the computational burden, measured in units
of necessary machine instructions, to the system parameters, and we obtain practical
working point values, in and below eq.(\ref{E:compload}).

(6) Universal maximal rate predictions, for a variety of quantum
cryptography scenarios, including Earth-to-LEO satellite in clear weather,
Earth-to-LEO satellite in poor weather, aircraft-to-LEO satellite,
Earth-to-GEO satellite, GEO-to-GEO satellite and fiber-optic cable links,
in Sections 5.3.1 and 5.3.2.


\subsection{Organization of the Paper}

The paper is organized as follows. In Section 3 we carry out a complete formal
derivation of the effective secrecy capacity and rate for practical
quantum cryptography systems. We obtain exact, closed-form results for the entire
system dynamics, including the calculation of
exact necessary and sufficient results, as well as useful practical results
for the required privacy amplification to ensure the unconditional secrecy of the
shared cipher. Our exact results allow us to explicitly
determine the requirements for
high speed quantum cryptography in practical implementations.
In Section 4 we perform a comprehensive analysis of all system losses and
loads, for both free-space-based and optical fiber cable-based quantum
cryptography systems, including
in particular the full classical communications bandwidth requirements
and the full computer machine instruction
requirements needed to support actual quantum cryptography
implementations. In Section 5 we analyze
precise requirements for and detailed methods to achieve successful practical
high speed quantum cryptography implementations in realistic environments.
Our conclusions and a discussion are contained in Section 6,
and Sections 7 and 8 contain acknowledgements and several appendices.

\subsection{Brief Description of Quantum Key Distribution Protocol}

Here we provide a very brief description of the basic elements of quantum
key distribution. We will illustrate this with the original four-state QKD protocol
developed by Bennett and Brassard in 1984 known as the ``BB84" protocol \cite{BB84}.
For definiteness in this illustration
we will assume that individual photons serve as the quantum bits for the protocol,
or more precisely, the polarization states of individual photons. To carry out the
protocol one of the parties transmits
a sequence of photons to the other party.
The parties publicly agree to make use of two distinct polarization bases which are
chosen to be maximally non-orthogonal. In a completely
random order, a sequence of photons are prepared in states of definite polarization
in one or the other of the two chosen bases and transmitted by one of the parties to the
other through a channel that preserves the polarization. The photons are measured by
the receiver in one or the other of the agreed upon bases, again chosen in a
completely random order. The choices of basis made by the transmitter and receiver
thus comprise two independent random sequences.
Since they are independent random sequences of binary numbers, about half of the basis
choices will be the same and are called the ``compatible" bases, and the other half
will be different and are called the ``incompatible" bases.
The two parties compare publicly, making use for this purpose of a classical communications
channel, the two independent random sets of polarization {\it bases} that were used,
without revealing the polarization {\it states} that they observed.
The bit values of those polarization states measured in the compatible bases
furnish the ``sifted key."
Note that, if the two parties used classical signals
to send the key, an eavesdropper 
could simply measure the signals to obtain complete knowledge of the key. If, on the
other hand, the two parties use 
single photons to transmit the key, the Heisenberg Indeterminacy Principle guarantees 
that an eavesdropper cannot measure the polarizations 
without being detected.
The sifted keys possessed by each of the parties will in general
be slightly different from each other due to errors caused by the use of imperfect
equipment.
A classical error correction procedure, carried out through the classical communication
channel, is executed in order to produce identical, error-free keys at both ends.
It is possible that an enemy may have obtained some information about the key during
the publicly-discussed error correction phase of the protocol.
In addition, it is also possible for the enemy to have obtained information due to the
presence in the sequence of quantum bits of multiple photon states.
The process of ``privacy amplification" is therefore applied to the sifted, error-free
key, which has the effect of reducing the information available to the enemy to less than
one bit, with extremely high probability.

These basic elements of quantum key distribution are discussed and analyzed in
detail in this paper.

\subsection{Secrecy and Security in Communications}

It is important to be clear about the ``security" advantage that does, and does not,
derive from the use of quantum key distribution, quantum cryptography
and the method of the Vernam cipher in
secret communications. For this purpose we introduce standard definitions and discuss
issues of context and application.

\subsubsection{Definitions}

``Secrecy" and ``security" do not have the same meaning: the former is included within
the
latter. Stated differently, all secure communications
systems provide secrecy, but not all secret communications
systems provide security.
In this paper we reserve the word {\it secrecy},
to mean what Shannon meant by the phrase ``perfect secrecy" in his
seminal work on the subject:
{\it Communication Theory of Secrecy Systems} ({\it cf} \cite{Shannon}).
The basic requirement for secrecy is that, in comparing the situation {\it before}
the enemy has intercepted the transmission with the situation
{\it after} any such interception (and analysis) has occurred,
the {\it a posteriori} and {\it a priori}
probabilities for the enemy to know the content of the transmission must
be identical.\footnote{
Of course, if the {\it a posteriori} and {\it a priori} probabilities are indeed
identical, but happen to be
identically equal to, say, unity, then we clearly don't have a secret system.
In the case of a string to be used as a Vernam cipher we
obviously {\it also} need that the probability for the enemy to know any specific bit
is equal to 50\%, independently of the ordering of the bits. Then perfect secrecy
in Shannon's sense means that the probability for
the enemy to know the entire string approaches zero
exponentially quickly with the number of bits.}
In an operational sense, we specifically intend the word ``secrecy" to characterize,
and apply solely to,
the protection provided {\it strictly} by the cryptographic protocol alone.
This operational meaning is best explained by placing secrecy in proper perspective
in the larger framework of ``security," or more precisely,
communications security~\cite{fs1037}. Communications {\it security},
(so called ``COMSEC"), may be naturally split into four separate categories (there are
other ways of organizing these concepts -- here we invoke the standard
scheme advocated by the U.S. National Security Agency in \cite{nsadoc}):

(1) cryptosecurity - [The] component of communications security that results from the
provision of {\it technically sound cryptosystems}
(emphasis added) and their proper use.

(2) emission security - Protection resulting from all measures taken to deny unauthorized
persons information of value which might be derived from intercept and analysis of
compromising emanations from crypto-equipment, computer and telecommunications systems.

(3) physical security - The component of communications security that results from all
physical measures necessary to safeguard classified equipment, material, and documents
from access thereto or observation thereof by unauthorized persons.

(4) transmission security - [The] component of communications security that results from
the application of measures designed to protect transmissions from interception and
exploitation by means other than cryptanalysis.

The word {\it secrecy} throughout this paper means no more and no less than {\it
cryptosecurity}, in the sense of definition (1) above. This is the secrecy protection
afforded purely {\it by the cryptographic protocol against purely cryptanalytic attacks}
only. {\it Unconditional secrecy} refers to secrecy that remains intact when the
cryptosystem is subjected to attacks by an enemy equipped with
unlimited time and - within the
constraints dictated by the laws of physics - unlimited computing machinery.

\subsubsection{Technically Sound Quantum Cryptosystem Design and Practice}

What {\it security} protection should unconditional {\it secrecy}
provide? Should the purview of cryptosecurity, the protection afforded specifically
by the cryptosystem {\it per s\'e}, be extended to include protection normally
provided by the other three elements of COMSEC? The answer to this question is
``no." Stated more precisely, if a so-called ``technically sound cryptosystem" is
properly operated, with the consequent
balance between the four elements of COMSEC that this implies,
there should be no {\it need} for such an extension of
purview. It is not the purpose of this paper to provide a detailed
analysis of proper cryptosystem {\it praxis}, but two observations should
suffice to illustrate the main point. 

As one example, we may imagine that a prefect cryptographic system that provides
unconditional secrecy has been set up and is in use. If the actual method of
use by the secret communicators, however, includes ``leaving
the door(s) open" at one (or both) of their facilities,
so that an eavesdropper can actually {\it gain access} to their system in some
way, the system {\it security} is obviously entirely lost, in spite of
the unconditional {\it secrecy} of the underlying
cryptosystem. Is it reasonable to insist that the cryptosystem, {\it per s\'e},
provide protection against such technically unsound cryptographic practice? The answer
is ``no." Before commenting on the technical
implications of this, let us consider a
different situation.

As another example, one might try to
argue (erroneously) that free-space {\it classical}
key distribution between a
satellite and a ground station is ``obviously" perfectly secure without the need for
making use of quantum bits, or even any cryptography at all.
If the transmission consists of classical bits,
encoded in optical pulses generated by a laser and propagated along a highly
collimated beam, wouldn't it be {\it very difficult} for an enemy to actually
physically intercept the beam at all? Wouldn't it be {\it almost impossible} for 
the enemy to somehow grab such a signal out of a highly collimated, thin beam?
Of course, the beam becomes broader as it propagates, and moreover, it is possible for
the enemy to exploit the scattering of such a beam
but nevertheless, the answer to this question is ``yes, it is difficult"
but that is irrelevant. 

What these two examples illustrate are two extremes in regard to what protection
the cryptosystem
should, and should not provide. We agree strongly with the philosophy
that cryptosecurity should be viewed as {\it only one part} of an overall system
for ensuring communications security ({\it cf} \cite{schneier}). Technically
sound quantum cryptosystem {\it design}, for
instance, dictates that if it is possible to trivially prevent the enemy from
modifying photon wavelengths in multiple photon pulses and thereby prevent the
``remote" adjustment of the quantum efficiency of the photon detector in a quantum
cryptography system by simply placing a narrow bandpass filter at the front
of the receiving apparatus, then such a technique, which falls {\it outside the
purview}
of pure cryptography and is instead an element of transmission security,
must be implemented.\footnote{
This particular issue is in fact of considerable
importance and is discussed in much more detail later in the paper.}
The point is that the fact that this is
being implemented
can be fully disclosed, without any loss of security whatsoever, to the enemy, as
there is nothing that can be done about it. Similarly, technically sound quantum
cryptosystem {\it practice} dictates that the communicating parties must
obtain an accurate measurement of the ambient
noise along the quantum channel prior to the use of the system.
On the other hand, potential
attacks on the secret communication cryptosecurity, as such, must be protected against,
{\it solely} on the basis of whatever features the cryptosystem itself provides.
It is not a valid argument that a
particular attack is ``difficult," since the technological capabilities of the
enemy may improve, and moreover, these
should never be underestimated. The consequences of the preceding qualitative
statements all translate into concrete mathematical implications for the detailed
analysis of the necessary and sufficient amount of privacy amplification (this is
introduced in the next section) required in a
practical quantum cryptography system, so we codify the
meaning of this below.

\vskip 10pt
\noindent {\it Quantum Cryptographic Conservative Catechism}

We propose the following ``doctrine of reasonableness" for analyzing
practical quantum cryptography systems: the
{\it Quantum Cryptographic Conservative Catechism} (QCCC), according to
which (1) it is presumed that both the physical hardware design {\it and} the
actual operation of any QC system will together furnish a technically
sound cryptosystem as determined both by the
precedents already established through the
history of cryptology and new features specific to quantum communications,
and (2) any proper theoretical analysis of the performance
characteristics of a practical QC system must incorporate the underlying
assumption that the enemy is limited solely by the laws of physics, relaxed {\it only}
to the extent that it is reasonable to take condition (1) into account. This is the
approach followed in our study.

\clearpage

\section{Theoretical Analysis of Effective Secrecy Capacity}

\noindent{In this chapter we will perform a careful derivation of the functions that
provide a full account of the operating characteristics of a general QKD system.
Our analysis is specifically constructed to characterize the Bennett-Brassard Four-State
(BB84) QKD protocol \cite{BB84}, or more precisely, a set of generalizations
that include the original BB84 protocol as a special case.\footnote{
In fact, the use of a source of data bits that
produces an admixture of single- and multiple-particle states means that the system does
{\it not} implement the original, pure BB84 protocol, which by definition requires pure,
idealized qubits represented by single particle states. However, our analysis is
sufficiently general to include all such implementations. In addition, although our
analysis does not focus specifically on quantum cryptography with entangled states,
such as the
Ekert protocol \cite{ekert} or
the recently demonstrated entangled state {\it variant} of the
original BB84 protocol \cite{BB84entangled}, our results can be modified
to apply to it as well. Work is in progress on the latter topic, which will
appear as Part Two of the current work~\cite{parttwo}.}
The figure-of-merit for the operating characteristics and secrecy of a QKD
system is provided by the effective secrecy capacity, ${\cal S}$, in terms of which
we may define the effective secrecy {\it rate} for the system, ${\cal R}$.\footnote{
In our calculation we shall adopt, and considerably expand, the notational scheme
introduced by Slutsky, {\it et. al.} in \cite{slutsky1} of the various system
characteristics and capacities. Our analysis is more complete than previous treaments
in accounting for {\it all} relevant system processes, and our use of previously
established
notation will in particular make it easy to identify the ways in which our analysis
extends
previously obtained results in this area.}
These quantities provide a full characterization of the operating characteristics of the
cryptographic communications system set up between the legitimate communicating parties,
traditionally referred to
as ``Alice" and ``Bob." The effective secrecy capacity is defined as the ratio of
the {\it final length} in bits,
of the secret shared cipher, to the number of bits initially transmitted by Alice to Bob
in order to establish the final cipher. The ``final length" is the length of the string
after the full execution of the protocol, including all of the required error correction,
privacy amplification and continuous authentication\footnote{
It is essential in any complete analysis of the characteristics of a QKD system to fully
account for what we refer to as ``continuous authentication." (The authors thank a
contact at the U.S. National Security Agency for suggesting to us the
phrase ``continuous
authentication.") Authentication is intended to ensure that only legitimate parties may
communicate via a cryptographic system.
We require the minimum amount of authentication, but no less than that,
in order to preserve the integrity of the QKD protocol. {\it Continuous} authentication,
for every single transmission between Alice and Bob, is absolutely required, but has not
been thoroughly studied before. It is not
sufficient to ``authenticate" once, as repeated attempts at system intrusion
may be made by the enemy. In this paper we carry out a full analysis of this process,
along with all other relevant system processes.}
has been applied to the original, ``raw" string, {\it i.e.}, the transmitted string that
has not yet been subjected to any processing at all.  We denote the number of ``raw"
pulses\footnote{
This number includes the ``empty" pulses -- those for which the filtering applied to the
output of the laser has resulted in the statistical extinction of the photon content.}
sent by Alice to Bob as $m$, the number of bits in the compatible polarization basis
that actually reach Bob as $n$, the number of {\it those} bits which are in error as
$e_T$, the total number of bits that must be subtracted\footnote{
The privacy amplification subtraction function, $s$, is defined here so
as {\it not} to include the privacy amplification security parameter $g_{pa}$.}
from the string in order to effect privacy amplification as $s$, the privacy amplification
security parameter as $g_{pa}$, and the number of bits required to be subtracted in
order to carry out continuous authentication as $a$. Then the effective secrecy capacity
is defined as}\footnote{
We choose a definition of the effective secrecy capacity appropriate for the QKD protocol
in which error correction is effected by identifying and {\it discarding}, rather than
identifying, correcting and {\it retaining}, the error bits. Our analysis can be easily
adapted to the case where error bits are identified, corrected and retained. This is
consistent with the conservative approach adopted throughout our
analysis and means that the various rate predictions we will make in fact constitute
{\it lower bounds} on achievable throughput values.}
%
\begin{equation}
{\cal S}\equiv{n-e_T-s-g_{pa}-a\over m}~.
\label{1}
\end{equation}
\noindent{The effective secrecy rate corresponding to the effective secrecy capacity
measures the effective throughput of secure Vernam cipher in bits per second
and is given by}
\be
{\cal R}=\tau^{-1}{\cal S},
\label{2}
\ee
\noindent{where $\tau$ is the bit cell period, the period of time that is required for the
system hardware to transmit one signal from Alice to Bob, ``reset" itself and become ready
to transmit the next signal.}

\subsection{Derivation of Effective Secrecy Capacity}

\noindent{We want to determine the conditions under which a fully realistic, practical
system implementation of the BB84 protocol for quantum key distribution can produce
unconditionally secret, shared key material between Alice and Bob. Moreover, we want to
discover those conditions under which we may obtain the highest possible data throughput
rate so that we can use the shared key as a real--time Vernam cipher and thus legitimately
speak of end-to-end quantum cryptography, {\it i.e.,} an unconditionally secret
communications system capable of supporting large volumes of data. For this purpose we
need to obtain an explicit, closed form expression for the effective secrecy capacity
that directly expresses $\cal S$ in terms of the actual operating parameters for a
realistic system implementation of quantum cryptography.}

\subsubsection{The Sifted Key and the Transmitted Errors}

We will deduce explicit expressions for the various quantities that
appear in the expression for the effective secrecy capacity, starting with the
length of the sifted key and the length of the transmitted error part. We will
assume that the ``Alice" system instrumentation principally includes a pulsed
laser which generates pulses of light in the form of coherent states. This
assumption can be made without loss
of generality since, as will be evident below, our formulation will include as a special
case the situation in which Alice instead utilizes a device that, through whatever means,
produces only single-photon states. The state function for a fiducial coherent state
produced by the laser is given by\footnote{
The enemy, in general, will not detect potential {\it intercepted} states
precisely as
coherent states, but will intstead (due to lack of a phase reference) detect a mixture of
Fock space states that are characterized by a Poisson distribution and described
by an appropriate density matrix \cite{lutkenhaus1}. This
fact has no bearing on the calculation of
the number of {\it sifted} bits shared between Alice and Bob, as they {\it do} possess
the necessary phase reference, and thus the use of explicit coherent states is
appropriate.}
%
\begin{equation}
\label{3}
\vert\phi\rangle=\sum_{l=0}^\infty {\sqrt {e^{-\mu}{\mu^l\over l!}}}e^{il\phi}\vert l
\rangle
\end{equation}
\noindent{where $\phi$ is the quantum mechanical phase and $\mu$ is defined as the
expectation value of the number operator, and in practice is the mean photon number
per pulse. The number of photons produced is thus characterized by a Poisson distribution.
We denote by $\hat\chi\left(\mu,l\right)$ the probability that a laser pulse (in a
stream of pulses characterized by $\mu$) will contain exactly $l$ photons and thus
\be
\label{4}
\hat\chi\left(\mu,l\right)=e^{-\mu}{\mu^l\over l!}~.
\label{poisson}
\ee
We will sometimes find it convenient to use the notation
\be
\label{5}
\psi_1\left(\mu\right)\equiv\hat\chi\left(\mu,1\right)=\mu e^{-\mu}
\ee
and
\be
\label{6}
\psi_2\left(\mu\right)\equiv\hat\chi\left(\mu,2\right)={\mu^2\over 2}e^{-\mu}
\ee
for the probabilities that exactly one and two photons, respectively, are in a pulse.

In the same manner we deduce that the probability, $\psi_{\ge 1}$, that a laser
pulse will contain one or more photons is given by}
\begin{equation}
\label{7}
\psi_{\ge 1}\left(\mu\right)=\sum_{l=1}^\infty e^{-\mu}{\mu^l\over l!}=1-e^{-\mu},
\label{p_one_or_more}
\end{equation}
and that the probability, $\psi_{\ge 2}$, that a laser pulse will contain two or more
photons is given by
\be
\label{8}
\psi_{\ge 2}\left(\mu\right)=\sum_{l=2}^\infty e^{-\mu}{\mu^l\over l!}=1-e^{-\mu}-\mu e^
{-\mu}.
\label{p_two_or_more}
\ee
In our analysis of quantum key distribution{\footnote {In strict accuracy one should
refer
to the QKD protocol as quantum key {\it expansion}, rather than distribution, since the
success of the entire process requires that Alice and Bob be in possession of a suitable
initial authentication string, which must be secret. This topic of authentication is
discussed in much greater detail below.}} we assume that Alice prepares and launches in
the direction of Bob a number, $m$, of laser pulses, referred to as the {\it raw bits}.
The time required to prepare, launch and ready the system to prepare another pulse is
the
{\it bit cell period}, $\tau$. The overall QKD event thus lasts for a duration
of $m\tau$.
Out of the full set of $m$ bit cells sent by Alice, a certain fraction only will
survive to
become potential bits in the secret key. In deducing the expression for the effective
secrecy capacity we must take into account the amount of attenutation,
$\alpha$,
that characterizes the propagation loss conditions of the trajectory connecting, and
including, the Alice and Bob systems. We also need to take account of the imperfect
intrinsic quantum efficiency, $\eta$, that characterizes Bob's detector, as well as the
intrinsic dark count probability, $r_d$. Alice
and Bob follow the standard protocol, whereby
the polarization {\it bases} (but not
the polarization {\it states}) of the bits collected by Bob are publicly discussed and
compared between Alice and Bob. The bases for which Alice and Bob find themselves in
agreement are referred to as {\it compatible} bases, and the remainder are referred to
as the {\it incompatible} bases. The random orientations of the polarizing and
polarization-discriminating apparatuses at Alice and Bob are assumed to comprise two
completely uncorrelated sequences, so that for about half of the bit cells about which
Alice and Bob conduct their discussion they will have noted compatible bases, and for
the
other half the bases will be incompatible. After taking into account the
various other effects that cause the bits shared between Alice and Bob to be
diminished in number, we can establish the number, $n$, of {\it sifted} bits for
the QKD problem.

In order to be as general as possible in our analysis
we will formulate the expression for the number of
sifted bits from first principles in terms of the various underlying probabilities
associated to the different processes that take place. Denoting the various relevant
probabilities by $\cal P$ with appropriate arguments, we have
\be
\label{9}
{\cal P}\left(l~{\rm photons~leave~Alice}\right)=e^{-\mu}{\mu^l\over l!}\equiv\hat\chi
\left(\mu,l\right),
\ee
\be
\label{10}
{\cal P}\left(l'~{\rm photons~reach~Bob}~{\Big\vert}~l~{\rm photons~leave~Alice}\right)=
\left(l\atop l'\right)\alpha^{l'}\left(1-\alpha\right)^{l-l'},
\ee
\be
\label{11}
{\cal P}\left(\lpp~{\rm photons~detected}~{\Big\vert}~\lp~{\rm photons~reach~Bob}
\right)=
\left(\lp\atop \lpp\right)\eta^{\lpp}\left(1-\eta\right)^{\lp-\lpp}\left(1-\delta_
{0,\lpp}
\right),
\ee
\be
\label{12}
{\cal P}\left({\rm no~dark~count~event}\right)=1-r_d,
\ee
and
\be
\label{13}
{\cal P}\left({\rm basis~compatibility}\right)={1\over 2}~.
\ee
In writing the expressions in eqs.(\ref{10}) and (\ref{11}) we are incorporating assumptions on the
statistical nature of the responses of both the qubit detector and the environmental
processes responsible for the line attenuation.  With the chosen form for the rhs of
eq.(\ref{10}) we are assuming that all attenuation processes act incoherently on a
$l$-photon pulse, and that there is no enhancement or suppression when $l$ photons
try to get through together. Similar assumptions apply in the case of eq.(\ref{11}), in
addition to which
the factor of $1-\delta_{0,l}$ enforces the condition that the detector apparatus may
not fire when zero photons are incident upon it (modulo dark count events, which
are described in a separate term as seen below).

These assumptions are quite reasonable, and they have evidently been implicitly adopted
in all previous work on this subject ({\it e.g.,} \cite{slutsky1,qcryptopapers}), but
we here
for the first time make them
explicitly clear. In fact, the final explicit form for $n$ obtained in
eq.(\ref{15}) below presumably {\it only}
follows upon making these two specific assumptions.\footnote{
Analysis of any specific forms for the number of sifted bits, $n$, that may arise
upon making {\it different} specific assumptions about the processes that underlie
eqs.(\ref{10}) and (\ref{11})
appears to have never been carried out. This is a worthwhile area for future research.}

We may now deduce the expression for the number of sifted bits by assembling the
appropriate probabilities, to yield
\bea
\label{sift1}
n &=& m{\Bigg \{}{\Bigg[}\sum_{l,l',l''}{\cal P}\left(l~{\rm photons~leave~Alice}\right)
\nonumber\\
&&\qquad\qquad\times{\cal P}\left(l'~{\rm photons~reach~Bob}~{\Big\vert}~l~
{\rm photons~leave~Alice}\right)
\nonumber\\
&&\qquad\qquad\times{\cal P}\left(l''~{\rm photons~detected}~{\Big\vert}~l'~
{\rm photons~reach~Bob}\right)\nonumber\\
&&\qquad\qquad\times{\cal P}\left({\rm no~dark~count~event}\right){\cal P}\left(
{\rm basis~compatibility}\right){\Bigg]}
\nonumber\\
&&\qquad+{\cal P}\left({\rm dark~count~event}\right){\cal P}\left({\rm basis~compatibility}
\right){\Bigg \}}
\nonumber\\
&=&m{\Bigg \{}\sum_{l,l',l''}{\Bigg[}e^{-\mu}{\mu^l\over l!}\left(l\atop l'\right)\alpha^
{l'}\left(
1-\alpha\right)^{l-l'}\left(l'\atop l''\right)\eta^{\lpp}\left(1-\eta\right)^{l'-l''}
\left(1-\delta_{0,\lpp}\right)
\left(1-r_d\right)\cdot {1\over 2}{\Bigg]}
\nonumber\\
&&\qquad\qquad+r_d\cdot {1\over 2}{\Bigg \}}
\nonumber\\
&=&{m\over 2}{\Bigg \{}\left(1-r_d\right)\sum_{l,l',l''}{\Bigg[}e^{-\mu}{\mu^l\over l!}
\left(l\atop l'\right)
\alpha^{l'}\left(1-\alpha\right)^{l-l'}\left(l'\atop l''\right)\eta^{\lpp}\left(1-\eta
\right)^{l'-l''}\left(1-\delta_{0,\lpp}\right)
{\Bigg]}+r_d{\Bigg \}}
\nonumber\\
&=&{m\over 2}{\Bigg \{}\left(1-r_d\right)\sum_{l=0}^\infty
e^{-\mu}{\mu^l\over l!}\sum_{\lp=0}^l
\left({l\atop\lp}\right)
\alpha^{\lp}\left(1-\alpha\right)^{l-\lp}
\sum_{\lpp=0}^{\lp}\left({\lp\atop\lpp}\right)\eta^{\lpp}\left(1-\eta
\right)^{\lp-\lpp}\left(1-\delta_{0,\lpp}\right)+r_d{\Bigg \}}
\nonumber\\
&=&{m\over 2}{\Bigg \{}\left(1-r_d\right)\sum_{l=0}^\infty
e^{-\mu}{\mu^l\over l!}{\Bigg [}1-\sum_{\lp=0}^l\left({l\atop\lp}\right)\alpha^{\lp}
\left(1-\alpha\right)^{l-\lp}\left({\lp\atop 0}\right)\left(1-\eta\right)^{\lp}{\Bigg ]}
+r_d{\Bigg \}}
\nonumber\\
&=&{m\over 2}{\Bigg \{}\left(1-r_d\right)\sum_{l=0}^\infty e^{-\mu}{\mu^l\over l!}
{\Bigg [}1-\left(1-\eta\alpha\right)^l{\Bigg ]}+r_d{\Bigg \}}
\nonumber\\
&=&{m\over 2}\llb\left(1-r_d\right){\Bigg \{}1-e^{-\mu}\sum_{l=0}^\infty{\left[\mu\left(
1-\eta\alpha\right)\right]^l\over l!}{\Bigg \}}+r_d\rrb
\nonumber\\
&=&{m\over 2}{\Bigg [}\left(1-r_d\right)\left(1-e^{-\eta\mu\alpha}\right)+r_d{\Bigg ]}
\nonumber\\
&=&{m\over 2}{\Bigg [}\left(1-r_d\right)\psi_{\ge 1}\left(\eta\mu\alpha\right)+r_d{\Bigg ]}
\\
&\simeq&{m\over 2}{\Bigg [}\psi_{\small{\ge 1}}\left(\eta\mu\alpha\right)+r_d{\Bigg ]}~.
\label{15}
\eea
where $\psi_{\ge 1}$, the probability of encountering one or more than one photon in a
pulse, is defined in eq.(\ref{7}). In the last step
above we have neglected $r_d$ in comparison to unity, which
means that we are ignoring the {\it dark
count coincidence} events, for which a dark count occurred in precisely the same bit cell
as an authentic photon detection event (this is a valid approximation for a good QKD
system equipped with a detector apparatus characterized by a small dark count rate).

As discussed below eq.(\ref{13}), the
form that we have derived for the number of sifted bits depends on
those assumptions that underlie eqs.(\ref{10}) and (\ref{11}).
In addition to these assumptions,
the result obtained in eq.(\ref{15}) above requires making the {\it further}
assumption on the intrinsic properties of the quantum channel that the Poisson
distribution of photon number that characterizes the output of the source laser
at the Alice end
{\it also} describes the state received at the Bob end of the quantum channel.

Let us consider the meaning of the terms in the square bracket in
eq.(\ref{15}) above.
The first term, $\psi_{\ge 1}\left(\eta\mu\alpha\right)$, is the
contribution to the number of sifted bits due to the bit cells comprised of single-photon
and multiple-photon pulses characterized by an {\it effective} mean photon number per
pulse of
$\eta\times\mu\times\alpha$, reflecting the fact that the stream is subjected to the
effects of both line attenuation and imperfect detection by Bob's apparatus.
The remaining term is simply the contribution to
the number of sifted bits due to those dark counts occurring in Bob's apparatus that do
not occur in a bit cell for which an authentic photon detection event takes place.

We also need to deduce the number of sifted bits that are in error, $e_T$. The
calculation is
similar to that for $n$, where we now take into account as well
the {\it intrinsic} quantum channel
error rate, $r_c$. In our parametrization of the QKD problem, $r_c$ measures only the
tendency of the system arrangement at Alice and Bob, along with the intrinsic properties
of the quantum channel itself, to cause polarization misalignment and dispersion of
photon arrival times between the instruments
at Alice and Bob, as we are assigning to other quantities ($\eta$, $r_d$ and most
significantly, $\alpha$) the role of measuring other system imperfections. In the case
of a free-space QKD system in which Alice and/or Bob are located on a moving platform,
such as a
satellite, $r_c$ may be a measure of the actual physical misalignment of the apparatuses at
the two ends. In the case of a fiber-optic cable QKD system, $r_c$ may be a measure of
certain dispersion effects intrinsic to the cable. In calculating $e_T$ we also
take account of the fact that the dark counts
produced by Bob's detector instrument will, by coincidence, be ``wrong" half of the time,
so that the factor of $r_d$ that appears in the expression for $n$ must be replaced by
$r_d/2$.

Following the same approach used to deduce eq.(\ref{15}) we may assemble
the necessary probability functions and carry out the required calculation
to obtain the expression
for the number of transmitted error bits, $e_T$, which is found to be given by
\bea
e_T&=&{m\over 2}{\Bigg [}\left(1-{r_d\over 2}\right)r_c\psi_{\ge 1}\left(\eta\mu\alpha
\right)+{r_d\over 2}{\Bigg ]}
\\
&\simeq&{m\over 2}{\Bigg [}r_c\psi_{\ge 1}\left(\eta\mu\alpha
\right)+{r_d\over 2}{\Bigg ]}~,
\label{17}
\eea
where in the second line above we have made the same approximation
utilized in deriving eq.(\ref{15}) and neglected $r_d$ in comparison to (in this
case, twice) unity.

\clearpage
\vskip 10pt
\noindent {\it Monitoring the Statistics of the Detection Events}

It may be advantageous for Alice and Bob to
specifically exclude from the sifting process those bit cells which are manifestly
associated with multiple photon pulses, since these bit cells provide an
opportunity for the enemy to obtain information on the final, shared key. For this
analysis, we envisage a generic, purely passive Bob apparatus suitable for use in
the BB84 protocol consisting of four photon detectors,\footnote{
The receiver design is discussed in detail in Section 5.2 below.}
for which there is a calculable probability
that, given a multi-photon pulse with $l$ photons incident on the apparatus, one and only
one of the four detectors will click
(we use the word ``click" to refer to the registration by the detector of an incident
photon). We will express this as the {\it confounding
probability} that Bob will not be able to distinguish a multi-photon pulse from a
single-photon pulse, and denote this probability by ${\hat z}\left(\eta,l\right)$,
where in general there is a dependence on both the photon number $l$ and the detector
efficiency $\eta$. Thus, Bob and
Alice can agree to discard those multi-photon pulses that manifestly produce more than a
single click at Bob's detector (leaving only those multi-photon pulses that happen to
produce a single click), and we can incorporate this condition quantitatively
in the expression for the effective secrecy capacity.
This will result in a shortened length for the sifted string, apparently thus
{\it reducing}
the possible value of the effective secrecy capacity. However, this procedure will
{\it also} reduce the size in bits of the associated privacy amplification subtraction
amount, thus
potentially {\it increasing} the effective secrecy capacity, and so the
two effects compete with each other. Because of the
complicated dependence on the various parameters that characterize the QKD problem, such
as the line attenuation, the mean photon number per pulse, the detector efficiency and
others (as derived in detail below), it is not clear {\it a priori} which term will
dominate. Such a scheme is a variant of
implementations of BB84 employing weak coherent pulses in which no distinction at all
is made between single- and multiple-photon pulses, and it is designed to  explore the
extent to which we can optimize the throughput rate as
well as guarantee unconditional secrecy identical to that achievable if only pure single
photon states are used.\footnote{
This extension of the BB84 protocol in fact may be further
generalized to include a {\it family} of generic extensions, distinguished from each other
by precisely how Bob monitors the statistics of the distribution of multiple clicks at his
detector. In Section 3.2 below we discuss this in more detail.}

To proceed, we return to the expression for the number of sifted bits derived above.
To emphasize the contributions to $n$ due separately to the single-photon and
multi-photon pulses we can rewrite
eq.(\ref{15}) as ({\it cf} eqs.(\ref{5}), (\ref{7}) and (\ref{8}))
\be
n={m\over 2}{\Bigg [}\psi_1\left(\eta\mu\alpha\right)+
\psi_{\ge 2}\left(\eta\mu\alpha\right)+r_d{\Bigg ]}~.
\label{18}
\ee
Inspection of the above expression would seem to indicate that to incorporate monitoring
of the click statistics it should be necessary to modify only the second term in the
square brackets, since, modulo dark count coincidence events, multiple clicks
can obviously never be produced when single-photon pulses (represented by the first
term in the square brackets) arrive at Bob's apparatus.
However, it turns out that the first term, {\it in addition to} the
second term, must be appropriately modified, as we show.

To understand this, we need to suitably modify the list of probabilities included in
eqs.(\ref{9}) to (\ref{13}), to allow for the characterization of
the single-click detection events. For this purpose we replace
\be
\label{19}
{\cal P}\left(l''~{\rm photons~detected}~{\Big\vert}~l'~
{\rm photons~reach~Bob}\right)=
\left(\lp\atop \lpp\right)\eta^{\lpp}\left(1-\eta\right)^{\lp-\lpp}\left(1-\delta_{0,
\lpp}\right)
\ee
with
\be
\label{20}
{\cal P}\left({\rm single~click~event}~{\Big\vert}~l'~{\rm photons~reach~Bob}\right)
\equiv{\hat z}_B\left(\eta,\lp\right)~,
\ee
where, as mentioned above, in general there is a dependence in ${\hat z}_B$
on both the number of received photons and the detector efficiency.

The explicit form of ${\hat z}_B\left(\eta,l\right)$, the confounding probability for
Bob that, given a laser pulse incident upon his apparatus
with $l$ photons in it, one and only one of
Bob's four detectors will click, will depend on (1) a model that specifies
the details of the detector apparatus,
{\it as well as on} (2) the particular click-monitoring scheme that is adopted. In the
following
we will take as a standard example a model of a purely passive setup with
four photon detectors
placed behind a pair of polarizing beamsplitters, which in turn are placed behind a
purely
passive 50/50 beamsplitter ({\it cf} Figure \ref{F:u_bob}). Having
thus picked the detector model, we
are still free to
specify the click-monitoring scheme. For instance, in one such scheme
we could require that Bob discard
all bit
cells in which any arrangement of simultaneous clicks occurs, while in another scheme
we could require that
only bit cells in which simultaneous clicks between, say, two detectors occur, but not
between three detectors, {\it etc}.\footnote{
Another possibility is that {\it all four} detectors simultaneously fire. This can only
occur if at least one of the four firings is due to a dark count event.}
To make the analysis as general as possible
we will make no specific assumption on this point. This still allows us to provide
explicit expressions for the cases of zero and one photons, so that we have
\be
\label{21}
{\hat z}_B\left(\eta,0\right)=0~,
\ee
\be
\label{22}
{\hat z}_B\left(\eta,1\right)=\eta~,
\ee
which indicates that there is a 100\% chance that a detector will fire if a single
photon impinges on the receiving apparatus, assuming a perfect detector (this may
be easily verified to follow from the assumption of a purely passive Bob apparatus,
as described above), and we define
\be
\label{23}
{\hat z}_B\left(\eta,\lp\right)\Big\vert_{\lp\ge 2}
\equiv
{\hat z}_{B,\ge 2}\left(\eta,\lp\right)~,
\ee
where ${\hat z}_{B,\ge 2}$ is a kernel to be operated on by a suitable probability
distribution that characterizes the distribution of the $\lp$ photons that have in
some manner propagated to the input of Bob's detector apparatus (the use of a general
expression for values of $\lp\ge 2$ allows a general treatment without specifying a
particular click-monitoring scheme).

We may now deduce the modified expression for the number of sifted
bits, $n_{mcs}$ (the subscript
stands for ``monitor click statistics") obtained with explicit
monitoring of the statistics of the detection clicks, by assembling the
appropriate probabilities, now supplemented by eq.(\ref{20}), to yield (note that we now
sum over two rather than three indices)
\bea
\label{sift1mcs}
n_{mcs} &=& m{\Bigg \{}{\Bigg[}\sum_{l,\lp}{\cal P}\left(l~{\rm photons~leave~Alice}
\right)
\nonumber\\
&&\qquad\qquad\times{\cal P}\left(l'~{\rm photons~reach~Bob}~{\Big\vert}~l~
{\rm photons~leave~Alice}\right)
\nonumber\\
&&\qquad\qquad\times
{\cal P}\left({\rm single~click~event}~{\Big\vert}~l'~{\rm photons~reach~Bob}\right)
\nonumber\\
&&\qquad\qquad\times{\cal P}\left({\rm no~dark~count~event}\right){\cal P}\left(
{\rm basis~compatibility}\right){\Bigg]}
\nonumber\\
&&\qquad+{\cal P}\left({\rm dark~count~event}\right){\cal P}\left({\rm basis~compatibility}
\right){\Bigg \}}
\nonumber\\
&=&m{\Bigg \{}\sum_{l,l'}{\Bigg[}e^{-\mu}{\mu^l\over l!}\left(l\atop l'\right)\alpha^
{l'}\left(1-\alpha\right)^{l-l'}
{\hat z}_B\left(\eta,\lp\right)
\left(1-r_d\right)\cdot {1\over 2}{\Bigg]}+r_d\cdot {1\over 2}{\Bigg \}}
\nonumber\\
&=&{m\over 2}{\Bigg \{}\left(1-r_d\right)\sum_{l,l'}{\Bigg[}e^{-\mu}{\mu^l\over l!}
\left(l\atop l'\right)
\alpha^{l'}\left(1-\alpha\right)^{l-l'}
{\hat z}_B\left(\eta,\lp\right){\Bigg]}+r_d{\Bigg \}}
\nonumber\\
&=&{m\over 2}{\Bigg \{}\left(1-r_d\right){\Bigg [}\sum_{l=0}^\infty e^{-\mu}{\mu^l
\over l!}
\sum_{\lp=0}^l
\left({l\atop\lp}\right)\alpha^{\lp}\left(1-\alpha\right)^{l-\lp}{\hat z}_B\left(
\eta,\lp\right){\Bigg ]}+r_d{\Bigg \}}
\nonumber\\
&=&{m\over 2}{\Bigg \{}\left(1-r_d\right)\sum_{l=0}^\infty e^{-\mu}{\mu^l\over
l!}{\Bigg [}l\alpha\left(1-\alpha\right)^{l-1}\eta+\sum_{\lp=2}^l\left({l\atop\lp}
\right)\alpha^{\lp}\left(1-\alpha\right)^{l-\lp}{\hat z}_B\left(\eta,\lp\right)
{\Bigg ]}+r_d{\Bigg \}}
\nonumber\\
&=&{m\over 2}\llb\left(1-r_d\right){\Bigg \{}\eta\mu\alpha e^{-\mu}\sum_{l=0}^\infty
{\Bigg [}{\mu^{l-1}\left(1-\alpha\right)^{l-1}\over\left(l-1\right)!}{\Bigg ]}+
{\Big\langle}\hat\chi\left(\mu,l\right){\hat {\cal Z}}_{\ge 2}\left(
\eta,\alpha,l\right){\Big\rangle}
{\Bigg \}}+r_d\rrb
\nonumber\\
&=&{m\over 2}{\Bigg \{}\left(1-r_d\right){\Bigg [}\eta\mu\alpha e^{-\mu\alpha}+
{\Big\langle}\hat\chi\left(\mu,l\right){\hat {\cal Z}}_{\ge 2}\left(
\eta,\alpha,l\right){\Big\rangle}
{\Bigg ]}+r_d{\Bigg \}}
\nonumber\\
&=&{m\over 2}{\Bigg \{}\left(1-r_d\right){\Bigg[}\eta\psi_1\left(\mu\alpha
\right)+
{\Big\langle}\hat\chi\left(
\mu,l\right){\hat {\cal Z}}_{\ge 2}\left(\eta,\alpha,l\right)
{\Big\rangle}{\Bigg]}+r_d{\Bigg \}}
\label{24}
\\
&\simeq&{m\over 2}{\Bigg [}\eta\psi_1\left(\mu\alpha\right)+
{\Big\langle}\hat\chi\left(
\mu,l\right){\hat {\cal Z}}_{\ge 2}\left(\eta,\alpha,l\right)
{\Big\rangle}+r_d{\Bigg ]}~,
\label{25}
\eea
where $\psi_1$, the probability of encountering exactly one photon in a pulse, is
defined in eq.(5), and we have defined as well
\bea
\label{26}
{\Big\langle}\hat\chi\left(
\mu,l\right){\hat {\cal Z}}_{\ge 2}\left(\eta,\alpha,l\right){\Big\rangle}
&\equiv &\sum_{l=0}^\infty\hat
\chi\left(
\mu,l\right){\hat {\cal Z}}_{\ge 2}\left(\eta,\alpha,l\right)
\nonumber\\
&=&\sum_{l=0}^\infty e^{-\mu}{\mu^l\over l!}{\hat {\cal Z}}_{\ge 2}
\left(\eta,\alpha,l\right)~,
\eea
and
\be
\label{27}
{\hat {\cal Z}}_{\ge 2}\left(\eta,\alpha,l\right)\equiv
\sum_{\lp=2}^l\left({l\atop\lp}\right)\alpha^{\lp}\left(1-\alpha
\right)^{l-\lp}{\hat z}_{B,\ge 2}\left(\eta,\lp\right)~,
\ee
and in eq.(\ref{25}) we have once again ignored $r_d$ in comparison to
unity.\footnote{
We employ a non-standard notation for averages, explicitly including inside the brackets
the specific distribution function with respect to which the average is defined, and
in particular including the argument of the distribution. This is
done to make clear which discrete variable is being summed over in every case.}
\footnote{
In deriving the results in eqs.(\ref{24}) and eq.(\ref{25})
we have also made use of the fact that, for {\it integer} values
of $l$, one has $\Gamma\left(-l\right)\rightarrow\infty\Rightarrow
1/\Gamma\left(-l\right)=0~\forall ~l\ge 1$.}

Let us consider the meaning of the three terms in the square bracket in the first equation
on the rhs of eq.(\ref{25}). The first term, $\eta\psi_1\left(\mu\alpha
\right)$ (to be compared with the first term in the square brackets in eq.(\ref{18})
which is $\psi_1\left(\eta\mu\alpha\right)$:
note the migration of the factor of $\eta$ out of the argument of $\psi_1$),
is the
contribution to the number of sifted bits due to the bit cells comprised of single-photon
pulses taken from a stream of pulses characterized by a mean photon number per pulse of
$\mu\times\alpha$ and further modified by the detector efficiency,
reflecting the fact that the stream is subjected to the
effects of both line attenuation and imperfect detection by Bob's apparatus, and
incorporating the effects of the click-monitoring procedure. The second
term, ${\Big\langle}\hat\chi\left(
\mu,l\right){\hat {\cal Z}}_{\ge 2}\left(\eta,\alpha,l\right)
{\Big\rangle}$, is
the contribution to the number of sifted bits due to those multi-photon bit cells that
cause only a single click to occur amongst the four detectors in Bob's apparatus.
(Those multi-photon pulses which cause multiple clicks in Bob's
device are watched for and discarded.) The remaining term is simply the contribution to
the number of sifted bits due to those dark counts occurring in Bob's apparatus that do
not occur in a bit cell for which an authentic photon detection event takes place.

Eq.(14) can in fact be recovered as a special case of
eq.(\ref{24}). It is easy to see this by rewriting $n_{mcs}$ instead as a sum to
manifestly include single photon pulses, so that we have
\bea
\label{28}
n_{mcs}&=&{m\over 2}{\Bigg [}\left(1-r_d\right){\Big\langle}\hat\chi\left(
\mu,l\right){\hat {\cal Z}}_{\ge 1}\left(\eta,\alpha,l\right)
{\Big\rangle}+r_d{\Bigg ]}
\nonumber\\
&\simeq&{m\over 2}{\Bigg [}{\Big\langle}\hat\chi\left(
\mu,l\right){\hat {\cal Z}}_{\ge 1}\left(\eta,\alpha,l\right)
{\Big\rangle}+r_d{\Bigg ]}~,
\eea
where
\bea
\label{29}
{\Big\langle}\hat\chi\left(
\mu,l\right){\hat {\cal Z}}_{\ge 1}\left(\eta,\alpha,l\right)
{\Big\rangle}&\equiv&
\sum_{l=0}^\infty\hat\chi\left(\mu,l\right)
{\hat {\cal Z}}_{\ge 1}\left(\eta,\alpha,l\right)
\nonumber\\
&=&\sum_{l=0}^\infty e^{-\mu}{\mu^l\over l!}
{\hat {\cal Z}}_{\ge 1}\left(\eta,\alpha,l\right)
\eea
and
\be
\label{30}
{\hat {\cal Z}}_{\ge 1}\left(\eta,\alpha,l\right)\equiv
\sum_{\lp=1}^l\left({l\atop\lp}\right)\alpha^{\lp}\left(1-
\alpha\right)^{l-\lp}{\hat z}_B\left(\eta,\lp\right)~.
\ee
The sum over $\lp$ in the above expression could just as well be allowed
to range from zero to infinity since ${\hat z}_B\left(\eta,0\right)$ vanishes
indentically. We have here marked the sum as beginning at $\lp=1$ merely
to emphasize specific value of the one-photon contribution in the result.

Clearly, making the replacement
\be
\label{31}
\sum_{\lp=0}^l\left({l\atop\lp}\right)\alpha^{\lp}\left(1-
\alpha\right)^{l-\lp}{\hat z}_B\left(\eta,\lp\right)
\Rightarrow
\sum_{\lp=0}^l\left({l\atop\lp}\right)\alpha^{\lp}\left(1-
\alpha\right)^{l-\lp}\sum_{\lpp=0}^{\lp}\left({\lp\atop\lpp}\right)\eta^{\lp}\left(1-
\eta\right)^{\lp-\lpp}\left(1-\delta_{0,\lpp}\right)
\ee
in eq.(\ref{29}) and carrying out the sums results in a specialization
from $n_{mcs}\Rightarrow n$, corresponding to our previous analysis
in which Bob makes no attempt whatsoever to distinguish single
photon pulses from multiple photon pulses. 

We also need to deduce the modification of the number of transmitted errors necessary
to account for the click statistics monitoring prescription by proceeding as in the
derivation of eq.(\ref{25}), from which we obtain
\bea
e_{T,mcs}&=&{m\over 2}{\Bigg \{}\left(1-{r_d\over 2}\right)r_c{\Bigg[}\eta\psi_1
\left(\mu\alpha\right)+
{\Big\langle}\hat\chi\left(
\mu,l\right){\hat {\cal Z}}_{\ge 2}\left(\eta,\alpha,l\right)
{\Big\rangle}{\Bigg]}+{r_d\over 2}{\Bigg \}}
\label{32}
\\
&\simeq&{m\over 2}{\Bigg \{}r_c{\Bigg [}\eta\psi_1\left(\mu\alpha\right)+
{\Big\langle}\hat\chi\left(
\mu,l\right){\hat {\cal Z}}_{\ge 2}\left(\eta,\alpha,l\right)
{\Big\rangle}{\Bigg ]}+{r_d\over 2}{\Bigg \}}~.
\label{33}
\eea
The functions $n_{mcs}$ and $e_{T,mcs}$ given in eqs.(\ref{25}) and (\ref{33}) are
quite general expressions that are
valid for any multi-click monitoring scheme provided one uses a detector model such that
there is unit probability (modified by $\eta$) that one detector will click given that
a single photon impinges on Bob's apparatus ({\it cf} eq.(22)).

\vskip 10pt
\noindent {\it A Digression on an Incorrect Approach to Calculating $\cal S$}

At first it might appear that a straightforward derivation of the effective secrecy
capacity $\cal S$, starting from the definition provided in eq.(\ref{1}), would consist
in a different development than that which was presented in the derivation of
eq.(\ref{15}). The
issue here is how to calculate the expression for the number of sifted bits, $n$ (and,
directly following on that, the number of transmitted error bits $e_T$). One might think
that the derivation of $n$ should consist in the following argument. One would first note
that the probability that a bit cell produced by a pulsed laser (generating a flux
of $\mu$) will contain one or more
photons is given by $\psi_{\ge 1}\left(\mu\right)$ ({\it cf} eq.(\ref{7})).
Since we
are interested in considering the fate of precisely those bit cells that contain one or
more photons, it might then seem that the construction of $n$ should consist merely in
multiplying $\psi_{\ge 1}\left(\mu\right)$ by the quantum efficiency $\eta$ of Bob's
detector
apparatus to account for the probability that the pulse will actually be detected, and
by the line attenuation $\alpha$ to account for the signal loss incurred in the passage
from Alice to Bob, adding to this product the dark count $r_d$, and finally multiplying
the entire expression by $m/2$ to account for the $50\%$ loss expected from incompatible
basis orientations. In this manner one would derive the quantity
\be
\label{34}
{\rm {``{\it n}"}}={m\over 2}\Big[\eta\psi_{\ge 1}\left(\mu\right)\alpha+r_d\Big]
\label{nono}
\ee
for the number of sifted bits, where we use the notation ``{\it n}" to distinguish this
incorrect expression from the correct expression for the number of sifted bits given by
$n$
in eq.(\ref{15}) above. Upon
comparing eqs.(\ref{34}) and (\ref{15}) we see that the difference between the
two
expressions is in the factors $\psi_{\ge 1}\left(\eta\mu\alpha\right)$ {\it versus}
$\eta\psi_{\ge 1}\left(
\mu\right)\alpha$. It is clear, however, that the quantity given in eq.(\ref{34})
certainly cannot be correct in general.

To see this, suppose that some number of sifted bits are established by using a particular
QKD setup. Now imagine that the intrinsic quantum efficiency, $\eta$, of the detector
apparatus is somehow doubled in value to $2\eta$. For instance, suppose that the original
value of $\eta$ is $45\%$, and we can double its value to $90\%$. If the {\it mean} number
of photons per pulse, $\mu$, is sufficiently small, say much less than unity, we would
expect that the corresponding number of detection clicks at the detector should double.
However, if instead $\mu$ was a very large number, say $\mu\approx 1000$, we would expect
that each bit cell that reached the detector would cause it to click anyway when we had
$\eta=45\%$, so that doubling $\eta$ to $90\%$ should {\it not} cause the number of
arriving bit cells that cause a click to become larger.

In fact it can easily be seen that the quantity $\eta\psi_{\ge 1}\left(\mu\right)\alpha$
furnishes
a lower bound to the quantity $\psi_{\ge 1}\left(\eta\mu\alpha\right)$. This is
illustrated
numerically in the figure below. Noting that the product $\eta\times\alpha$ satisfies the
inequality $0\le\eta\times\alpha\le 1$ since both $0\le\eta\le 1$ and $0\le\alpha\le 1$,
we plot curves that compare the values of $\eta\psi_{\ge 1}\left(\mu\right)\alpha$ with
$\psi_{\ge 1}\left(\eta\mu\alpha\right)$ for four different values of $\mu$.

\begin{figure}[htb]
\vbox{
\hfil
\scalebox{0.66}{\rotatebox{0}{\includegraphics{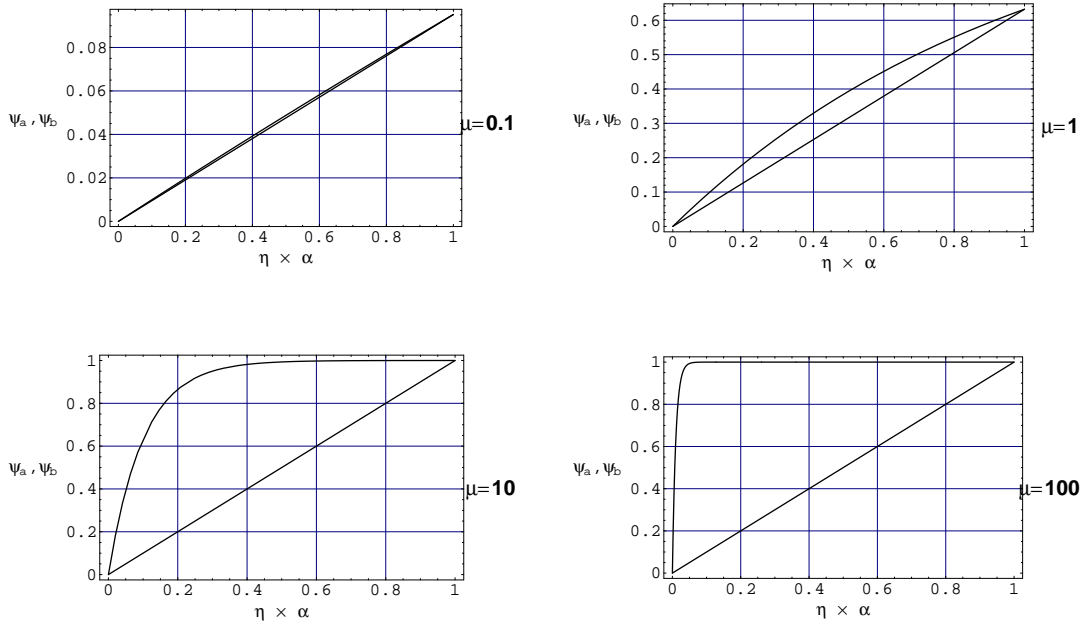}}}
\hfil
\hbox to -1.25in{\ } 
}
\bigskip
\caption{%
Comparison of $\eta\psi_{\ge 1}\left(\mu\right)\alpha$ with $\psi_{\ge 1}\left(\eta
\mu\alpha\right)$
}
\label{F:lower_bound}
\end{figure}

In each of the four graphs, the upper curve is the
function $\psi_{\ge 1}\left(\eta\mu\alpha\right)$ and the lower curve the function
$\eta\psi_{\ge 1}\left(\mu\right)\alpha$. It might
{\it still} appear that it would be sufficient to make use of the latter function in the
expression for the the number of sifted bits, $n$, since, as a lower bound on the correct
expression, one would at worst be underestimating the effective secrecy capacity $\cal S$.
After all, if $\eta\psi_{\ge 1}\left(\mu\right)\alpha$, provides a lower bound to
$\psi_{\ge 1}\left(\eta\mu\alpha\right)$ then we have $``n"\le n$ as well, so one might
think that the effective
secrecy capacity $\cal S$ is lower-bounded in this way as well. However, that is not in
general guaranteed to be true.
In fact {\it each term} in the numerator of $\cal S$
in eq.(\ref{1}), with the exception of $g_{pa}$, is a function of $\mu$.
(We will be developing the explicit $\mu$-dependence of the quantities contained in
the term $s$ in $\cal S$, as
well as that of $a$, in the sections below.) Each of these functions is, moreover, a
function as well of other quantities such as $\alpha$, $\eta$, {\it etc.}, which give
rise to a complicated parametric behavior.

\subsubsection{Privacy Amplification: General Remarks}

In the defining expression for the effective secrecy capacity, $\cal S$, given
in eq.(\ref{1}), $s$ is the number of bits of the sifted, error--corrected key that must
be discarded to implement privacy amplification. This number should be no less (and,
ideally, no more) than the number of bits that are ``at risk," in the sense that the
eavesdropper may have been able to obtain information as to their values. We refer to
$s$ as the privacy amplification subtraction function. According to
the {\it privacy amplification theorem} \cite{bbcm}, assuming
the use of a representative of the appropriate class
of hash function, if $s$ bits are removed from the shared
key and upon removing an additional $g_{pa}$ bits,\footnote{
The symbol $g_{pa}$, which will be discussed in more detail in Section 4.4.1, is
referred to as the {\it privacy amplification security parameter}.}
it is guaranteed that
the probability, $P$, that the eavesdropper can know one, or more than one bit of the
remaining key is given by\footnote{
Privacy amplification is described in much greater detail in Sections 3.1.2,
3.1.3, 3.1.4 and 3.1.5 below.}
%
\be
\label{35}
P\le{2^{-g_{pa}}\over\ln 2}~.
\ee
There are three possible ways in which the eavesdropper might gain partial information
about the shared key: (1) The execution of any particular error correction protocol
requires that Alice and Bob exchange information via the public channel. Although this
channel is assumed to be secured against {\it spoofing} by the eavesdropper through the
use of a suitable authentication protocol (discussed fully in
Section 4.4.1), the channel is
nevertheless assumed to be
completely open to eavesdropping and monitoring, and leakage of information is
accordingly
possible. (2) The ``pure" BB84 protocol, by which we mean the original idealized
protocol in
which {\it only} authentic qubits, {\it i.e.}, single-particle states are transmitted,
is provably perfectly secret in the sense defined by
Shannon \cite{bb84proofs1,bb84proofs2,bb84proofs3,bb84proofs4}.
However, unless the quantum
channel is perfect and completely free of any noise whatsoever, it is possible for the
eavesdropper to obtain some information from the single particle transmitted states by
exploiting the presence of the noise in the channel. Thus, in any {\it practical} system
implementation it
is necessary to account for the possibility of information leakage due to measurements
performed even on the single particle states. (3) Much attention has been devoted to the
use of weak coherent pulses generated by pulsed lasers in implementing QKD. In this
case,
and indeed in {\it any} case in which any sort of imperfect source at all\footnote{
This applies both to attenuated lasers and nonlinear crystals.}
is employed, it is in fact possible in principle for the eavesdropper to obtain
information from suitable
attacks on the multi-photon pulses in the stream, and this potential information leakage
must be accounted for in deducing the amount of required privacy amplification
subtraction
that should be carried out. 

The quantity $s$ is thus given by
\begin{equation}
\label{36}
s\equiv q+t+\nu,
\end{equation}
where $q$ is the Renyi information (in bits) leaked via error
correction, $t$ is the Renyi information leaked via measurements on the single-photon
pulses, and $\nu$ is the corresponding quantity associated to attacks performed on the
multiple-photon pulses.\footnote{
We restrict attention to privacy amplification carried out with classical computing
machines. So-called ``quantum privacy amplification" \cite{qpa} implemented with quantum
computing machines will be considered in Part Two of this paper.}

\subsubsection{Privacy Amplification: Error Correction}

Since $q$ should provide a bound on possible
information leakage caused by eavesdropping on the
process of error correction, it is natural to define $q$ so that it is measured in units
of error bits. We therefore write
\begin{equation}
\label{37}
q=Q\cdot e_T,
\end{equation}
so that we must deduce the appropriate form for the bounding function $Q$, which in all
generality should satisfy the equation of state given by
\begin{equation}
\label{38}
Q=Q\left(x,n,e_T\right)~,
\end{equation}
where, in addition to the dependence on $n$ and $e_T$ we also introduce and define a
parameter $x\ge 1$ to measure the degree to which Alice and Bob approach the Shannon
bound for perfect error correction in whatever error correction protocol that they
utilize. We will refer to $x$ as the {\it Shannon deficit parameter}, where $x=1$
corresponds to prefect error correction. It is clear that the bounding function $Q$
should depend on $n$ and $e_T$ through the ratio $e_T/n$, the error fraction, so that the
equation of state becomes
\begin{equation}
\label{39}
Q=Q\left(x,{e_T\over n}\right)~.
\end{equation}
Note that in the limit of vanishing dark count, $r_d=0$, the ratio $e_T/n$ reduces to the
intrinsic channel loss, so that we have
\be
\label{40}
{e_T\over n}{\Bigg\vert}_{r_d=0}=r_c~.
\ee
To deduce the explicit forms for $Q$ and $q$, we need to determine the entropy associated
with the information potentially leaked due to eavesdropping on error correction. This is
provided by the Shannon entropy function for binary information states, which is given by
\begin{equation}
\label{41}
h\left(\zeta\right)\equiv -\zeta\log_2\zeta-\left(1-\zeta\right)\log_2\left(1-\zeta
\right)
\end{equation}
where $\zeta$ is the transmitted bit error fraction. Since the bit error fraction
associated to the $n$-bit sifted string is given by $e_T/n$, the {\it minimum} amount
of information that will be leaked will be for the case of perfect error correction
in which the Shannon limit is attained, corresponding to $x=1$, and is given by
\be
\label{42}
q_{min}=nh\left({e_T\over n}\right)~.
\ee
In practice, perfect error correction cannot be attained, so an additional fractional
amount of information measured by the Shannon deficit parameter will be leaked. Thus, the
total information leakage, $q$, due to error correction is given by
\be
\label{43}
q=xnh\left({e_T\over n}\right)~.
\ee
Comparing (\ref{43}) with (\ref{37}) we deduce
\begin{equation}
\label{44}
Q\left(x,\zeta\right)\equiv{xh\left(\zeta\right)\over\zeta}~.
\end{equation}
Since $Q$ depends on $n$ and $e_T$ only through the ratio of the two, we see that the
$m$-dependence ({\it i.e.}, the dependence on the number of raw bits) drops out
entirely ({\it cf} eqs.(\ref{15}) and (\ref{17})).
This will be a useful
fact in determining important characteristics of the behavior of
QKD systems. In particular, the form of $Q$ is such that the expression is identically
exact in the limit of an arbitrarily long cipher {\it and} a cipher of finite length.

\subsubsection{Privacy Amplification: Single Photon Pulses}

Although in the idealized case of a noiseless channel
the quantum mechanical properties of the single-photon states
guarantee the perfect secrecy of the transmission against any and all
attacks by ``Eve" (the conventional name used to denote the enemy), the fact
that there is noise in a practical quantum channel provides an
opportunity to nevertheless carry out measurements which may provide some information to
the eavesdropper. The eavesdropper can generally attempt to be clever and simply not
measure {\it too much}, hoping
thereby to not generate too much noise, by ``flying under the radar" of the noise
present in the quantum channel.\footnote{
Terminology aside, this is in fact a rather apt analogy, since in flying under the
radar a pilot tries to
mask the radar signature of his aircraft in the radar clutter that is copiously present
near the surface of the earth.}
Following the nomenclature of \cite{slutsky2}, we refer to the
function that counts the number of required privacy amplification subtraction bits
associated to this possibility as the {\it defense frontier function},
as it maps out the safe ``frontier" for a sufficient amount of privacy amplification.
The specific form of this function that is appropriate to the special, limiting case
of the distribution of a cipher of infinite
length was first obtained by L\"utkenhaus in \cite{lutkenhaus2}.
The generalized form appropriate to actual ciphers
of finite length (which includes the previously obtained version applicable to infinite
length ciphers as a special
case) was later obtained by Slutsky, {\it et.al.} in \cite{slutsky2}.

In the approach adopted in \cite{slutsky2}, a defense frontier function was constructed
so
as to ensure that a ``successful" attack against the sifted, error-corrected bits
could not be carried out with a probability
any larger than a selectable infinitesimal quantity, $\epsilon$.\footnote{
Quoting the analysis provided in \cite{slutsky2}, we
define a {\it successful} attack
as one which introduces some number of errors $e_T$ into an $n$-bit sifted
data string resulting from an $m$-bit transmission, while yielding the enemy an
amount of Renyi information $I>t\left(n,e_T,\epsilon\right)$, where
$t\left(n,e_T,\epsilon\right)$
is the defense frontier function displayed below. As we discuss below, our treatment
departs somewhat from the analysis given in \cite{slutsky2} in explicitly restricting
consideration here solely to the {\it single-photon pulse part} of the entire
transmission. This point was left unclear in the original treatment.}
The detailed
calculation produced a quantity that provides
sufficient, but not necessary and sufficient, privacy amplification subtraction to
guarantee the desired result. In our application of the defense frontier function,
unlike in the original derivation given in \cite{slutsky2},
we explicitly restrict the arguments of
the defense frontier function, $t$, to
the {\it single photon parts only} of the
numbers of sifted and error bits, which we define, respectively, as
({\it cf} eqs.(\ref{15}) and (\ref{17}))
\bea
\label{45}
n_1&=&n_1\left(\eta,\mu,\alpha,r_d\right)
\nonumber\\
&\equiv&{m\over 2}{\Bigg [}\psi_1\left(\eta\mu\alpha
\right)+r_d{\Bigg ]}~,
\eea
and
\bea
\label{46}
e_{T,1}&=&e_{T,1}\left(\eta,\mu,\alpha,r_c,r_d\right)
\nonumber\\
&\equiv&{m\over 2}{\Bigg [}r_c\psi_1\left(\eta\mu
\alpha\right)+{r_d\over 2}
{\Bigg ]}~.
\eea
The reason for this restriction, discussed more fully in Section 3.1.5 below, is that in
our analysis we additively split into completely separate terms the privacy
amplification subtraction functions associated to the single- and multiple-photon
pulses. 

With the proviso that we restrict the functional arguments as indicated above,
we may adapt the derivation of \cite{slutsky2} to obtain
as the explicit expression for the defense frontier function
\be
\label{47}
t\left(n_1,e_{T,1},\epsilon\right)=\left(n_1-e_{T,1}
\right){\bar I}_{max}^R\left({e_{T,1}\over n_1}+
\xi\left(n_1,\epsilon\right)\right)+
\xi\left(n_1,\epsilon\right){\Big [}n_1\left(n_1-e_{T,1}\right){\Big ]}^{1/2}~,
\ee
\noindent{where ${\bar I}_{max}^R$ is the maximum average amount of Renyi information
leaked to the eavesdropper, with ${\bar I}_{max}^R$ calculated to be}
\be
\label{48}
{\bar I}_{max}^R\left(\zeta\right)\equiv 1+\log_2\left[1-{1\over 2}\left({1-3\zeta\over 1-
\zeta}\right)^2\right]~,
\ee
and $\xi$ is defined by
\be
\label{49}
\xi\left(n_1,\epsilon\right)\equiv{1\over{\sqrt {2n_1}}}{\rm {erf}}^{-1}
\left(1-\epsilon\right)~.
\ee

We note here that the expression for ${\bar I}_{max}^R$ given above is in fact precisely
the same as an
associated expression derived by L\"utkenhaus in \cite{lutkenhaus2,lutkenhaus3},
although it doesn't look like at it first sight.\footnote{
The two versions of the maximum average Renyi entropy become manifestly equal
through the rescaling $\zeta_{rescaled}\rightarrow {\zeta\over 1-\zeta}$, as pointed
out in \cite{lutkenhaus3}.}

Recall that in the previous section we chose a form for the
quantity $q=Qe_T$ (the bound on the amount of Renyi information
that may be leaked to the eavesdropper) that is measured in units of error bits. This
is
natural in view of the fact that this information leakage is associated with
eavesdropping on the error
correction process. In the same way, we express the defense frontier function in the
form
$t=Te_T$, introducing thereby an explicit dependence on both $e_T$ and $e_{T,1}$,
since the result of {\it any} measurements on the single photon pulses is to
{\it necessarily} generate some
number of errors in the transmitted string. Here we have introduced the new
function $T$ which plays a role similar to that of $Q$,
in that it is a bounding function on the privacy amplification subtraction amount.
Upon writing this out explicitly we find
\be
\label{50}
t\left(n_1,e_T,e_{T,1},\epsilon\right)=T\left(n_1,e_T,e_{T,1},\epsilon\right)\cdot e_{T}~,
\ee
where\footnote{
In the approximate form of eq.(\ref{51}) we are neglecting terms of the first order
of smallness as given
by ${e_{T,1}\over e_T}\simeq 1-\left[r_c\psi_{\ge 2}/\left(r_c\psi_1+{r_d\over 2}\right)
\right]+\cdots$, with a similar expression for $n_1/e_{T,1}$. Recall
({\it cf} eqs.(\ref{18}), (\ref{45}), {\it etc.})
that the argument of both of the functions $\psi_1$ and $\psi_{\ge 2}$ is
$\eta\mu\alpha<<1$.}
%
\bea
\label{51}
T\left(n_1,e_T,e_{T,1},\epsilon\right)&=&
\left({n_1\over e_{T}}-{e_{T,1}\over e_T}\right)
{\bar I}_{max}^R\left({e_{T,1}\over n_1}
+\xi\left(n_1,\epsilon\right)\right)+
\xi\left(n_1,\epsilon\right){n_1\over e_{T}}\left(1-{e_{T,1}\over n_1}\right)^{1/2}
\nonumber\\
&\simeq&\left({n_1\over e_{T,1}}-1\right)
{\bar I}_{max}^R\left({e_{T,1}\over n_1}
+\xi\left(n_1,\epsilon\right)\right)+
\xi\left(n_1,\epsilon\right){n_1\over e_{T,1}}\left(1-{e_{T,1}\over n_1}\right)^{1/2}~.
\nonumber\\
\eea
Unlike $Q$, $T$ does not depend on $n$ and $e_T$ (restricted here, of course, solely
to the single-photon pulse parts) {\it only} through terms that are
functions of the ratio of the two, for which the $m$-dependence identically drops out
entirely. In addition, due to the presence of $\xi$, $T$ includes a dependence on $m$
that does not intrinsically cancel out. Thus, the infinite-cipher
limit and finite-length
cipher version of $T$ are not identical. A straightforward calculation reveals that
\be
\label{52}
\lim_{m\rightarrow\infty}\xi\left(n_1,\epsilon\right)=0
\ee
and
\be
\label{53}
\lim_{m\rightarrow\infty}{\Bigg \{}\xi\left(n_1,\epsilon\right){n_1\over e_{T,1}}
\left(1-{e_{T,1}\over n_1}\right)^{1/2}{\Bigg \}}=0~,
\ee
so that we have
\bea
\label{54}
\lim_{m\rightarrow\infty}T\left(n_1,e_T,e_{T,1},\epsilon\right)&=&\left({n_1
\over e_{T,1}}-1\right)
{\bar I}_{max}^R\left({e_{T,1}\over n_1}\right)
\nonumber\\
&\equiv& T_\infty~,
\eea
which is independent of both $m$ and $\epsilon$. As stated above, this form recovers
the expression for ciphers of infinite length first
obtained by L\"utkenhaus.

\subsubsection{Privacy Amplification: Multiple Photon Pulses}

There is reason to be concerned about the effect on the secrecy, and hence the
security, of QKD systems when pulsed lasers are used to generate weak coherent
pulses in place of ideal, perfect qubits.
Indeed, the use of {\it any} imperfect source, such as excited nonlinear crystals,
must be considered to be potentially problematic in this regard. This
concern stems from the fact that, unlike the case for single-photon pulses, multi-photon
pulses offer the opportunity in principle for an eavesdropper to obtain full information
on the bit value encoded in the polarization state. However, this problem can be
{\it completely neutralized} by employing a sufficient amount of privacy amplification
in the processing of the shared key. It is essential in the
analysis of this problem to use the proper tool, which is the complete, comprehensive
form for the effective secrecy capacity of the system. Thus must include {\it all} the
contributing terms, and in particular must include a complete characterization of the
full
amount of privacy amplification due to all causes (as well as the full amount
of shared key material that must be removed to implement continuous authentication).
As always in this
analysis, it is {\it required} to determine at least the minimum number of subtraction
bits for privacy amplification; it is {\it acceptable} (although undesirable) to
overestimate this number, but it is strictly {\it unacceptable} to underestimate
it.\footnote{
Any claim of a QKD system rate based on an underestimated numerical value for the
privacy amplification subtraction function is potentially untenable and should be
rejected as characterizing a potentially vulnerable communications system.}

\begin{figure}[htb]
\vbox{
\hfil
\scalebox{0.6}{\rotatebox{270}{\includegraphics{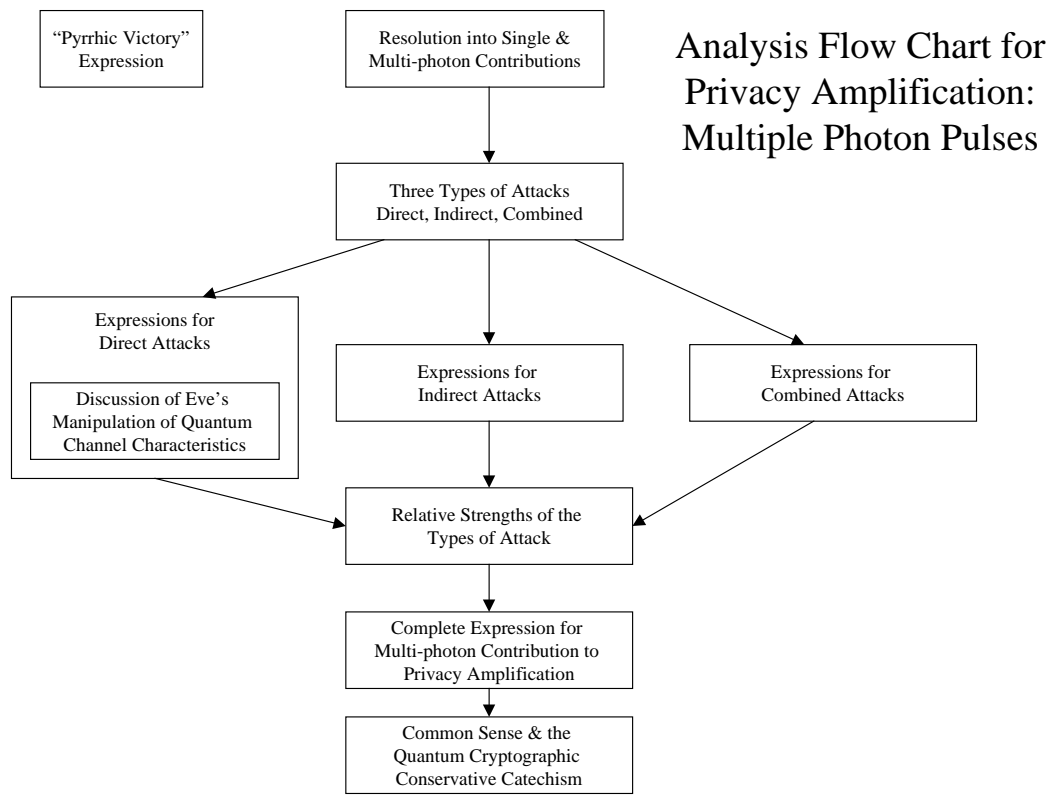}}}
\hfil
\hbox to -1.25in{\ } 
}
\bigskip
\caption{%
Flow Chart of Analysis of Multi-Photon Pulse Privacy Amplification
}
\label{F:flow}
\end{figure}

The detailed analysis of the privacy amplification associated to multi-photon
pulses is rather complicated. As we cannot know
in advance precisely how the enemy will choose to attack these pulses, it is
necessary to enumerate all possibilities by providing a
complete taxonomy of all attacks, after which the various possibilities may be
compared against each other
to ascertain which are the strongest in various circumstances. It is then
possible to deduce the
expressions for the requisite amounts of privacy amplification.
To provide an overall perspective of the various steps in the logic we
illustrate the structure of the analysis with a flow
chart in Figure \ref{F:flow}.

\vskip 10pt
\noindent {\it The ``Pyrrhic Victory" Approach to Privacy Amplification}

It would of course be possible to guarantee absolutely that none of the information
resident in the
multi-photon portion of the stream be available to an enemy, by simply carrying out
sufficient privacy amplification subtraction to discard {\it all} of the bit values
associated to {\it all} of the multi-photon pulses. This effectively and
completely solves the problem
of the vulnerability of the information in the multi-photon pulses, and moreover
considerably shortens the analysis!
The entire information content of
the multi-photon part of the transmission is given by $\psi_{\ge 2}\left(\mu\right)$,
so that we could denote the total privacy amplification subtraction function by
$\nu_{{\rm Pyrrhic}}$ and simply write
\be
\label{55}
\nu_{{\rm Pyrrhic}}\equiv\psi_{\ge 2}\left(\mu\right)
\ee
and be done with it, confident that the eavesdropper cannot carry out {\it any}
useful attack on the multi-photon pulses: even if she attacks every single multi-photon
pulse in any manner whatsoever and is completely successful in every single case, she
gains nothing because we
will have removed all the associated information. The remaining shared key would be
as secret as
any key generated by a source of pure single particle states.\footnote{
That is, such a key will be unconditionally secret in the sense of privacy amplification,
assuming as one must, that privacy amplification would be always required for {\it any
practical} system employing even solely single particle states, as they are subject to
the effects of machine-induced errors that must be corrected, for which privacy
amplification is required to mitigate the effect of possible eavesdropping on the
error correction protocol.}
However, numerical analysis based on the complete expression for the effective secrecy
capacity derived in this paper demonstrates
that the achievable throughput rates in this
case are unacceptably low. Thus the use of $\nu_{{\rm Pyrrhic}}$ as a privacy
amplification amount serves to defeat Alice and Bob as surely as it defeats Eve. We need
to try to find a better, not merely sufficient
bound that will result in acceptable throughput values.

In the remainder of this section, as depicted in the flow chart in Figure \ref{F:flow},
we therefore construct for the first time the general functions
that measure the {\it necessary and sufficient}
amount of privacy amplification subtraction bits required to account for information loss
associated to the multiple-photon pulses, in addition to which we also
construct an explicit expression that
provides a practical, universal bound that is {\it always} at least as large as the
minimum number or required subtraction bits.

\vskip 10pt
\noindent {\it Splitting the Privacy Amplification Function into Single- and Multi-Photon
Parts}

In our analysis we explicitly, {\it additively} separate into two pieces the privacy
amplification associated to single- and multi-photon pulses. The two kinds of
cryptanalytic\footnote{
Strictly speaking, we are not referring to cryptanalytic attacks as such, as this
is traditionally defined to mean attacks on enciphered {\it data}, whereas here we are
discussing attacks designed to determine the {\it key}.}
attacks on these two kinds of pulses, that necessitate carrying out
privacy amplification in the first place, are each of a very different character.
Attacks on single-photon pulses {\it necessarily} generate errors, while attacks on
multi-photon pulses, when properly performed by the enemy, generate no detectable
errors at all. Thus, as we have emphasized in Section 3.1.4 above, it is natural to
mathematically express the privacy amplification  subtraction function associated
to single-photon pulses in units of error bits, as in eq.(\ref{50}), while
it is clearly not physically correct to do so in the
case of the privacy amplification subtraction function associated to multi-photon
pulses. Taking this point further, we note that in our view it is neither necessary nor
physically meaningful to lump together functionally (as done for instance,
in \cite{lutkenhaus3}) the privacy amplification subtraction
amounts for these two very different kinds of cryptanalytic attacks. Moreover, the form
for
the privacy amplification subtraction amount associated to single-photon pulses derived
in Section 3.1.4 above is appropriate for, and allows us to analyze quantitatively,
the dynamics
of the transmission of {\it actual} ciphers of finite length, as opposed to merely
idealized ciphers of infinite length. Lumping togther the privacy amplification
contribution for single- and muli-photon pulses into a functional form which is
only applicable to idealized ciphers of infinite length obscures this possibility.

Thus, we advocate cleanly additively splitting into two distinct parts the separate
contributions,
which serves both to emphasize the different physical characteristics of the two types
of attack and allows us to properly analyze the dynamics of ciphers of finite length.

\vskip 10pt
\noindent {\it Three Types of Individual Cryptanalytic Attacks on Multiple Photon
Pulses}

In this paper we consider three
distinct kinds of attack that can be
carried out against the multiple-photon pulses in the stream:\footnote{
As noted above in Section 2.2 we again point out that in Part One of this paper we
are only considering so-called ``individual
attacks," {\it i.e.,} those attacks that do not require that the enemy apply unitary
transformations to the intercepted state with a quantum computing device. We will
address quantum computer-based attacks in Part Two.}
(1) {\it direct} attacks, (2) {\it indirect} attacks and (3)
{\it combined direct and indirect} attacks.

We define the {\it direct} attacks on
multi-photon pulses as attacks in which Eve intercepts the stream and attempts to directly
determine the polarization of the coherent state by performing a suitable measurement.
This attack requires a pulse with three or more photons in it. The
direct attack necessarily destroys the state as received by Eve, but, if she is successful
in determining the polarization she can attempt to send another state with identical
polarization on to Bob.\footnote{
So-called ``quantum non-demolition" measurements, which have been experimentally
demonstrated to be able to repeatedly count photons without destroying them, play
no role here. To profit from the direct attack it is necessary for Eve to determine
the state of polarization. In order to measure the state of polarization (more precisely,
the eigenvalues of the helicity operator) of the photon, it is necessary to select a
particular basis.}
We refer to the state prepared by the enemy and sent on to Bob as the {\it surrogate}
pulse. If he receives and detects this state it will have arrived just as if it been
sent by
Alice and was untouched by Eve, and in this way the information will be known to Eve.
There is a quantifiable probability that this attack will be successful, which we discuss
below.

We define {\it indirect} attacks on multi-photon pulses as attacks in which Eve
``splits the beam," as a result of which she ``keeps and preserves" one or more of the
photons
in the pulse, without measuring their state, and allows the remnant pulse to propagate on
to Bob. This attack requires a pulse with two or more photons in it.
In the indirect attack Eve knows that she must not interfere in any way with the
remnant pulse that is allowed to continue on to Bob: it must arrive at Bob's instrument in
the polarization state that it left Alice's system, only differing from the original
multi-photon pulse sent by Alice in that it contains some smaller number of photons than
when it left Alice, unbeknownst to Bob. He then measures the state of the pulse, and
carries out the public discussion phase of the QKD protocol as per usual. Eve eavesdrops on
the public discussion and learns thereby the particular {\it basis}, but not the {\it
state}, of the pulse. Since she has advanced technology at her disposal and has preserved
the photons that she split off in their original state, she merely measures the
polarization in the announced basis in order to precisely determine the actual state
of polarization. There is a quantifiable probability that this attack will be successful
which we discuss below.

Finally, the {\it combined attack} occurs when a pulse with
five or more
photons in it (the reason for this requirement on the number of photons is explained
below) is intercepted and split up by the enemy, allowing both a direct and indirect
attack to be performed. This particular individual attack appears not to have been
discussed previously in the literature of this subject.

In any of these cases Eve will have succeeded in determining the state of polarization of
the pulse in question, without Alice and Bob noticing that anything has happened: in
particular, the enemy will have obtained the information without having induced
{\it any}
elevation in the error rate. Without an increase in the error rate to indicate that Eve
has compromised the system, there will be no way for Alice and Bob to know that Eve has
the same information on those particular bits that they do. Of course, strictly
speaking this is not a weakness of {\it ideal} quantum key distribution, but rather
of {\it practical} quantum key distribution. However, this is a distinction without a
difference, as we must concern ourselves with the actual features of a realistic
implementation in any serious study of this subject. {\it Ideal} quantum key
distribution is actually almost irrelevant here: pure quantum bits propagating along
a noiseless quantum channel between perfect Alice and Bob instruments comprise
a fiction that has very little to do with anything that can be implemented in practice.

We emphasize again that the potential secrecy weakness of practical systems employing
weak
coherent pulses produced by a laser is in general {\it shared} by those systems
employing nonlinear crystals and parametric downconversion as a source of raw bits
for Alice. As in the case of an attenuated pulsed laser, nonlinear crystals in actual
system implementations will also sometimes produce multiple photon pulses, which in
principle may be exploited by an enemy equipped with suitable technology.

In our analysis of the requisite privacy amplification function associated to
multiple-photon pulses, we will make use of the results we have obtained for the 
general case in which Bob explicitly monitors for, and eliminates from the sifting
process, those bit cells which manifestly contain more than one photon. We also
consider the special case in which no such monitoring of click statistics is carried
out.

We now consider in more detail the three types of attack that can be carried out on the
multi-photon pulses.

\vskip 10pt
\noindent {\it Direct Attacks}

The logical structure of the analysis carried out in this section is illustrated with the
flow chart shown in Figure \ref{F:da} below.
\begin{figure}[htb]
\vbox{
\hfil
\scalebox{0.6}{\rotatebox{270}{\includegraphics{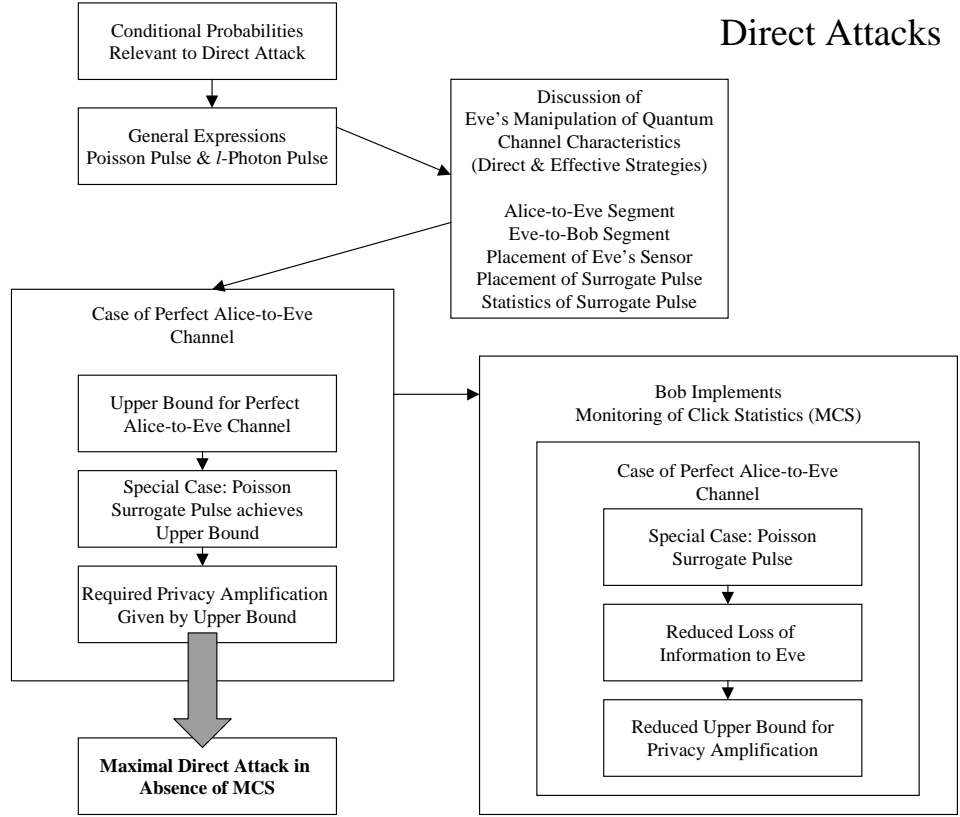}}}
\hfil
\hbox to -1.25in{\ } 
}
\bigskip
\caption{%
Flow Chart for Analysis of Direct Attacks
}
\label{F:da}
\end{figure}

In the direct attack the enemy intercepts the pulse transmitted by Alice and measures it
with her
apparatus. It becomes increasingly {\it more} likely to determine with complete knowledge
the polarization state of a multi-photon pulse as the number of photons in the pulse
increases. At the same time, owing to the Poisson distribution that governs the output
of
the pulsed laser used by Alice, and the fact that Alice will in general adjust
the flux to be
suitably weak through the use of appropriate intensity filters, it becomes
increasingly {\it less} likely to encounter a
multi-photon pulse as the number of photons in the pulse increases. Thus, there is a
competition between these two effects, and to ensure the secrecy of the shared cipher it
is essential to analyze the balance between them to carefully deduce precisely the
maximum amount of information that may be obtained by Eve in measuring these states.

An important point, discussed in more detail
below, is that the direct attack can only succeed if the pulse received by the enemy
apparatus contains three or more photons in it.

To proceed, we note that in \cite{lutkenhaus1} it is shown
that one may make use of the results of \cite{cheflesbarnett} to deduce an explicit
expression for the maximum probability to unambiguously determine the polarization of an
incident Fock state comprised of $l$ photons distributed according to a Poisson
distribution. (Such an incident state includes as a special case in particular a {\it
coherent} state comprised of $l$ photons.) We shall refer to this probability
as ${\hat z}_E\left(l\right)$, the maximum probability that Eve may with complete
knowledge determine the state of polarization of an incoming $l$-photon pulse.
In \cite{cheflesbarnett,lutkenhaus1} this was found to be given by
\be
\label{56}
{\hat z}_E\left(l\right) = \left\{
\begin{array}{l@{\;\;\;}l}
0 & l \leq 2\\
1-2^{1-l/2} & \mbox{$l$ even}\\
1-2^{(1-l)/2}& \mbox{$l$ odd.}
\end{array}
\right.
\ee
We know that irrespective of the apparatus utilized by Eve, it will be the case that the
pulse she intercepts from Alice will be characterized by a Poisson
distribution,\footnote{
As mentioned above, we {\it assume} that the intrinsic characteristics of the quantum
channel are such that the Poisson number distribution produced by Alice is preserved (of
course, as we discuss explicitly below, it is entirely possible that {\it the enemy}
may somehow alter the distribution on its way to being received by Bob).}
so the
appropriate average maximum probability, $z_E\left(\mu\right)$, for Eve to be able to
determine the polarization state with complete knowledge is
\bea
\label{57}
z_E\left(\mu\right)&\equiv&\sum_{l=0}^\infty e^{-\mu}{\mu^l\over l!}{\hat z}_E\left(l
\right)
\nonumber\\
&\equiv&\Big\langle\hat\chi\left(\mu,l\right){\hat z}_E\left(l\right)\Big\rangle_{l\ge 0}
\nonumber\\
&=&\Big\langle\hat\chi\left(\mu,l\right){\hat z}_E\left(l\right)\Big\rangle_{l\ge 3}
\nonumber\\
&=&1-e^{-\mu}{\Bigg (}\sqrt 2\sinh{\mu\over\sqrt 2}+2\cosh{\mu\over\sqrt 2}-1{\Bigg )}
~.\eea
It is obvious that, insofar as determining the polarization {\it state} with complete
knowledge is concerned, a two-photon pulse is no more useful than a single-photon pulse
is: with only two photons in the pulse it is not possible to determine the state of
polarization, although it is possible to determine the polarization {\it basis} in this
case. However, a direct measurement of the pulse that furnishes the identity of the basis
necessarily destroys the polarization of the two photons, making the pulse unsuitable for
a further measurement to determine the state. (Moreover, the basis, in any event, would
have been publicly revealed to the enemy
in the discussion between Alice and Bob.) As expected, and as
noted in \cite{lutkenhaus1}, the
leading order behavior of $z_E\left(\mu\right)$ varies as the cube of
the mean photon number, and is specifically given ({\it cf} eq.(\ref{57}))
by $z_E\left(\mu\right)={1\over 12}\mu^3+O\left(\mu^4\right)$, reflecting the fact that
three or more photons are required in order to
unambiguously determine the state of polarization of the pulse. We note that there is no
manifest appearance of the efficiency, $\eta_E$, of Eve's detector apparatus in the
quantities ${\hat z}_E\left(l\right)$ or $z_E\left(\mu\right)$, as we are implicitly
assuming that Eve is equipped with
perfect detection equipment, so that we have implicitly set $\eta_E=1$.

The function $z_E\left(\mu\right)$ is ``universal," in the sense that it provides an
upper
bound for all choices of apparatus on Eve's probability to determine the polarization of
an intercepted multi-photon pulse with complete knowledge. With any
particular measurement apparatus, Eve may typically in practice realize a {\it lower}
probability of
polarization determination: in \cite{bbbss,yuen} a
particular setup was described which provides Eve
with a probability of ${1\over 32}\mu^3+O\left(\mu^4\right)$.
In our analysis we will employ the maximal value
given in eq.(\ref{57}) above, to allow for the strongest possible enemy attack.

To proceed in as general a manner as possible, we will formulate the expressions
for privacy amplification subtraction from first principles in terms of the various
underlying probabilities associated to the different processes that take place. This
is analogous to our deduction of the numbers of sifted and transmitted error bits
previously carried out in Section 3.1.1. There we considered all
processes that characterize the propagation
of signals in the absence of an eavesdropper. An important difference here is that we must
{\it explicitly} take into account the various types of activities that an eavesdropper
may conduct. Denoting as before the various relevant probabilities by $\cal P$ with
appropriate arguments, we have
\be
\label{58}
{\cal P}\left(l~{\rm photons~leave~Alice}\right)=\hat\chi\left(\mu,l\right)~,
\ee
\be
\label{59}
{\cal P}\left(\lp~{\rm photons~reach~Eve}~{\Big\vert}~l~{\rm photons~leave~Alice}\right)=
\left({l\atop\lp}\right)\left(\alphaAE \rhoAE \right)^{\lp}\left(1-\alphaAE 
\rhoAE \right)^{l-\lp}~,
\ee
\be
\label{60}
{\cal P}\left({\rm polarization~determined~with~certainty}~{\Big\vert}~\lp~
{\rm photons~reach~Eve}\right)={\hat z}_E\left(\lp\right)~,
\ee
\be
\label{61}
{\cal P}\left(l_E~{\rm photons~leave~Eve~in~some~distribution~\hat\Xi}\right)=
\hat\Xi\left(\mu_E,l_E\right)~,
\ee
\be
\label{62}
{\cal P}\left(\lpp~{\rm photons~reach~Bob}~{\Big\vert}~l_E~{
\rm photons~leave~Eve}\right)=\left({l_E\atop\lpp}\right)\left(\alphaEB \rhoEB 
\right)^{\lpp}\left(1-\alphaEB \rhoEB\right)^{l_E-\lpp}
\ee
and
\be
\label{63}
{\cal P}\left(\lppp~{\rm photons~detected}~{\Big\vert}~\lpp~{\rm photons~reach~Bob}\right)
=\left({\lpp\atop\lppp}\right)\eta^{\lppp}\left(1-\eta\right)^{\lpp-\lppp}\left(1-
\delta_{0,\lppp}\right)~.
\ee
Note that, as discussed above, ${\hat z}_E\left(\lp\right)$ in eq.(\ref{60}) embodies
the requirement that the pulses vulnerable to the direct attack must contain three
or more photons, in effect it includes a factor of the Heaviside function
$\theta\left(l-3\right)$ in its definition ({\it cf} eq.(\ref{56})).\footnote{
Indeed, one way to proceed in evaluating sums over products of ${\hat z}_E\left(l
\right)$ with generic $l$-dependent quantities $y_l$ is to write such sums as
$\sum_l{\hat z}_E\left(l\right)y_l=\sum_{k=1}\left(1-2^{1-k}\right)y_{2k}+
\sum_{k=1}\left(1-2^{-k}\right)y_{2k+1}$. The implicit factor
of $\theta\left(l-3\right)$ in ${\hat z}_E\left(l\right)$
ensures that the indicated ranges of summation over $k$, in each case beginning
with $k=1$, are in fact correct as written for both the even and odd terms
in the two sums.}

In eqs.(\ref{59}) and (\ref{62}) we have introduced
the quantities $\alphaAE $, $\rhoAE $,
$\alphaEB $ and $\rhoEB $. The quantities $\alphaAE $ and $\alphaEB $ are,
respectively,
the line attenuation amounts along the quantum channel for the Alice-Eve and Eve-Bob link
segments. Note that the product of the two {\it partial} line attenuations gives the
{\it total} line attenuation, $\alpha$, along the entire quantum channel from Alice
to Bob (we assume that Eve is between Alice and Bob), so that we have
\be
\label{64}
\alphaAE\alphaEB=\alpha~.
\ee

The quantities $\rhoAE $ and $\rhoEB $ each satisfy the
inequalities $0\le\rhoAE \le\alphaAE ^{-1}$ and $0\le\rhoEB \le\alphaEB ^{-1}$.
These are parameters that
measure the degree to which the enemy can somehow ``adjust" the transparency of the
quantum channel so as to increase the amount of information
that can be obtained
on the transmitted bits. Thus, a value of $\rho_{AE,EB}=0\Rightarrow\alpha_{AE,EB}
\rho_{AE,EB}=0$ corresponds to a totally degraded quantum channel,\footnote{
This case constitutes denial of service, since no signals of any kind can propagate
through the channel when $\rho_{AE,EB}=0$, and therefore falls outside the purview
of an analysis of eavesdropping attacks.}
a value of $\rho_{AE,EB}=1\Rightarrow\alpha_{AE,EB}\rho_{AE,EB}=\alpha_{AE,EB}$
corresponds to the case when the enemy does not modify the transparency, thus leaving
the fiducial amount of line attenuation in the channel, and a value
of $\rho_{AE,EB}=\alpha_{AE,EB}^{-1}\Rightarrow\alpha_{AE,EB}\rho_{AE,EB}=1$
corresponds to the case in which the enemy has made the quantum channel perfectly
transparent.

There are different cases for us to consider. In the case of a fiber-optic implementation
of QKD, it {\it may} be
reasonable to analyze the case in which the {\it entire} quantum
channel is surreptitiously replaced by the enemy with an ``ideal," lossless cable.
In this case, we would have $\rhoAE =\alphaAE ^{-1}$ and $\rhoEB =\alphaEB ^{-1}$
after the cable replacement, resulting in the elimination of the line attenuation
along the
cable. In the case of a free space implementation, it is {\it unreasonable}
to imagine that the enemy can replace the channel with one of perfect, or even improved,
transparency. This
is entirely well-motivated physically: in this case a replacement of the quantum channel
with one of better transmission characteristics amounts to imagining that Eve can replace
the atmosphere with one that {\it she} prefers. Even if this were possible, it would
presumably not go unnoticed by Alice, Bob and the rest of the population of the planet.
Note that this effect {\it cannot} be mimiced by adjusting the frequency of the photons
in
the reconstructed pulses, as proper ``technically sound cryptosystem"
operating procedure dictates the use
of a narrow bandpass wavelength filter in the front of Bob's receiver which will
physically exclude any such wavelength-modified incoming photons.

However, in the case of a free space QKD implementation it may nevertheless
be possible for the
enemy to anyhow effect a partial, and sometimes even a complete, {\it effective}
improvement of the transparency of the
channel. For instance, we may imagine that the interception apparatus of the enemy
is secretly located immediately adjacent to the Alice site (we ascribe to the enemy
superb powers of camouflage and technical skill),\footnote{
As discussed below in Sections 4 and 5, we envisage for the free space case an
implementation in which Alice is located on
an orbiting satellite and Bob is located on an aircraft or the ground, so
that the undetected placement of an eavesdropping interception
device immediately adjacent to Alice is almost impossible to imagine for any actual,
practical situation unless Eve can literally make herself invisible. Again, we are
assuming
that the entire praxis of {\it communications security}, in addition to the narrower
requirement
of {\it cryptosecrecy}, is properly implemented, so that physical access of the enemy
to the Alice and Bob devices is (1) assumed to be prevented, and (2) in any event is not
reasonably within the purview of the QKD protocol {\it per se}.}
thereby effectively producing the value
$\rhoAE =\alpha_{AE} ^{-1}\Rightarrow\rhoAE \alphaAE =1$.
(Strictly speaking we
would presumably actually have the condition $\rhoAE\simeq\alphaAE^{-1}$ rather than
$\rhoAE=\alphaAE^{-1}$ in this case, since Eve is presumed {\it adjacent to}, but not
{\it physically coincident with}, Alice.) In the case of the direct
attack, in which a surrogate pulse is sent on to Bob, we may {\it further}
imagine that the
physical location of the site from which the surrogate pulse is launched is likewise
placed immediately adjacent to (and somehow undetected by) Bob. This allows the two
enemy collaborators to
entirely circumvent the attenuation of the free space quantum channel
by simply communicating instructions to each other clasically, which
has the effect of
causing $\rhoEB =\alphaEB ^{-1}\Rightarrow\rhoEB \alphaEB =1$. Note that, in the
case of the {\it indirect} attack (to be discussed in great detail below), the same
options are not {\it simultaneously} available to the enemy in the case of the free space
quantum
channel. This is because, unlike in the direct attack, it is necessary that the pulse
that is allowed to travel on to Bob not be modified, and its polarization state remains
unknown to the enemy.
Hence, no ``conspirator" can participate in the transmission of the pulse. Thus, Eve
can be located either immediately adjacent to Alice, in which case we have
$\rhoAE =\alphaAE ^{-1}\Rightarrow\rhoAE \alphaAE =1$, or Eve can be located
immediately adjacent to Bob, in which case we have
$\rhoEB =\alphaEB ^{-1}\Rightarrow\rhoEB \alphaEB =1$, (or somewhere in between)
but both endpoint conditions
cannot simultaneously be realized: there cannot be a ``second Eve" located at the other
end.\footnote{
We emphasize that all detailed discussion of entanglement in quantum
cryptography, including analysis of
entanglement-assisted attacks, is performed in Part Two
of this paper. We here merely point out that inclusion of
entanglement-assisted attacks against
{\it individual} quantum bits in particular, which would allow a second Eve adjacent
to Bob to effectively eliminate the line attenuation even for the indirect attack (if
prior entanglement is shared between her and the first Eve located
adjacent to Alice), does {\it not}
change the functional form of the final expressions for multi-photon
privacy amplification obtained in eqs.(\ref{141}) through (\ref{151}) below. The only
change is that the distinction between optical-fiber and free-space implementations
is eliminated, with a consequent modification of the associated throughput
rates \cite{parttwo}.}

Whether for a free space or fiber optic cable implementation,
after imposing the condition $\rhoAE=\alphaAE^{-1}$ the
expression for the privacy amplification that results will still retain a dependence on
$\alphaEB$. In the case
of a free space implementation only, one may without loss of generality always
interpret this residual
$\alphaEB$-dependence in fact as dependence on the {\it entire} line
attenuation, $\alpha$, of
the quantum channel, since in this case the value of $\alphaEB$ is due to the conditions
along the entire propagation path from Eve (who is adjacent to Alice) to Bob. However, in
the case of a fiber-optic cable system, for which it might be possible for Eve
to achieve the condition $\rhoAE=\alphaAE^{-1}$ by directly replacing the channel
between Alice and herself {\it without} necessarily having to place an interception device
immediately adjacent to Alice (although this might also be done), it obviously
need not be true in general that the residual $\alphaEB$-dependence corresponds to the
entire line attenuation $\alpha$.

Note that we do {\it not} introduce a parameter analogous to $\rhoAE$ or $\rhoEB$ to
describe
the degree to which the enemy can remotely ``control" or adjust the quantum efficiency,
$\eta$, of Bob's detector device. We are assuming, in accord with the discussion in
Section 2.4.2, that
proper cryptographic procedure and design is followed in the system
implementation, so that the enemy {\it cannot} enforce the condition $\eta\rightarrow 1$,
nor even cause any non-negligible change in the value of $\eta$ at all.
The only physically reasonable method whereby such remote ``control" could succeed is
through the modification of the wavelength of those photons that propagate successfully
to Bob's detector. Clearly this cannot work in the case of the indirect attack, as
its success {\it requires} that Eve not prepare the state allowed to continue on to
Bob in any way. Moreover, even for the direct attack such wavelength modification can,
as mentioned above, be trivially
neutralized in any event by placing a narrow bandpass filter in front of Bob's
detector apparatus,
thereby preventing (with high probability) any photons of modified wavelength from
entering the device.

After successfully determining the polarization state of the intercepted pulse, the enemy
may prepare an identically polarized state in any way that is deemed to be advantageous
and send the surrogate pulse on to Bob. Although the enemy may send
{\it any} (properly polarized) pure state or mixture to Bob, these states may
practically be regarded as mixtures of number states\footnote{
This has been noted as well in \cite{lutkenhaus1}.}
owing to the facts that (1) Bob is employing a
photon number detector, and (2) anything {\it else} will provide a signature that the
eavesdropper has tampered with the signal, alerting Alice and Bob who will then discard
the bit cell. Therefore, although the distribution function $\hat\Xi$
in eq.(\ref{61}) is
not necessarily identical to the Poisson distribution, we can without loss of
generality {\it always} take it to be some {\it discrete} distribution,
$\hat\Xi=\hat\Xi\left(\mu_E,l_E\right)$, that is characterized for each
bit cell by both a mean, $\mu_E$, and a particular number, $l_E$, of photons. Of course,
there need be no particular {\it a priori} relation between the number of photons, $\lp$,
in the pulse that Eve intercepted and measured in order to detemine the
polarization state, and the number of photons, $l_E$, that is sent on by Eve to Bob.

We may now deduce the explicit expression for the privacy amplification subtraction
function associated to individual direct attacks, which we denote by $\nu_d$,
by assembling the appropriate probabilities from eqs.(\ref{58}) through (\ref{63}). We
find
\bea
\label{65}
\nu_d&=&{m\over 2}\sum_{l,\lp,\lpp,\lppp,l_E}
{\cal P}\left(l~{\rm photons~leave~Alice}\right)
\nonumber\\
&&\qquad\qquad\times
{\cal P}\left(\lp~{\rm photons~reach~Eve}~{\Big\vert}~l~{\rm photons~leave~Alice}\right)
\nonumber\\
&&\qquad\qquad\times
{\cal P}\left({\rm polarization~determined~with~certainty}~{\Big\vert}~\lp~
{\rm photons~reach~Eve}\right)
\nonumber\\
&&\qquad\qquad\times
{\cal P}\left(l_E~{\rm photons~leave~Eve~in~some~distribution~\hat\Xi}\right)
\nonumber\\
&&\qquad\qquad\times
{\cal P}\left(\lpp~{\rm photons~reach~Bob}~{\Big\vert}~l_E~{
\rm photons~leave~Eve}\right)
\nonumber\\
&&\qquad\qquad\times
{\cal P}\left(\lppp~{\rm photons~detected}~{\Big\vert}~\lpp~{\rm photons~reach~Bob}\right)
\nonumber\\
&=&{m\over 2}\sum_{l=0}^\infty\hat\chi\left(\mu,l\right)\sum_{\lp=0}^l
\left({l\atop\lp}\right)\left(\alphaAE\rhoAE \right)^{\lp}\left(1-\alphaAE \rhoAE
\right){\hat z}_E\left(\lp\right)
\nonumber\\
&&\qquad\sum_{l_E=0}^\infty\hat\Xi\left(\mu_E,l_E\right)\sum_{\lpp=0}^{l_E}
\left({l_E\atop\lpp}\right)\left(\alphaEB \rhoEB 
\right)^{\lpp}\left(1-\alphaEB \rhoEB \right)^{l_E-\lpp}
\nonumber\\
&&\qquad\qquad\sum_{\lppp=0}^{\lpp}
\left({\lpp\atop\lppp}
\right)\eta^{\lppp}\left(1-\eta\right)^{\lpp-\lppp}\left(1-\delta_{0,\lppp}\right)
\nonumber\\
&=& {m\over 2}\sum_{l=0}^\infty\hat\chi\left(\mu,l\right)\sum_{\lp=0}^l\left({l\atop
\lp}\right)\left(\alphaAE \rhoAE \right)^{\lp}\left(1-\alphaAE \rhoAE 
\right)^{l-\lp}{\hat z}_E\left(\lp\right)
\nonumber\\
&&\qquad\sum_{l_E=0}^\infty\hat\Xi\left(\mu_E,l_E
\right)\sum_{\lpp=0}^{l_E}\left({l_E\atop\lpp}\right)\left(\alphaEB \rhoEB 
\right)^{\lpp}\left(1-\alphaEB \rhoEB \right)^{l_E-\lpp}{\Big [}1-\left(1-
\eta\right)^{\lpp}{\Big ]}
\nonumber\\
&=& {m\over 2}\sum_{l=0}^\infty\hat\chi\left(\mu,l\right)\sum_{\lp=0}^l\left({l\atop
\lp}\right)\left(\alphaAE \rhoAE \right)^{\lp}\left(1-\alphaAE \rhoAE 
\right)^{l-\lp}{\hat z}_E\left(\lp\right)
\nonumber\\
&&\qquad\sum_{l_E=0}^\infty\hat\Xi\left(\mu_E,l_E\right)
{\Big [}1-\left(1-\eta\alphaEB \rhoEB \right)^{l_E}{\Big ]}~.
\eea
The general form of this quantity is simple to explain in physical terms: it
is the probability that
the enemy can with certainty determine the polarization of a multi-photon pulse
given that it has been intercepted, multiplied by the probability that a surrogate
pulse in the same state of polarization can arrive at and be detected by Bob, measured
in a distribution $\hat\Xi$ chosen solely by the eavesdropper.\footnote{
We note in passing that the sum $\sum_{\lp=0}^l\left({l\atop\lp}\right)\left(\alphaAE
\rhoAE \right)^{\lp}\left(1-\alphaAE \rhoAE \right)^{l-\lp}{\hat z}_E\left(\lp
\right)$ can be explicitly evaluated to a closed form, but as it is not particularly
illuminating we have not displayed the result here.}

This expression provides the amount of privacy amplification subtraction required
in order to compensate for direct attacks on {\it all} of the multi-photon pulses that
contain three or more photons. It is clearly not the most general expression for the
total amount of privacy amplification subtraction required in order to ensure
unconditional secrecy, as such a uniform attack on all multi-photon pulses with three
or more photons is only one possible cryptanalytic strategy that may be employed by
the enemy. In general, we also need the expression for the amount of privacy
amplification
required in order to protect against a direct attack on any {\it particular}
multi-photon pulse with
$l$ photons in it, which we denote by $\nu_{d,l}$. We may obtain the relevant
quantity by direct inspection of eq.(\ref{65}), to find
\bea
\label{66}
\nu_{d,l}&=&{m\over 2}\hat\chi\left(\mu,l\right)\sum_{\lp=0}^l
\left({l\atop\lp}\right)\left(\alphaAE \rhoAE \right)^{\lp}\left(1-\alphaAE 
\rhoAE \right)^{l-\lp}{\hat z}_E\left(\lp\right)
\nonumber\\
&&\qquad\sum_{l_E=0}^\infty\hat\Xi\left(\mu_E,l_E\right)
{\Big [}1-\left(1-\eta\alphaEB \rhoEB \right)^{l_E}{\Big ]}
\nonumber\\
&=&{m\over 2}\psi_l\left(\mu\right)
\sum_{\lp=0}^l\left({l\atop\lp}\right)\left(\alphaAE \rhoAE \right)^{\lp}\left(
1-\alphaAE 
\rhoAE \right)^{l-\lp}{\hat z}_E\left(\lp\right)
\nonumber\\
&&\qquad\sum_{l_E=0}^\infty\hat\Xi\left(\mu_E,l_E\right)
{\Big [}1-\left(1-\eta\alphaEB \rhoEB \right)^{l_E}{\Big ]}~,
\eea
where we have introduced the notation $\psi_l\left(\mu\right)\equiv\hat
\chi\left(\mu,l\right)$.
Inspection of the above expression reveals the fact that, irrespective of
the particular form of $\hat\Xi$, the magnitude of the privacy amplification function
associated to direct attacks {\it increases} as the quantum efficiency of Bob's
photon detector increases (recall that the product $\eta\alphaEB\rhoEB$ satisfies
the inequality $\eta\alphaEB\rhoEB\le 1$). Thus, we have the seemingly
paradoxical situation that, due to the
special nature of the direct attack, {\it more} information is potentially at risk
to compromise by Eve when Bob's detector apparatus is characterized by a better
detector efficiency than when characterized by a poorer efficiency.

Now we note that, if $\rhoAE =\alphaAE ^{-1}$, which means either that in some way
the enemy has arranged that the quantum channel between Alice and herself is free of
any attenuation (in the case of a fiber-optic cable system), or effectively done the
same thing by situating the interception apparatus immediately adjacent to Alice (in
either a cable- or free space-based implementation), the cofactor of ${\hat z}_E$ in
the summand of the sum over $\lp$ becomes a Kronecker delta\footnote{
Upon specifically setting $\rhoAE =\alphaAE ^{-1}$ in the general sum $\sum_{\lp=0}^l
\left({l\atop\lp}
\right)\left(\alphaAE\rhoAE\right)^{\lp}\left(1-\alphaAE\rhoAE\right)^{l-\lp}y_{\lp}$
we obtain $\sum_{\lp=0}^l\left({l\atop\lp}
\right)\left(1\right)^{\lp}\left(1-1\right)^{l-\lp}y_{\lp}=
\sum_{\lp=0}^l\left({l\atop\lp}\right)\delta_{l,\lp}y_{\lp}=y_l$
for any $\lp$-dependent quantity $y_{\lp}$.}
enforcing $\lp\rightarrow l$,
so that we have (note that the sums over $\lp$ and $l_E$ are completely functionally
independent of each other)
\bea
\label{67}
\nu_d\Big\vert_{\rhoAE =\alphaAE ^{-1}}
&=&{m\over 2}\sum_{l=0}^\infty\hat\chi\left(\mu,l\right){\hat z}_E\left(l\right)
\sum_{l_E=0}^\infty\hat\Xi\left(\mu_E,l_E\right){\Big [}1-{\Big (}1-\eta\alphaEB 
\rhoEB {\Big )}^{l_E}{\Big ]}
\nonumber\\
&=&{m\over 2}z_E\left(\mu\right)\sum_{l_E=0}^\infty\hat\Xi\left(\mu_E,l_E\right){\Big [}
1-{\Big (}1-\eta\alphaEB \rhoEB {\Big )}^{l_E}{\Big ]}~,
\eea
in the case that the enemy has performed a direct attack on {\it all} the multi-photon
pulses with three or more photons (where $z_E\left(\mu\right)$ is given
in eq.(\ref{57})), and
\be
\label{68}
\nu_{d,l}\Big\vert_{\rhoAE =\alphaAE ^{-1}}
={m\over 2}\psi_l\left(\mu\right){\hat z}_E\left(l\right)
\sum_{l_E=0}^\infty\hat\Xi\left(\mu_E,l_E\right)
{\Big [}1-\left(1-\eta\alphaEB \rhoEB \right)^{l_E}{\Big ]}~,
\ee
in the case of a direct attack on any {\it particular} pulse with $l$ photons
in it.

We now confront an important fact about the direct attack. Alice and Bob may {\it never}
be able to learn the detailed functional form  of $\hat\Xi\left(\mu_E,l_E\right)$, and
certainly will not if we simply (conservatively) assume that the enemy is {\it always}
capable of witholding this information from them.
Without knowing the explicit form of $\hat\Xi\left(\mu_E,l_E\right)$ chosen by the enemy
for the preparation of the surrogate pulse to be sent on to Bob, we are to a certain
extent
limited as to what we can predict about the effect of this function on the operating
characteristics of a practical QKD system. However, we may note that the sum over
$l_E$ is a probability function, and thus its value is constrained to range
between $0$ and $1$ only. Thus from the point of view of Alice and Bob,
the {\it worst case}, or maximum values possible for either $\nu_d$ or $\nu_{d,l}$ are
\be
\label{69}
\nu_d\Big\vert_{\rhoAE =\alphaAE ^{-1}}^{max}={m\over 2}z_E\left(\mu\right)
\ee
and
\be
\label{70}
\nu_{d,l}\Big\vert_{\rhoAE =\alphaAE ^{-1}}^{max}={m
\over 2}\psi_l\left(\mu\right){\hat z}_E\left(l\right)~,
\ee
respectively. We note that these worst case results, which have been defined to be
independent of
$\alphaAE$, are clearly also independent of both $\eta$ and $\alphaEB$ (and therefore
also independent of $\alphaAE\alphaEB=\alpha$). Thus, if the enemy can choose a
suitable distribution function $\hat\Xi$ in which to prepare the surrogate
pulses such that sum over $l_E$ in eq.(\ref{65}) (or eq.(\ref{66})) becomes
\be
\label{71}
\sum_{l_E=0}^\infty\hat\Xi\left(\mu_E,l_E\right)
{\Big [}1-\left(1-\eta\alphaEB \rhoEB \right)^{l_E}{\Big ]}=1~,
\ee
the enemy can gain full effective control over both the total line attenuation {\it and}
the quantum efficiency of Bob's detector without having to physically tamper
with either the quantum channel or the detector! In order for her to achieve this,
though, it is essential that Bob {\it not}
monitor the click statistics of his detector. If he does monitor click statistics, as we
discuss below, he can partly prevent Eve from gaining such control: he can prevent her
from controlling his detector efficiency, but cannot prevent her from gaining control
over the line attenuation.

We can do more if we assume a particular form for $\hat\Xi\left(\mu_E,l_E\right)$. If
as before, we take $\rhoAE =\alphaAE ^{-1}$ and we {\it further} assume that the
enemy in particular prepares the surrogate photon states
in a Poisson distribution, so that we have $\hat\Xi=\hat\chi$, we find
\bea
\label{72}
\nu_d\Bigg\vert_{{\rhoAE =\alphaAE ^{-1}\atop
\!\!\!\!\!\!\!\!\!\!\!\!\!\!\!\!\!\!\!\!\hat\Xi=\hat\chi}}
&=&{m\over 2}z_E\left(\mu\right)\sum_{l_E=0}^\infty\hat\chi\left(\mu_E,l_E\right)
{\Big [}1-{\Big (}\eta\alphaEB \rhoEB {\Big )}^{l_E}{\Big ]}
\nonumber\\
&=&{m\over 2}z_E\left(\mu\right){\Bigg \{}1-e^{-\mu_E}\sum_{l_E=0}^\infty
{\left[\mu_E\left(1-\eta\alphaEB \rhoEB \right)\right]^{l_E}\over
l_E!}{\Bigg \}}
\nonumber\\
&=&{m\over 2}z_E\left(\mu\right){\Bigg [}1-e^{-\mu_E}e^{\mu_E\left(1-\eta\alphaEB 
\rhoEB \right)}{\Bigg ]}
\nonumber\\
&=&{m\over 2}z_E\left(\mu\right){\Bigg (}1-e^{-\eta\mu_{\small E}\alphaEB \rhoEB }
{\Bigg )}
\nonumber\\
&=&{m\over 2}z_E\left(\mu\right)\psi_{\ge 1}\left(\eta\mu_E\alphaEB \rhoEB \right)~,
\eea
in the case of a direct attack on all of the multi-photon pulses containing three or
more photons, and
\be
\label{73}
\nu_{d,l}\Bigg\vert_{{\rhoAE =\alphaAE ^{-1}\atop
\!\!\!\!\!\!\!\!\!\!\!\!\!\!\!\!\!\!\!\!\hat\Xi=\hat\chi}}
={m\over 2}\psi_l\left(\mu\right){\hat z}_E\left(l\right)
\psi_{\ge 1}\left(\eta\mu_E\alphaEB \rhoEB \right)~,
\ee
in the case of a direct attack on any {\it particular} pulse with $l$ photons in it.
This result is easily interpreted as the probability that Alice sends a pulse with $l$
photons
in it, multiplied by the probability that Eve can with complete knowledge determine the
state of polarization of that pulse, multiplied by the probability that Bob will observe
a pulse with one or more photons in it, taken from a stream sent by Eve characterized
by an effective mean photon number per pulse of $\eta\mu_E\alphaEB \rhoEB$.

We now see, quite explicitly in the case of the direct attack based on the use by Eve
of a
Poisson distribution for the surrogate pulses allowed to go on to Bob, that the value
of the fact that Eve cannot
``remotely" control the value of the quantum efficiency, $\eta$, of Bob's
detector\footnote{
As discussed previously, this is due to the presumed use by Bob of a narrow bandpass
wavelength filter in front of his apparatus.}
is {\it completely taken away}. In other words, it doesn't matter that Eve
can't directly control the quantum efficiency of Bob's detector: as long as Bob
chooses {\it not} to monitor the click statistics of his detector (to be discussed
below), Eve can effectively mimic control over $\eta$. An appropriate tuning by the
enemy of the value of the
statistical mean flux $\mu_E$ such that the product $\eta\mu_E\alphaEB\rhoEB$ is as large
as required to produce a value of $\psi_{\ge 1}\left(\eta\mu_E\alphaEB\rhoEB\right)$
that is arbitrarily close to unity, can evidently have the effect of achieving the same
result of maximizing the amount of
compromised bit information.\footnote{
Note that Eve does not need to know in advance the value of $\eta$
to achieve this
(in general Eve won't know the value of $\eta$
if Alice and Bob follow proper technically sound cryptosystem
practice and withhold this from her).
If Eve doesn't know the value of $\eta$ in advance she can infer the value
as follows:
During a period of the transmission in which she does not carry out any attacks as such,
she may perform quantum non-demolition photon number measurements from which she can
determine the value
of the mean photon number $\mu$ characterizing Alice's source. 
She can also determine the fraction $n/m$ by listening to the public discussion pertaining
to this portion of the transmission.
Using the relationship $n/m={1\over 2}\left[\psi_{\ge 1}\left(\eta\mu\alpha\right)+
r_d\right]\simeq{1\over 2}\psi_{\ge 1}\left(\eta\mu\alpha\right)\approx\eta\mu\alpha$
({\it cf} eqs.(\ref{7}) and (\ref{15})) she can then deduce the approximate
value of the product $\eta\alpha$, from which she can then infer the value of $\eta$.
Alternatively she can simply reasonably assume that Bob is using a detector for which
the value of $\eta$ is not too small to be useful and adjust $\mu_E$ accordingly.}
By the same token we see that Eve does not have to arrange
for a collaborator to be located next to Bob: the tuning of $\mu_E$ also results in the
effective replacement of $\alphaEB$ by unity.

Thus, to sum up, {\it in the absence of explicit monitoring of click statistics by Bob},
whether a
free space {\it or} optical fiber quantum channel is used is immaterial: in either
case, with
or without an enemy collaborator located at another position along the quantum channel,
and with or without any capability to physically adjust the
transparency of the quantum channel in any way, and without ascribing to Eve the
physically
nonsensical ability to ``remotely control" the quantum efficiency of Bob's detector,
in carrying out the direct attack the enemy can anyhow entirely ``tune away" the values
of both
$\alpha$ and $\eta$ to the values that allow for maximal vulnerability of the
transmitted multi-photon pulses. This has explicitly been proved to be true
({\it cf} eq.(\ref{72})) in the
case that the enemy prepares the surrogate pulse in a Poisson distribution, and is
probably true for many other distributions that might be chosen as well. However, the
fact that we have found at least one distribution for which this is clearly possible
dictates that we must assume that Eve can always choose to achieve this maximal,
worst-case possibility. Thus, we can replace the expression given in
eq.(\ref{69}) with a
universal maximal privacy amplification amount for the direct attack, for which
we no longer impose {\it only} the condition $\rhoAE=\alphaAE^{-1}$, but
also $\rhoEB=\alphaEB^{-1}$ (which amounts to allowing that Eve has completely
eliminated the attenuation by effecting the replacement $\alpha\rightarrow 1$):

\bea
\label{74}
\nu_d\Bigg\vert_{{\rhoAE =\alphaAE ^{-1}\atop\rhoEB =\alphaEB ^{-1}}}^{max}&=&
{m\over 2}z_E\left(\mu\right)
\nonumber\\
&\equiv&\nu_d^{max}
\eea

and

\bea
\label{75}
\nu_{d,l}\Bigg\vert_{{\rhoAE =\alphaAE ^{-1}\atop\rhoEB =\alphaEB ^{-1}}}^{max}&=&
{m\over 2}\psi_l\left(\mu\right){\hat z}_E\left(l\right)
\nonumber\\
&\equiv&\nu_{d,l}^{max}~.
\eea

However, we have assumed thus far that Bob adamantly
does {\it not}
monitor the statistics of his detector clicks in the above analysis.
As we shall now show, the very strong capabilities of the enemy reflected by
the above two equations are somewhat reduced if
Bob explicitly {\it does} monitor click statistics.

Now we examine the more general case in which Bob explicitly monitors the click statistics
and discards those bit cells which manifestly contain more than one photon by having
produced simultaneous clicks. We do this by following the procedure used in making the
replacement of eq.(\ref{15}) by eq.(\ref{25}), in which case we find (as before, the
subscript {\it mcs} stands for ``monitor click statistics")
\bea
\label{76}
\nu_{d,mcs}&=&{m\over 2}\sum_{l=0}^\infty\hat\chi\left(\mu,l\right)\sum_{\lp=0}^l
\left({l\atop\lp}\right)\left(\alphaAE\rhoAE \right)^{\lp}\left(1-\alphaAE \rhoAE \right)
{\hat z}_E\left(\lp\right)
\nonumber\\
&&\qquad\sum_{l_E=0}^\infty\hat\Xi\left(\mu_E,l_E\right)\sum_{\lpp=0}^{l_E}
\left({l_E\atop\lpp}\right)\left(\alphaEB \rhoEB 
\right)^{\lpp}\left(1-\alphaEB \rhoEB \right)^{l_E-\lpp}
{\hat z}_B\left(\eta,\lpp\right)
\nonumber\\
&=& {m\over 2}\sum_{l=0}^\infty\hat\chi\left(\mu,l\right)\sum_{\lp=0}^l\left({l\atop
\lp}\right)\left(\alphaAE \rhoAE \right)^{\lp}\left(1-\alphaAE \rhoAE 
\right)^{l-\lp}{\hat z}_E\left(\lp\right)
\nonumber\\
&&\qquad\sum_{l_E=0}^\infty\hat\Xi\left(\mu_E,l_E\right)
{\Bigg [}l_E\eta\alphaEB \rhoEB \left(1-\alphaEB \rhoEB 
\right)^{l_E-1}
\nonumber\\
&&\qquad\qquad\qquad+\sum_{\lpp=2}^{l_E}\left({l_E\atop\lpp}\right)
\left(\alphaEB \rhoEB \right)^{\lpp}\left(1-\alphaEB \rhoEB 
\right)^{l_E-\lpp}{\hat z}_{B,\lpp\ge 2}\left(\eta,\lpp\right){\Bigg ]}~,
\nonumber\\
\eea
in the case that all multi-photon pulses with three or more photons are subjected to
a direct attack, and
\bea
\label{77}
\nu_{d,l, mcs}&=& {m\over 2}\hat\chi\left(\mu,l\right)\sum_{\lp=0}^l
\left({l\atop
\lp}\right)\left(\alphaAE \rhoAE \right)^{\lp}\left(1-\alphaAE \rhoAE 
\right)^{l-\lp}{\hat z}_E\left(\lp\right)
\nonumber\\
&&\qquad\sum_{l_E=0}^\infty\hat\Xi\left(\mu_E,l_E\right)
{\Bigg [}l_E\eta\alphaEB \rhoEB \left(1-\alphaEB \rhoEB 
\right)^{l_E-1}
\nonumber\\
&&\qquad\qquad\qquad+\sum_{\lpp=2}^{l_E}\left({l_E\atop\lpp}\right)
\left(\alphaEB \rhoEB \right)^{\lpp}\left(1-\alphaEB \rhoEB 
\right)^{l_E-\lpp}{\hat z}_{B,\lpp\ge 2}\left(\eta,\lpp\right){\Bigg ]}
\nonumber\\
&=&{m\over 2}\psi_l\left(\mu\right)\sum_{\lp=0}^l
\left({l\atop
\lp}\right)\left(\alphaAE \rhoAE \right)^{\lp}\left(1-\alphaAE \rhoAE 
\right)^{l-\lp}{\hat z}_E\left(\lp\right)
\nonumber\\
&&\qquad\sum_{l_E=0}^\infty\hat\Xi\left(\mu_E,l_E\right)
{\Bigg [}l_E\eta\alphaEB \rhoEB \left(1-\alphaEB \rhoEB 
\right)^{l_E-1}
\nonumber\\
&&\qquad\qquad\qquad+\sum_{\lpp=2}^{l_E}\left({l_E\atop\lpp}\right)
\left(\alphaEB \rhoEB \right)^{\lpp}\left(1-\alphaEB \rhoEB 
\right)^{l_E-\lpp}{\hat z}_{B,\lpp\ge 2}\left(\eta,\lpp\right){\Bigg ]}~,
\nonumber\\
\eea
in the case of a direct attack on any {\it particular} pulse with $l$ photons in it.

As in our previous analysis leading to eq.(\ref{67}), if
we again (very conservatively) assume that
the enemy has the capability of somehow arranging for the removal of any line attenuation
between the location of Alice and the interception site, so that we have
$\rhoAE =\alphaAE ^{-1}$, we find
\bea
\label{78}
\nu_{d, mcs}\Big\vert_{\rhoAE =\alphaAE ^{-1}}
&=&{m\over 2}\sum_{l=0}^\infty\hat\chi\left(\mu,l\right){\hat z}_E\left(l\right)
\sum_{l_E=0}^\infty\hat\Xi\left(\mu_E,l_E\right)
{\Bigg [}l_E\eta\alphaEB \rhoEB \left(1-\alphaEB \rhoEB 
\right)^{l_E-1}
\nonumber\\
&&\qquad\qquad+\sum_{\lpp=2}^{l_E}\left({l_E\atop\lpp}\right)
\left(\alphaEB \rhoEB \right)^{\lpp}\left(1-\alphaEB \rhoEB 
\right)^{l_E-\lpp}{\hat z}_{B,\lpp\ge 2}\left(\eta,\lpp\right){\Bigg ]}
\nonumber\\
&=&{m\over 2}z_E\left(\mu\right)\sum_{l_E=0}^\infty\hat\Xi\left(\mu_E,l_E\right)
{\Bigg [}l_E\eta\alphaEB \rhoEB \left(1-\alphaEB \rhoEB 
\right)^{l_E-1}
\nonumber\\
&&\qquad\qquad+\sum_{\lpp=2}^{l_E}\left({l_E\atop\lpp}\right)
\left(\alphaEB \rhoEB \right)^{\lpp}\left(1-\alphaEB \rhoEB 
\right)^{l_E-\lpp}{\hat z}_{B,\lpp\ge 2}\left(\eta,\lpp\right){\Bigg ]}~,
\nonumber\\
\eea
and
\bea
\label{79}
\nu_{d,l, mcs}\Big\vert_{\rhoAE =\alphaAE ^{-1}}
&=&{m\over 2}\psi_l\left(\mu\right){\hat z}_E\left(l\right)
\sum_{l_E=0}^\infty\hat\Xi\left(\mu_E,l_E\right)
{\Bigg [}l_E\eta\alphaEB \rhoEB \left(1-\alphaEB \rhoEB 
\right)^{l_E-1}
\nonumber\\
&&\qquad\qquad+\sum_{\lpp=2}^{l_E}\left({l_E\atop\lpp}\right)
\left(\alphaEB \rhoEB \right)^{\lpp}\left(1-\alphaEB \rhoEB 
\right)^{l_E-\lpp}{\hat z}_{B,\lpp\ge 2}\left(\eta,\lpp\right){\Bigg ]}~.
\nonumber\\
\eea
If we now also examine the case in which the enemy in particular prepares the
surrogate states in a Poisson distribution so that $\hat\Xi=\hat\chi$,
one obtains

\bea
\label{80}
\nu_{d, mcs}\Bigg\vert_{{\rhoAE =\alphaAE ^{-1}\atop
\!\!\!\!\!\!\!\!\!\!\!\!\!\!\!\!\!\!\!\!\hat\Xi=\hat\chi}}
&=&{m\over 2}z_E\left(\mu\right){\Bigg \{}e^{-\mu_E}\mu_E\eta\alphaEB \rhoEB 
\sum_{l_E=0}^\infty{\left[\mu_E\left(1-\alphaEB \rhoEB \right)\right]^{l_E-1}\over
\left(l_E-1\right)!}
\nonumber\\
&&\qquad\qquad\qquad\qquad\qquad\qquad\qquad+{\Big\langle}\hat\chi\left(\mu_E,l_E\right)
{\cal Z}_{\ge 2}\left(
\eta,\alphaEB \rhoEB ,l_E\right){\Big\rangle}{\Bigg \}}
\nonumber\\
&=&{m\over 2}z_E\left(\mu\right){\Bigg [}\eta\mu_E\alphaEB \rhoEB e^{-\mu_E}
e^{\mu_E\left(1-\alphaEB \rhoEB \right)}
+{\Big\langle}\hat\chi\left(\mu_E,l_E\right){\cal Z}_{\ge 2}\left(
\eta,\alphaEB \rhoEB ,l_E\right){\Big\rangle}{\Bigg ]}
\nonumber\\
&=&{m\over 2}z_E\left(\mu\right){\Bigg [}\eta\mu_E\alphaEB \rhoEB 
e^{-\mu_E\alphaEB \rhoEB }
+{\Big\langle}\hat\chi\left(\mu_E,l_E\right){\cal Z}_{\ge 2}\left(
\eta,\alphaEB \rhoEB ,l_E\right){\Big\rangle}{\Bigg ]}
\nonumber\\
&=&{m\over 2}z_E\left(\mu\right){\Bigg [}\eta
\psi_1\left(\mu_E\alphaEB \rhoEB \right)
+{\Big\langle}\hat\chi\left(\mu_E,l_E\right){\cal Z}_{\ge 2}\left(
\eta,\alphaEB \rhoEB ,l_E\right){\Big\rangle}{\Bigg ]}
\nonumber\\
\eea
and
\be
\label{81}
\nu_{d,l, mcs}\Bigg\vert_{{\rhoAE =\alphaAE ^{-1}\atop
\!\!\!\!\!\!\!\!\!\!\!\!\!\!\!\!\!\!\!\!\hat\Xi=\hat\chi}}
={m\over 2}\psi_l\left(\mu\right){\hat z}_E\left(l\right){\Bigg [}\eta
\psi_1\left(\mu_E\alphaEB \rhoEB \right)
+{\Big\langle}\hat\chi\left(\mu_E,l_E\right){\cal Z}_{\ge 2}\left(
\eta,\alphaEB \rhoEB ,l_E\right){\Big\rangle}{\Bigg ]}~.
\ee
Comparison of eqs.(\ref{81}) with (\ref{73}) reveals
the benefit of incorporating the monitoring of
click statistics into the QKD protocol. First we rewrite eq.(\ref{73}) as
\bea
\label{82}
\nu_{d,l}\Bigg\vert_{{\rhoAE =\alphaAE ^{-1}\atop
\!\!\!\!\!\!\!\!\!\!\!\!\!\!\!\!\!\!\!\!\hat\Xi=\hat\chi}}
&=&{m\over 2}\psi_l\left(\mu\right){\hat z}_E\left(l\right)\psi_{\ge 1}\left(
\eta\mu_E\alphaEB\rhoEB\right)
\nonumber\\
&=&{m\over 2}\psi_l\left(\mu\right){\hat z}_E\left(l\right){\Bigg [}
\psi_1\left(\eta\mu_E\alphaEB\rhoEB\right)+\psi_{\ge 2}
\left(\eta\mu_E\alphaEB\rhoEB\right){\Bigg ]}
\eea
and compare with eq.(\ref{81}). Looking first at the {\it first terms} in
each of the square brackets, we see that by tuning the mean flux $\mu_E$
appropriately, the products
$\mu_E\alphaEB\rhoEB$ and $\eta\mu_E\alphaEB\rhoEB$ (for the cases of
$\nu_{d,l,mcs}$ and $\nu_{d,l}$, respectively) can assume values such that the
functions $\psi_1\left(\eta\mu_E\alphaEB\rhoEB\right)$ and
$\psi_1\left(\mu_E\alphaEB\rhoEB\right)$ reach their maximal,
optimal values (from the perspective of the enemy), thereby maximizing the
amount of information obtainable by the enemy.
However, we
see that, by employing monitoring of the click statistics, Alice and
Bob can {\it force} the reduction of this maximum amount of vulnerable information on bit
encodings otherwise available to Eve by an amount $\eta$: this is where significance
of the fact that Eve cannot ``remotely" control the value of $\eta$ comes into full
force. An analogous reduction in the amount of vulnerable information will arise from
the remaining terms in the square brackets, which will in general depend on the details
of the
functional form of ${\hat z}_B\left(\eta,l\right)$.

\vskip 10pt
\noindent {\it Indirect Attacks}

The logical structure of the analysis carried out in this section is illustrated with the
flow chart shown in Figure \ref{F:ia} below.
\begin{figure}[htb]
\vbox{
\hfil
\scalebox{0.6}{\rotatebox{270}{\includegraphics{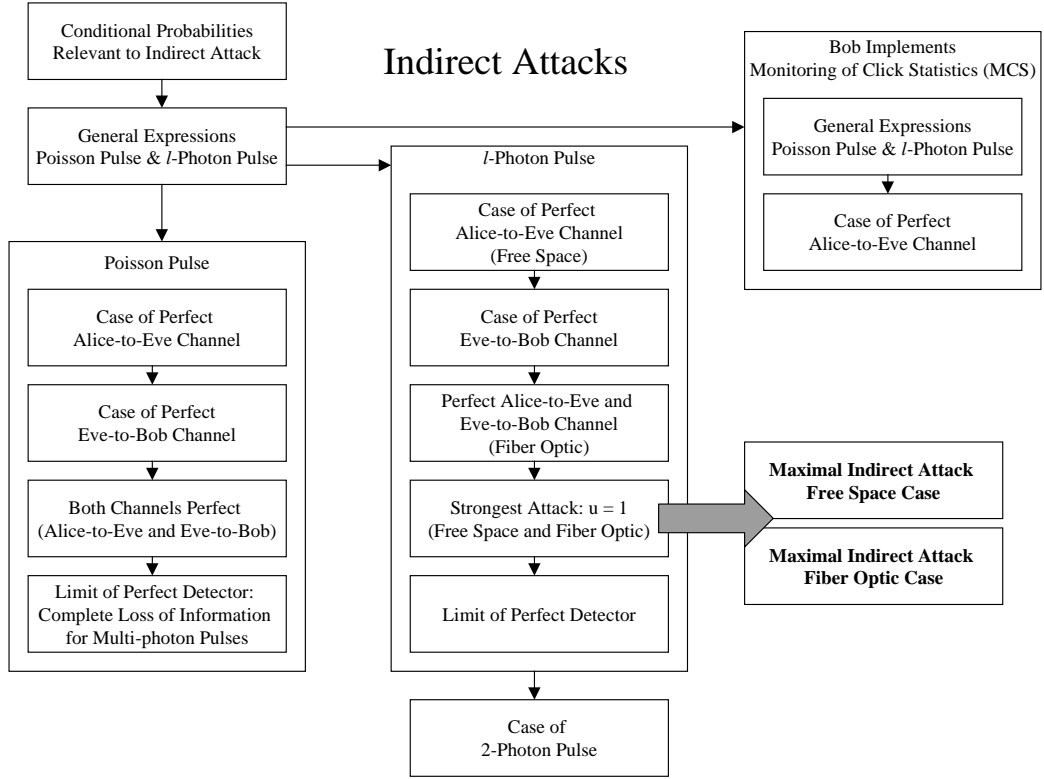}}}
\hfil
\hbox to -1.25in{\ } 
}
\bigskip
\caption{%
Flow Chart for Analysis of Indirect Attacks
}
\label{F:ia}
\end{figure}

As described above, in the indirect attack the enemy
receives a multi-photon pulse, ``splits the beam" and retains one or more
photons - unmeasured - in an appropriate quantum memory while allowing the remaining
photons in the pulse to propagate on to Bob, without disturbing them in any way. As always
in our analysis, we ascribe to the enemy superior technological capabilities, and do
not delve into the methods whereby the photon or photons retained in quantum memory
can actually be so preserved. We also assume as before that the enemy possesses perfect
photon
detection equipment, so that $\eta_E$, the intrinsic quantum efficiency of Eve's
detector apparatus, may be once and for all set equal to unity.

We proceed to deduce the form of the privacy amplification function appropriate to
indirect attacks on multi-photon pulses. As before we work from first principles by
listing the relevant probabilities for the various processes that make up the
dynamics, which are given by
\be
\label{83}
{\cal P}\left(l\ge 2~{\rm photons~leave~Alice}\right)=
\hat\chi\left(\mu,l\right)\theta\left(l-2\right)~,
\ee
\be
\label{84}
{\cal P}\left(\lp~{\rm photons~reach~Eve}~{\Big\vert}~l\ge 2~{
\rm photons~leave~Alice}\right)=
\left({l\atop\lp}\right)\left(\alphaAE \rhoAE \right)^{\lp}\left(1-\alphaAE 
\rhoAE \right)^{l-\lp}~,
\ee
\be
\label{85}
{\cal P}\left(\lpp~{\rm photons~reach~Bob}~{\Big\vert}~\lp-u~{
\rm photons~pass~Eve}\right)=\left({\lp-u\atop\lpp}\right)\left(\alphaEB \rhoEB 
\right)^{\lpp}\left(1-\alphaEB \rhoEB\right)^{\lp-u-\lpp}
\ee
and
\be
\label{86}
{\cal P}\left(\lppp~{\rm photons~detected}~{\Big\vert}~\lpp~{\rm photons~reach~Bob}\right)
=\left({\lpp\atop\lppp}\right)\eta^{\lppp}\left(1-\eta\right)^{\lpp-\lppp}\left(1-
\delta_{0,\lppp}\right)~.
\ee
In eq.(\ref{83}) the $\theta$-function enforces the
condition that there must be at least two
photons in the pulse received by the enemy in order to carry out the indirect attack.
In eq.(\ref{85}) $u$ is the number of photons split off (and preserved in
some suitable quantum memory) by the enemy, and it satisfies the inequality $u\le u_
{max}$, where $u_{max}$ is (one less than) the number of photons that were contained in
the pulse as received by Eve. We refer to the part of the pulse that is allowed to
propagate on to Bob as the ``remnant" pulse.
Unlike the case in the direct attack,
the remnant pulse must not be amplified or prepared
in any way by Eve, lest she give herself away. In the general implementation scenario
in which Bob monitors the statistics of the detector clicks, it is not obvious {\it a
priori} which value for $u$ is
optimal for the enemy. For instance, if Eve splits off and retains {\it one} photon, this
allows the maximum strength remnant pulse to go on to Bob, which is presumably
advantageous
in the case of a channel with strong attenuation, but in this case there will be some
admixture of multiple clicks observed and discarded by Bob. Alternatively, if Eve splits
off and retains {\it all but one} of the photons in the pulse, then the single photon in
the remnant pulse will defninitely {\it not} cause a multiple click to occur in Bob's
detectors, and thus the factor $z_B\left(\eta,l\right)$ will not cause Eve to lose some
of her advantage, although the received signal will be fully subjected to the effects of
line attenuation and Bob's detector inefficiency. This type of puzzle can be resolved
through the use of numerical methods. In the case of the indirect attack on
a multi-photon pulse with $l$ photons, there will in general be $l-1$ distinct
possible values for $u$ that may be chosen by the enemy.

We now deduce the explicit expression for the privacy amplification subtraction function
associated to indirect attacks, which we denote by $\nu_i^{(u)}$, where the superscript
``$(u)$" indicates the number of photons that the enemy chooses to remove from the
multi-photon pulse and retain untouched in quantum memory. Upon assembling the
appropriate probabilities from eqs.(\ref{83}) through (\ref{86}) we find
\bea
\label{87}
\nu_i^{(u)}&=&{m\over 2}\sum_{l,\lp,\lpp,\lppp}{\cal P}\left(l\ge 2~{
\rm photons~leave~Alice}\right)
\nonumber\\
&&\qquad\qquad\times{\cal P}\left(\lp~{\rm photons~reach~Eve}~{\Big\vert}~l\ge 2~{
\rm photons~leave~Alice}
\right)
\nonumber\\
&&\qquad\qquad\times{\cal P}\left(\lpp~{\rm photons~reach~Bob}~{\Big\vert}~\lp-u~{
\rm photons~pass~Eve}\right)
\nonumber\\
&&\qquad\qquad\times{\cal P}\left(\lppp~{\rm photons~detected}~{\Big\vert}~\lpp~{
\rm photons~reach~Bob}
\right)
\nonumber\\
&=&{m\over 2}\sum_{l=0}^\infty\hat\chi\left(\mu,l\right)
\theta\left(l-2\right)\sum_{\lp=0}^l\left(
{l\atop\lp}\right)\left(\alphaAE \rhoAE \right)^{\lp}\left(1-\alphaAE\rhoAE
\right)^{l-\lp}
\nonumber\\
&&\qquad\qquad\sum_{\lpp=0}^{\lp-u}\left({\lp-u\atop\lpp}\right)\left(
\alphaEB \rhoEB \right)^{\lpp}\left(1-\alphaEB \rhoEB \right)^{\lp-u-\lpp}
\nonumber\\
&&\qquad\qquad\qquad\sum_{\lppp=0}^{\lpp}\left({\lpp\atop\lppp}\right)\eta^{\lppp}
\left(1-\eta\right)^{\lpp-\lppp}\left(1-\delta_{0,\lppp}\right)
\nonumber\\
&=&{m\over 2}\sum_{l=0}^\infty\hat\chi\left(\mu,l\right)\theta\left(l-2\right)
\sum_{\lp=0}^l\left(
{l\atop\lp}\right)\left(\alphaAE \rhoAE \right)^{\lp}\left(1-\alphaAE \rhoAE
\right)^{l-\lp}
\nonumber\\
&&\qquad\qquad\sum_{\lpp=0}^{\lp-u}\left({\lp-u\atop\lpp}\right)\left(
\alphaEB \rhoEB \right)^{\lpp}\left(1-\alphaEB \rhoEB \right)^{\lp-u-\lpp}
{\Big [}1-\left(1-\eta\right)^{\lpp}{\Big ]}
\nonumber\\
&=&{m\over 2}\sum_{l=0}^\infty\hat\chi\left(\mu,l\right)\theta\left(l-2\right)
\sum_{\lp=0}^l\left(
{l\atop\lp}\right)\left(\alphaAE \rhoAE \right)^{\lp}\left(1-\alphaAE \rhoAE
\right)^{l-\lp}{\Big [}1-\left(1-\eta\alphaEB \rhoEB \right)^{\lp-u}{\Big ]}
\nonumber\\
&=&{m\over 2}\sum_{l=0}^\infty\hat\chi\left(\mu,l\right)\theta\left(l-2\right)
{\Bigg [}1-\sum_{\lp=0}^l\left(
{l\atop\lp}\right)\left(\alphaAE \rhoAE \right)^{\lp}\left(1-\alphaAE \rhoAE 
\right)^{l-\lp}\left(1-\eta\alphaEB \rhoEB \right)^{\lp-u}{\Bigg ]}
\nonumber\\
&=&{m\over 2}\sum_{l=0}^\infty\hat\chi\left(\mu,l\right)\theta\left(l-2\right)
{\Bigg [}1-\left(1-\eta\alphaEB \rhoEB \right)^{-u}\left(1-\eta\alphaAE 
\rhoAE \alphaEB \rhoEB \right)^l{\Bigg ]}
\nonumber\\
&=&{m\over 2}\sum_{l=2}^\infty\hat\chi\left(\mu,l\right)
{\Bigg [}1-\left(1-\eta\alphaEB \rhoEB \right)^{-u}\left(1-\eta\alphaAE 
\rhoAE \alphaEB \rhoEB \right)^l{\Bigg ]}
\nonumber\\
&=&{m\over 2}{\Bigg [}\psi_{\ge 2}\left(\mu\right)-\left(1-\eta\alphaEB \rhoEB 
\right)^{-u}\sum_{l=2}^\infty\hat\chi\left(\mu,l\right)\left(1-\eta\alphaAE \rhoAE 
\alphaEB \rhoEB \right)^l{\Bigg ]}
\nonumber\\
&=&{m\over 2}{\Bigg \{}\psi_{\ge 2}\left(\mu\right)-\left(1-\eta\alphaEB \rhoEB 
\right)^{-u}e^{-\mu}\sum_{l=2}^\infty{\left[\mu\left(1-\eta\alphaAE \rhoAE 
\alphaEB \rhoEB \right)\right]^l\over l!}{\Bigg \}}
\nonumber\\
&=&{m\over 2}{\Bigg \{}\psi_{\ge 2}\left(\mu\right)-\left(1-\eta\alphaEB \rhoEB 
\right)^{-u}e^{-\mu}{\Bigg [}e^{\mu\left(1-\eta\alphaAE \rhoAE \alphaEB \rhoEB 
\right)}-1-\mu\left(1-\eta\alphaAE \rhoAE \alphaEB \rhoEB \right){\Bigg ]}
{\Bigg \}}
\nonumber\\
&=&{m\over 2}\llb\psi_{\ge 2}\left(\mu\right)-\left(1-\eta\alphaEB \rhoEB 
\right)^{-u}{\Bigg \{}e^{-\eta\mu\alphaAE \rhoAE \alphaEB \rhoEB }-e^{-\mu}
{\Bigg [}1+\mu{\Bigg (}1-\eta\alphaAE \rhoAE \alphaEB \rhoEB {\Bigg )}{\Bigg ]}
{\Bigg \}}\rrb
\nonumber\\
\eea
The physical description of this result is straightforward. This quantity
is the total amount of the information contained in the multi-photon pulses
(the first term, $\psi_{\ge 2}$, in the double square brackets), diminished by a
complicated expression that
takes into account the effects of the imperfect nature of both the quantum channel itself
and the quantum efficiency of Bob's detector. This form clearly shows that simply
subtracting the {\it entire} amount of information contained in the multi-photon
pulses (as has apparently been done in all previous analyses)
in order to protect against eavesdropping attacks is sufficient, but obviously
not necessary in the presence of attenuation and/or imperfect detector efficiency in
Bob's apparatus.

The above result provides the amount of privacy amplification subtraction required to
compensate for indirect attacks on {\it all} multi-photon pulses. As in the case
of the direct attack, we also need the expression for the
amount of privacy amplification required in order to protect against an indirect attack
on any {\it particular}
multi-photon pulse with $l$ photons in it, which we denote by
$\nu_{i,l}^{(u)}$. Reading off the relevant quantity from the above calculation, we have
\bea
\label{88}
\nu_{i,l}^{(u)}&=&{m\over 2}\hat\chi\left(\mu,l\right)\theta\left(l-2\right)
{\Bigg [}1-\left(1-\eta\alphaEB \rhoEB \right)^{-u}\left(1-\eta\alphaAE 
\rhoAE \alphaEB \rhoEB \right)^l{\Bigg ]}
\nonumber\\
&=&{m\over 2}\psi_l\left(\mu\right)\theta\left(l-2\right)
{\Bigg [}1-\left(1-\eta\alphaEB \rhoEB \right)^{-u}\left(1-\eta\alphaAE 
\rhoAE \alphaEB \rhoEB \right)^l{\Bigg ]}~.
\eea
These results may be explored in a number of limits. If we specialize the above to the
case $\rhoAE =\alphaAE ^{-1}$, we have
\be
\label{89}
\nu_i^{(u)}\Big\vert_{\rhoAE =\alphaAE ^{-1}}
={m\over 2}\llb\psi_{\ge 2}\left(\mu\right)-\left(1-\eta\alphaEB \rhoEB 
\right)^{-u}{\Bigg \{}e^{-\eta\mu\alphaEB \rhoEB }-e^{-\mu}{\Bigg [}1+\mu{\Big (}
1-\eta\alphaEB \rhoEB {\Big )}{\Bigg ]}{\Bigg \}}\rrb~.
\ee
In the case that we have $\rhoEB =\alphaEB ^{-1}$, we find
\be
\label{90}
\nu_i^{(u)}\Big\vert_{\rhoEB =\alphaEB ^{-1}}
={m\over 2}\llb\psi_{\ge 2}\left(\mu\right)-\left(1-\eta
\right)^{-u}{\Bigg \{}e^{-\eta\mu\alphaAE \rhoAE }-e^{-\mu}{\Bigg [}1+\mu{\Big (}
1-\eta\alphaAE \rhoAE {\Big )}{\Bigg ]}{\Bigg \}}\rrb~.
\ee
If we consider the case in which both $\rhoAE =\alphaAE ^{-1}$ {\it and}
$\rhoEB =\alphaEB ^{-1}$, corresponding to the situation for a fiber-optic cable
implementation\footnote{
As discussed above, in the case of a free space implementation it is not necessary to
analyze the case in which the transparency of the quantum channel is modified. At most,
in this case we can imagine that {\it either} $\rhoAE =\alphaAE ^{-1}$ (meaning that
Eve is physically next to Alice) {\it or} $\rhoEB =\alphaEB ^{-1}$ (meaning that Eve
is physically next to Bob), but both conditions together are not possible for Eve to
impose.}
in which the enemy has somehow replaced the cable with an ideal ``lossless"
channel, we have
\bea
\label{91}
\nu_i^{(u)}\Bigg\vert_{{\rhoAE =\alphaAE ^{-1}\atop\rhoEB =\alphaEB ^{-1}}}
&=&{m\over 2}\llb\psi_{\ge 2}\left(\mu\right)-\left(1-\eta
\right)^{-u}{\Bigg \{}e^{-\eta\mu}-e^{-\mu}{\Bigg [}1+\mu\left(
1-\eta\right){\Bigg ]}{\Bigg \}}\rrb
\nonumber\\
&\equiv&
\nu_i^{(u),max}~.
\eea
We note that, unlike the case for the direct attack, in the indirect attack
there is no parameter that
multiplies $\eta$ that is under the control of the enemy that can be used by the enemy
to ``remotely control" or adjust the
value of the quantum efficiency of Bob's detector to a value that is optimal for Eve,
so that the above expression is indeed the worst case (from the perspective of Alice and
Bob), or maximal value of the amount of privacy amplification subtraction that needs
to be carried out to ensure a secret shared cipher.

Note that if we anyway examine the (artificial) limit of perfect detector
efficiency, $\eta\rightarrow 1$, the quantity
$\nu_i^{(u),max}$ in this ``more-than-maximal" case (denoted by the superscript
``{\it max}+") becomes
\bea
\label{92}
\nu_i^{(u),max+}&\equiv&\lim_{\eta\rightarrow 1}\nu_i^{(u),max}
\nonumber\\
&=&\lim_{\eta\rightarrow 1}\nu_i^{(u)}\Bigg\vert_{{\rhoAE =\alphaAE ^{-1}\atop\rhoEB =
\alphaEB ^{-1}}}
\nonumber\\
&=&{m\over 2}{\Bigg \{}\psi_{\ge 2}\left(\mu\right)-\lim_{\eta\rightarrow 1}
{e^{-\eta\mu}-e^{-\mu}\left[1+\mu\left(1-\eta\right)\right]\over\left(1-\eta
\right)^u}{\Bigg \}}
\nonumber\\
&=&{m\over 2}\psi_{\ge 2}\left(\mu\right)~,
\eea
so that, as expected, in the limit of a perfectly lossless
channel (so that all of the multi-photon pulses reach Eve and that all of the split-off
pieces she lets pass go on to reach Bob), perfect detector efficiency (ensuring that all
of the split-off pulses that reach Bob are in fact detected) and complete, indirect
attack
compromise of {\it all} of the multi-photon pulses (effected by the sum over all $l$),
there is a corresponding loss to the enemy of {\it all} of the information
contained in those pulses.

The results in eqs.(\ref{87}) through (\ref{92}) apply
to the case in which all multi-photon pulses
are subjected to indirect cryptanalytic attack. In case of an indirect attack on a
particular multi-photon pulse with $l$ photons in it, for the situation in which
$\rhoAE =\alphaAE ^{-1}$, we have
\be
\label{93}
\nu_{i,l}^{(u)}\Big\vert_{\rhoAE =\alphaAE ^{-1}}
={m\over 2}\psi_l\left(\mu\right)\theta\left(l-2\right){\Big [}1-\left(
1-\eta\alphaEB \rhoEB \right)^{l-u}{\Big ]}~,
\ee
and in the case that $\rhoEB =\alphaEB ^{-1}$ we have
\be
\label{94}
\nu_{i,l}^{(u)}\Big\vert_{\rhoEB =\alphaEB ^{-1}}
={m\over 2}\psi_l\left(\mu\right)\theta\left(l-2\right){\Big [}
1-\left(1-\eta\right)^{-u}\left(1-\eta\alphaAE\rhoAE\right)^l{\Big ]}~.
\ee

The situation described by eq.(\ref{93}) for the
case that $\rhoAE=\alphaAE^{-1}$, which
arises either if Eve is located immediately adjacent to Alice (in the case of a
free-space or
fiber-optic cable implementation) or if Eve has somehow been able to replace the cable
between herself and Alice with a perfect one, is particularly interesting. We will
see that
it is {\it always} advantageous, irrespective of the values of $\rhoAE$ or $\rhoEB$
(and, importantly, in the absence of click statistics monitoring by Bob), for
the enemy to choose
the value $u=1$, which means that only {\it one} photon in the multi-photon pulse is
split off and retained in quantum memory, with the other $l-1$ photons allowed to travel
on to Bob in the remnant pulse. For example, there are two possible values
for $u$ in the case of a
three-photon pulse: $u=1$ and $u=2$. We find
\be
\label{95}
\nu_{i,l=3}^{(2)}\Big\vert_{\rhoAE =\alphaAE ^{-1}}
={m\over 2}\psi_3\left(\mu\right)\eta\alphaEB\rhoEB~,
\ee
and
\bea
\label{96}
\nu_{i,l=3}^{(1)}\Big\vert_{\rhoAE =\alphaAE ^{-1}}
&=&{m\over 2}\psi_3\left(\mu\right){\Bigg [}1-\left(1-\eta\alphaEB
\rhoEB\right)^2{\Bigg ]}
\nonumber\\
&=&{m\over 2}\psi_3\left(\mu\right)\left(-\eta^2\alphaEB^2\rhoEB^2+2\eta\alphaEB\rhoEB
\right)
\nonumber\\
&=&{m\over 2}\psi_3\left(\mu\right)\eta\alphaEB\rhoEB\left(2-\eta\alphaEB\rhoEB\right)
\nonumber\\
&=&\nu_{i,l=3}^{(2)}\Big\vert_{\rhoAE =\alphaAE ^{-1}}\left(2-\eta\alphaEB\rhoEB\right)
\nonumber\\
&\ge&\nu_{i,l=3}^{(2)}\Big\vert_{\rhoAE =\alphaAE ^{-1}}~,
\eea
since $2-\eta\alphaEB\rhoEB\ge 1$ due to the fact that $\eta\alphaEB\rhoEB\le 1$,
and using eq.(\ref{88}) it is easy to show that in general one has
\be
\label{97}
\nu_{i,l}^{(1)}
\ge \nu_{i,l}^{(u>1)}~,
\ee
and finally we note that this inequality remains true for all allowed values of
$\rhoEB$, and in particular for the case $\rhoEB=\alphaEB^{-1}$. When
{\it both}
$\rhoAE=\alphaAE^{-1}$ and $\rhoEB=\alphaEB^{-1}$ we have modeled the worst case
scenario (from the perspective of Alice and Bob) in which the enemy has completely
replaced the quantum channel with one of perfect transparency,\footnote{
As noted before, this is perhaps reasonable to discuss in the case of a fiber-optic
cable implementation, but is not possible in the case of a free-space impementation.
Moreover, since we are now analyzing the indirect rather than the direct attack,
it is not possible for the enemy to circumvent the physical line attenuation by
employing a collaborator located adjacent to Bob. Accordingly, when predictions
derived from the use of $\nu_{i,l}^{(u),max}$, based as it is on the absence of
any line attenuation whatsoever, are applied to the case of a free
space implementation, any such results can be safely understood to be overly
conservative.}
so that
\bea
\label{98}
\nu_{i,l}^{(u)}\Bigg\vert_{{\rhoAE =\alphaAE ^{-1}\atop\rhoEB =
\alphaEB ^{-1}}}&=&
{m\over 2}\psi_l\left(\mu\right)\theta\left(l-2\right){\Big [}1-\left(1-\eta\right)
^{l-u}{\Big ]}
\nonumber\\
&\equiv&
\nu_{i,l}^{(u),max}~,
\eea
and we have
\be
\label{99}
\nu_{i,l}^{(1),max}\ge\nu_{i,l}^{(u>1),max}
\ee
in complete generality. Since in the above case we assume that there is no explicit
monitoring
of multiple click statistics, this result is easily explained. The enemy retains only
one photon, maximizing the number of photons in the remnant pulse and
thereby increasing the chance that the remnant pulse will be able to
propagate through to Bob in spite of the presence of some amount of line
attenuation. Of course, if Bob is actively monitoring click statistics,
the enemy faces the risk that a larger number of photons in the remnant pulse will
cause the bit cell to be identified as carrying a multi-photon pulse and thus be
discarded from the sifting process.

As with eq.(\ref{92}), where we studied for the case of the
indirect attack on {\it all}
the multi-photon pulses the articifial (and unenforceable by the enemy) but
theoretically interesting limit in
which $\eta\rightarrow 1$, we may examine this for the
``more-than-maximal" strength indirect
attack on a {\it particular} multi-photon pulse (denoted as before by the
superscript ``{\it max}+"). From eq.(\ref{98}) we have
\bea
\label{100}
\nu_{i,l}^{(u),max+}&\equiv&\lim_{\eta\rightarrow 1}\nu_{i,l}^{(u),max}
\nonumber\\
&=&\lim_{\eta\rightarrow 1}\nu_{i,l}^{(u)}\Bigg\vert_{{\rhoAE =
\alphaAE ^{-1}\atop\rhoEB = \alphaEB ^{-1}}}
\nonumber\\
&=&
{m\over 2}\lim_{\eta\rightarrow 1}{\Bigg \{}\psi_l\left(\mu\right)\theta\left(l-2\right)
{\Big [}1-\left(1-\eta\right)^{l-u}{\Big ]}{\Bigg \}}
\nonumber\\
&=&{m\over 2}\psi_l\left(\mu\right)\theta\left(l-2\right)\left(1-\delta_{l,u}
\right)~.
\eea
This result means that as before, in the limit of a perfectly lossless channel and
perfect detector efficiency in Bob's apparatus,
as long as the enemy doesn't make the mistake of keeping
{\it all} of the intercepted photons (which corresponds to setting $u=l$, in which
case none of the information is compromised since Bob doesn't receive anything,
effected by the factor $1-\delta_{l,l}=0$), {\it all}
of the information contained in the $l$-photon pulse is compromised
in this artificial and unrealizable more-than-maximal version of the indirect attack.

The case of pulses with precisely two photons is of considerable importance, since
{\it only} indirect attacks are possible for these. In this case
the only allowed value for $u$ is
$u=1$, and we find
\be
\label{101}
\nu_{i,l=2}^{(1)}={m\over 2}\psi_2\left(\mu\right)
{\Big [}1-\left(1-\eta\alphaEB \rhoEB \right)^{-1}\left(1-\eta\alphaAE 
\rhoAE \alphaEB \rhoEB \right)^2{\Big ]}~.
\ee
If we specialize to the case that $\rhoAE =\alphaAE ^{-1}$ this becomes
\bea
\label{102}
\nu_{i,l=2}^{(1)}\Big\vert_{\rhoAE =\alphaAE ^{-1}}
&=&{m\over 2}\psi_2\left(\mu\right)
{\Big [}1-\left(1-\eta\alphaEB \rhoEB \right){\Big ]}
\nonumber\\
&=&{m\over 2}\psi_2\left(\mu\right)\eta\alphaEB \rhoEB~.
\eea
If we now {\it also} set $\rhoEB=\alphaEB^{-1}$, which as above amounts to assuming
that the
enemy has completely replaced the quantum channel with one of perfect transparency,
we see that the {\it worst} case (again, from the perspective of Alice and Bob), or
maximum value of required privacy amplification is
\be
\label{103}
\nu_{i,l=2}^{(1),max}={m\over 2}\psi_2\left(\mu\right)\eta~.
\ee
Thus, the fact that the enemy cannot remotely control and alter the value of
$\eta$ is very significant,
as it implies that the enemy {\it cannot} obtain the full information
content of the two-photon pulses in the transmission. Unlike the case of the direct
attack, for which the quantity $\mu_E$ provides the enemy with a parameter (that is
beyond the control of Alice and Bob) that can ``tune" the quantum efficiency of
Bob's detector to that value which is optimal for Eve, in the indirect attack the
enemy can {\it at most} obtain
a fraction $\eta$ of the information content of the two-photon pulses.

We may reconsider the entire analysis of indirect attacks for the case
corresponding to explicit monitoring of click statistics. Carrying through the algebra
for this yields
\bea
\label{104}
\nu_{i,mcs}^{(u)}&=&{m\over 2}\sum_{l=0}^\infty\hat\chi\left(\mu,l\right)\theta\left(l-2
\right)
{\Bigg \{}\eta\alphaEB \rhoEB {\left(1-\alphaAE \rhoAE \alphaEB
\rhoEB\right)^l\over\left(1-\alphaEB \rhoEB \right)^{u+1}}{\Bigg [}{l\alphaAE 
\rhoAE \left(1-\alphaEB \rhoEB \right)\over 1-\alphaAE \rhoAE 
\alphaEB \rhoEB }-u{\Bigg ]}
\nonumber\\
&&\qquad+\sum_{\lp=0}^l\left({l\atop\lp}\right)\left(\alphaAE \rhoAE \right)^{\lp}
\left(1-\alphaAE \rhoAE \right)^{l-\lp}
\nonumber\\
&&\qquad\qquad\sum_{\lpp=2}^{\lp-u}\left({\lp-u\atop\lpp}
\right)\left(\alphaEB \rhoEB \right)^{\lpp}\left(1-\alphaEB \rhoEB \right)^
{\lp-u-\lpp}{\hat z}_{B,\ge 2}\left(\eta,\lpp\right){\Bigg \}}~,
\eea
for the case that {\it all} the multi-photon pulses are subjected to the indirect
attack in the presence of click statistics monitoring, and
\bea
\label{105}
\nu_{i,l,mcs}^{(u)}&=&{m\over 2}\hat\chi\left(\mu,l\right)\theta\left(l-2
\right)
{\Bigg \{}\eta\alphaEB \rhoEB {\left(1-\alphaAE \rhoAE \alphaEB
\rhoEB\right)^l\over\left(1-\alphaEB \rhoEB \right)^{u+1}}{\Bigg [}{l\alphaAE 
\rhoAE \left(1-\alphaEB \rhoEB \right)\over 1-\alphaAE \rhoAE 
\alphaEB \rhoEB }-u{\Bigg ]}
\nonumber\\
&&\qquad+\sum_{\lp=0}^l\left({l\atop\lp}\right)\left(\alphaAE \rhoAE \right)^{\lp}
\left(1-\alphaAE \rhoAE \right)^{l-\lp}
\nonumber\\
&&\qquad\qquad\sum_{\lpp=2}^{\lp-u}\left({\lp-u\atop\lpp}
\right)\left(\alphaEB \rhoEB \right)^{\lpp}\left(1-\alphaEB \rhoEB \right)^
{\lp-u-\lpp}{\hat z}_{B,\ge 2}\left(\eta,\lpp\right){\Bigg \}}~,
\eea
for the case that a particular $l$-photon pulse is attacked.

If we assume that $\rhoAE =\alphaAE ^{-1}$ we find the considerably simplified forms
\bea
\label{106}
\nu_{i,mcs}^{(u)}\Big\vert_{\rhoAE =\alphaAE ^{-1}}
&=&{m\over 2}\sum_{l=0}^\infty\hat\chi\left(\mu,l\right)\theta\left(l-2
\right)
{\Bigg [}\eta\alphaEB \rhoEB \left(1-\alphaEB \rhoEB \right)^{l-u-1}\left(
l-u\right)
\nonumber\\
&&\qquad+\sum_{\lpp=2}^{l-u}\left({l-u\atop\lpp}\right)\left(\alphaEB \rhoEB 
\right)^{\lpp}\left(1-\alphaEB \rhoEB \right)^{l-u-\lpp}{\hat z}_{B,\ge 2}\left(
\eta,\lpp\right){\Bigg ]}~,
\nonumber\\
\eea
and
\bea
\label{107}
\nu_{i,l,mcs}^{(u)}\Big\vert_{\rhoAE =\alphaAE ^{-1}}
&=&{m\over 2}\hat\chi\left(\mu,l\right)\theta\left(l-2
\right)
{\Bigg [}\eta\alphaEB \rhoEB \left(1-\alphaEB \rhoEB \right)^{l-u-1}\left(
l-u\right)
\nonumber\\
&&\qquad+\sum_{\lpp=2}^{l-u}\left({l-u\atop\lpp}\right)\left(\alphaEB \rhoEB 
\right)^{\lpp}\left(1-\alphaEB \rhoEB \right)^{l-u-\lpp}{\hat z}_{B,\ge 2}\left(
\eta,\lpp\right){\Bigg ]}~.
\nonumber\\
\eea

\vskip 10pt
\noindent {\it Combined Direct {\it and} Indirect Attacks}

The logical structure of the analysis carried out in this section is illustrated
with the flow chart shown in Figure \ref{F:ca} below.
\begin{figure}[htb]
\vbox{
\hfil
\scalebox{0.6}{\rotatebox{270}{\includegraphics{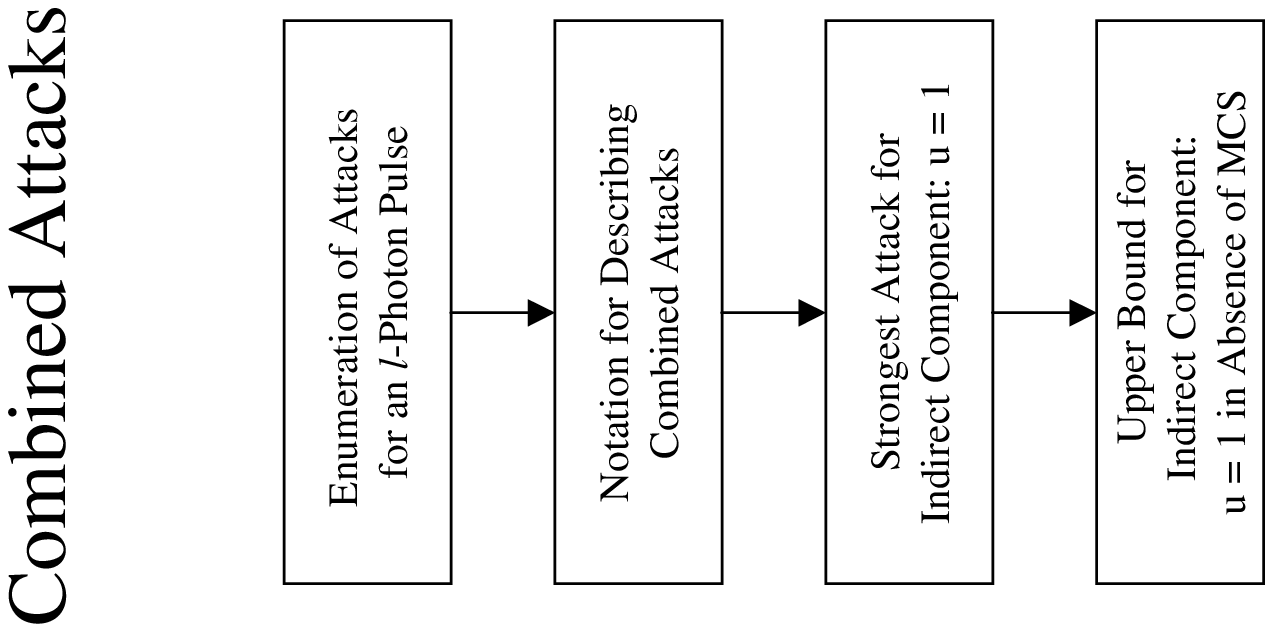}}}
\hfil
\hbox to -1.25in{\ } 
}
\bigskip
\caption{%
Flow Chart for Analysis of Combined Direct and Indirect Attacks
}
\label{F:ca}
\end{figure}

Until now we have considered the situation in which, having intercepted a
multi-photon pulse containing $l$ photons, the enemy carries out {\it either}
the direct attack {\it or} the indirect attack, and we have considered this in the
cases that the attacks are performed either on a particular pulse or on some number
of them (including all of them). However, in
addition to carrying out a general admixture of {\it purely} direct and {\it purely}
indirect attacks, distributed in some way
amongst the various multi-photon pulses, it is also
possible for the enemy to perform what we shall refer to as
a {\it combined} direct and indirect attack on any
given intercepted multi-photon pulse, as long as the pulse contains {\it five or more}
photons. This type of attack appears to have never been previously analyzed.
The requirement for five or more photons arises as follows. To carry out the
{\it direct} part of the attack, the enemy requires at least three photons in order to
determine the state of polarization with complete knowledge; to carry out the {\it
indirect} part of the attack, the enemy must split the beam and retain at least one
(unmeasured) photon in a suitable quantum memory, and allow a remnant pulse of at least
one (unmeasured) photon to propagate on to Bob. Of course, any
given attack on any given intercepted
multi-photon pulse is either successful in providing the enemy the identity of the
state or it isn't: the purpose for the enemy in carrying out the
combined attack would be to try to
increase the likelihood that the information extracted on the
pulse can be increased to a higher value than possible with either a purely direct or
purely indirect attack. The question is whether or not this occurs.
As we will show, in the general case the combined attack
is not as strong as either a particular purely direct or purely indirect
attack.\footnote{
For completeness
it is nevertheless of value to understand the combined attack in detail.}
The combined attack furnishes, for any given multi-photon pulse,
a ``continuum region" of success outcomes for the enemy connecting the purely direct
and purely indirect attacks.
The analysis is complicated by the competing
effects of the quantum efficiency of Bob's detector and any residual
line attenuation on the
quantum channel that the enemy has not managed to in some way eliminate.

There are a variety of ways in which the photon
content of a
given multi-photon pulse with five or more photons in it can be disassembled by the
enemy to carry out the combined attack.
The number of distinct types of combined attack grows rapidly with the number of
photons in the multi-photon pulses. For example, with a multi-photon pulse that
contains $l=5$ photons, there is only {\it one} possible combined attack. In this case
the enemy can split off three photons from the pulse to carry out the direct attack
and subject the remaining two photons to the indirect attack, for which the number
of photons in the remnant pulse is necessarily unity so that $u=1$ photon is
retained in quantum memory by the enemy. When there are
$l=6$ photons in the pulse, there are {\it three} distinct combined attacks possible:
(1) the enemy can split off three photons for the direct attack, which leaves three
photons for the indirect attack, of which $u=1$ is retained in quantum memory with
two photons in the remnant pulse, or
(2) the enemy can split off three photons for the direct attack, which leaves three
photons for the indirect attack, of which $u=2$ are retained in quantum memory with
one photon in the remnant pulse, or (3) the enemy can split off four photons for the
direct attack, which leaves two photons for the indirect attack, of which $u=1$ is
retained in quantum memory with one photon in the remnant pulse.

The analysis of the various possibilities is governed by the fact that, for any
given multi-photon pulse, the enemy can choose to perform any of the allowed
combined attacks, or any of the allowed purely direct or indirect attacks. To
organize the different possibilities,
we introduce the following notation to represent the particular way in which a
given $l$-photon pulse has been disassembled by the enemy
in order to carry out a chosen attack:
for an $l$-photon pulse, we designate by
$(l_d,l_i)^{u+(l_i-u)}$, subject to the
constraint $l_d+l_i=l$, the situation in
which $l_d$ of the photons in the pulse are subjected to a direct attack and
$l_i$ of the photons are subjected to an indirect attack, with $u$ photons being
intercepted and stored in quantum memory by the enemy for the indirect part.
This {\it kinematical symbol} notation is completely general and is very useful
in describing any possible attack. For example, in the
case of a pulse containing $l=8$ photons, the symbol $(3,5)^{1+4}$ means that 3 of
the photons are subjected to a direct attack, and the remaining 5 photons are subjected
to the an indirect attack, with Eve choosing to split off and retain 1
of those 5 photons, letting the
remaining 4 photons go on to Bob. In the event that so many of
the photons are taken for one kind of attack that there are not enough left to
carry out the {\it other} kind
of attack, we place the remaining number inside parentheses:
thus, the symbol $((1),7)^{4+3}$ means that 7 of the 8 photons in the intercepted pulse
are taken by the enemy
for the indirect attack (here with $u=4$ of these retained by Eve and 3 allowed
to go on to Bob), which only leaves 1 additional
photon, which is not enough to carry out a direct attack, and so it is placed
inside parentheses. 

To illustrate the different types of attack that are possible, in
Table \ref{table1} we enumerate in full the 12 distinct attacks that are possible to
carry out on a
multi-photon pulse containing $l=5$ photons, and in Table \ref{table2} we do the same
for the 30 distinct attacks that are possible on a multi-photon pulse containing
$l=8$ photons.

\begin{table}
\halign {

# &  \hskip 2.0 in # \hfil & \hskip 0.5 in # \hfil & \hskip 0.5 in # \hfil \cr
&&$(5,(0))$ &   \cr
&&$(4,(1))$ &   \cr
&&$(3,2)^{1+1}$ &   \cr
&&$((2),3)^{1+2}$ &   \cr
&&$((2),3)^{2+1}$ &   \cr
&&$((1),4)^{1+3}$ &   \cr
&&$((1),4)^{2+2}$ &   \cr
&&$((1),4)^{3+1}$ &   \cr
&&$((0),5)^{1+4}$ &   \cr
&&$((0),5)^{2+3}$ &   \cr
&&$((0),5)^{3+2}$ &   \cr
&&$((0),5)^{4+1}$ &   \cr
}
\bigskip
\caption{
Set of Distinct Purely Direct, Purely Indirect and Combined Attacks for $l=5$.
}
\label{table1}
\end{table}


\begin{table}
\halign {

# &  \hskip 1.5 in # \hfil & \hskip 0.5 in # \hfil & \hskip 0.5 in # \hfil \cr
&&$(8,(0))$ & $((1),7)^{1+6}$   \cr
&&$(7,(1))$ & $((1),7)^{2+5}$   \cr
&&$(6,2)^{1+1}$ & $((1),7)^{3+4}$   \cr
&&$(5,3)^{1+2}$ & $((1),7)^{4+3}$   \cr
&&$(5,3)^{2+1}$ & $((1),7)^{5+2}$   \cr
&&$(4,4)^{1+3}$ & $((1),7)^{6+1}$   \cr
&&$(4,4)^{2+2}$ & $((0),8)^{1+7}$   \cr
&&$(4,4)^{3+1}$ & $((0),8)^{2+6}$   \cr
&&$(3,5)^{1+4}$ & $((0),8)^{3+5}$   \cr
&&$(3,5)^{2+3}$ & $((0),8)^{4+4}$   \cr
&&$(3,5)^{3+2}$ & $((0),8)^{5+3}$   \cr
&&$(3,5)^{4+1}$ & $((0),8)^{6+2}$   \cr
&&$((2),6)^{1+5}$ & $((0),8)^{7+1}$   \cr
&&$((2),6)^{2+4}$ &   \cr
&&$((2),6)^{3+3}$ &   \cr
&&$((2),6)^{4+2}$ &   \cr
&&$((2),6)^{5+1}$ &   \cr
}
\bigskip
\caption{
Set of Distinct Purely Direct, Purely Indirect and Combined Attacks for $l=8$.
}
\label{table2}
\end{table}
%
The entries in each
list comprises a continuum of attacks ranging from purely direct to purely indirect.
In the $l=5$ case, of the 12 possibilities there are two distinct,
purely direct attacks (represented by $(5,(0))$ and $(4,(1))$, one combined
attack (represented by $(3,2)^{1+1}$), and a total of 9 purely indirect attacks
(represented by the ordered pairs in which the first element is contained within
parentheses). Similarly, in the $l=8$ case, of the 30 possibilities
there are two distinct, purely direct
attacks (represented by $(8,(0))$ and $(7,(1))$), ten different combined attacks
(represented by the various
entries for which neither element inside the ordered pair is contained within
parentheses), and a total of 18 purely indirect attacks (as before, represented by the
ordered pairs in which the first element is contained within parentheses). As always, it
is simple to identify an attack that is optimal from the perspective of the enemy.
The attack for which the net value of the associated privacy amplification function
is maximal
is the strongest attack, as it requires the largest number of bits to be subtracted in
order to ensure that the remaining bits shared between Alice and Bob will be secret.
As we shall see, most of the attacks displayed in the two Tables
are {\it not} optimal from the perspective of the enemy, and
are listed here only for completeness.

In the absence of explicit click statistics monitoring by Bob we may continue to
assume for the combined attack, just as was shown in eqs.(\ref{96}) and (\ref{97})
for
the case of the purely indirect attack, that the indirect attack {\it part} of a
given combined attack is optimal for the enemy if the value $u=1$ is selected, so that
only one photon is retained in quantum memory and thus the largest possible number of
photons are allowed to go on to Bob in the remnant pulse, thereby increasing
the likelihood of overcoming whatever line attenutuation may be present in the
quantum channel. This assumption, which is only demonstrably valid when {\it no}
click monitoring is in effect, considerably reduces the number of distinct types
of combined attack that need to be considered in our analysis: now, if
the intercepted multi-photon
pulse contains $l$ photons, it is easy to see that there are a total of $l-4$
possible distinct {\it combined} attacks available to the enemy. More generally, if we
restrict consideration of indirect attacks to those for which $u=1$ there are a
total altogether of $l-1$ distinct attacks of {\it any} kind. Thus, in the case
of a multi-photon pulse with $l=5$ photons, out of the 12 possible attacks, we see
from Table \ref{table1} that there is $5-4=1$ combined attack for which $u=1$ (in this
case this is also the {\it only} combined attack), and we see that there are $5-1=4$
attacks in general
in which the indirect part is characterized by $u=1$. The reduction in the
number of attacks that need to be considered is much more dramatic in the case of
$l=8$: here we see from Table \ref{table2} that, of the 30 attacks that are possible
in total, only $8-4=4$ of them are combined attacks for which $u=1$, and only
$8-1=7$ of them include indirect attacks, combined or not, for which $u=1$.

It is useful to study this reduced set of attacks for which $u=1$, since in
this case we know (in the absence of click statistics monitoring) that the attack
is optimal for the enemy, {\it both} due to having $u=1$ {\it and} due
to the absence of click statistics monitoring, since the modifying the former
condition can only have the effect of reducing the strength of the enemy's attack,
and the same is true if the latter activity is implemented.
Thus, the associated amount of privacy
amplification subtraction is guaranteed to provide sufficient protection against the
strongest possible cryptanalytic attacks, as Bob is needlessly weakening his protection
({\it i.e.}, Bob is not executing a ``technically sound quantum cryptosystem")
against the enemy by not monitoring click statistics.

Having sketched a taxonomy and thus delineated the ``kinematics" of the various
types of
multi-photon attacks we now turn our attention to the ``dynamics," {\it i.e.}
an assessment of the relative strengths of the purely direct, purely indirect
and combined attacks by making use of the closed form expressions for the
privacy amplification that we have derived.

\vskip 10pt
\noindent {\it Comparison of Strengths of Direct, Combined and Indirect Attacks}

The logical structure of the analysis carried out in this section is illustrated with the
flow chart shown in Figure \ref{F:cs} below.
\begin{figure}[htb]
\vbox{
\hfil
\scalebox{0.6}{\rotatebox{270}{\includegraphics{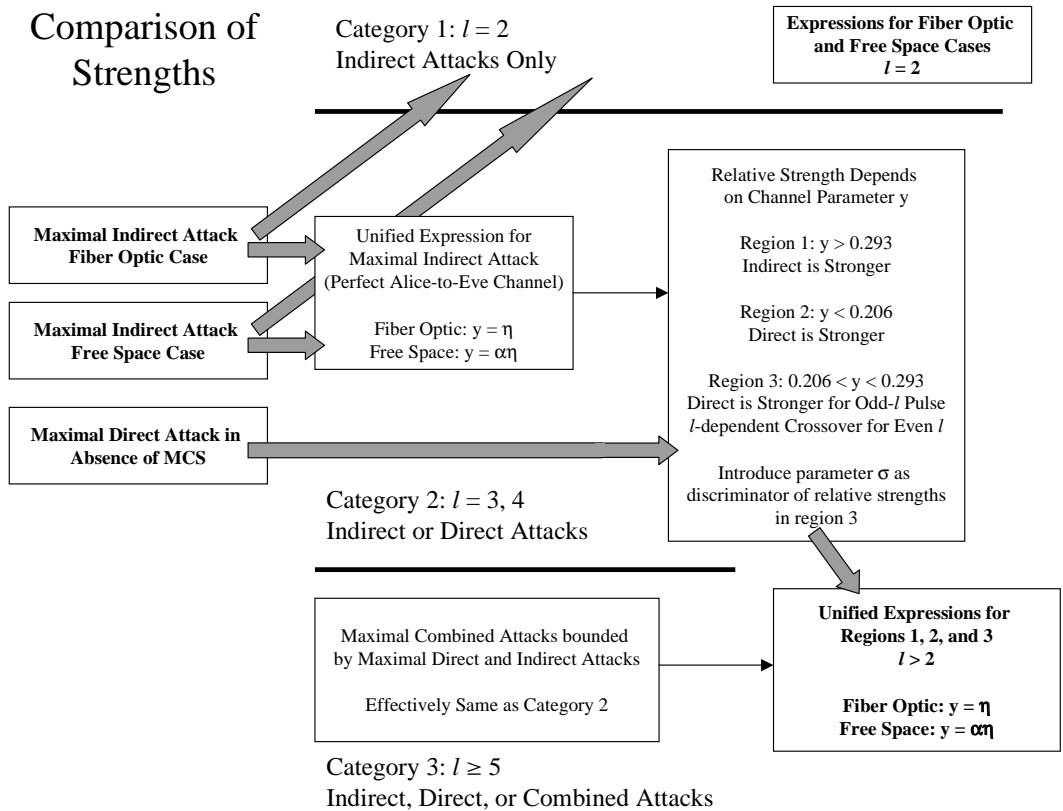}}}
\hfil
\hbox to -1.25in{\ } 
}
\bigskip
\caption{%
Flow Chart for Analysis of Comparison of Attack Strengths
}
\label{F:cs}
\end{figure}

The enemy can choose to carry out any admixture of purely direct and purely indirect
attacks, or to carry out simultaneous combinations of the two types,
and we must assume that the relative proportion chosen for each is unknown to Alice and
Bob, who must therefore implement sufficient privacy amplification subtraction to
protect
against the worst case scenario. By an ``admixture of purely direct and purely
indirect attacks" we mean that, for a given transmission from Alice to Bob
that includes some fixed number of multi-photon
pulses that contain $l$ photons each, the enemy can choose to carry out the purely
indirect attack
on a fraction $j_l$ of the $l$-photon pulses and subject the remaining fraction of
$1-j_l$ of the $l$-photon pulses in the
stream to the purely direct attack (as long as $l\ge 3$, since otherwise
the direct attack
cannot succeed). There is no reason that the $j_l$ values need to be the same for
different values of $l$, and in general we need to consider the case that they are
not, in order to assess what the strongest enemy attack might be.
To understand this it is important to compare directly against each other the
strengths of the direct and indirect attacks. 

We first note that there are three different categories of multi-photon
pulses to consider in this analysis, based on the associated value of
$l$: Category 1: $l=2$,
Category 2: $l=3$ and
$l=4$ and Category 3: $l\ge 5$. We address each category in turn.

$\bullet$ Category 1, $l=2$: For two-photon pulses, {\it only} indirect attacks
are possible.\\
{}$~~~~~~$We conservatively
assume that {\it all} two-photon pulses are subjected to indirect attack. Thus
no comparison of strengths of different types of attacks is necessary or possible: the
only question to address is the appropriate amount of privacy amplification to
apply (necessary and sufficient, or sufficient). This is addressed in the following
section in which we write down the complete expression for multi-photon pulse
privacy amplification.

$\bullet$ Category 2, $l=3$ and $l=4$: For
three- and four-photon pulses, {\it either} a direct {\it or}
an indirect attack, but not both, are possible on any given pulse, and
thus no combined attack is possible.\\
{}$~~~~~~$We conservatively
assume that {\it all} three- and four-photon pulses are subjected to one or
the other of the two types of attack, denoting
the fraction of three-photon pulses subjected to indirect attack by $j_3$, and the
fraction of three-photon pulses subjected to direct attack by $1-j_3$ (with
similar meanings for $j_4$ and $1-j_4$, respectively, for the case of
four-photon pulses). 

$\bullet$ Category 3, $l\ge 5$: For multi-photon pulses with five or more photons
direct, indirect or combined direct and indirect attacks are possible.

In the absence of special intelligence information provided
through espionage or other means,
Alice and Bob cannot in general expect to know the values that
the enemy will choose for the various $j_l$ that will be used against pulses from
categories 2 and 3, and they will also not know which combined attacks, if any, may
be carried out against pulses from category 3, so
that the only way to ensure secrecy
is to determine the strength of the
worst case attack. For this purpose we explicitly compare the
closed form expressions for the privacy amplification functions deduced above.

As written above, no comparison of attack strengths is required for
Category 1 pulses since only indirect attacks are possible.
We will proceed by first comparing the strengths of the purely direct and purely
indirect attacks, which are the {\it only} kind possible in Category 2, and then
show that (as mentioned above) for any given multi-photon pulse it is always the
case that the strongest possible attack is {\it either} a purely direct attack
or a purely indirect attack and never a combined attack,
so that for {\it both} Category 2 and Category 3 it suffices to consider direct
and indirect attacks only.

In addition to not knowing the values of the $j_l$, as discussed
above Alice and Bob will also not in general know the identity of the distribution
function $\hat\Xi$ chosen by the enemy for the preparation of the surrogate pulse in
the case of the direct attack. However, we found above
that if Eve chooses in particular to prepare the surrogate pulse in
the Poisson
distribution, $\hat\Xi\left(\mu_E,l_E\right)=\hat\chi\left(\mu_E,l_E\right)$, she
can in principle tune the value of $\mu_E$ such that the entire line attenuation and
detector inefficiency of Bob's apparatus are effectively
eliminated, resulting in the strongest
possible direct attack, $\nu_{d,l}^{max}$, given in eq.(\ref{75}) as
\be
\label{108}
\nu_{d,l}^{max}={m\over 2}\psi_l\left(\mu\right){\hat z}_E\left(l\right)~.
\ee
Since we have explicitly demonstrated that it is {\it possible} for the enemy to
achieve the maximal strength direct attack, we must assume that the
maximal strength direct attack {\it will} be achieved (assuming
that the direct attack has been chosen).
Of course, as we found in eq.(\ref{77}), the strength
of this attack can be controlled and
reduced by Bob through active monitoring of the click statistics,
resulting in a leading order diminution ({\it cf} eq.(\ref{81}))
of the strength of the attack by a factor of $\eta$. We will proceed at
first by assuming that Bob does {\it not} employ click statistics monitoring, which
means we will be working with the absolutely
worst case scenario from the perspective of Alice and Bob, the
strongest possible version of the direct attack, {\it i.e.,} the form given by
eq.(\ref{108}) above. This form equally applies to the cases of a free space or fiber
optic
cable quantum channel since the difference between them, {\it i.e.,} the fact that
it is not possible for the enemy to physically improve the transparency of the
atmosphere, is effectively eliminated since the enemy can tune the value of
$\mu_E$ to achieve the same result.

Similarly, we found in eq.(\ref{98}) the form of the
maximal strength indirect attack that can
be carried out on a multi-photon pulse,
\be
\label{109}
\nu_{i,l}^{(u),max}={m\over 2}\psi_l\left(\mu\right)\theta\left(l-2\right){\Big [}
1-\left(1-\eta\right)^{l-u}{\Big ]}~,
\ee
which is the applicable form if the enemy has somehow managed to surreptitiously
replace the quantum channel
with another one of perfect transparency, which
is only reasonable to suppose is possible (if at all!)
in the case of a fiber optic cable implementation. In the case of a free space
implementation we should instead use eq.(\ref{88}),
\be
\label{110}
\nu_{i,l}^{(u)}={m\over 2}\psi_l\left(\mu\right)\theta\left(l-2\right){\Bigg [}
1-\left(1-\eta\alphaEB\rhoEB\right)^{-u}\left(1-\eta\alphaAE\rhoAE\alphaEB
\rhoEB\right)^l{\Bigg ]}~.
\ee
The interception apparatus of the enemy must of course
be located {\it somewhere}, and for this analysis we will
take as this location a position immediately adjacent to the Alice
site, which has the effect of setting
$\rhoAE=\alphaAE^{-1}$, and this becomes ({\it cf} eq.(\ref{93}))
\bea
\label{111}
\nu_{i,l}^{(u)}\Big\vert_{\rhoAE =\alphaAE ^{-1}}&=&
{m\over 2}\psi_l\left(\mu\right)\theta\left(l-2\right){\Big [}
1-\left(1-\eta\alphaEB\rhoEB\right)^{l-u}{\Big ]}
\nonumber\\
&=&{m\over 2}\psi_l\left(\mu\right)\theta\left(l-2\right){\Big [}
1-\left(1-\eta\alpha\right)^{l-u}{\Big ]}~,
\eea
where in the second equation above we have simply set $\alphaEB\rhoEB=\alpha$,
since in the case of the generic free space system that we are discussing,
the residual amount $\alphaEB$ of line attenuation
appearing in the first equation above is simply the total line attenuation
$\alpha$, and
the condition $\rhoEB=1$ is imposed by the physical impossibility of replacing the
atmosphere with one of improved transmissivity.
The difference between eqs.(\ref{109}) and (\ref{111}) for
the maximal strength fiber optic cable and
free space implementation versions of the indirect attack privacy amplification amount
is then just the presence of the factor of
$\alpha$ that multiplies $\eta$.

Thus, the general, worst case (for Alice and Bob) combination of attacks for
multi-photon pulses with $l=3$ or $l=4$ photons is given by
\be
\label{112}
j_l\nu_{i,l}^{(u),max}+\left(1-j_l\right)\nu_{d,l}^{max}
\ee
where $\nu_{i,l}^{(u),max}$ is understood to be given by eq.(\ref{109}) in the case of a
fiber-optic cable implementation and by eq.(\ref{111}) in the case of a free space
implementation. Of course, the above expression denotes the most general
possible mix between
purely direct and purely indirect attacks for {\it any} value of $l\ge 3$, not just
the cases $l=3$ and $l=4$, but for the latter two values this form
encompasses all possibilities since no combined attacks are allowed.
The question now is: what values of $j_l$
will optimize this for the enemy, thereby
prescribing for Alice and Bob the corresponding
amount of needed privacy amplification?

We consider the situation in which Bob does
not actively monitor the click statistics of his detector. In this case we found
in eq.(\ref{97}) that the maximum strength indirect attack takes place when the enemy
selects the value $u=1$, retaining only one photon from the split beam.
To be as conservative as possible we
utilize this value of $u$ in the following analysis, thus ensuring that we are
considering worst case results.
To measure the relative strengths of the two types of maximal attack
we examine functions
of their differences and ratios. We first note that
the quantities $\nu_{i,l}^{(u),max}$ and $\nu_{d,l}^{max}$ are each
proportional to the factor ${m\over 2}\psi_l\left(\mu\right)$.
We therefore construct the normalized
difference function $\Delta^{max}$ between the two maximal
privacy amplification amounts as
\be
\label{113}
\Delta^{max}\equiv{\nu_{i,l}^{(1),max}-\nu_{d,l}^{max}\over{m\over 2}\psi_l\left(
\mu\right)}~.
\ee

More generally, we note that
with the help of eqs.(\ref{66}) and (\ref{88})
we can define the universal difference function that is {\it always}
valid (not just for maximal attacks) in the absence of click statistics monitoring
as
\be
\label{114}
\Delta\equiv{\nu_{i,l}^{(u)}-\nu_{d,l}\over{m\over 2}\psi_l\left(
\mu\right)}~,
\ee
and, if click statistics monitoring is executed, we may write using eqs.(\ref{77})
and (\ref{105})
\be
\label{115}
\Delta_{mcs}\equiv{\nu_{i,l,mcs}^{(u)}-\nu_{d,l,mcs}\over{m\over 2}\psi_l\left(
\mu\right)}~.
\ee
Proceeding with the analysis of $\Delta^{max}$,
since ${m\over 2}\psi_l\left(\mu\right)>0$, we see that when
the condition $\Delta^{max}>0$ is satisfied the maximal
{\it indirect} attack is the stronger of the
two for the enemy, and when $\Delta^{max}<0$ the maximal {\it direct} attack is
superior.
Owing to the form of the function ${\hat z}_E\left(l\right)$ contained within the
quantity $\nu_{d,l}^{max}$, we must write the difference function
out separately for multi-photon pulses with even and odd numbers
of photons. Upon defining
$y\equiv\eta\alpha$ (satisfying $0\le y\le 1$) in order to consider
both the fiber-optic cable and free space implementation cases with the same notation,
we have for
the even integer case, when $l=2k$ with $k\ge 2$
(due to the chosen range for $k$ we have not needed to manifestly
include either the explicit $\theta$-function $\theta\left(l-2\right)$
contained in $\nu_{i,l}^{(1),max}$
or the implicit $\theta$-function $\theta\left(l-3\right)$
contained in $\nu_{d,l}^{max}$ as they are both equal to unity)
\bea
\label{116}
\Delta_e^{max}&=&\Delta_e^{max}\left(k,y\right)
\nonumber\\
&=&1-\left(1-y\right)^{2k-1}-\left(1-2^{1-k}\right)
\nonumber\\
&=&2^{1-k}-\left(1-y\right)^{2k-1}~,
\eea
and for the odd integer case, when $l=2k+1$ with $k\ge 1$ we
have (for the same reason as above we have not needed to write out the
$\theta$-functions)
\bea
\label{117}
\Delta_o^{max}&=&\Delta_o^{max}\left(k,y\right)
\nonumber\\
&=&1-\left(1-y\right)^{2k}-\left(1-2^{-k}\right)
\nonumber\\
&=&2^{-k}-\left(1-y\right)^{2k}~.
\eea

To determine the boundary separating the regions for which $\Delta^{max}>0$
and $\Delta^{max}<0$ we can solve the equations
\be
\label{118}
0=\Delta_e^{max}\left(k,y\right)\Big\vert_{y=y_e}
\ee
and
\be
\label{119}
0=\Delta_o^{max}\left(k,y\right)\Big\vert_{y=y_o}
\ee
for $k$ and invert the solutions to obtain
\bea
\label{120}
y_e&=&y_e\left(k\right)
\nonumber\\
&=&
1-2^{-{1-k\over 1-2k}}\quad\forall ~k\ge 2
\eea
for the even photon number case with $l=2k$, and
\bea
\label{121}
y_o&=&1-{1\over\sqrt 2}\quad\forall ~k\ge 1
\nonumber\\
&\simeq& 0.292893
\eea
for the odd photon number case with $l=2k+1$. In solving the equations to obtain
$y_e$ and $y_o$ all other solutions than the two that are listed here were discarded
since they either do not yield real values for $y_{e,o}$
or do not satisfy the unitarity constraints $0\le y_{e,o}\le 1$.
We also note in
particular that $y_o$ is manifestly independent of any specific value of $k$.

We note that the solution $y_e$ is a monotonic function\footnote{
The function $y_e$ is of course strictly
only defined at discrete, integer values of its argument $k$.}
of $k$, with its smallest value given by
\be
\label{122}
y_e\left(2\right)=1-{1\over\sqrt[3] {2}}~,
\ee
and we find that it asymptotically achieves its maximum in the limit
\be
\label{123}
\lim_{k\rightarrow\infty}y_e\left(k\right)\uparrow 1-{1\over\sqrt 2}~.
\ee
We have thus found the answer to the question: ``Which is stronger for a
given multi-photon pulse, the maximal purely direct attack or the maximal
purely indirect attack?" The answer is completely determined, in all generality, by
the value of the quantity $\eta\alpha$ in the case of a free space implementation
of quantum cryptography, or by the value of the quantity $\eta$ in the case of a
fiber optic cable implementation (if we wish to be conservative and allow for the
possibility that the enemy might surreptitiously replace the cable with a ``lossless"
one). Rather than writing expressions first with
$\eta\alpha$ and then again with $\eta$, we will continue to employ the symbol $y$
with the appropriate substitution understood for free space or fiber optic cable
systems. The relative strength of maximal purely direct and maximal purely indirect
attacks is determined by locating $y$ in one of the following regions:
\clearpage
$\bullet$ Region 1  - when the condition

\be
\label{124}
y>1-{1\over\sqrt 2}\simeq 0.293
\ee

is satisfied, it is {\it always true that the
indirect attack is stronger than the direct attack
for multi-photon pulses with any number of photons},

$\bullet$ Region 2 - when the condition

\be
\label{125}
y<1-{1\over\sqrt[3] 2}\simeq 0.206
\ee

is satisfied, it is {\it always true that the
direct attack is stronger than the indirect attack
for multi-photon pulses with any number of photons},

$\bullet$ Region 3 - when the condition

\be
\label{126}
1-{1\over\sqrt[3] 2}~<~y~<~1-{1\over\sqrt 2}
\ee

{\it i.e.},

\be
\label{127}
0.206~\lsim~y~\lsim~0.293
\ee

is satisfied, for multi-photons pulses with an odd number $l=2k+1$
of photons it
is always true that the direct attack is stronger than than the indirect
attack, and for multi-photon pulses with an even number $l=2k$ of photons the
particular
value of $y$ which separates the two cases is determined for a particular
value of $k$ by the expression given on the rhs in eq.(\ref{120}).\footnote{
Note that the special value $\eta\alpha\simeq 0.293$
was also noted in \cite{lutkenhaus1}, but there the full
significance of this number was not inferred (nor was the critical value
$\eta\alpha\simeq 0.206$ discovered at all). There it was concluded that as long
as the condition $\eta\alpha\gsim 0.293$ is satisfied it is not always
possible for the direct attack to succeed. This result is contained in our result.
We have inferred the universal conclusion
that if $\eta\alpha\gsim 0.293$ it is {\it always} the case that some purely
indirect attack is stronger than any purely direct attack
for any given multi-photon pulse where both attacks are possible ({\it i.e.}, for
$l\ge 3$).}

We are able to unambiguously deduce these results since we are directly comparing
the {\it maximal} values of the explicit privacy amplification functions for the
two types of attack. Recall in particular that, from eq.(\ref{66}), the largest possible
value of $\nu_{d,l}$ occurs when, for a given value of $\mu$,
the coefficient in $\nu_{d,l}$ of ${m\over 2}\psi_l\left(\mu\right)
\sum_{\lp=0}^l\left({l\atop\lp}\right)\left(\alpha_{AE} \rho_{AE} \right)^{\lp}\left(
1-\alpha_{AE}\rho_{AE} \right)^{l-\lp}{\hat z}_E\left(\lp\right)$
is as large as it can be, which is unity since that coefficient is itself a probability
function. The critical value and region
$\eta\alpha\gsim 0.293$ determine the condition for
which the {\it strongest possible} purely direct attack is not as strong as {\it some}
purely indirect
attack, and the critical value and region $\eta\alpha\lsim 0.206$ determine
the condition for which the {\it strongest possible} purely
indirect attack is not as strong
as {\it some} purely direct attack.

\begin{figure}[htb]
\vbox{
\hfil
\scalebox{0.66}{\rotatebox{0}{\includegraphics{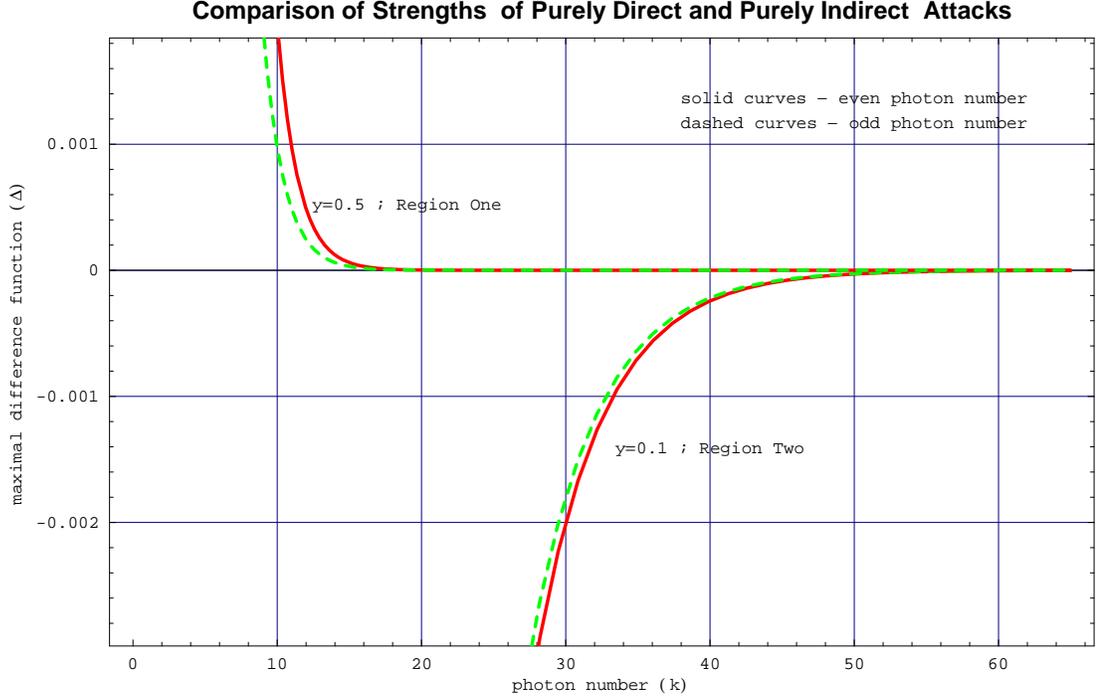}}}
\hfil
\hbox to -1.25in{\ } 
}
\bigskip
\caption{%
Comparison of Strengths of Purely Direct and Purely Indirect Attacks
}
\label{F:strength_pure}
\end{figure}

These analytical results are illustrated numerically
in Figure \ref{F:strength_pure} where we display a graph of curves
of the maximal difference function $\Delta^{max}$, resolved
into the even and odd parts $\Delta_e^{max}$ and $\Delta_o^{max}$. This is done
for Region 1 and Region 2, represented by $y$ values of $y=0.5$ and $y=0.1$,
respectively.

The preceding analysis was carried out in order to establish whether, for a given
value of $l$, the maximal purely direct or maximal purely indirect is stronger. We
have answered this in all generality for all multi-photon pulses in Regions 1 and 2,
as well as for pulses with an odd number of photons in Region 3. To complete the
analysis and
determine this for the case of even values of $l$ in Region 3 where
$1-1/\sqrt[3] 2\le y\le 1-1/\sqrt 2$ we need to introduce the quantity
\be
\label{128}
\sigma^{max}\equiv{\nu_{i,l}^{(1),max}\over\nu_{d,l}^{max}}
\ee
which measures the relative strength of the two types of attacks.
Although we will only need this function for the case of even values of $l$,
for completeness we can as above separately consider both the cases when $l$ is an
even integer and when $l$ is
an odd integer (and as before we do not display the $\theta$-functions due to the
chosen ranges for $k$):
\bea
\label{129}
\sigma_e^{max}&=&\sigma_e\left(k,y\right)
\nonumber\\
&=&{1-\left(1-y\right)^{2k-1}\over 1-2^{1-k}}~,
\eea
and
\bea
\label{130}
\sigma_o^{max}&=&\sigma_o\left(k,y\right)
\nonumber\\
&=&{1-\left(1-y\right)^{2k}\over 1-2^{-k}}~.
\eea
We will use $\sigma_e^{max}$ in the next section where we
assemble the complete expression for multi-photon pulse privacy amplification.

Having performed a direct comparison of the strengths of direct {\it versus} indirect
attacks, which are the only possible attacks for pulses with $l=3$ and $l=4$ photons,
we now consider pulses with five or more photons.
In this case the full set of attacks include the combined as well as purely direct
and purely indirect attacks.
Inspection of the kinematical symbol
entries in Tables 1 and 2 reveals that the full set of
allowed attacks for any given multi-photon pulse fills out a continuum of
attack strengths. To see this
explicitly, we consider as an example
the entry for $l=8$ displayed in Table 2, and restrict our
analysis to the case that $u=1$ for all indirect attacks (both for purely indirect
attacks and the indirect {\it parts} of the allowed combined attacks), and as before
assume that the enemy carries out the {\it maximal} attack always.
To analyze the privacy amplification function for a generic combined attack, we
need to introduce the additional
notation $\nu_{c(d),l,l_d}$ and $\nu_{c(i),l,l_i}^{(u)}$. The
first symbol denotes the privacy amplification function associated to the {\it
direct part} of a combined attack (this is indicated by the ``$c(d)$" in the subscript)
on a multi-photon pulse with a total of $l$
photons out of which $l_d$ photons have been taken by the enemy for the direct attack
part. Similarly, the second symbol denotes the privacy amplification function
associated to the {\it indirect part} of a combined attack on a multi-photon pulse
with a total of $l$ photons out of which $l_i$ photons have been taken by the enemy
for the indirect attack part. 

Note that in the cases that $l_d=l$ and $l_i=l$ the
quantities $\nu_{c(d),l,l_d}$ and $\nu_{c(i),l,l_i}^{(u)}$ should
reduce, respectively, to the expressions for $\nu_{d,l}$ and $\nu_{i,l}^{(u)}$, so
that we have
\be
\label{131}
\nu_{c(d),l,l}=\nu_{d,l}
\ee
and 
\be
\label{132}
\nu_{c(i),l,l}^{(u)}=\nu_{i,l}^{(u)}~.
\ee
The important point to observe is that the constraint $l=l_d+l_i$ must always be
satisfied in the combined attack, and this means in particular that the photon number
arguments of different factors
that appear in the associated privacy amplification functions will always be different
from each other in the case of any combined attack, only becoming equal to each other
in the limit that either $l_d=l$ or $l_i=l$, in which case the combined attack reduces
to a purely direct or purely indirect attack. For example, in the case of
the direct attack part of a generic combined attack on a pulse with $l$ photons, we
have (for this example
we are assuming that the direct attack part of this combined attack is
maximal)
\be
\label{133}
\nu_{c(d),l,l_d}^{max}={m\over 2}\psi_l\left(\mu\right){\hat z}_E\left(l_d\right)~,
\ee
where the subscript for $\psi_l$ is indeed
different than the argument for ${\hat z}_E$,
with an analogous splitting amongst the appropriate arguments
between $l$ and $l_i$ in the case of the
privacy amplification function for the indirect part.

In the above expression the constraint $l_d\le l$ must always be satisfied. With
equality between $l_d$ and $l$ we now see explicitly that
the above expression goes over to the privacy
amplification function for the purely direct attack on {\it all} the photons in a
multi-photon pulse with $l$ photons:

\bea
\label{134}
\nu_{c(d),l,l}^{max}&=&{m\over 2}\psi_l\left(\mu\right){\hat z}_E\left(l\right)
\nonumber\\
&=&\nu_{d,l}^{max}~.
\eea

Thus, to compare the strengths of any of the direct attack parts of
a combined attack on a multi-photon pulse with $l$ photons, it suffices to compare
the magnitudes of ${\hat z}_E\left(l_d\right)$ and ${\hat z}_E\left(l\right)$ for
all $l_d$ satisfying $l_d<l$. Inspection of eq.(\ref{56}) for ${\hat z}_E\left(l\right)$
reveals that one has

\be
\label{135}
{\hat z}_E\left(l_d\right)<{\hat z}_E\left(l\right)~\forall~l_d<l
\ee

and therefore

\be
\label{136}
\nu_{c(d),l,l_d}^{max}<\nu_{d,l}^{max}~\forall~l_d<l~.
\ee

Thus, for a given value of $l$, there is no direct attack part of any combined attack
that is stronger than the purely direct attack carried out on the full set of $l$
photons contained in the pulse. A similar argument can easily be made to show the
analogous
result in the case of the relation between the indirect part of any combined attack
and the associated purely indirect attack. Not surprisingly, as a result
it turns out that for any fixed value of $l$,
the various allowed combined attacks {\it always} are
characterized by maximal privacy amplification function values ({\it i.e.}, worst
case attack
strengths) that are less than those for the maximal purely direct and maximal
purely indirect attacks, the
``endpoint" symbols in lists such as in Tables 1 and 2.
We can also motivate this result numerically as follows. Going down the list
of entries in Table 2 from first to last, let us examine four representative
attacks denoted by the kinematical
symbols $(8,(0))$, $(6,2)^{1+1}$, $(3,5)^{1+4}$ and
$((0),8)^{1+7}$. We see that:

$(8,(0))$ corresponds to the single privacy
amplification function $\nu_{d,8}^{max}$,

$(6,2)^{1+1}$ corresponds to the {\it two} privacy
amplification functions $\nu_{c(d),8,6}^{max}$ and $\nu_{c(i),8,2}^{(1),max}$,

$(3,5)^{1+4}$
corresponds to the two privacy amplification functions $\nu_{c(d),8,3}^{max}$
and $\nu_{c(i),8,5}^{(1),max}$,

$((0),8)^{1+7}$ corresponds to the single privacy
amplification function $\nu_{i,8}^{(1),max}$.

\begin{figure}[htb]
\vbox{
\hfil
\scalebox{0.66}{\rotatebox{0}{\includegraphics{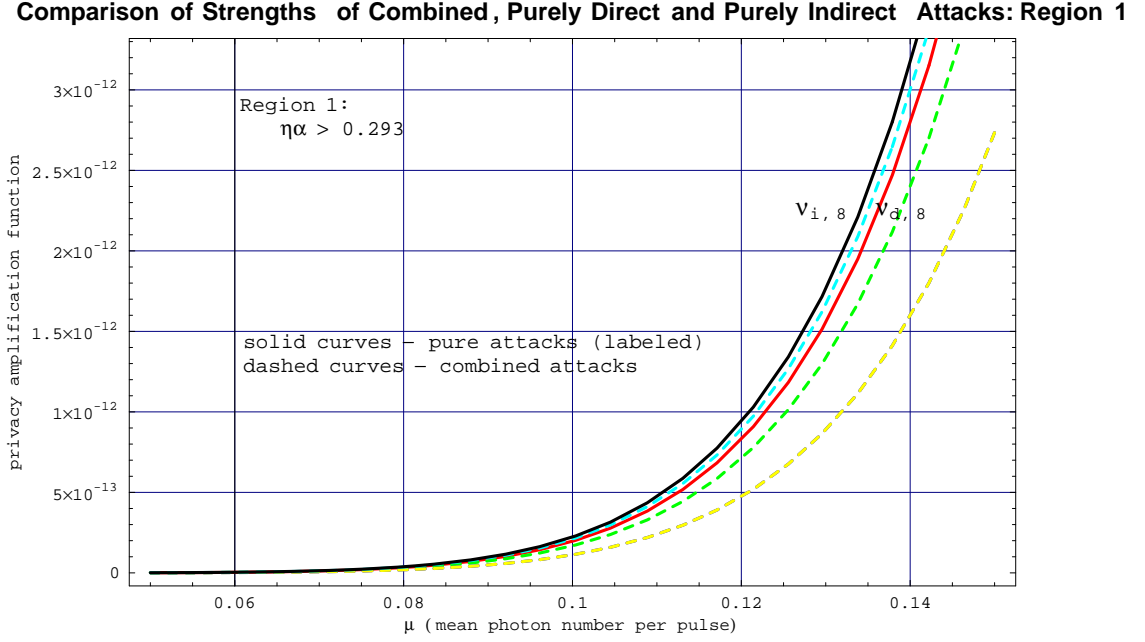}}}
\hfil
\hbox to -1.25in{\ } 
}
\bigskip
\caption{%
Comparison of Strengths of Combined, Purely Direct and Purely Indirect Attacks:\newline
Region One
}
\label{F:strength_combined1}
\end{figure}

\begin{figure}[htb]
\vbox{
\hfil
\scalebox{0.66}{\rotatebox{0}{\includegraphics{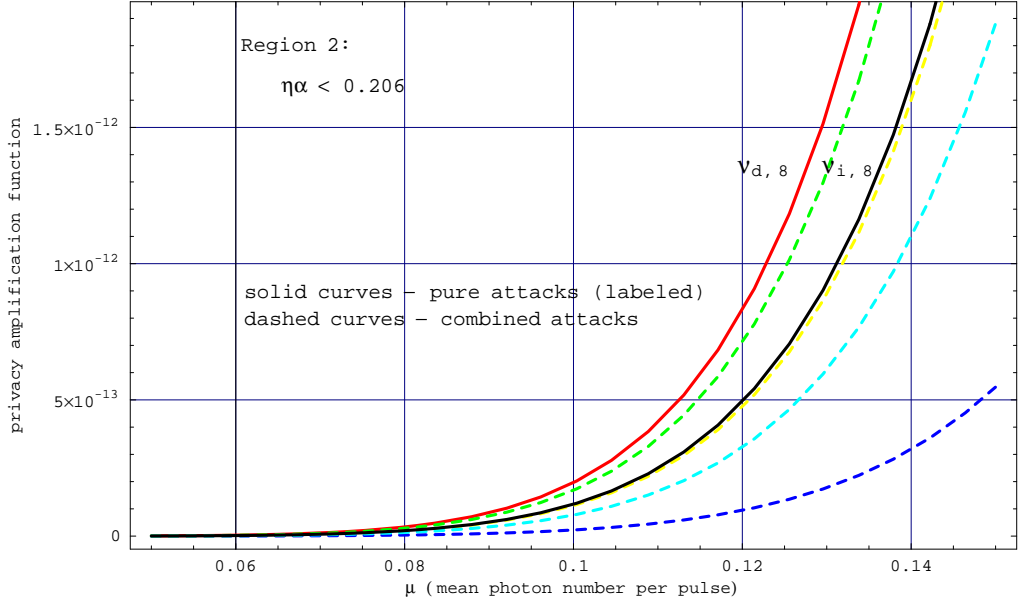}}}
\hfil
\hbox to -1.25in{\ } 
}
\bigskip
\caption{%
Comparison of Strengths of Combined, Purely Direct and Purely Indirect Attacks:\newline
Region Two
}
\label{F:strength_combined2}
\end{figure}

In Figures \ref{F:strength_combined1}
and \ref{F:strength_combined2}
we have plotted for Regions 1 and 2, respectively,
the six privacy amplification functions corresponding to the
different attacks identified by the four kinematical symbols listed above.
The two solid curves in each graph are the privacy amplification functions for
the purely direct and purely indirect attacks,
and the four dashed curves in each graph correspond to the various combined attacks.
Inspection of the curves in the graphs reveals that, depending
on the value of $y=\eta\alpha$ in precisely the way determined by $y_e$ and $y_o$
given in eqs.(\ref{120}) and (\ref{121}), the
purely direct or purely indirect attacks are {\it
always} stronger than any of the combined attacks.
We are led to conclude that we can therefore
bound the worst possible effect of any combined attacks on multi-photon pulses
with $l\ge 5$ photons by carrying out the privacy amplification analysis as if {\it
only} purely direct or purely indirect attacks were available to the enemy.
Putting this together with the previous analysis for the multi-photon pulses with
$l=3$ and $l=4$ photons, we see that for {\it all} multi-photon pulses with $l\ge 3$
photons we can determine the strongest combination of direct and indirect attacks
in a universal manner by making use of the critical values for $\eta\alpha$, for
even or odd photon number, as determined by eqs.(\ref{120}) and (\ref{121}). The
generic expression for privacy amplification, now for all $l\ge 3$, is
therefore given by
\be
\label{137}
j_l\nu_{i,l}^{(u)}+\left(1-j_l\right)\nu_{d,l}
\ee
as in (\ref{112}) above.

\vskip 10pt
\noindent {\it The Complete Expression for Multi-Photon Pulse Privacy Amplification}

The logical structure of the analysis carried out in this section is illustrated with
the flow chart shown in Figure \ref{F:ce} below.
\begin{figure}[htb]
\vbox{
\hfil
\scalebox{0.6}{\rotatebox{270}{\includegraphics{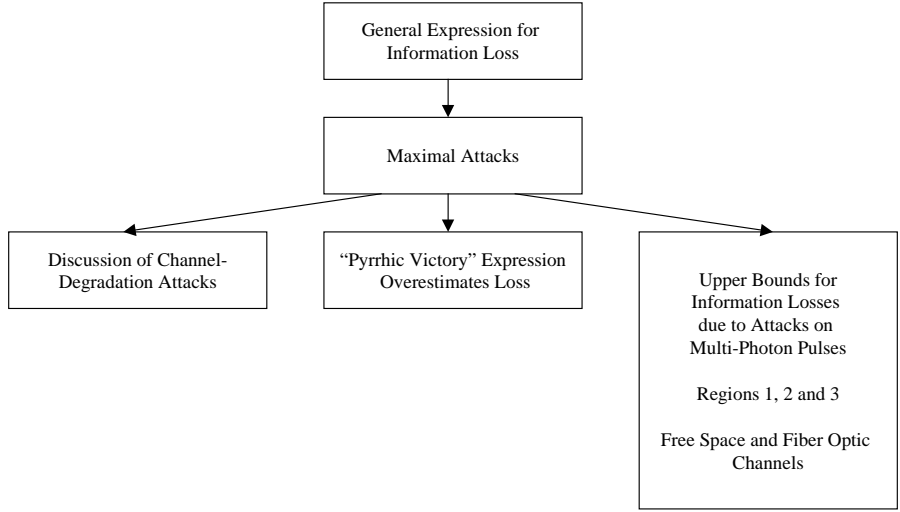}}}
\hfil
\hbox to -1.25in{\ } 
}
\bigskip
\caption{%
Flow Chart for Analysis of Complete Expressions for Multi-Photon Pulse
Privacy Amplification
}
\label{F:ce}
\end{figure}

Here we assemble the complete expression that provides both
necessary and sufficient privacy
amplification to ensure unconditional secrecy (in the sense of the privacy amplification
theorem) against attacks on the multi-photon pulse part of the transmission from Alice
to Bob. We adopt the very conservative assumption that {\it all} multi-photon pulses
are intercepted and subjected to some form of attack. We at first proceed by
considering the possible attacks on a photon number-by-photon number
basis.

$\bullet$ For two-photon pulses, {\it only} indirect attacks are possible.\\
{}$~~~~~~$We assume that all two photon pulses are subjected to the maximal
indirect attack.

$\bullet$ For three- and four-photon pulses each, {\it either} a direct {\it or}
an indirect attack, but not both, are possible.\\
{}$~~~~~~$We assume that all three- and four-photon pulses are subjected to
the maximal version of one or the other of the two types of attack,
whichever is stronger.

$\bullet$ For pulses with five or more photons, direct, indirect or
combined direct and indirect attacks are possible.\\
{}$~~~~~~$In the previous section we showed that, for any fixed number of photons
in a multi-photon pulse, the strength of the generic combined
attack is always less than a particular maximal purely direct or maximal
purely indirect
attack. We therefore proceed by assuming that all pulses with five or more photons
are subjected to one or the other of the two types of pure attack, whichever
is stronger, just as for three- and four-photon pulses.

Putting together our various results, the
overall expression for the {\it necessary and sufficient} amount of privacy
amplification $\nu$, in all generality,
for the entire multi-photon pulse part of the transmission from
Alice to Bob is finally given by

\be
\label{138}
\nu=\nu_{i,2}^{(u)}+\sum_{l=3}^\infty{\Big [}j_l\nu_{i,l}^{(u)}+\left(1-j_l
\right)\nu_{d,l}{\Big ]}~,
\ee

where 

\be
\label{139}
\nu_{i,l}^{(u)}={m\over 2}\psi_l\left(\mu\right)\theta\left(l-2\right)
{\Bigg [}1-\left(1-\eta\alphaEB
\rhoEB\right)^{-u}\left(1-\eta\alphaAE\rhoAE\alphaEB\rhoEB\right)^l{\Bigg ]}
\ee

and

\bea
\label{140}
\nu_{d,l}&=&{m\over 2}\psi_l\left(\mu\right)
\sum_{\lp=0}^l\left({l\atop\lp}\right)\left(\alphaAE \rhoAE \right)^{\lp}\left(
1-\alphaAE 
\rhoAE \right)^{l-\lp}{\hat z}_E\left(\lp\right)
\nonumber\\
&&\qquad\sum_{l_E=0}^\infty\hat\Xi\left(\mu_E,l_E\right)
{\Big [}1-\left(1-\eta\alphaEB \rhoEB \right)^{l_E}{\Big ]}~.
\eea

Although this is indeed the most general, necessary and sufficient expression for the
privacy amplification function $\nu$, its use by Alice and Bob in practical
application to actual quantum cryptography is problematic at best, and typically
this expression cannot in fact be used. It provides the formally necessary and
sufficient amount of privacy amplification, presuming that Alice and Bob have somehow
ascertained various facts that are in principle, entirely under the control of the
enemy, such as the particular distribution $\hat\Xi$ chosen by the enemy for the
preparation of surrogate pulses, the value of $u$ chosen by the enemy, {\it etc.}

Since such information will in general not be available, the only alternative is
to utilize the maximal versions of the privacy amplification amounts that we have
derived above. This has the virtue of ensuring that Alice and Bob will share secret
bits irrespective of the attacks carried out by the enemy, while at the same time
avoiding the situation of the ``Pyrrhic victory" that results when {\it too many}
bits are subtracted, as would occur if eq.(\ref{55}) were used instead. Thus, we have
as the {\it best practical expression} for the required amount of privacy
amplification on multi-photon pulses:

\be
\label{141}
\nu^{max}=\nu_{i,2}^{(1),max}+\sum_{l=3}^\infty{\Big [}j_l\nu_{i,l}^{(1),max}+
\left(1-j_l\right)\nu_{d,l}^{max}{\Big ]}~.
\ee

The natural way to organize this practical expression for
multi-photon pulse privacy amplification is in terms of location in
$\eta\alpha$-space, rather than in terms of photon numbers, and
moreover to do so separately for the cases of free space and optical
fiber cable implementations. An issue of importance, over which Alice and Bob
have no control, is how much modification, if any, the enemy will impose on the
transparency of the quantum channel. For {\it both} the free space and optical fiber
implementations of QC we will ignore the possibility that the enemy might be
able to, and will decide to, {\it degrade} the transmissivity of the channel. Thus
we do not consider the situation in which Eve causes a reduction in the value of
$\alpha$ from the fiducial value initially measured by Alice and Bob to some
lower value. This is because,
as our explicit derivations of the various privacy amplification functions show,
such a decrease in the value of $\alpha$ leads to a {\it decrease} in the number of
bits that can be compromised by the enemy, who will presumably not want this outcome
to occur, and moreover any degradation of channel transparency
constitutes an attempt at denial of service rather than information
compromise in any event.\footnote{
Strictly speaking, even though a reduction in the value
of $\alpha$ reduces the number of bits that may be compromised,
one should allow for the possibility that, by reducing the value
of $\alpha$ and thus reducing the value of the product $\eta\alpha$, the enemy will
drive the attack dynamics from Region 1 to either Region 3 or to Region 2 (or
drive the attack
dynamics from Region 3 to Region 2). Presumably the enemy will attempt to degrade
the quantum channel
surreptitiously, so that if Alice and Bob don't notice this they
will not know that they should adjust the privacy amplification function
accordingly.
This might {\it possibly} be advantageous for
the enemy since, in Region 1, the privacy amplification should be
optimized for the indirect attack, while in Region 2, say, the privacy
amplification should be optimized for the direct attack. The outcome
will depend on whether
the specific number of fewer bits that may be compromised by the enemy as
a result of having
reduced the value of $\alpha$ is greater or smaller than the {\it difference}
between the attack strength of the indirect attack evaluated at the {\it original} value
of $\alpha$ and the attack strength of the direct attack
evaluated at the {\it reduced} value of
$\alpha$. This analysis will be performed elsewhere.}

Thus we are left with two possibilities: (1) In the case of a free space 
implementation we can assume that the value of the product $\eta\alpha$ is
given once and for all
by the fiducial value initially measured by Alice and Bob.\footnote{
As always, we are
presuming that proper, ``technically sound cryptosystem" technique is being executed
by Alice and Bob, and they have thus obtained an accurate initial measurement for 
the {\it a priori} value of $\alpha$, and we also assume that they  know very well
what the value of $\eta$ is.}
They will thus adjust their privacy amplification appropriately according to the
$\eta\alpha$ regions mapped out below. (2) In the case of an optical fiber cable
implementation, Alice and Bob cannot be confident that the fiducial value of $\alpha$
that they believe characterizes the cable will not be increased by the enemy. They
must therefore compare the value of $\eta\alpha$, based on the fiducial value of
$\alpha$, with the value of $\eta$. If both endpoints of this set of values lie within
one and only one of the three regions mapped out below, then they should simply
{\it set} $\alpha=1$ in the appropriate privacy amplification function. Otherwise,
the supremum of the privacy amplification functions between the different regions,
calculated for the two endpoint values, should
be used in a practical system implementation.
In the following list, we will assume for simplicity
in the case of the optical fiber cable implementations that both endpoints
($\eta$ and $\eta\alpha$) lie within one and only one of the designated regions.
We then have the following requirements for appropriate privacy amplification
processing associated to the multi-photon pulses:

Region 1: $\eta\alpha~>~1-{1\over\sqrt 2}~~{\Big (}i.e.,~\eta\alpha~
\gsim~0.293{\Big )}~\Rightarrow~j_l=1$

$\qquad\bullet$~Free space implementation of quantum cryptography
\bea
\label{142}
\nu^{max}&=&\nu_{i,2}^{(1),max}+\sum_{l=3}^\infty{\Big [}j_l\nu_{i,l}^{(1),max}+
\left(1-j_l\right)\nu_{d,l}^{max}{\Big ]}
\nonumber\\
&=&\nu_{i,2}^{(1),max}+\sum_{l=3}^\infty\nu_{i,l}^{(1),max}
\nonumber\\
&=&\sum_{l=2}^\infty\nu_{i,l}^{(1),max}
\nonumber\\
&=&{m\over 2}\llb\psi_{\ge 2}\left(\mu\right)-\left(1-\eta\alpha
\right)^{-1}{\Bigg \{}e^{-\eta\mu\alpha}-e^{-\mu}{\Bigg [}1+\mu\left(
1-\eta\alpha\right){\Bigg ]}{\Bigg \}}\rrb~.
\eea

$\qquad\bullet$~Optical fiber cable implementation of quantum cryptography
\be
\label{143}
\nu^{max}={m\over 2}\llb\psi_{\ge 2}\left(\mu\right)-\left(1-\eta
\right)^{-1}{\Bigg \{}e^{-\eta\mu}-e^{-\mu}{\Bigg [}1+\mu\left(
1-\eta\right){\Bigg ]}{\Bigg \}}\rrb~.
\ee

Region 2: $\eta\alpha~<~1-{1\over\sqrt[3] 2}~~{\Big (}i.e.,~\eta\alpha~
\lsim~0.206{\Big )}~\Rightarrow~j_l=0$

$\qquad\bullet$~Free space implementation of quantum cryptography
\bea
\label{144}
\nu^{max}&=&\nu_{i,2}^{(1),max}+\sum_{l=3}^\infty{\Big [}j_l\nu_{i,l}^{(1),max}+
\left(1-j_l\right)\nu_{d,l}^{max}{\Big ]}
\nonumber\\
&=&\nu_{i,2}^{(1),max}+\sum_{l=3}^\infty\nu_{d,l}^{max}
\nonumber\\
&=&{m\over 2}\psi_2\left(\mu\right)\eta\alpha+{m\over 2}z_E\left(\mu\right)
\nonumber\\
&=&{m\over 2}{\Bigg [}\psi_2\left(\mu\right)\eta\alpha+
1-e^{-\mu}{\Bigg (}\sqrt 2\sinh{\mu\over\sqrt 2}+2\cosh{\mu\over\sqrt 2}-1{\Bigg )}
{\Bigg ]}~.
\eea

$\qquad\bullet$~Optical fiber cable implementation of quantum cryptography
\be
\label{145}
\nu^{max}={m\over 2}{\Bigg [}\psi_2\left(\mu\right)\eta+
1-e^{-\mu}{\Bigg (}\sqrt 2\sinh{\mu\over\sqrt 2}+2\cosh{\mu\over\sqrt 2}-1{\Bigg )}
{\Bigg ]}~.
\ee

Before proceeding to the case of Region 3 we recall that we must now separately
consider the cases of multi-photon pulses with even and odd numbers of photons.
Eq.(\ref{121}) above implies that for {\it all} multi-photon pulses with an
odd number of
photons the strongest attack is a maximal purely direct attack in Region 3, and thus
in this case Alice and Bob should always choose $j_l=0$. However,
in the case of multi-photon pulses with an even number of photons
it is necessary to determine which of the two maximal attacks is strongest on the
basis of the solution to eq.(\ref{120}). For this purpose we now make use of the
strength ratio function $\sigma_e\left(k,y\right)$ introduced
in eq.(\ref{129}) to define the appropriate value of $j_l$, for even $l$ only, as
\be
\label{146}
j_l=\theta{\Big (}\sigma_e\left(k,y\right)-1{\Big )}\qquad l=2k~~,~~k\ge2~.
\ee
We observe
that this form ensures that the value of $j_l$ correctly identifies in Region
3 (actually, it could also be used in the other two regions, but it is
not needed there) the optimal attack for multi-photon pulses with even numbers
of photons, yielding $j_l=1$ when a maximal indirect attack is the
strongest, and yielding $j_l=0$ when
a maximal direct attack is the strongest. This expression can be easily
computed for any
particular value of $k=l/2$, and thus provides a practical method of determining for
the multi-photon pulses with an even number of photons in Region 3 what the
correct amount of privacy amplification is. We then have:

Region 3: $1-{1\over\sqrt[3] 2}~<~\eta\alpha~<~1-{1\over\sqrt 2}~~~~{
\Big (}i.e.,~0.206~\lsim~\eta\alpha~\lsim~0.293{\Big )}$

$\qquad\bullet$~Free space implementation of quantum cryptography
\bea
\label{147}
\nu^{max}&=&\nu_{i,2}^{(1),max}+\sum_{l=3}^\infty{\Big [}j_l\nu_{i,l}^{(1),max}+
\left(1-j_l\right)\nu_{d,l}^{max}{\Big ]}
\nonumber\\
&=&
\nu_{i,2}^{(1),max}+\sum_{{l=3\atop\left(l~{\rm even}\right)}}
^\infty{\Big [}j_l\nu_{i,l}^{(1),max}+
\left(1-j_l\right)\nu_{d,l}^{max}{\Big ]}+
\sum_{{l=3\atop(l~{\rm odd})}}^\infty\nu_{d,l}^{max}
\nonumber\\
&=&
\nu_{i,2}^{(1),max}+{\sum}_e+{\sum}_o~,
\eea

where
\bea
\label{148}
{\sum}_o&\equiv&\sum_{{l=3\atop(l~{\rm odd})}}^\infty\nu_{d,l}^{max}
\nonumber\\
&=&\sum_{{l=3\atop(l~{\rm odd})}}^\infty{m\over 2}\psi_l\left(\mu\right){\hat z}_E\left(
l\right)
\nonumber\\
&=&{m\over 2}e^{-\mu}\sum_{k=1}^\infty{\mu^{2k+1}\over\left(2k+1\right)!}\left(1-
2^{-k}\right)
\nonumber\\
&=&{m\over 2}e^{-\mu}{\Bigg (}\sinh\mu-{\sqrt 2}\sinh{\mu\over\sqrt 2}{\Bigg )}
\eea
and
\bea
\label{149}
{\sum}_e&\equiv&\sum_{{l=3\atop\left(l~{\rm even}\right)}}
^\infty{\Big [}j_l\nu_{i,l}^{(1),max}+
\left(1-j_l\right)\nu_{d,l}^{max}{\Big ]}
\nonumber\\
&=&{m\over 2}\sum_{k=2}^\infty\psi_{2k}\left(\mu\right){\Bigg \{}\theta
{\Big (}\sigma_e\left(k,y\right)-1{\Big )}{\Big [}1-\left(1-y\right)^{2k-1}{\Big ]}
+{\Big [}1-\theta{\Big (}\sigma_e\left(k,y\right)-1{\Big )}{\Big ]}\left(1-
2^{1-k}\right){\Bigg \}}
\nonumber\\
&\equiv&{\sum}_e\left(y,\mu\right)~,
\eea

so that we finally have
\bea
\label{150}
\nu^{max}&=&{m\over 2}\llb\psi_2\left(\mu\right)\eta\alpha+
e^{-\mu}{\Bigg (}\sinh\mu-{\sqrt 2}\sinh{\mu\over\sqrt 2}{\Bigg )}
\nonumber\\
&&+
\sum_{k=2}^\infty\psi_{2k}\left(\mu\right){\Bigg \{}\theta
{\Big (}\sigma_e\left(k,y\right)-1{\Big )}{\Big [}1-\left(1-y\right)^{2k-1}{\Big ]}
+{\Big [}1-\theta{\Big (}\sigma_e\left(k,y\right)-1{\Big )}{\Big ]}\left(1-
2^{1-k}\right){\Bigg \}}\rrb{\Bigg\vert}_{y=\eta\alpha}
\nonumber\\
\eea

$\qquad\bullet$~Optical fiber cable implementation of quantum cryptography
\bea
\label{151}
\nu^{max}&=&{m\over 2}\llb\psi_2\left(\mu\right)\eta+
e^{-\mu}{\Bigg (}\sinh\mu-{\sqrt 2}\sinh{\mu\over\sqrt 2}{\Bigg )}
\nonumber\\
&&+
\sum_{k=2}^\infty\psi_{2k}\left(\mu\right){\Bigg \{}\theta
{\Big (}\sigma_e\left(k,y\right)-1{\Big )}{\Big [}1-\left(1-y\right)^{2k-1}{\Big ]}
+{\Big [}1-\theta{\Big (}\sigma_e\left(k,y\right)-1{\Big )}{\Big ]}\left(1-
2^{1-k}\right){\Bigg \}}\rrb{\Bigg\vert}_{y=\eta}
\nonumber\\
\eea

It should be pointed out that, as will be shown in Section 4, a
practical system implementation will typically be characterized by a large amount
of line attenuation, corresponding to a small numerical value for $\alpha$.
Moreover the
quantum efficiencies of available detectors are usually smaller than one would
desire. Thus realistic system values of $\eta\alpha$ will
typically lie well within Region 2 as defined above, so that
eqs.(\ref{144}) and (\ref{145})
will usually be the appropriate values for multi-photon pulse
privacy amplification.


\vskip 10pt
\noindent {\it Common Sense and the Quantum Cryptographic Conservative Catechism}

As presented in the Introduction, the Quantum Cryptographic Conservative
Catechism (QCCC) provides a ``doctrine of reasonableness" that serves as a guide in
analyzing the various cryptanalytic attacks that the enemy may perform. Consistent
with this, we have determined in this section
the proper amount of privacy amplification subtraction required to
ensure that Alice and Bob share bits that are secret, based on the assumption that the
enemy is essentially only constrained by the laws of physics (modified, though,
by point (1)
in the definition of QCCC in Section 2.5.2). Thus,
we have {\it not} presumed that the enemy is
limited by currently perceived
difficulties of practical engineering that may indeed constrain the possibility of
actually {\it carrying out} the attacks that we study in this paper.

Having said that, however, it ought to at least be mentioned in passing that practical
engineering issues are in fact {\it highly} constraining today.
The attacks that can realistically be carried out by the enemy are greatly limited as
a result. As one example, in the case of the direct
attack, it is necessary that the {\it time} required for the
enemy to perform all physical manipulations to
intercept the pulse, measure the state of the pulse, prepare a chosen surrogate
pulse {\it and} have the surrogate pulse propagate through whatever distance is
required in order to reach Bob, be {\it no greater than the bit cell period}. If this
time constraint is violated then Alice and Bob will be able to detect a corresponding
error. Related
constraints also apply in the case of the indirect attack (such as a constraint
on the time
required to place the retained part of the pulse in an appropriate quantum memory). For
a high speed quantum cryptosystem characterized by a small bit cell period
this basic constraint may be so difficult
to satisfy that the associated attack might as
well be forbidden by the laws of physics.

\subsubsection{Continuous Authentication}

As has been stressed several times above, it is important to ensure that the
quantum cryptography system remains protected against possible spoofing for the
{\it entire duration} of the
transmission. ``Spoofing" occurs when the enemy gains access to the public channel,
interposes herself between the legitimate transmitter and receiver and attempts to
misrepresent her identity in order to gain information, interfere with the system or
both. It must be assumed {\it on each and every use} of the public channel
that the enemy will attempt to carry out a spoofing attack, which explains the
need for continuous authentication. The detailed
derivations of the explicit functional forms of the complete
continuous authentication cost functions are provided in
Section 4.4.1 below. Here we anticipate those results and
list what the cost functions are so that we may
incorporate
them into the complete expressions for the effective secrecy capacity and
effective secrecy rate of QKD
systems. Although the need for {\it initial} authentication of the public channel in
quantum key
distribution has been mentioned by many authors, previous analyses have not
included explicit derivations of the exact functions that describe the full and
precise cost in bits of {\it continuous} authentication.

The complete analytical expression for the cost function for continuous authentication
obtained in Section 4.4.1 is found to be given by
\begin{equation}
\label{152}
a=\tilde g_{EC}+\sum_{j=1}^5w_j\left(g_j,c_j\left(\mu\right)\right)~,
\end{equation}
where we define the important {\it Wegman-Carter function},
$w\left(g,c_i\right)$, as\footnote{
The full and complete expression for the quantity that we denote by $w$ and refer to as
the Wegman-Carter function, which is of crucial importance
in practical quantum cryptography, does not appear to have been properly analyzed
previously in the context of QC (nor apparently even {\it named}
by any authors). Surprisingly,
the closed-form function, as
such, doesn't appear as a numbered equation in \cite{wc}. In fact, it must be obtained
instead by combining quantities that appear in lines 3 and 17 in the first paragraph of
section 3 in \cite{wc}.}
%
\be
\label{153}
w\left(g_i,c_i\right)=4{\Big [}g_i+\log_2\left(\log_2c_i\right){\Big ]}\log_2c_i~.
\ee
The authentication cost function $a$ is the sum of six terms: two of the terms in the
sum represent the authentication cost associated with the sifting process, and the
remaining four terms are the cost associated with the error correction process.

In Sections 4.4.1 and 4.4.2
we explicitly derive the complete distinct costs, in bits, of the
various
communications exchanges required to support the continuous authentication of
the public channel. We list there that
\begin{equation}
\label{154}
c_1=2n\left(1+\log_2m\right)~,
\end{equation}
\begin{equation}
\label{155}
c_2=2n~,
\end{equation}
\begin{equation}
\label{156}
c_3=n~,
\end{equation}
\begin{equation}
\label{157}
c_4=g_{EC}
\end{equation}
and
\be
\label{158}
c_5=\tilde g_{EC}~,
\ee
and all of the security parameter constants $g_i$ that appear in the summand
in eq.(\ref{152}), which includes the quantities
$g_{EC}$ and $\tilde g_{EC}$, are (as explained in Section 4.4.1)
taken to be equal to 30.

These quantities characterize the amount of communications required to effect continuous
authentication for the sifting and error correction phases of the QC protocol. We note
that no communication, and hence no authentication at all,
is required to execute the privacy amplification phase of
the protocol. This may at first appear surprising and appears not to have been discussed
in detail before in the literature.\footnote{
We thank J. Guttman for emphasizing the fact
that there need be no communication between Alice and
Bob to carry out privacy amplification.}
The bit values that must be identically shared between Alice and Bob in explicitly carrying
out privacy amplification consist of a random set to be used to compute the privacy
amplification hash function. This set can be obtained without any communication at all
between Alice and Bob, and as indicated above,
since there is nothing to be communicated via the public channel,
there is obviously no need to authenticate that channel for this purpose. The trick is
for
Alice and Bob to exploit the {\it untapped randomness} resident in the processes used to
execute the protocol. During the public sifting discussion Alice and Bob keep a record
not
only of the identities of the compatible bases, but should as well keep a record of the
index position within the overall bit cell stream of those compatible basis events. They
will be able to generate, in real time, two random strings of bits, each of
length $m/2$ (modulated by the system losses),
by first simply recording the index positions, respectively, of the compatible and
incompatible basis events. Then Alice and Bob may compute the {\it parities} of these two
strings to obtain two completely random bit strings of length $m/2$ each. Either of these
two strings (this choice can be made by public agreement between Alice and Bob) can be
used to compute the privacy amplification hash function, and {\it no} information will
have been communicated between Alice and Bob for this purpose. Although Eve can {\it also}
perform this exercise, since the particlar random sequence generated in this way for use in
the privacy amplification hash function doesn't {\it exist} prior to the sifting
discussion, it is of no use at all to Eve in deducing any information about the shared
key whatsoever.

As discussed in the next section, we will
need to solve an extremization equation in order to deduce the optimal values for both
the effective secercy capacity $\cal S$ and the effective secrecy rate $\cal R$.
To solve the optimization equation we need an explicit expression for $2a_{,\mu}/m$,
which we find is given by

\begin{equation}
\label{159}
{-2a_{,\mu}\over m}={-8\psi_{\ge 1,\mu}\over\psi_{\ge 1}+r_d}\cdot{1
\over m}\cdot{\Big \{}3\left(g+1\right)+\log_2\Big[\left(\log_2c_1\right)\left(\log_2c_2
\right)\left(\log_2c_3\right)\Big]{\Big \}},
\end{equation}

where the $c_i$ are the costs listed above, in classical bits, of
the various communication links required for continuous authentication.
In the next section we will also consider the effective secrecy capacity in the
limit of an infinitely long cipher, for which purpose we will need to use the fact
that
\be
\label{160}
\lim_{m\rightarrow\infty}{a\over m}=0~,
\ee
which is straightforward to verify using the expression for $a$ given in
eq.(\ref{152}) above.

\subsubsection{The Complete Expressions for the Effective Secerecy Capacity and Rate}

We are now in a position to put together the results we have obtained on
the numbers of sifted bits, error bits, privacy amplification subtraction bits
and continuous authentication bits to obtain the complete expressions for the
effective secrecy capacity, $\cal S$, and the effective secrecy rate, $\cal R$,
of general, practical QKD systems implementations.
For this purpose we first introduce the useful function, $f$, which we define as
\be
\label{161}
f\equiv 1+Q+T~
\ee
in the case of a QC system implementation in which Alice and Bob
identify and {\it discard} error bits in the sifted string.\footnote{
In the case of a QC system implementation in which Alice and Bob identify, correct and
{\it retain} error bits in the sifted string, we would instead have $f\equiv Q+T$.
As stated in Section 3.1.1 above, unless explicitly
otherwise mentioned, in this paper we will adopt the
``error discard" approach, with the consequence that the various rate predictions
based on it will furnish universal {\it lower bounds} on achievable throughput rates.}
The purpose of the function $f$ is as follows.
Of the total amount of information that must be removed from the sifted string in order
to achieve a secret shared key, the function $f$ groups together and
measures just that portion that is
natural to measure directly in units of error bits. (This leaves in distinct terms
those portions of the information that is to be removed that are due to multi-photon
pulses, continuous authentication and the privacy amplification security parameter.)
For instance, in the case of $f$ as defined in eq.(\ref{161}), appropriate to the error
correction procedure in which error bits are discarded, the first term of
unity indicates that
the entire fiducial set of error bits are indeed subtracted, and the second and third
terms of $Q$ and $T$ represent the subtractions for error correction information leakage
and single-photon pulse measurements, respectively.

We now collect the results of the above sections for the number of sifted bits, the
number of error bits, the total amount of privacy amplification and the cost of
continuous authentication and substitute them into eq.(\ref{1}) to deduce the
form of the complete, effective secrecy capacity as
\bea
\label{162}
{\cal S}&\equiv&{n-e_T-s-g_{pa}-a\over m}
\nonumber\\
&=&
{n-fe_T-\nu-g_{pa}-a\over m}
\nonumber\\
&=&{1\over 2}{\Bigg [}\psi_{\ge 1}+r_d-f\left(r_c\psi_{\ge 1}+{r_d\over 2}\right)
-\tilde\nu{\Bigg ]}-{g_{pa}+a\over m}
\nonumber\\
&=&{1\over 2}{\Bigg [}\psi_{\ge 1}\left(1-fr_c\right)+\left(1-{f\over 2}\right)r_d
-\tilde\nu{\Bigg ]}-{g_{pa}+a\over m}~,
\eea
where we have defined 
\be
\label{163}
\tilde\nu\equiv 2\nu /m
\ee
so that the rescaled quantity $\tilde\nu$ is
independent of the number of raw bits, $m$. In the above expression for $\cal S$ the
argument of the first term in the square brackets, {\it i.e.} the argument of
the function
$\psi_{\ge 1}$, is equal to $\eta\mu\alpha$ as derived and
discussed in Section 3.1.1 above.

The $m$-dependence in $\cal S$ is important, as it allows us to study the dynamics
of actual ciphers of finite length in addition to studying properties of abstract
ciphers of infinite length. Note that in addition to the manifest $m$-dependence
that appears in the term
${g_{pa}+a\over m}$, there is also $m$-dependence contained within the function
$f$ through {\it its} dependence on $T$ ({\it cf} eqs.(\ref{49}) and (\ref{50})),
but not on $Q$ ({\it cf} eq.(\ref{39})).

Making use of eq.(\ref{160})
from Section 3.1.6, we see that the expression for the effective secrecy
capacity in the limit of a cipher of infinite length becomes
\bea
\label{164}
\lim_{m\rightarrow\infty}{\cal S}
&=&
{1\over 2}\lim_{m\rightarrow\infty}
{\Bigg [}\psi_{\ge 1}\left(1-fr_c\right)+\left(1-{f\over 2}\right)r_d
-\tilde\nu{\Bigg ]}-\lim_{m\rightarrow\infty}{g_{pa}\over m}-
\lim_{m\rightarrow\infty}{a\over m}
\nonumber\\
&=&
{1\over 2}{\Bigg [}\psi_{\ge 1}\left(1-f_\infty r_c\right)+\left(1-{f_\infty
\over 2}\right)r_d
-\tilde\nu{\Bigg ]}
\nonumber\\
&\equiv&{\cal S}_\infty~,
\eea
where
\be
\label{165}
f_\infty\equiv 1+Q+T_\infty
\ee
and $T_\infty$ is given in eq.(\ref{54}).

The effective secrecy rate is given by
\be
\label{166}
{\cal R}={\cal S}/\tau~,
\ee
where $\tau$ is the bit cell period of the QKD system implementation.

We stress that the various quantities $\psi_{\ge 1}$, $f$, $\tilde\nu$ and $a$ appearing
in $\cal S$ depend in a complicated way on a large number of parameters. Rather than
writing this out in full, we display the complete parametric dependence of the effective
secrecy capacity and rate with the following equations of state:
\be
\label{167}
{\cal S}={\cal S}\left(\eta,\mu,\alpha,r_c,r_d,m,{\vec g},\epsilon,{\vec\rho},{\vec j},x
\right)~,
\ee
and
\be
\label{168}
{\cal R}={\cal R}\left(\eta,\mu,\alpha,r_c,r_d,m,{\vec g},\epsilon,{\vec\rho},{
\vec j},x,\tau\right)~.
\ee

\vskip 10pt
\noindent {\it Optimization of the Effective Secrecy Capacity and Rate}

\noindent{The mean photon number per pulse, $\mu$, is the one system parameter that can
by assumption always be directly controlled and adjusted by Alice. This is
accomplished by adding or removing
neutral density filters, as appropriate, to achieve the desired value of emitted
intensity. This desired value should optimize the effective secrecy capacity
and rate of
the system. The optimization equation for determining the optimal value,
$\mu_{{\rm opt}}$,
of the mean number of photons per pulse is given by}
\begin{equation}
\label{169}
0={\cal S}_{,\mu}{\Big\vert}_{\mu=\mu_{{\rm opt}}}~,
\end{equation}
\noindent{which is explicitly written as}
\begin{equation}
\label{170}
0=\psi_{\ge 1,\mu}\left(1-fr_c\right)-\psi_{\ge 1}f_{,\mu}r_c-{f_{,\mu}\over 2}r_d-
\tilde\nu_{,\mu}-{2a_{,\mu}\over m}{\Bigg\vert}_{\mu=\mu_{{\rm opt}}}.
\end{equation}
The resulting optimal value $\mu_{{\rm opt}}$ satisfies the equation of state
\be
\label{171}
\mu_{{\rm opt}}=\mu_{{\rm opt}}\left(\eta,\alpha,r_c,r_d,m,{\vec g},\epsilon,{\vec\rho},
{\vec j},x\right)~.
\ee
The optimal value of the effective secrecy capacity, ${\cal S}_{{\rm opt}}$, is
obtained by evaluating $\cal S$ at $\mu_{{\rm opt}}$, so that we have
\be
\label{172}
{\cal S}_{{\rm opt}}={\cal S}\left(\mu_{{\rm opt}}\right)~,
\ee
and the corresponding expression for the optimal effective secrecy rate is given by
\be
\label{173}
{\cal R}_{{\rm opt}}={\cal S}_{{\rm opt}}/\tau~.
\ee
The optimal effective secrecy capacity and rate, ${\cal S}_{{\rm opt}}$
and ${\cal R}_{{\rm opt}}$, are the quantities that
should be used in practice
to make predictions about and study the performance characteristics of
any particular quantum cryptograhy system.

When the {\it complete} explicit expressions for the
functions $\psi_{\ge 1}$, $\psi_{\ge 1,\mu}$, $f$, $f_{,\mu}$,
$\tilde\nu_{,\mu}$ and $a_{,\mu}$ are written out in full
and substituted into the optimization
equation (eq.(\ref{170})), it
becomes apparent that numerical methods must be used to obtain an
answer for the optimal value of $\mu$.
The {\it general} problem of practical quantum cryptography exhibits
such a complicated dependence on the many parameters that are required
to provide a complete system characterization that a full
mathematical description evidently does not admit closed form analytical
solutions for $\mu_{{\rm opt}}$, except in
special limiting cases.

\vskip 10pt
\noindent {\it Effective Secrecy Capacity and Rate with Click Statistics Monitoring}

In the general case for which Bob monitors the click statistics
and discards those bit cells that manifestly contain multiple photon pulses, the
expression for the effective secrecy capacity becomes
\bea
\label{174}
{\cal S}_{mcs}&\equiv&{n-e_T-s-g_{pa}-a\over m}\Bigg\vert_{mcs}
\nonumber\\
&=&
{n-fe_T-\nu-g_{pa}-a\over m}\Bigg\vert_{mcs}
\nonumber\\
&=&
{1\over 2}{\Bigg [}{\Bigg (}\eta\psi_1
+
{\Big\langle}\hat\chi{\hat {\cal Z}}_{\ge 2}
{\Big\rangle}
{\Bigg )}
\left(1-f_{mcs}r_c\right)+\left(1-{f_{mcs}\over 2}\right)r_d-\tilde\nu_{mcs}{\Bigg ]}
-{g_{pa}+a_{mcs}\over m}~.
\nonumber\\
\eea
In the above expression for ${\cal S}_{mcs}$, the argument of $\psi_1$ is equal to
$\mu\alpha$. The functions $f_{mcs}$ and $a_{mcs}$ are defined in terms of
the quantities $n_{mcs}$ and $e_{T,mcs}$
in place of the corresponding quantities $n$ and $e_T$.
For example, $f_{mcs}$ is given explicitly by
\bea
\label{175}
f_{mcs}&\equiv& \left(1+Q+T\right)\Big\vert_{mcs}
\nonumber\\
&=&1+Q\left(x,{e_{T,mcs}\over n_{mcs}}\right)+T\left(n_{mcs},e_{T,mcs},\epsilon\right)~,
\eea
and $a_{mcs}$ is given by
\be
\label{176}
a_{mcs}=\tilde g_{EC}+\sum_{j=1}^5w_j
\left(g_j,c_j\left(\mu\right)\right)\Bigg\vert_{n=n_{mcs}}~.
\ee
The quantity $\tilde\nu_{mcs}\equiv 2\nu_{mcs}/m$ is formed from the appropriate
expressions for $\nu_{d,l,mcs}$ 
and $\nu_{i,l,mcs}^{(u)}$, respectively, given in eqs.(\ref{77}) and (\ref{105}).

\vskip 10pt
\noindent {\it Effective Secrecy and Capacity and Rate in Special Limits}

\noindent{It is worthwhile to examine this result in the special case that there is
no eavesdropping activity but for which there {\it is} attenuation in the quantum
channel and
loss at Bob's detector, in which circumstance we may define the associated secrecy
capacity, ${\cal S}_{\rm no~enemy}$. If there is no eavesdropping activity
we have $Q=T=0$ (since no information is lost in particular due to eavesdropping
on either error correction or single-photon pulses), so that
we also have $f=1$. We may set $\tilde\nu=0$, since none of the multi-photon pulses
are at risk in this scenario. In the absence of an eavesdropper it should also not
be necessary to undertake any
authentication, so that we can impose the condition $a=0$. Moreover, in this case
we can
safely set the privacy amplification security parameter, $g_{pa}$, equal
to zero, so that we finally have}
\begin{equation}
\label{177}
{\cal S}_{\rm no~enemy}={1\over 2}\Big[\psi_{\ge 1}\left(1-r_c\right)+{1
\over 2}r_d\Big]
\end{equation}
\noindent{Inspection of the above expression reveals the basic physical effects that
are responsible for the effective bit rate: the factor $\left(1-r_c\right)$ gives the
proportion of the bit cells that will reach Bob without being subjected to
depolarization or other intrinsic channel errors,
the factor $\psi_{\ge 1}$ gives the probability that a particular
laser pulse will contain at
least one photon, and in the argument of $\psi_{\ge 1}$ the
factor $\alpha$ gives the proportion of qubit photons that will
reach
Bob in spite of attenuation due to atmospheric losses, the factor $\eta$ gives the
fraction of photons that will actually be detected at Bob's receiving instrument in spite
of the intrinsic inefficiency of his apparatus, the overall factor of $1/2$ gives
the fraction
of photons lost as a result of the statistically independent, random choices of
polarization basis made between Alice and Bob, and finally the term ${1\over 2}r_d$
gives the contribution to the effective secrecy capacity due to the presence of dark
count activity.}

\vskip 10pt
\noindent {\it Finite Length {\it versus} Infinite Length Ciphers}

It is crucially important to obtain, as we have done,
closed form, analytical expressions for the effective
secrecy capacity and rate that are valid for actual ciphers of finite length, as
opposed to expressions that are only valid in the abstract limit of infinitely long
ciphers. Why is this?
One of our principal objectives in this study is to determine how and under what
conditions we may achieve high data throughput rates for practical quantum
cryptograhy systems. As we shall discuss in detail in Section 5.2.6, one of the
techniques that may be used to achieve this objective is to assemble a collection of
transmitters and multiplex their outputs together in a common transmission stream.
However, the totality of computing resources (measured by
the number of basic computer machine
instructions) required to {\it actually carry out} the QKD protocol furnishes an
important practical
constraint on any quantum cryptography system. As this has never been analyzed
before, we work out the details in full in this paper in Section 4.4.3 below.
We find for the first time
a closed form expression (eq.(\ref{E:compload}) below)
relating the computing resources required for
carrying
out the protocol to the processing block size taken from the transmitted bit stream.
With the help of this functional relationship, and also making use of the optimal effective
secrecy capacity to numerically determine the dependence of the throughput rate on any
reduction or increase in the number of transmitted raw bits assigned to each processing
block, it is possible to deduce
the rate of any proposed multiplexing scheme while satifying the important constraint that
there are sufficicent computing resources to achieve it. {\it This is not possible to
do without closed form expressions for the secrecy capacity and rate that are valid for
ciphers of finite length.} With such functions at our disposal,
however, we will determine in Section 5.3 below what are the highest possible
rates than can actually be achieved for practical quantum cryptography.

\subsection{An Extended Family of Four-State Quantum Key Distribution Protocols}

We saw in the discussion in Section 3.1.1 above that there are a number of
choices of schemes that may be adopted by Bob and Alice (and fully disclosed to Eve)
to effect the monitoring of click statistics.
As described there, to fully define any such scheme it is necessary to both select
a particular model that specifies
the details of the detector apparatus {\it and} to select
a particular click-monitoring scheme to be used in that model. For each
such choice one obtains different numerical values for the higher-order terms\footnote{
These are terms ({\it cf} eq.(\ref{26})) such as ${\Big\langle}\hat\chi\left(
\mu,l\right){\hat {\cal Z}}_{\ge 2}\left(\eta,\alpha,l\right){\Big\rangle}
\equiv \sum_{l=0}^\infty\hat
\chi\left(
\mu,l\right){\hat {\cal Z}}_{\ge 2}\left(\eta,\alpha,l\right)$.}
in the
explicit expressions for the number of sifted bits, the number of transmitted error
bits, the cost of continuous authentication
and the various privacy amplification subtraction functions. Each of these choices will
generate different specific numerical results for overall system performance, in
particular affecting the total integrated cipher throughput rates that can be
achieved. Each of these may be thought of as an element in an extended family of
BB84-like protocols. In fact, there are a denumerable infinity of different versions
of these click monitoring-based schemes, distinguished from each
other according to how Bob chooses to distribute any click monitoring he carries out
amongst the bit cells. He can choose to monitor click statistics for the entire
transmission, for certain fractions of the transmission, for certain fractions of
the transmission for different amounts of time {\it etc.} Although these different
variations will in the general case of the strongest possible attack by Eve
be suboptimal from the perspective of Alice and Bob
compared to simply carrying out the maximum amount of click monitoring, {\it i.e.},
executing the monitoring all of the time, for the entire transmission, and discarding
{\it all} mulliple-click event bit cells, there are situations for which it is preferable
for Alice and Bob to choose one of the other options. If Alice and Bob happen to have,
through whatever means, access to privileged information regarding the set of
attacks that
the enemy can or will carry out, it is possible to tailor an appropriate click
monitoring scheme specifically against that set of attacks - such a
specially ``tailored" click statistics monitoring scheme may result in a greater
overall throughput of secret bits. A full anaysis and discussion of this sensitive
topic is beyond the scope of the present paper, and will be treated elsewhere.

\subsection{Secrecy in the Presence of Weak Coherent Pulses}

\noindent{The four--state quantum cryptography protocol (the BB84 protocol) in the ideal
situation -- {\it i.e.,} in the absence of any system noise and with a source
of perfect quantum bits -- is unconditionally
secret in the presence of any cryptanalytic attacks by Eve. Our purpose
here, though, is in considering the
case of a {\it realistic} system for quantum cryptography. Any actual implementation of
the
BB84 protocol obviously requires the use of actual physical hardware: we know that in
{\it
any} actual implementation the real system will be such that the the intended
communication
will be characterized by both transmission losses and errors. It is the inevitable
presence
of the errors, in particular, that absolutely forces us to effectively re-define the
``pure" BB84 protocol ({\it i.e.}, generation of the sifted key only)
to include sufficient error correction and privacy amplification in order to
assure that Alice and Bob share a secret Vernam cipher at the end of the communication.
The privacy amplification functions calculated in great detail in the previous section
serve the purpose of ensuring that any information possibly
obtained by Eve through whatever means is removed from the final string shared between
Alice and Bob. Of course, if Alice and Bob somehow knew with certainty that Eve did not
exist, or that if she did exist was not present, or if present would not eavesdrop now
or in the future, it
would not be necessary to implement privacy amplification, although it would still be
necessary to carry out error correction, due to the presence of system errors. However,
in
this unrealistic case -- the absence of present or future adversaries -- it would not be
necessary to make use of {\it any} cryptography, quantum or classical, in
the first place. In {\it real} circumstances we must assume that Eve might be present,
either actively attempting to decode our communications, or attempting to
intecept,
record and store them for possible future cryptanalysis. In this case, which describes
the
real world of communications in the presence of adversaries, we {\it must} implement
privacy amplification. In other words, as a {\it practical} protocol for secret key
distribution the ``pure" BB84 protocol is not complete: in any actual application
the noise
inherent to the system hardware dictates that the protocol actually used must be BB84
supplemented by sufficient error correction and privacy amplification. In several extant
implementations quantum key distribution has been implemented by generating the signal
states with an intensity--filtered pulsed laser. As discussed in great detail in the
previous section, the output of the laser in this case is
in the form of weak coherent pulses of laser light: some of the pulses contain no more
than
a single photon, and some of the pulses contain two or more than two photons. The need
to
carry out privacy amplification, however, applies equally for the formal BB84 protocol
implemented solely with proper quantum bits, which would consist solely of single photon
states, {\it or} the weak coherent pulse implementation, which would include both single
and multiple photon states. Even if there are no multi-photon pulses {\it at all}
amongst
the signals sent from Alice to Bob, the fact that the physical hardware generates
errors,
combined with the fact that Eve may be present, requires privacy amplification. This
means
that, even if the pure BB84 protocol involving solely proper single-particle qubits is
implemented on a real system, the effect of the required execution of privacy
amplification
dictates that the probability $P$ that Eve will be able to know more than one bit of the
final shared key sequence is given by $P\le{2^{-g_{pa}}\over\ln 2}$, where $g_{pa}$ is
the
privacy amplification security parameter (the length in bits of the tag for the hash
function utilized in effecting the privacy amplification). {\it Precisely the same
degree of secrecy} is realized if the BB84 protocol is implemented with pulsed lasers
generating weak coherent pulses that include amongst them multi--photon states, as long
as sufficient additional privacy amplification is performed to account for the maximum
amount of Renyi information that may have possibly been obtained by Eve as a result of
any physically allowed attacks on such states.}

\clearpage

\section{Comprehensive Analysis of System Losses and Loads}

In order to apply the expressions for the effective secrecy capacity and rate
that we have derived
to a realistic system it is essential to supply accurate values for the various parameters
that characterize the losses and loads which result in a reduction of the throughput of
secret key material. This must of course include the various quantities that specify the
actual losses suffered by the signal as it propagates from Alice to Bob, but it is also
important to include in the analysis those {\it ancillary} costs associated to the
supporting classical communication required to actually carry out the QKD protocol. In
addition, it is crucial to estimate and include in the analysis the amount of computing
power that is required to carry out the various operations including error correction,
computation of authentication and privacy amplification hash functions, real-time data
record keeping, {\it etc.}, that must go on ``behind the scenes" in order for a practical
system to actually work. Such costs can only be determined for actual ciphers of finite
length, as abstract limits for infinitely long ciphers are inapplicable to the
practical situation. It is only after complete account is taken of {\it all} these
effects that one can accurately estimate the actual throughput rates and other operating
characteristics that describe a real QKD system.

\vskip 10pt
\noindent {\it Some Practical System Considerations}

In Section 5 of this paper we consider in detail
the practical requirements for achieving QKD at a high throughput rate. Looking
ahead to that development, in the various sections below of this chapter we will
illustrate the general analytical results that we obtain for system losses by making use
of certain system parameter values to calculate sample numerical results. In these
illustrations we will typically consider a system in which Alice uses a laser that
produces light at a wavelength of 1550 nanometers (consistent with many modern
telecommunications and laser communications systems). Moreover for the case in which
a free space quantum channel is utilized, such as for QKD between a satellite
and a ground station, we in general assume a QKD configuration in which {\it Alice is
elevated} and {\it Bob is on the ground} (or possibly in an aircraft). As we will see,
a consequence of the various system losses for a QKD system operating through
the atmosphere is that it is advantageous in optimizing the throughput rate to {\it
increase} the size of Bob's receiving instrument aperture compared to Alice's, and it
is easier and less expensive to do that by placing the Bob system on the ground (or in
an airplane) and the Alice system on the satellite. In these cases we will principally
consider numerical examples for two different satellite altitudes: a low earth orbit
(LEO) satellite located at 300 kilometers altitude, and a geosynchronous (GEO) satellite
located at 35783 kilometers (22236 miles) altitude.

\vskip 10pt
We now turn to an analysis of all of the above issues.

\subsection{System Losses: The Line Attenuation - Free Space}
\label{SS:LA}

The line attenuation, $\alpha$, is {\it defined} to include the loss suffered by the
signal
due to four distinct causes: (1) the diffraction loss, {\it i.e.,} the geometrical
vacuum
beam spreading loss, (2) the static atmospheric losses, due to atmospheric scattering
and
absorption, (3) the turbulent atmospheric losses, due to several causes as enumerated in
Section ~\ref{SS:turbloss} below, and (4) the ``optics package" losses due to
the imperfect nature of the various components present in the system. Note that the line
attenuation $\alpha$ is {\it not} the ``total" attentuation suffered by the qubit signal.
Specifically not included in the definition of the line attenuation are: the loss,
$\eta$, associated with photon detector efficiency, the intrinsic quantum channel loss,
$r_c$, the basic, $50\%$ sifting loss due to the definition of the BB84 protocol, the dark
count loss, $r_d$, associated to the photon detector, and the loss associated with the use
of weak laser pulses, described by the probabilistic distributions
$\psi_l\left(\mu\right)$, $\psi_{\ge 1}$, {\it etc}.,
in the effective secrecy capacity ${\cal S}$.\footnote{
Our line attenuation, $\alpha$, is defined to be equal (in linear units)
to unity when there is no signal
loss, and equal to zero when there is complete loss of signal. This is often referred to
as the transmittance.}

\subsubsection{Diffraction Vacuum Beam Spreading Losses}
\label{SS:VBSL}

The use of finite optics dictates that the beam generated at Alice and transmitted to Bob
will become a spread beam due to diffraction. The radius, $\rho_d$, of the purely
diffraction--limited
spot size of the beam incident upon a flat receiving plane at the location of Bob's
apparatus is found from a straightforward calculation to be given by
\begin{equation}
\label{178}
\rho_d=\Bigg[{4L^2\over\left(kD_A\right)^2}+\left({D_A\over 2}\right)^2\Bigg]^{1/2},
\end{equation}
\noindent{where $L$ is the path length over which the signal propagates, $D_A$ is the
diameter of the aperture of Alice's transmitting instrument and $k$ is the wavenumber of
the photons in the beam. The calculation of loss associated with this spreading of the
beam will be deferred to allow for the inclusion of the additional beam spreading loss
caused by atmospheric turbulence as deduced in Section ~\ref{SS:turbloss} below.}

\subsubsection{Static Atmospheric Losses}

Even in the absence of any turbulent motions at all, the atmosphere will induce a
variety
of scatterings and absorptions of the pulses in the beam, leading to a reduction in the
received signal intensity at Bob. We have made use of the FASCODE (``Fast Atmospheric
Signature Code") ~\cite{fascode1,fascode2,hitran} computer code developed
by the U.S. Air Force Research
Laboratory to numerically compute typical examples of such losses in a wide variety of
operating environmental conditions. In our analysis many computations were carried out
for a wide range of boundary conditions. As a representative example some of our
computations were done with the following assumptions input to the code: (1) 1550
nanometer wavelength light, using the high-resolution version of FASCODE, (2) $45$
degrees slant angle (the slant angle is defined to be equal to $90$ degrees
at zenith), (3) inputted geographic coordinates for Hanscom Air Force Base,
Massachusetts,(4) minor sunspot actvity, (5) azimuth angle equal to $0$ degrees
({\it i.e.,} looking northward), (6) clear conditions (this defined as
yielding 23 kilometers visibility),\footnote{
This is only representative, as FASCODE computations were carried
out for a variety of less favorable weather conditions, as reflected
in the analysis presented in Section 5.3.2 of this paper.}
(7) no significant recent weather or volcanic activity, (8) date and time for computer
run: 21 March, 2000, noon.

We will denote the losses defined by the output of FASCODE runs by
\be
\label{179}
{\cal L}_{{\rm static~atmospheric}}=10\cdot\log_{10}\left({
\rm normalized~FASCODE~signal~output}\right)~,
\ee
and incorporate this in Section 4.1.5 below in a complete account of the total
free space line attenuation.

A summary of the physical meaning of these numerical computations follows:

$\bullet$ The typical attenuation obtained for a path length of $300$ kilometers,
representing the distance from mean sea level (MSL) to a low-earth-orbit (LEO) satellite,
in the direction of propagation from satellite to ground against clear weather
conditions for 1550 nanometer laser light indicates a static atmospheric attenuation
of order $-1$ dB.\footnote{
In this paper we adopt the convention of denoting attenuation values (when measured
in decibels) as negative quantities.}

$\bullet$ The static atmospheric attenuation effectively disappears when the two ends of
the link are located at elevations of 10 kilometers and 300 kilometers, respectively.

$\bullet$ Rain, and even light drizzle, will severely attenuate the beam to the extent
that in many cases useful signal cannot be transmitted at all.

\begin{figure}[htb]
\vbox{
\hfil
\scalebox{0.69}{\rotatebox{0}{\includegraphics{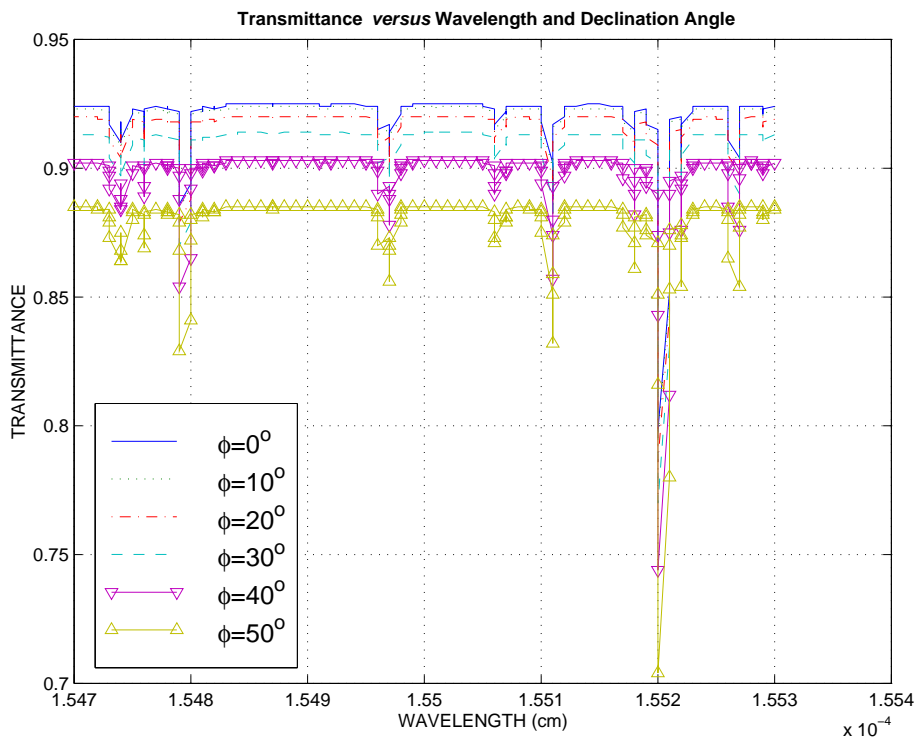}}}
\hfil
\hbox to -1.25in{\ } 
}
\bigskip
\caption{%
Sample FASCODE Results for Static Atmospheric Attenuation
}
\label{F:fascode1}
\end{figure}

The results of typical FASCODE computer runs are illustrated in Figure \ref{F:fascode1}
with a set of numerical curves that depict characteristic static atmospheric
losses. For the computations in this example we have assumed that clear weather
conditions obtain, defined as above to yield 23 kilometers visibility.
In this graph we display curves of the atmospheric transmission as a function
of the declination angle with respect to zenith, with 0 degrees
corresponding to the zenith position. Inspection of the
curves reveals the expected functional dependence, with the static atmospheric
transmission loss increasing as
the declination angle increases from 0 degrees to 50 degrees.

(The computer analysis was specifically carried out using the Air Force Research
Laboratory's PLEXUS system \cite{plexus}, which provides an interface to FASCODE.)

\subsubsection{Turbulent Atmospheric Losses}
\label{SS:turbloss}

Atmospheric turbulence will potentially induce a variety of signal losses in the
propagating beam.
These are: (1) turbulence-induced beam spreading (this is beam spreading {\it in addition
to} that beam spreading due to purely geometrical diffraction effects), (2)
turbulence-induced beam wander, (3) turbulence-induced coherence loss, (4)
turbulence-induced scintillation, and (5) turbulence-induced pulse distortion
and/or broadening. (Another well-known type of turbulence-related loss, {\it thermal
blooming}, is not relevant here as the filled bit cells in the beam comprise principally
a sequence of single photons which can only heat the atmosphere to a negligible degree.)

\begin{figure}[htb]
\vbox{
\hfil
\scalebox{0.6}{\rotatebox{0}{\includegraphics{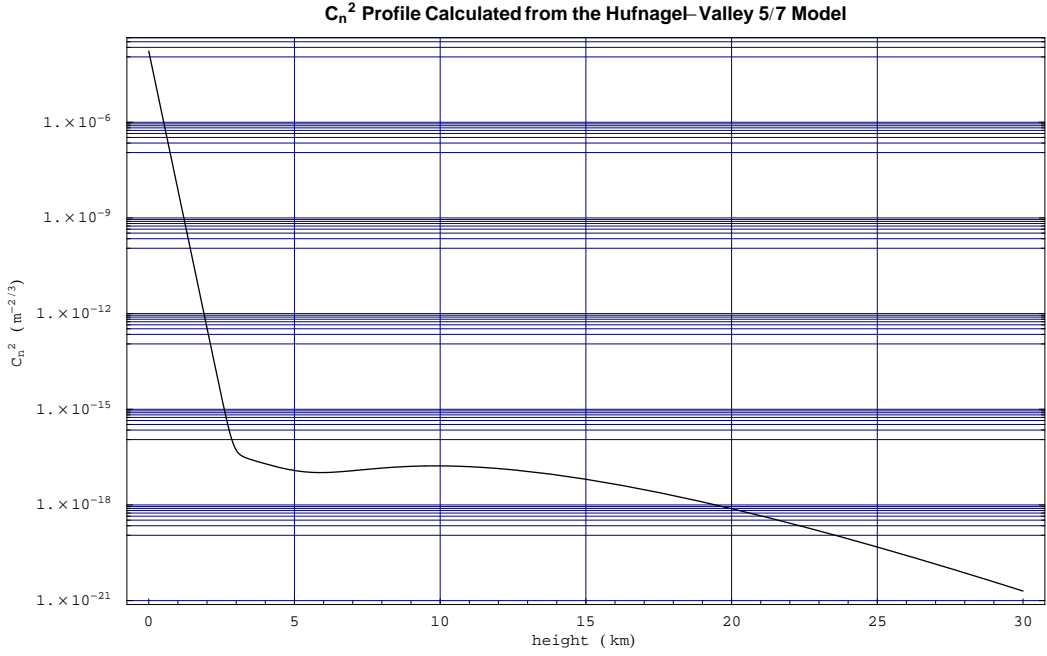}}}
\hfil
\hbox to -1.25in{\ } 
}
\bigskip
\caption{%
The Hufnagel-Valley 5/7 Model for Atmospheric Turbulence
}
\label{F:hufnagel}
\end{figure}

In analyzing turbulence-induced losses, it is necessary to adopt a particular model for
the refractive index structure function $C_n^2$ in order to characterize the turbulent
motions in the atmosphere. In our analysis we have made use of two standard models of
atmospheric turbulence, the
Hufnagel-Valley $5/7$ model \cite{hufnagel1,hufnagel2,hufnagel3} and the CLEAR I
model \cite{clearI1,clearI2,clearI3}. The
dependence on altitude of $C_n^2$ in the former model is illustrated in
Figure \ref{F:hufnagel}.

We now consider in turn each of the five types of turbulence-associated losses enumerated
above.

\vskip 10pt
\noindent {\it Turbulence-Induced Beam Spreading}

\noindent{From standard results in turbulence theory \cite{fante,EO} the
transverse coherence length is given by}
\begin{equation}
\label{180}
\rho_0=\Bigg[ 1.46k^2\sec\varphi\int_0^L d\eta C_n^2\left(\eta\right)\left(1-{\eta
\over L}\right)^{5/3}\Bigg]^{-3/5},
\end{equation}
where $k$ is the wavenumber, $L$ is the path length over which the signal
propagates, $\varphi$ is the declination angle with respect to the zenith direction
and $C_n^2$ is the refractive index structure function. The (mean squared value of the)
``short-time beam spread radius" due to {\it both} turbulence and vacuum geometrical
effects is given by
\begin{equation}
\label{181}
\langle\rho_s^2\rangle=\rho_d^2+{4L^2\over\left(k\rho_0\right)^2}\Bigg[1-0.62\left(
{\rho_0\over D_A}\right)^{1/3}\Bigg]^{6/5},
\end{equation}
\noindent{where $D_A$ is the diameter of the aperture of Alice's transmitting
instrument and}
\begin{equation}
\label{182}
\rho_d=\Bigg[{4L^2\over\left(kD_A\right)^2}+\left({D_A\over 2}\right)^2\Bigg]^{1/2}
\end{equation}
\noindent{is the vacuum beam spread radius presented in
Section \ref{SS:VBSL} above.\footnote{\label{fff}
The expressions used in eq.(\ref{181})
for $\langle\rho_s^2\rangle$ and in eq.(\ref{184}) below
for $\langle\rho_c^2\rangle$ are valid \cite{fante} in the region of the system
parameter space characterized
by the inequalities
$\rho_0<<D_A<L_o$ and $L\lsim kX^2$, where $L_o$ furnishes a measure of the largest
distances over which fluctuations in the index of refraction are correlated (typically
$L_o\approx 100$ m), $X$ is the smaller of $D_A$ and $\rho_0$, and $L$
is the signal path length. In the
cases considered here it is straightforward
to show numerically that both inequalities are satisifed.}
Then the
associated loss, {\it i.e.} the {\it total} beam spread loss due to both turbulence and
vacuum geometrical effects, is given by}
\begin{equation}
\label{183}
{\cal L}_{\rm{beam \ spread}}={\cal L}_{\rm{beam \ spread}}\left(k,\varphi,L,D_A,D_B
\right)=10\cdot\log_{10}\left({D_B^2\over 4\langle\rho_s^2\rangle}
\right),
\end{equation}
\noindent{where $D_B$ is the diameter of Bob's receiving instrument. The above
expression is defined to give a negative value ({\it i.e.,} indicate a loss) when
the area of Bob's receiving instrument is less than the effective area of the
turbulence- and diffraction-limited spot of the fully spread laser beam. (In this
approach it is assumed that there is uniform illumination across the turbulence- and
diffraction-limited received spot. This is acceptable because a different situation in
which, say, there is a Gaussian pattern, will generate {\it less} loss, so that we are
at worst {\it overestimating} the loss in this case.)}

As a particular example, we assume that the diameter of the aperture of Alice's
transmitting instrument is $D_A=30~\rm cm$ for laser light at 1550 nanometers wavelength.
The largest amount of diffraction- and turbulence-induced beam spread loss will be realized
in systems with receiving apertures that are too small. We consider two cases: (1) Alice is
on a LEO satellite orbiting at 300 kilometers altitude above Bob who is at mean sea level.
If the diameter of Bob's receiving instrument is $D_B=30~\rm cm$, the total beam spread
loss is $-17.3237$ dB with Alice located at zenith above Bob, and increases
to $-19.0205$ dB
with Alice at a slant angle of 45 degrees. If the diameter of Bob's receiving instrument
is increased to $D_B=1~\rm m$, these losses are reduced
to $-6.86612$ dB and $-8.56288$ dB,
respectively. (2) Alice is on a GEO satellite orbiting at 22236 miles altitude and Bob
is assumed to be located either on a mountain at high altitude, or on an aircraft, in
either case such that most of the turbulent effects are effectively very small. In this
case the signal loss due to beam spreading can be determined from the effects of
diffraction: if $D_B=1~\rm m$ and Bob is on a mountain at an altitude of 13500 feet
(such as at the Mauna Kea Observatory location) the loss is $-41.4144$ dB; if Bob is on
an airplane at 35000 feet altitude the loss is $-41.4128$ dB. If $D_B=10~\rm m$, which is
the size of the effective aperture of the Keck Telescope on Mauna Kea, these losses
decrease to $-21.4144$ dB and $-21.4128$ dB, respectively.

\vskip 10pt
\noindent {\it Turbulence-Induced Beam Wander}

The presence of atmospheric turbulence will cause the beam to appear to ``wander" around
a bit on its passage through the space between Alice and  Bob. The mean squared value of
the radius of this beam wander region is given by turbulence
theory \cite{fante,EO} as
\be
\label{184}
\langle\rho_c^2\rangle={2.97L^2\over k^2\rho_0^{5/3}D_A^{1/3}}~.
\ee
Existing engineered devices that apply active closed-loop feedback control between Alice
and Bob are available to generate in excess of $30$ dB rejection of turbulence-induced
beam
wander ~\cite{geolite}. These systems employ fast steering mirrors that scan the
incoming
tracking beam to correct for lower frequency wander ($\le 100$ Hz).

The explicit loss in signal associated specifically with turbulence--induced beam wander
is given by
\begin{equation}
\label{185}
{\cal L}_{\rm{beam \ wander}}={\cal L}_{\rm{beam \ wander}}\left(k,\varphi,L,D_A,D_B
\right)=10\cdot\log_{10}\left({D_B^2\over 4\langle\rho_c^2\rangle}
\right)~.
\end{equation}
As a particular example, we again assume that the diameter of the aperture of Alice's
transmitting instrument is $D_A=30~\rm cm$ for laser light at 1550 nanometers wavelength.
As with the loss due to beam spreading, the largest amount of turbulence-induced beam
wander loss will be realized in systems with receiving apertures that are too small. We
again consider two cases: (1) Alice is on a LEO satellite orbiting at 300 kilometers
altitude. Even if the diameter of Bob's receiving instrument is as small as $D_B=30~\rm
cm$, the same size as for Alice, the beam wander loss is $-16.4917$ dB with
Alice located at
zenith above Bob, and increases to $-17.9969$ dB with Alice at a slant angle
of 45 degrees.
(2) Alice is on a GEO satellite orbiting at 22236 miles altitude. As long as
$D_B\ge 2.11 \rm ~meter$ the beam wander loss is no worse than $-30$ dB if
the slant angle
is not taken into account: this is appropriate in the case of the earth-GEO satellite
link, as it is assumed that the receiving platform will be located either on an aircraft
or a mountain (such as the Mauna Kea Observatory location) in order to minimize atmospheric
effects in general.

Thus, we can arrange that the total signal loss associated with turbulence-induced beam
wander can be suppressed to a level less than a magnitude of 30 dB in the cases of
interest, for which
purpose there are engineering solutions available to completely mitigate this loss.

Therefore, it is possible to construct a QKD system in which beam wander loss is
effectively eliminated for both LEO and GEO satellite links.

\vskip 10pt
\noindent {\it Turbulence-Induced Coherence Loss}

In this section we consider two types of coherence loss that will affect the
state of 
the pulse as it arrives at Bob's detector.  In principle, these coherence
losses could 
affect the probability that photons arriving at the Bob's apparatus will be
detected, thus 
reducing the rate at which secret bits can be produced.  We will show that, if
the 
effective cross section of the detector is sufficiently large, these effects
do not reduce the count rate of Bob's apparatus.  

The first effect is due to the 
loss of spatial coherence. Consider first what happens when a classical
optical signal 
impinges on a telescope \cite{optics}. If the lateral coherence length of the
signal is 
less than the diameter of the receiving optics, the intensity 
pattern at the focus becomes spread
out over a larger area. This is due to a diffraction effect in which the
effective aperture 
is given by the coherence length of the signal rather than the diameter of the
telescope objective.  In applying this to the case of a weak pulse containing
small 
numbers of photons, we note that the classical intensity corresponds to the
probability 
of measuring a photon in a given region. As long as the detector is designed 
to capture any signal that appears in the diffraction disk, it can be expected
to 
collect individual photons that propagate through the same optical device.
Provided 
this condition is met, there is no loss of photon counts due to loss of
spatial coherence 
in the incoming pulse, and we then have
\be
\label{186}
{\cal L}_{{\rm spatial \ coherence}}=0~.
\ee

The second effect is the decoherence of the quantum mechanical phases
associated with the 
initial coherent state sent by Alice.  The initial state is described by a
coherent 
superposition of photon number states:
\be
\label{187}
\vert \phi \rangle = \sum_{k=0}^\infty \sqrt {{e^{-\mu} \mu^k} \over k!} 
        e^{ik\phi} \vert k \rangle~,
\ee
\noindent
where $\phi$ is the semi-classical phase of the coherent state.  Note that we
idealized 
the pulse as a pure monochromatic state, and that we have suppressed
polarization and 
wavevector indices.  The actual physical pulse will be a superposition of such
states 
summed over a region of wavevectors to produce a wave packet that is localized
in 
space and time.  This linear superposition is irrelevant to our argument, as
we will 
only be concerned with the relative phases of the terms appearing in the
monochromatic 
coherent state.  The 
density matrix corresponding to this state is
\be
\label{188}
\rho^{(coh)} = \sum_{k,l=0}^\infty e^{-\mu} \sqrt {{\mu^{k+l}}\over {k!l!}}
       e^{i\phi_{kl}} \vert k \rangle \langle l \vert~,
\ee
\noindent
where we define the phase factor
\be
\label{189}
\phi_{kl} \equiv \left( k-l \right) \phi~.  
\ee
\noindent
Since the phases for the on-diagonal elements are identically zero, this can
be rewritten 
as
\bea
\label{rhocoh}
\rho^{(coh)} &=&
\sum_{k=0}^\infty e^{-\mu} {{\mu^k} \over k!} \vert k \rangle \langle k
\vert
     \nonumber\\
   &&+\sum_{k \neq l}^\infty e^{-\mu} \sqrt {{\mu^(k+l)}\over {k!l!}}
      e^{i\phi_{kl}}  \vert k \rangle \langle l \vert~.  
\eea
\noindent
The density matrix for the incoherent state is obtained by averaging over the 
phases $\phi_{kl}$.  The resulting density matrix is
\bea
\label{rhoincoh}
\rho^{(incoh)} &=&
\sum_{k=0}^\infty e^{-\mu} {{\mu^k} \over k!} \vert k \rangle \langle k
\vert 
     \nonumber\\
   &&+\sum_{k \neq l}^\infty e^{-\mu} \sqrt {{\mu^(k+l)}\over {k!l!}}
       \langle e^{i\phi_{kl}} \rangle \vert k \rangle \langle l \vert~.  
\eea
We now use these expressions to find the response of the detector in the
coherent 
and incoherent cases.  An idealized detector is a device which produces 
a count if it measures in state $\vert k \rangle$ for $k \geq 1$.  This 
measurement corresponds to 
the projection operator
\bea
{\cal M} &=& \sum_{k=1}^\infty \vert k \rangle \langle k \vert \nonumber\\
         &=& {\unit} - \vert 0 \rangle \langle 0 \vert~,
\eea
\noindent
so that the expectation value of the measurement on a mixed state
characterized 
by the density matrix $\rho$ is
\bea
\langle {\cal M} \rangle 
   &=& Tr \left( \rho {\cal M} \right) \nonumber\\
   &=& 1-\rho_{00}~.  
\eea
\noindent
But from equations (\ref{rhocoh}) and (\ref{rhoincoh}), we see that
\be
\rho_{00}^{(coh)} = \rho_{00}^{(incoh)}~,
\ee
\noindent
and
\be
\langle {\cal M} \rangle_{(coh)} = \langle {\cal M} \rangle_{(incoh)}~,
\ee
so that the result of the measurement is independent of whether or not the 
quantum mechanical phases have lost coherence on their way from Alice to Bob,
and we thus have
\be
\label{196}
{\cal L}_{{\rm quantum \ coherence}}=0~.
\ee
It is clear from the discussion that this insensitivity to the coherence
properties of the received signal is due to two factors.  First, the loss of
quantum coherence only affects the off-diagonal terms in the density matrix,
and, second, the response of the detector is dependent only on the diagonal
terms.  This property of the detector is due to the fact that it is essentially
a photon counting device, which performs measurements that projects the measured
states onto a set of states described in terms of the photon number basis.  Our
conclusions apply to any detector that can be so described, including the imperfect
$(\eta < 1)$ detectors of a real implementation.

\vskip 10pt
\noindent {\it Turbulence-Induced Scintillation}

Atmospheric turbulence will cause the received value of the signal intensity $I$ to
fluctuate about its average value. This will manifest itself as a distinct scintillation
of the laser beam. We will calculate the magnitude of the scintillation in the weak
turbulence regime, for which the Rytov approximation holds. (There is little point in
QKD applications in considering the regime of stronger turbulence as we would not expect
sufficient signal to survive the transit to even a LEO satellite in this case.)

The magnitude of the normalized variance of the signal intensity that is responsible
for the intensity scintillations is given by~\cite{fante,EO}
\be
\label{197}
\sigma_I^2={\langle\left(I-\langle I\rangle\right)^2\rangle\over\langle I
\rangle^2}\simeq 4\sigma_\chi^2,
\ee
where
\be
\label{198}
\sigma_\chi^2=0.56k^{7/6}\int_0^LdzC_n^2\left(z\right)z^{5/6}~,
\ee
so that we have for the associated loss
\be
\label{199}
{\cal L}_{\rm{scintillation}}={\cal L}_{\rm{scintillation}}\left(k,L\right)
=10\cdot\log_{10}\left(1-\sqrt{\sigma_I^2}\right)~.
\ee
The Rytov approximation appropriate for the regime of weak turbulence is specified by
the inequality
\be
\label{200}
\sigma_I^2\le 0.3.
\ee
As a numerical example, if we assume that Alice transmits laser pulses at a wavelength of
1550 nanometers and we employ the Hufnagel-Valley 5/7 model for the refractive index
structure function we obtain $\sigma_ \chi^2=0.0158$, corresponding to $\sigma_I^2\approx
0.06305$, which indicates that we are well within the regime of validity for the Rytov
theory. This implies a signal loss due to intensity scintillations of $-1.26$ dB. Note
that, due to the rapid decay with increasing altitude of the Hufnagel-Valley 5/7 model,
this result holds almost identically for the cases of Alice located on a LEO satellite
{\it and} on a GEO satellite.

\vskip 10pt
\noindent {\it Turbulence-Induced Pulse Distortion and/or Broadening}

For pulses of sufficiently short duration the consequences of atmospheric turbulence can
include inducing distortion and/or broadening of the shape of the pulses through
the generation of dispersion. In order to mitigate this problem we would like to find
the conditions that ensure that the spectrum of a short pulse {\it in} a turbulent
medium be equal to that of the same pulse when it was {\it incident upon} and entered
the medium. These conditions have been worked out by Fante and consist of the two
inequalities (eq.(106) and eq.(107) in ~\cite{fante}):
\be
\label{201}
{0.91\Omega C_n^2L^2\over cl_o^{1/3}}<<1
\ee
and
\be
\label{202}
{0.39\Omega^2C_n^2L_o^{5/3}L\over c^2}<<1.
\ee
Here $c$ is the speed of light, $\Omega$ is the bandwidth of the pulse, $L$ is the path
length and $L_o$ and $l_o$ are, respectively, the outer and inner scale sizes of typical
turbulent eddies: $L_o$ (as mentioned in footnote \ref{fff} above)
furnishes a measure of the largest distances over which
fluctuations in the index of refraction are correlated, and $l_o$ is a measure of the
smallest correlation distances.

\begin{figure}[htb]
\vbox{
\hfil
\scalebox{0.6}{\rotatebox{270}{\includegraphics{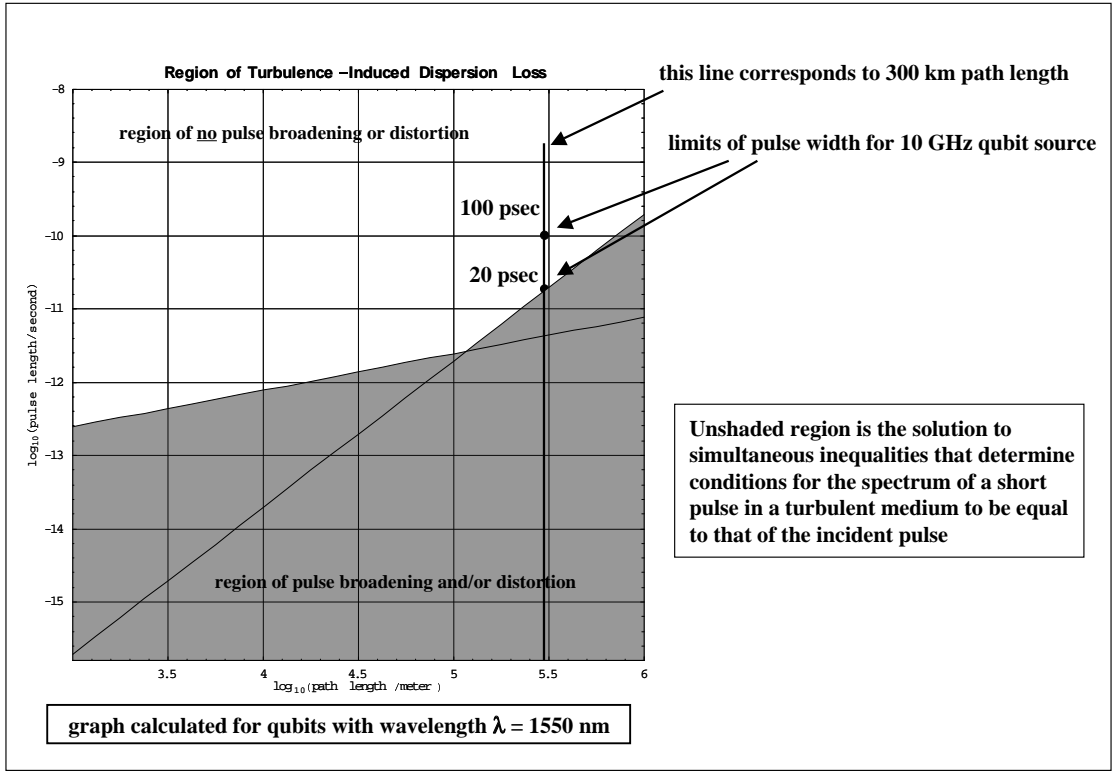}}}
\hfil
\hbox to -1.25in{\ } 
}
\bigskip
\caption{%
Pulse Distortion and/or Broadening Graph
}
\label{F:pulse_distort}
\end{figure}

The above inequalities can be numerically solved and inspection of the plotted results
allows one to identify parameter regions where pulse distortion and/or broadening do
occur and where they do not. We have plotted in Figure \ref{F:pulse_distort} below
a numerical solution for the
case when the laser pulses are at a frequency of 1550 nanometers, and we have taken
typical values~\cite{fante} of $L_o=100\rm~m$ and $l_o=0.001\rm~m$, making use of the
Hufnagel-Valley 5/7 model for the refractive index structure function. We see that
when Alice is orbiting on a LEO satellite at 300 kilometers altitude, in order to
avoid incurring any pulse distortion and/or broadening loss, the width of the pulse
must be greater than 19.8 picoseconds, corresponding to a laser pulse repetition
frequency (PRF) of no greater than about 50 GHz.\footnote{
Note that the bit cell period, $\tau$, which is by definition
equal to the reciprocal of the PRF of the laser, is in general larger than the width of
the pulse. The calculation of a lower bound on the pulse width thus enables us
to deduce an upper bound on the value of the PRF.}
This means that, ignoring for the moment all other
concerns regarding operating a QKD system at a basic clock rate of 50 GHz (such as
photon detection, real-time data recording, {\it etc.}), it is preferable to utilize
lasers with PRFs of less than 50 GHz for an Earth-LEO satellite link, and even slower
laser pulse rates for links to higher altitude satellites. When these conditions
are met we have for the associated loss
\be
\label{203}
{\cal L}_{{\rm pulse~distortion/broadening}}=0~.
\ee
These are not necessarily
rigid constraints on system design, however. In the case of an Earth-LEO satellite
link, for example, the use of a laser with a larger PRF than 50 GHz {\it might} be
acceptable: the
effect of the addition to the line attenuation, and the corresponding decrease in
system throughput caused by the resulting noise would need to balanced against the
increase in throughput caused by the shorter bit cell period. Such an analysis would
be a fruitful area for future numerical research.

In Section 5 below we will discuss the design of a practical high speed QKD system in
which a laser with a PRF of 10 GHz is employed, corresponding to a bit cell period of
100 picoseconds. In Figure \ref{F:pulse_distort} we have indicated the two limits on
the pulse width provided by the 100 picosecond and 19.8 picosecond points,
respectively.

\subsubsection{Optics Package Losses}

\begin{figure}[htb]
\vbox{
\hfil
\scalebox{0.66}{\rotatebox{270}{\includegraphics{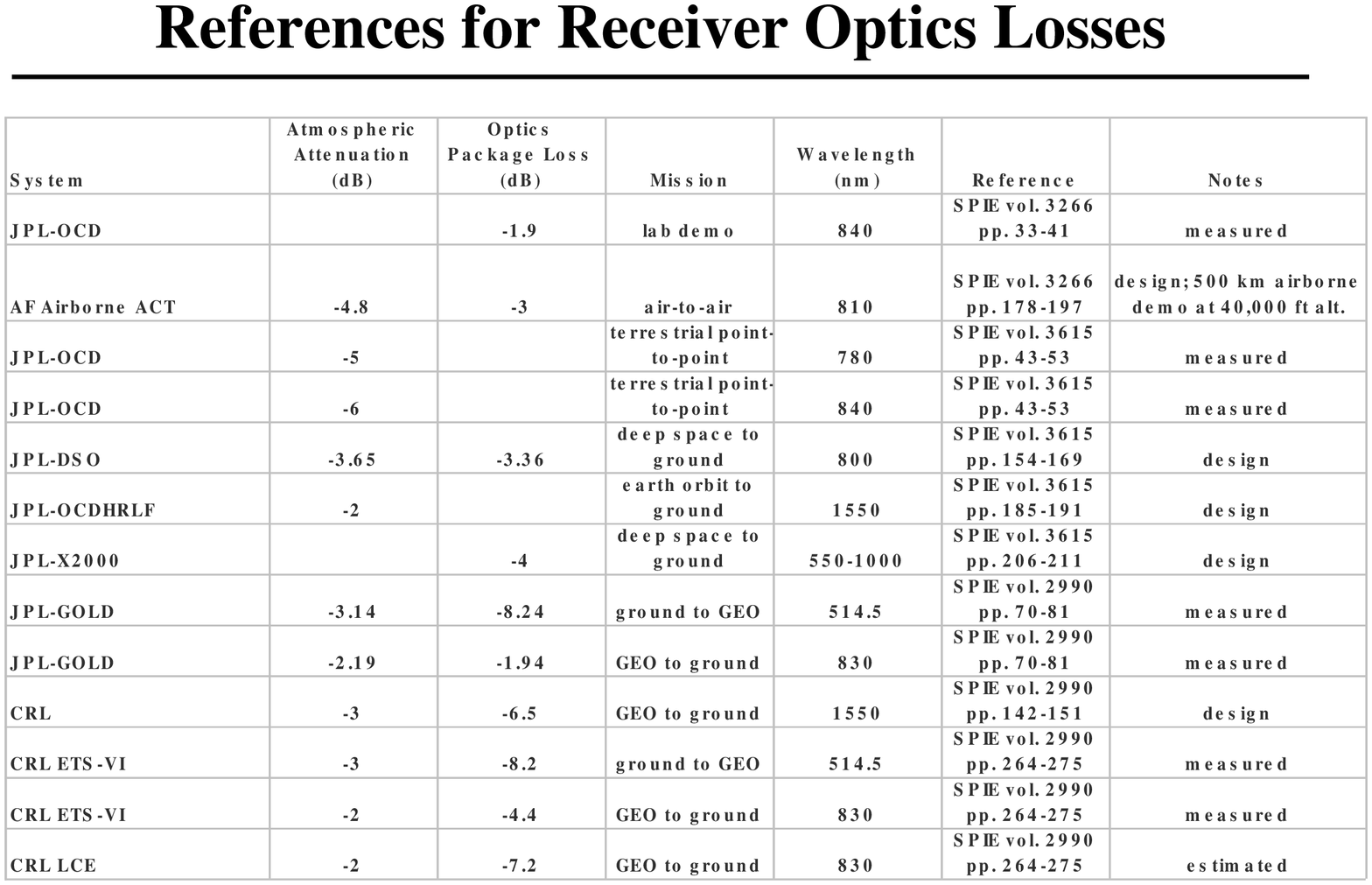}}}
\hfil
\hbox to -1.25in{\ } 
}
\bigskip
\caption{%
Optics Package Loss Comparison Table
}
\label{F:opticspackage}
\end{figure}

The optics package losses can be estimated by comparing a proposed system design for the
Alice and Bob apparatuses with demonstrated optical communications systems of comparable
complexity. This complexity is assessed in terms of approximately equal numbers of
system
components of corresponding quality and characterstics appropriate for a QKD system
setup.
In Figure \ref{F:opticspackage}
below we compare the demonstrated losses for a variety of laser
communications terminals as reported in the literature.\footnote{
We thank C.P. McClay for assembling this information.}

Based on this analysis of extant systems of comparable complexity, it is reasonable
to take a value of -5 dB for the expected optics package loss associated to the QC
system proposed in Section 5 below. We will employ this value of loss for the optics
package in computing the total line attenuation for a free space implementation.

\vskip 10pt
\noindent {\it ``Behind-the-Telescope Loss"}

We briefly mention here the loss that may arise between the ``back" of Bob's telescope
and front of his detector apparatus, deferring to Section 5.2.1 below a more detailed
discussion.
We have already calculated and discussed the beam spreading loss associated to the
passage
of the laser beam from Alice to Bob. The consequence of this effect is that the laser
``spot" that is incident upon the front of Bob's telescope is larger than we would like
it to be - this is a problem of received beam size. There is {\it another} beam size
problem that can develop behind Bob's telescope, if the received beam incident upon the
surface of Bob's photon detection apparatus is too large. We propose in Section 5.2.1
below a method of mitigating such loss for the novel type of fast photon detector that
is a central and unique feature of the high-speed QKD system we decribe. It is important,
in general, to take careful account of this source of loss, as it can be a very important
contribution to the overall optics package loss unless successfully mitigated.

\subsubsection{Complete Line Attenuation Losses - Free Space}

We collect here the results of the above sections to display the total line
attenuation-associated losses on the strength of the signal in the case of
free-space propagation. This analysis allows us to understand the operating
characteristics of an actual quantum cryptography implementation set up as a free
space communications system. The various losses that contribute to the complete
free space line
attenuation $\alpha$ can be assembled into a single function that can be numerically
plotted. Putting together the expressions from
eqs.(\ref{179}), (\ref{183}), (\ref{185}), (\ref{186}), (\ref{196}), (\ref{199}),
and (\ref{203}),
we form the complete line attenuation function $\alpha_{free~space}$ as the sum
of all losses:
\bea
\label{204}
\alpha_{free~space}&=&\alpha_{free~space}\left(k,\varphi,L,D_A,D_B\right)
\nonumber\\
&=&\sum\cal L
\nonumber\\
&=&
{\cal L}_{{\rm static~atmospheric}}+
{\cal L}_{{\rm beam~spread}}+
{\cal L}_{{\rm beam~wander}}+{\cal L}_{{\rm spatial~coherence}}
\nonumber\\
&&+
{\cal L}_{{\rm quantum~coherence}}+{\cal L}_{{\rm scintillation}}+
{\cal L}_{{\rm pulse~distortion/broadening}}+{\cal L}_{{\rm optics~package}}~,
\nonumber\\
\eea
where the equation of state in the first line above displays only some of the
various functional dependences that characterize the total line attenuation. As
only one
example of several additional ones that could be discussed, the dependence
on fog conditions that is implicit in the
term ${\cal L}_{{\rm static~atmospheric}}$ is not {\it explicitly} indicated as part
of the functional dependence, although it certainly needs to be specified in order
to obtain a concrete numerical result (and in practice this is done by providing
fog data in an appropriate input file for the FASCODE runs discussed in Section 4.1.2
above).

\begin{figure}[htb]
\vbox{
\hfil
\scalebox{0.6}{\rotatebox{0}{\includegraphics{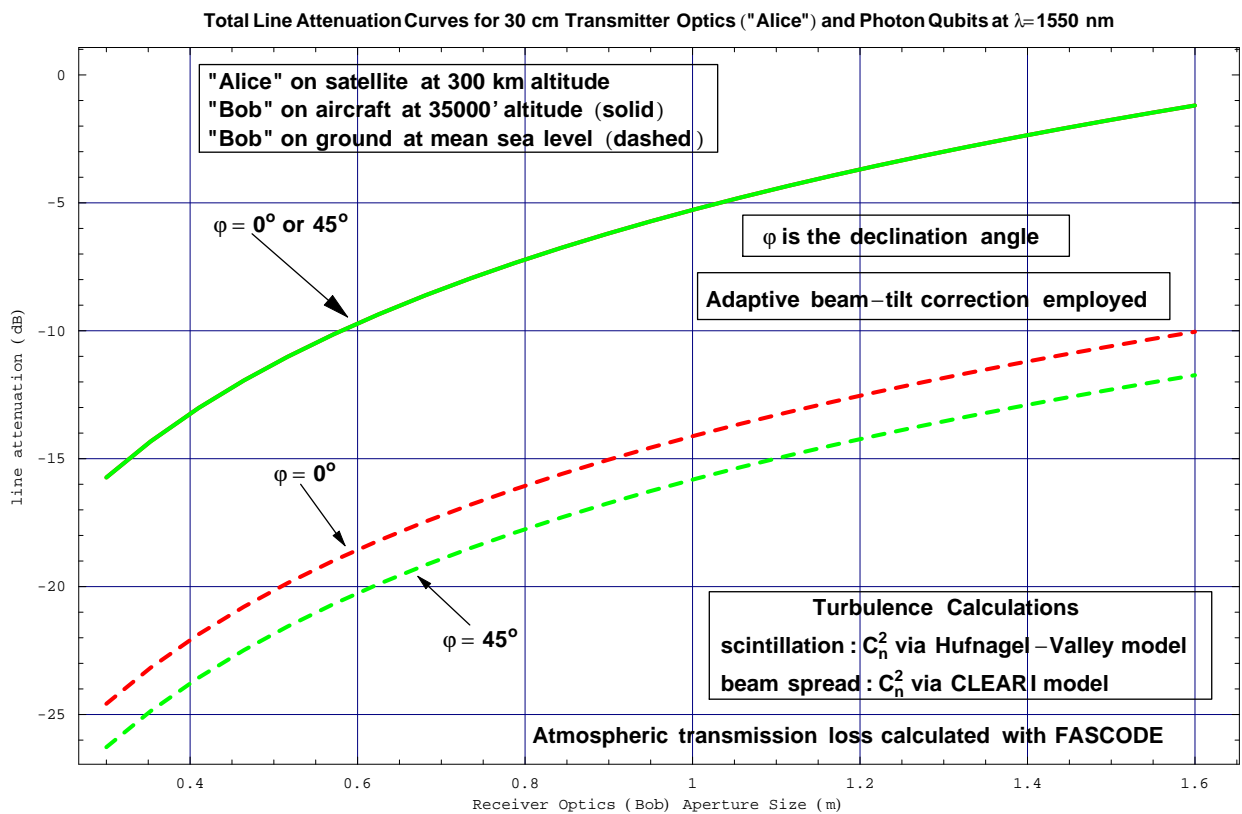}}}
\hfil
\hbox to -1.25in{\ } 
}
\bigskip
\caption{%
Line Attenuation for $\lambda=1550~\rm {nm}$
}
\label{F:attenuation1}
\end{figure}

\begin{figure}[htb]
\vbox{
\hfil
\scalebox{0.6}{\rotatebox{0}{\includegraphics{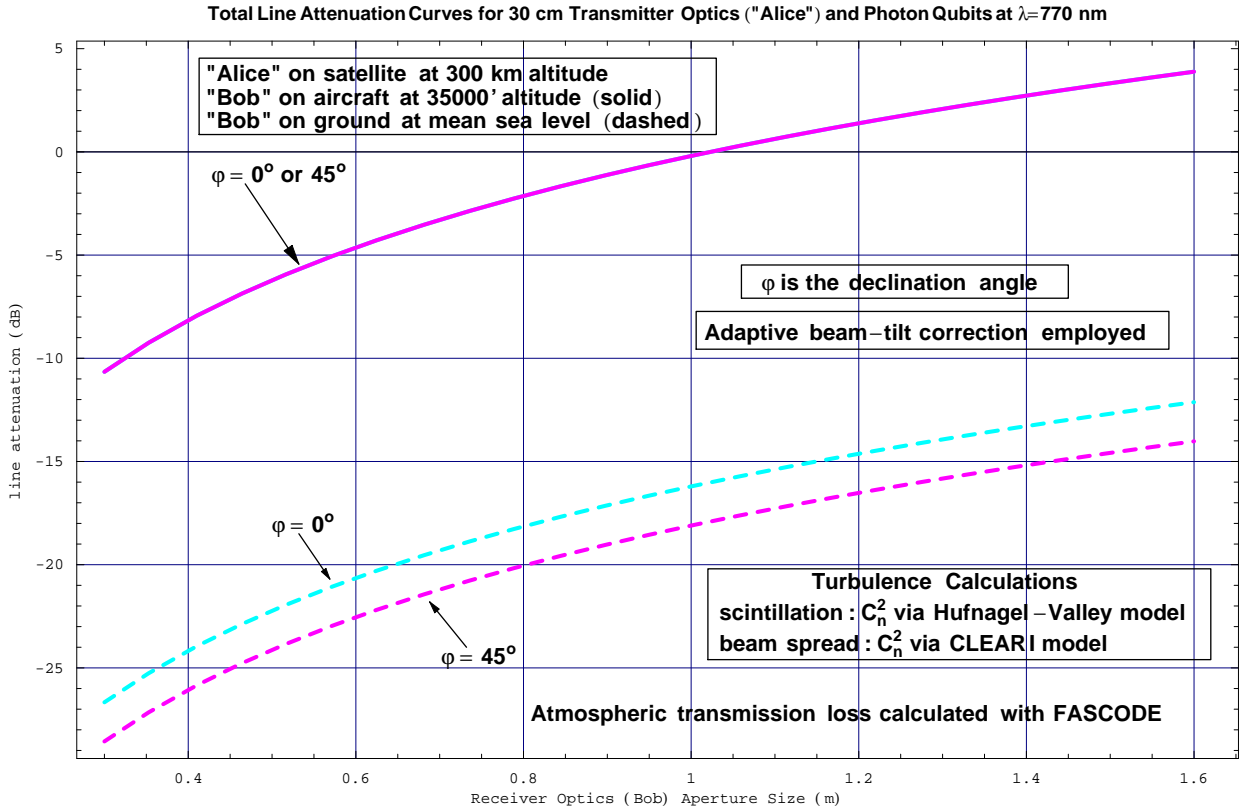}}}
\hfil
\hbox to -1.25in{\ } 
}
\smallskip
\caption{%
Line Attenuation for $\lambda=770~\rm {nm}$
}
\label{F:attenuation2}
\end{figure}

In Figures \ref{F:attenuation1}
and \ref{F:attenuation2} below
we plot the dependence of this loss on the diameter of
the aperture of Bob's receiving instrument, $D_B$, assuming that the transmitting
instrument used by Alice has a diameter of $D_A=30~\rm cm$. These curves are computed
for the cases of light at wavelengths of
$\lambda=1550~\rm nanometers$ and $\lambda=770~\rm nanometers$,
respectively. As discussed in Section 4.1.3 above, the
use of sufficient adaptive beam tilt correction to have effectively
mitigated the effects of beam wander has been assumed. The static atmospheric
transmission losses have been computed using the FASCODE computer program as described
in Section 4.1.2 above. The
scintillation and beam spread losses were calculated using the
Hufnagel-Valley 5/7 and CLEAR I atmospheric turbulence models, respectively.
For all the curves it has been assumed that Alice is located on a LEO satellite
at 300 kilometers altitude.
The solid curves are for Bob located at an altitude of 35000' for arbitrary slant
angle, and the dashed curves are for Bob located at mean sea level for different slant
angles of 0 degrees and 45 degrees. We have assumed that clear weather
conditions apply for the curves in both Figure \ref{F:attenuation1} and
Figure \ref{F:attenuation2}.\footnote{
As mentioned in Section 4.1.2 above ``clear weather" conditions are defined as
yielding 23 kilometers visibility.}

As an example, inspection of Figure \ref{F:attenuation1} reveals that,
at $\lambda=1550~\rm nanometers$ for a value of $D_A=30~\rm
cm$ and with Bob on an airborne platform at 35000 feet, in order to achieve a line
attenuation amount of -10 dB it is necessary for Bob to use a receiving instrument with
an aperture of 58 cm. If we switch to a wavelength of 770 nanometers, inspection of
Figure \ref{F:attenuation2} reveals that roughly the same size of aperture for Bob
will produce a line attenuation value of -5 dB.\footnote{
An analysis is given in \cite{GOLD} of
the Ground/Orbiter Lasercomm Demonstration (GOLD) free-space
optical communications experiment carried out between Table Mountain Facility near
the Jet Propulsion Laboratory in Pasadena, California and the (geosynchronous)
Japanese ETS-VI satellite. Both the predicted values and the measured values given
there
for the losses are larger than the values predicted in our analysis.
This is due to the fact that the authors of \cite{GOLD} have
defined and calculated system
losses by making use of the generic radar equation, in particular
modelling optical elements as antennas. When the appropriate
antenna gain quantities are taken into account,
the predicted and measured values given in \cite{GOLD} become
consistent with the predicted values for the line attenuation given here,
after substituting LEO for GEO.}

\subsection{System Losses: The Line Attenuation - Optical Fiber}
\label{SS:LAFC}

The line attenuation for a QKD system with optical fiber as the quantum channel is
expressed in a different way from that for a free space quantum channel. The general
expression is given in this case as
\bea
\label{205}
\alpha_{fiber}&=&\alpha_{fiber}\left(L_{fiber},A,\kappa\right)
\nonumber\\
&=&10^{-{AL_{fiber}+\kappa\over 10}},
\eea
where $L_{fiber}$ is the length of the fiber cable connecting Alice and Bob. In this
expression $A$ is a parameter that measures the intrinsic loss characteristic, per unit
length, of the fiber (this is a quantity essentially ``built in" at the factory where the
cable is manufactured) and $\kappa$ is the ``bulk loss" constant associated with the
fiber that includes, for example, splicing
losses due to the presence of spliced links of fiber.
At the present time it is possible to obtain optical fiber cable of high quality
for the transmission of 1550 nanometer wavelength light that has an intrinsic
attenuation constant of $A=0.2$ dB/kilometer.
In addition to the attenuation,
{\it per s\'e}, along the fiber, one must also take into account the intrinsic
channel error, $r_c$, which is caused by the phenomenon of dispersion
of photon arrival times as the pulse propagates from
the Alice to Bob sites. As we will discuss below, this
problem can be mitigated by including appropriate segments of so-called ``compensating"
fiber in the channel. Each such inserted link, however, introduces a contribution to
the value of the bulk loss constant $\kappa$, thereby affecting the net value of
$\alpha_{fiber}$. Thus, in the case of an optical fiber implementation of
quantum cryptography, the intrinsic channel error $r_c$ and the line attenutuation
$\alpha_{fiber}$ are ``connected" to each other in a way that is not the case for
a free space implementation, as we discuss below.

\subsection{The Intrinsic Channel Error}
\label{SS:ICE}

The effective secrecy capacity $\cal S$ exhibits a sensitive dependence on the intrinsic
quantum channel error parameter $r_c$. Relatively small changes in the
value of the intrinsic channel error
can have a significant effect on the magnitude of the
effective secrecy capacity and total throughput rate
of a practical quantum cryptography system, even if all other sources of system loss
have been mitigated. In this section we analyze the
characteristics and causes of this source of error.

\subsubsection{Free Space Quantum Channel}

The intrinsic channel error, $r_c$, is the parameter that measures the tendency of the
free space QKD
system characteristics to cause the states transmitted by Alice to suffer polarization
misalignment by
the time they are detected by Bob. For a free-space quantum channel the
depolarization rate will be determined principally by the actual angular mismatch between
the Alice and Bob instruments as the platforms supporting one or the other or both of them
move. This will occur, for instance, if Alice is on a satellite and Bob is located at a
ground station, or perhaps on another moving platform.

In Appendix A it is demonstrated that, if the relative angular mismatch between the Alice
and Bob instruments is denoted by $\delta$, the fractional error rate due to polarizer
misalignment is given by
\be
\label{206}
r_c=\sin^2\left(\delta\right)~.
\ee
Thus, even if $\delta$ is as large as $1/10$ radian (5.7 degrees) the probability of error
is less than $1\%$. Since $\delta$ is {\it relative} angular mismatch between Alice and
Bob, the solid angle cone within which relative motion is allowed (for a given value of
$r_c$) is $2\delta$. Thus if the platforms on which Alice and Bob are located can be
subjected to attitude control such that corrections can be applied that allow motion
through a cone of solid angle of no more than 11.5 degrees, the intrinsic channel error
can be made to satisfy $r_c\le 0.01$. This constraint on the necessary real-time
control of the attitude of a satellite on which the Alice device could be placed
and on telescope adjustment requirements is well within the currently achievable
state-of-the-art \cite{pointingacc1,pointingacc2}, and
it is therefore reasonable for us
to employ a value of $r_c=0.01$ in the effective secrecy capacity in
computing the operating characteristics of a free space implementation of quantum
cryptography.

\subsubsection{Optical Fiber Quantum Channel}

In the case of a fiber-optic quantum channel, the dispersion characteristics of
the fiber will generate an intrinsic error rate. Dispersion causes the shape of the
transmitted pulse to spread as it travels along the fiber. We envisage an optical fiber
cable-based QKD system built using single mode telecom fiber, used to transmit signals
at a wavelength of 1550 nanometers. In this situation we are primarily concerned with
two types of dispersion that appear in available optical fiber cables: (1) chromatic
pulse dispersion (CD), and (2) polarization-mode pulse dispersion (PMD). Because we
neither advocate nor analyze a QKD system built from multi-mode fiber we will not be
concerned with other types of dispersion effects that can arise, such
as modal dispersion.

If the effects of dispersion along the cable are not mitigated this will specifically
appear as a certain amount of dispersion of the photon arrival time of the
transmitted signal as received by Bob.
The figure-of-merit in assessing this is provided by comparing the dispersion delay
time with the appropriate characteristic
{\it critical time}, which is given by the bit cell
period $\tau$ (the reciprocal of the pulse repetition frequency) of the data source
laser at Alice. In the special
case of an inteferometry-based quantum cryptography system as
described in \cite{plugandplay1,plugandplay2}, one
should compare the dispersion delay time with {\it two} generally
different characteristic critical times, the bit cell period $\tau$
as before, and the coherence time, $\tau_{coh}$ associated with the data source laser.

\clearpage
\vskip 10pt
\noindent {\it Chromatic Pulse Dispersion}

Chromatic pulse dispersion stems from the dependence of the index of refraction on the
wavelength. In more detail, chromatic dispersion in a fiber is partly due to ``material
dispersion," the dependence of the fiber core index of refraction on the wavelength,
and to ``waveguide dispersion," the dependence of the constant propagation mode on the
wavelength. For single mode fibers that transmit light at 1550 nanometers wavelength,
the amount of chromatic dispersion, $d_{CD}$, is given by
\be
\label{207}
d_{CD}\approx 4~{{\rm picoseconds}\over{\rm nanometer}\cdot{\rm kilometer}}~,
\ee
which corresponds to a chromatic dispersion pulse delay time, $\tau_{CD}$, of
\bea
\label{208}
\tau_{CD}&=&d_{CD}\cdot L\cdot\Delta\lambda
\nonumber\\
&\approx& 160 ~{\rm picoseconds}
\eea
for a cable link of $L=50~{\rm kilometers}$, assuming a laser source with a linewidth
of $\Delta\lambda=0.8 ~{\rm nanometer}$. To compare this with the two characteristic
times we assume that the source laser has a coherence time $\tau_{coh}$ of at least 1
nanosecond, or $\tau_{coh}=1000~{\rm picosecond}$. We therefore have
\be
\label{209}
\tau_{CD}<\tau_{coh}~,
\ee
and no mitigation of chromatic pulse dispersion is required on this basis. However
we see that the requirement 
\be
\label{210}
\tau_{CD}<\tau
\ee
means that we must mitigate chromatic pulse dispersion if we use a fast data source
laser with a bit cell period shorter than 160 picoseconds, corresponding to a pulse
repetition frequency greater than 6.25 GHz. As we discuss below, we envisage the use
of high speed lasers with pulse repetition frequencies of 10 GHz. We conclude that
chromatic pulse dispersion needs to
be mitigated if we utilize a high speed laser with a pulse repetition frequency of
10 GHz (corresponding to a bit cell period of 100 picoseconds). Mitigation of this
problem can be accomplished by making use of appropriate lengths of dispersion
compensation fiber spliced into the line to reduce the photon arrival time dispersion
to less than approximately 50 picoseconds. (As discussed in Section 4.2 above, each
additional inserted cable link will increase the line attenuation by increasing the
splicing loss.)

\vskip 10pt
\noindent {\it Polarization-Mode Pulse Dispersion}

Polarization-mode dispersion due to intrinsic and induced birefringence will cause
the pulses to broaden as they propagate along the fiber \cite{PMD}. In
currently available
good quality single mode fiber operating at a wavelength of 1550 nanometers,
the amount of polarization-mode dispersion, $d_{PMD}$, is typically given by\footnote{
In this example we have used representative sample numerical values appropriate for
Lucent ${\rm TrueWave}^{\rm TM}$ fiber.}
%
\be
\label{211}
d_{PMD}\approx 0.1~{{\rm picoseconds}\over{\sqrt{{\rm kilometer}}}}~.
\ee
The corresponding polarization-mode dispersion pulse delay time, $\tau_{PMD}$,
is given for a cable link of $L=50~{\rm kilometers}$ by
\bea
\label{212}
\tau_{PMD}&=&d_{PMD}\cdot\sqrt L
\nonumber\\
&\approx& 0.7~{\rm picoseconds}~.
\eea
We see that, again assuming a laser coherence time of 1 nanosecond, this polarization
arrival time dispersion strongly satisfies the requirement
\be
\label{213}
\tau_{PMD}<\tau_{coh}~.
\ee
Furthermore we see that the requirement
\be
\label{214}
\tau_{PMD}<\tau
\ee
is satisfied unless the pulse repetition frequency of the laser is greater than
$\tau_{PMD}^{-1}=1.43~{\rm THz}$. The current state-of-the-art in photon detection and
data correlation is such that this value does not furnish a constraint. Thus,
polarization-mode dispersion will not be a practical problem.

\subsection{System Loads}

In the previous section we analyzed the various {\it losses} that together reduce the
effective throughput of a QKD system. In this section we consider the complete set of
system {\it loads} that together must be taken into account as well to establish a bound
on the achievable data rate. Our analysis includes careful calculations of the cost in
bits of maintaining
{\it continuous} authentication of the QKD protocol, the throughput requirements on the
classical communications channel, and the computational requirements measured in
machine instructions that must be satisfied in order to carry out real-time processing of
the key. 

Although these processes are crucial to the implementation of any QKD system, there has
been essentially no explicit, quantitative analysis of any of them presented heretofore
in the literature on the
subject. In the case of the authentication cost this seems to stem from the erroneous
notion that an initially supplied, ``short" authentication string shared between Alice
and Bob will suffice to protect the protocol from spoofing. This is mistaken: (1) either
the initially supplied, shared authentication string is indeed ``short," in which case
in due course it will be used up and must be replaced by removing bits from the
key generated by the QKD protocol itself, {\it or} (2) the initially supplied, shared
authentication string is in fact ``long," in which case a primary justification
for the use of quantum key distribution in the first place is severely weakened. In
the case of the
communications throughput load, the potential use of QKD on satellite systems in
particular, clearly
requires a detailed analysis of the constraints that must be satisfied by the
specialized satcom equipment that will be used, and similar importance attaches to
the computational burden, again with special consideration for the satellite problem,
owing to obvious space constraints on the computing hardware that can in practice be
installed on a spaceborne platform.

\subsubsection{The Cost of Continuous Authentication}

\vskip 10pt
\noindent {\it General Remarks}

\indent{Prior to the error correction phase, the BB84 protocol furnishes Alice
and 
Bob with a 
block of $n$ bits, of which Eve has managed to obtain an amount of information
equivalent 
to at most $t+\nu$ bits by making undetected measurements of single photon
pulses and by 
using some combination of direct and indirect attacks on the multiple photon pulses.   
The blocks obtained by Alice and Bob are not identical due to the 
errors introduced by imperfections in the physical apparatus.  By listening to
the 
public transmissions by which Alice and Bob eliminate the errors from their
blocks, 
Eve is able to obtain additional information about Alice and Bobís bits,
giving her a total of at most $q+t+\nu$ bits of information about the error-corrected
block.  As the 
last step of the process, Alice and Bob use a privacy amplification technique
as described in \cite{bbcm} 
to arrive at a block of bits that is shared identically between them and about
which Eve is expected to have no information to a high degree of confidence.  
This section describes the details of a specific implementation of the error
correction and privacy amplification process.  It is assumed throughout that
Eve 
has complete knowledge of the protocols, as well as unrestricted access to all
communications between Alice and Bob that occur during this period.} 

\indent{Our description of these protocols is intended to achieve two goals.
The first 
is to determine the authentication cost of 
the protocols.  This authentication cost is the number of shared secret bits
that 
need to be sacrificed 
in order to guarantee that the protocols perform correctly, that is, that 
execution of the protocols results in some predictable 
amount of secret key material that is shared identically between Alice and Bob
but 
about which Eve has no information with a high degree of confidence.  Since
the 
authentication cost represents a sacrifice of previously existing key material
in order to guarantee the generation of a new block of key material, this cost
has 
a direct impact on the rate at which keys can be generated.}

\indent{The second goal is to estimate the burdens these protocols place on
computational and communications resources.  Unlike the authentication cost,
these costs have no direct effect on the rate of key generation, but they do
provide an estimate of the computational and communications resources required
to support key generation, resulting in constraints on the rate of key generation
for given set of resources. In addition, the fact that the computational complexity
is quadratic in the block size of the key material results in a practical upper bound
on the block size, again for a given set of computational resources.}

We begin by giving a brief overview of the protocol. The first phase is the production 
of a sifted string of bits shared, except for some errors, by Alice and Bob. This is 
achieved by applying the BB84 protocol previously described. If the equipment
were perfect and there were no possibility of errors, the sifted strings would be identical
and any attempt by Eve to obtain more than a few bits would be revealed by the
presence of errors in the sifted string.  Since this is not the case for a practical
implementation, it is essential to include mechanisms to correct the errors and to
eliminate information leaked to Eve in the process. The second phase is thus error
correction. Alice and Bob agree on a systematic scheme of computing and comparing
parities for subsets of the sifted string in order to identify and correct the errors.
Since Eve can eavesdrop on this discussion, an additional amount of data is leaked.
The third phase is privacy amplification, during which Alice and Bob apply a hash
transformation to the error-corrected string which results in a shorter string about
which Eve's expected information is vanishingly small. At various points during
these three phases, Alice and Bob must authenticate their communications to ensure
that Eve is not making a man-in-the-middle attack.

\vskip 10pt
\noindent {\it Sifting Phase}

\indent{The first phase of the key distribution protocol is the generation of
an initial sifted string that is shared between Alice and Bob, but which may
contain errors and about which Eve may have partial information.  We describe a
specific implementation of the BB84 protocol.
Alice generates two blocks of $m$ random bits.  The first block is the
raw key material, and the second block determines the choice of basis she uses
to transmit the bits over the quantum channel. Bob generates a single block of
$m$ bits that reflect his choice of basis in measuring the incoming qubits. Bob
must now identify to Alice those pulses for which he detected a qubit and inform
her of his choice of basis for those pulses.  Bob has several choices available in
deciding how he wants to encode this information. The simplest approach is to send
two bits corresponding to each of Alice's pulses. The first bit tells whether a photon
was detected, the second describes the choice of basis. This means that Bob must send
$2m$ bits to Alice for each block of key material. A more efficient version of this
scheme is to send the second bit only when the first bit indicates that a photon was
detected. In this case, Bob only sends $m+2n$ bits on average. (The factor of 2 in $2n$
comes from the fact that Bob's choice of basis agrees with Alice's on average half of
the time, so that there are twice as many detected photons as there are bits in the
sifted key.) Since we always have $2n\leq m$ this alternative is no less efficient
than the first, and usually it is more efficient. A third alternative is to send two
pieces of information for each detected photon, the first indicating for which of the
$m$ bit cells the photon was detected, and the second giving Bob's choice of basis.
This requires that Bob send $2n\left(1+\log_2m\right)$ bits for each block of key
material. This alternative is more efficient than the others when}
\be
2n\log_2m <m
\ee
\noindent{Since we are primarily concerned with situations in which $m>>n$,
this is normally the most efficient of the three alternatives thus far discussed.  
More efficient encodings are certainly possible.  For instance one might imagine sending
differences between successive indices instead of the entire index.  For purposes of
obtaining reasonable estimates of the communications cost without overcomplicating the
analysis, the third alternative is a reasonable choice, and we will proceed by
restricting the protocol to this case in subsequent discussions.}

Once Alice has received Bob's information, she compares Bob's basis choices
with her own and informs Bob of the results.  Alice can accomplish this by
sending Bob a single bit corresponding to each of the photons Bob detected,
giving a total of $2n$ bits.  

We now augment the protocol with provisions that will prevent Eve from making
the so called man-in-the-middle attack.  In this attack, Eve interposes
herself between Alice and Bob, measuring Alice's pulses on the quantum channel as
though she were Bob, and generating a distinct set of pulses to send to Bob as though
she were Alice.  In all her subsequent correspondence with Alice over the
classical channel, she responds just as Bob would, and in all correspondence
with Bob she plays the role of Eve. After the first phase of the protocol, Eve
has two blocks of sifted keys, one of which she shares with Alice and the other
with Bob.  Assuming she can continue this attack through the error correction and
privacy amplification phases, she will have completely compromised Alice and Bob's
ability to use the keys to transmit secret information. In fact, Eve will be able to
decipher any encrypted information sent between Alice and Bob, always passing the
ciphertext to the intended recipient so that neither Alice nor Bob is any the wiser.  

In order to prevent this state of affairs, it is necessary to provide an
authentication mechanism to guarantee that the transmissions received by
Bob were sent by Alice, not Eve, and that the transmissions received by
Alice were sent by Bob.  Wegman and Carter \cite{wc} describe an authentication
technique that is well suited to this problem.  The authentication works
as follows.  Alice and Bob first agree upon a suitable space of hash functions
to be used for authentication.  All details of their agreement may be 
revealed to Eve without compromising the authentication.  For each message
that is to be authenticated, Alice picks a hash function from the space that is known
to Bob, but not to Eve.  She does this by using a string of secret bits that is
known only to herself and Bob as an index to select the hash function.  She then
applies the hash function to the block of raw data to produce an authentication key.
This authentication key is transmitted to Bob along with the message.  Bob uses the
same string of secret bits to pick the same hash function, applies it to the
message, and compares the result with the authentication key sent by Alice.
If they match, Bob  concludes that Alice, and not Eve was the sender of the message.
Wegman and Carter describe a class of hash functions such that the probability that
Eve can generate the correct authentication key without knowing the index used is
vanishingly small. Let ${\cal M}_1$ denote the precondition that Eve has obtained
a copy of the message to be authenticated and ${\cal T}^{(E)}_1$ denote the set of
outcomes in which Eve guesses the tag for the message. The probability of such
an outcome is 
\be
\label{E:authprob}
{\cal P}({\cal T}^{(E)}_1|{\cal M}_1) = 2^{-g_{auth}}~,
\ee
\noindent where $g_{auth}$ depends on the space of hash functions Alice and
Bob have chosen to use for the protocol.  It can be made as large as desired
by making the space sufficiently large.  Alice and Bob do pay a price for increased
confidence. A larger space of functions requires a larger set of indices, and thus a
longer string of secret bits must be sacrificed to perform the authentication.  The
other restriction on the protocol is that a new hash function, and thus a new index,
must be used for each message to be authenticated if we desire to maintain this
upper bound on Eve's ability to spoof the authentication process. If we allow Eve
to obtain one prior message and tag, denoted as ${\cal M}_1{\cal T}_1$, and then
allow her to obtain the next message, denoted as ${\cal M}_2$, as well as the
information that Alice and Bob intend to use the same hash function 
for both, her chances of guessing the second tag improve only slightly to
\be
{\cal P}({\cal T}^{(E)}_2|{\cal M}_1{\cal T}_1{\cal M}_2) = 2^{1-g_{auth}}~.
\ee
\noindent If we allow additional messages to be authenticated using the same
hash function, Wegman and Carter's analysis provides no upper bound on Eve's
ability to produce a correct authentication tag.  Although it would be more
efficient to allow the same hash function to be applied exactly twice, we will
consider the simpler case in which a new hash function is picked for each message.  

We now consider which transmissions need to be authenticated. We will not
attempt to authenticate the communications on the quantum channel. Any
man-in-the-middle attack by Eve on the quantum channel will become evident
when the error correction process reveals that there is no correlation between
the Alice and Bob's sifted strings. Eve also gains no advantage from a selective
attack on a subset of the pulses sent by Alice. Suppose Alice and Bob predict an
expected number of errors $\langle e_T\rangle$ based on the known physical properties
of the channel and their equipment.  They then select a maximum threshold value 
$e_T^{max}>\langle e_T\rangle$ so that, if the measured error rate is greater
than the threshold, that is, if 
\be
e_T^{meas}>e_T^{max}
\ee
\noindent then they will conclude that Eve has interfered with the quantum
channel, perhaps by making a man-in-the-middle attack, and will terminate
processing for that block of data.  There is still the possibility that
$e_T^{meas}<e_T^{max}$, but that Eve has nevertheless corrupted the quantum
channel. In this case, the protocol proceeds as usual, 
the additional errors are identified and corrected, and the information leaked
to Eve is removed during privacy amplification.  This type of attack falls under the
category of an attack on secrecy that has no effect other than to reduce the overall
generation rate of key material.  

Authentication is required for the classical discussion of Alice and Bob's
choice of bases and the identification of the pulses received by Bob. If there is no 
authentication of this step, Eve can successfully mount the man-in-the-middle
attack which results in two sets of keys, one shared between Alice and herself, the
other between Bob and herself. Authentication guarantees Alice and Bob that they are
working with the same subset of the pulses sent by Alice and that any 
remaining errors are due to physical imperfections of the equipment or
attempts by Eve to measure, and therefore disturb, the pulses sent by Alice.  

The authentication of the classical discussion results in a cost to the
overall rate of quantum key generation, since some of the secret bits
produced by previous iterations of the protocol must be sacrificed to
generate an authentication tag that Alice or Bob can validate but that
Eve cannot forge.  Wegman and Carter \cite{wc} show that 
the size of the secret index required to select a hashing function is
\be
\label{E:hashkeysize}
w\left( g,c\right) = 4\left( g+\log_2 \log_2 c\right) \log_2 c
\ee
\noindent
where $c$ is the length in bits of the message to be authenticated and $g$ is
the length in bits of the authentication tag. Note that $g$ determines the degree of 
confidence in the authentication according to eq. (\ref{E:authprob}), in which $g$ is 
denoted $g_{auth}$. 
We define the parameter 
$g_{auth}$ as the length of the authentication tags used in this protocol.
The first message to be authenticated is Bob's transmission of the indices of the
detected pulses and his choice of basis for each. The length of the message is
\be
\label{E:c1}
c_1=2n\left( 1 + \log_2m\right)~,
\ee
\noindent
giving an authentication cost of 
\be
\label{E:w1}
w_1=4{\Big \{} g_{auth} + \log_2\log_2{\Big [} 2n{\Big (}
1+\log_2m{\Big )}{\Big ]}{\Big \}}
      \log_2{\Big [} 2n{\Big (} 1+\log_2m{\Big )}{\Big ]}~.
\ee
\noindent The second message is Alice's transmission of her choice of basis
for the pulses that Bob detected. The length of the message is 
\be
\label{E:c2}
c_2=2n~,
\ee
\noindent and the corresponding authentication cost is
\be
\label{E:w2}
w_2=4{\Big [} g_{auth}+\log_2\log_2\left( 2n\right){\Big ]} \log_2\left(
2n\right)~.
\ee
\noindent
We consider next the load this phase imposes on the classical communication
channel. We assume that the communication protocol employs some form of
error-correction coding that increases the length of the message by a factor
$\chi_{EC}$ \cite{commec}. The protocol then breaks the message into packets of length
$m_p$ and adds an amount of frame 
overhead $f_o$ to each packet.  Finally, the authentication tag is sent as a
single  packet, on the assumption that the tag size after encoding for error
correction is less than $m_p$:
\be
\chi_{EC}g_{auth}\leq m_p
\ee
\noindent This is a reasonable assumption, as we generally will take
$g_{auth}<50$,
$\chi_{EC}\approx 2$, so that we only require $m_p>100$, which is easily
achieved for 
typical optical ground-to-satellite links or terrestrial fiber 
optic channels.  The number of packets sent from Bob to Alice is then 
\be
{\cal N}_{B\rightarrow A}^{sift}=\Big\lceil{\chi_{EC}2n\left( 1+\log_2m\right)\over 
                         m_p}\Big\rceil+1~,
\ee
\noindent and the load in bits carried by the channel is, approximately, 
\bea
\label{E:cbasift}
  {\cal C}_{B\rightarrow A}^{sift} &\simeq&
      \left( 1+{f_o\over m_p}\right) 
          {\Big [}\chi_{EC}2n\left( 1+\log_2m\right){\Big ]}\nonumber\\
      &&+ \left(\chi_{EC} g_{auth} + f_o\right)~,
\eea
\noindent where we have used the packetization approximation described in Appendix B.  
The communication from Alice to Bob 
required for sifting is given similarly by
\bea
  {\cal C}_{A\rightarrow B}^{sift} &\simeq&
      \left( 1+{f_o\over m_p}\right) 
          {\Big (}\chi_{EC}2n{\Big )}\nonumber\\
      &&+ {\Big (}\chi_{EC} g_{auth} + f_o{\Big )}~.
\eea

\vskip 10pt
\noindent {\it Error Correction Phase}

At this point Bob and Alice move on to the error correction phase.  We will
estimate the 
authentication, communication, and computational costs for a modified version
of the 
error correction 
protocol described by Bennett {\it et. al.\/}, \cite{bbbss}.  More 
efficient techniques have been developed, for example the 
``Shell" and ``Cascade" protocols described in \cite{brassardsalvail}, 
but the method described here is more suitable for
our purposes 
since it is simpler to analyze and can be expected to provide a practical upper
bound 
for the communications cost.  

At the beginning of the error correction phase, Alice and Bob each have a
string of $n$ 
bits.  The strings are expected to be nearly identical, but they will also
contain errors 
for which Alice and Bob disagree on the value of the bit.  It is the goal of
error 
correction to identify and remove all of these errors, so that Alice and Bob
can 
proceed with a high degree of certainty that the strings are identical.  
Error correction consists of three steps.  The first 
step is the error detection and correction step, which eliminates all or
almost all 
of the errors.  The validation step which follows eliminates any residual
errors and 
iteratively tests randomly chosen subsets of the string to generate a high
degree of 
confidence that the strings are identical.  The final step is authentication,
which 
protects against a man-in-the-middle attack by Eve during the error correction
process.  

At the beginning of the error detection and correction step, Alice and Bob
each 
shuffle the bits in their string using a random shuffle upon which they have
previously 
agreed.  The purpose of this shuffle is to separate bursts of errors so that
the 
errors in the shuffled string are uniformly distributed.  Alice and Bob may
use 
the same shuffle each time they process a new string of sifted bits, and 
security is not compromised if Eve has complete prior knowledge of the
shuffling 
algorithm, even including any random numbers used as parameters.  

The error detection and correction step is an iterative process.  Alice and
Bob 
begin each iteration $i$ by breaking their strings into shorter blocks.  The
block 
length is chosen so that the expected number 
of errors in each block is given by a parameter $\varrho$.  The number of 
blocks in the string is then
\be
J^{\left( i\right)} = {\Big\lceil}{e_T^{\left( i-1\right)}\over\varrho}{\Big\rceil}~.
\ee
\noindent
and the average number of bits per block is

\be
k^{\left( i\right)} = {n\over J^{\left( i\right)}}~,
\ee

\noindent 
where $e_T^{\left( i\right)}$ is the expected number of errors remaining 
after the $i$th, or at the beginning of the $i+1$st, iteration.  
In principle the parameter $\varrho $ could 
change from iteration to iteration.  We assume that it is a constant to
simplify 
the analysis.  Alice and Bob compute the parity of each 
of the blocks and exchange their results.  Blocks for which the parities match
necessarily contain at least one error.  For each of the blocks in which Alice
and 
Bob have detected an error, they isolate the erroneous bit by a bisective
search, 
which proceeds as follows.  Alice and Bob bisect one of the blocks containing
an error, 
that is, they divide it as evenly as 
possible into 2 smaller blocks.  Alice and Bob each compute the parity of one
of the blocks,
say the one that lies closer to the beginning of
the shuffled string, and exchange the results.  
If the parities do not match, the error is in the lower block.  If they match,
the error is 
in the upper block.  Alice and Bob then bisect the block that contains the 
error and proceed 
recursively until they find an erroneous bit.  Bob then inverts that bit in
his 
string, and in so doing 
the error is removed.  

We have described the bisective search as though the search were 
completed for any block containing a detected error before beginning the
bisection on 
the next block.  In fact, it is more efficient from a communications
standpoint to apply 
each bisection to all the blocks with detected errors at the same time,
exchange parities 
for all of the sub-blocks, and then to proceed recursively to the next
bisection.  This 
results in fewer, but larger, packets of data for each exchange between Bob
and Alice, 
thus reducing the overall frame overhead.  

When the bisective search is completed for all blocks in which an error is
detected, 
a new blocksize is computed based on the expected number of errors remaining,
the string 
is broken up into a new set of larger blocks, parity checks are compared for
the blocks, 
and bisective searches are made in those blocks containing detected errors.  
This process is repeated until there would be only one or two blocks in the
string 
for the next iteration, that is, until

\be
J^{\left( N_1+1\right)} \leq 2~,
\ee

\noindent
where $N_1$ is the number of iterations in the error correction and detection
step.  
An equivalent stopping criterion is that the blocksize for the subsequent
iteration is more 
than half the length of the string, that is

\be
k^{\left( N_1+1\right)} \geq {n\over 2}~,
\ee

\noindent
and the expected number 
of errors remaining after the final iteration satisfies

\be
e_T^{\left( r\right)}\equiv e_T^{\left( N_1\right)} \leq 2\varrho~.
\ee

We summarize here some important results that are needed for an analytical
description of 
the communications required to support this part of the error correction
phase.  
As shown in the appendix ``Statistical Results for Error Correction,"
the expected number of errors remaining after the $i$th
iteration is

\be
e_T^{\left( i\right)} \simeq \beta^i e_T^{\left( 0\right)}~,
\ee

\noindent
and the expected number of errors found and corrected in the $i$th iteration
is

\be
e_f^{\left( i\right)} \simeq \left( 1-\beta\right) \beta^{i-1} e_T^{\left(
0\right)}~,
\ee

\noindent
where $\beta $ is defined by the expression

\be
\beta\equiv{2\varrho - 1 + e^{-2\varrho} \over 2 \varrho}~.
\ee

\noindent
We obtain the number of iterations in the error detection and correction step
by setting
$J^{\left( N_1+1\right)} \leq 2$, which gives
\be
N_1 = {\Bigg\lceil}{\log_2{2\varrho \over e_T^{\left( 0\right)}}\over
\log_2\beta}{\Bigg\rceil}~,
\ee
\noindent
and the expected number of remaining errors becomes
\be
e_T^{\left( r\right)} \simeq \beta^{N_1} e_T^{\left( 0\right)} \simeq 2\varrho~.
\ee

The second step in the error correction phase, validation, 
is also iterative.  During each iteration, Alice and Bob 
select the same random subset of their blocks.  They compute the parities and
exchange 
them.  If the parities do not match, Alice and Bob execute a bisective search
to 
find and eliminate the error.  Iterations continue until $N_2$ consecutive 
matching parities are found.  At this point, Alice and Bob conclude that their
strings 
are error free.  As shown in the appendix ``Statistical Results for Error
Correction," the probability of one or more errors remaining is 

\be
{\cal P}({\rm errors~after~validation}) \leq 
   e^{2 \varrho} \left( {1 \over 2} \right)^{N_2}
\ee

\noindent
the expected number of iterations in which no error is found is given, to a
good 
approximation, by

\be
N_2^{\left( n\right)} \simeq N_2 + e_T^{\left( r\right)} \simeq N_2 +
2\varrho~,
\ee

\noindent
and the expected number of iterations in which an error is found is 

\be
N_2^{\left( f\right)} \simeq e_T^{\left( r\right)} \simeq 2\varrho~.
\ee

The selection of the same random subsets for validation can be accomplished by
using a 
deterministic random number generator \cite{randgen} and resetting the random seed to a
predetermined 
value at the beginning of the validation phase.  We will assume that Eve has
complete 
knowledge of the algorithm and the random seed as well.  Note that it is to
Alice and Bob's 
advantage to keep the algorithm and seed secret, since Eve can make use of
this information to interpret 
the parities she intercepts on the public channel, but it is not essential to
the 
secrecy of the overall result, since we eliminate all the information Eve can
have 
obtained in privacy amplification.    

The last step in the error correction phase is authentication.  Up until now,
Alice and Bob 
have made no attempt to authenticate their exchange of parity information on
the classical  
channel.  Eve could mount a man-in-the-middle attack during the error
correction phase 
that would fool Alice and Bob into correcting the wrong set of bits.  This
would not give 
Eve any additional information about the secret string, but it could result in
Alice and 
Bob believing that their strings are identical when in fact they are not.
Even if one bit 
is different, the privacy amplification phase will produce strings that are
completely 
uncorrelated, and Alice and Bob will still believe that their strings are
identical.  
The solution to this problem is for Alice and Bob to verify that their strings
are the same 
at the end of the error correction phase.  This effectively authenticates
their prior 
communications, since any successful attempt by Eve to steer the error
correction 
process will be immediately apparent.  

This approach presupposes that Alice and Bob can verify that their strings are
the same 
without leaking too much additional information to Eve.  This can be
accomplished if 
Alice and Bob apply the same hash function to their strings and compare the
resulting 
tag.  This does not provide an absolute guarantee that the strings are the
same, but 
if the hash function is chosen as described in \cite{wc}, the probability that two 
different strings will yield the same tag is 

\be
{\cal P}\left({\rm same~tag,~two~strings}\right) = 2^{-g_{EC}}~,
\ee

\noindent
where $g_{EC}$ is the length of the tag.  This gives a high degree of
confidence that 
the strings are identical even for relatively short ($g_{EC}\sim 30$) tags.  
The price Alice and Bob have to pay for this 
is that they must use a portion of the secret bits obtained from previous
iterations 
of the protocol to select the hash function, indicate whether the keys match,
and 
authenticate their transmissions.  

We introduce a specific protocol for Alice and Bob to carry out the
authentication 
step for purposes of estimating the costs associated with this step.  Alice
and Bob 
agree to set aside a portion of the secret bits derived from each block of the
quantum 
transmission for use in processing subsequent blocks.  Some of these bits are
used for 
authentication during the sifting phase as previously discussed.   Some
additional bits 
are required to use as a key to select the hash function for the equivalence check.  
The size of this
key is given 
by eq. (\ref{E:hashkeysize}), where 

\be
\label{E:c3}
c_3=n~,
\ee

\noindent
is the length of the string to be hashed and $g_{EC}$
is the length of the tag, so that the authentication cost for equivalence
checking is

\be
\label{E:w3}
w_3=4\left( g_{EC} + \log_2\log_2 n\right)
      \log_2 n~.
\ee

\noindent
Alice and Bob both compute an equivalence tag using this hash function, and
Bob 
sends his to Alice.  Bob must also authenticate his message, since otherwise
Eve 
can mount a man-in-the-middle attack in which she simply sends an arbitrary
tag 
to Alice, convincing her that her string doesn't match Bob's string when, in
fact, it 
does.   Although this is only a denial-of-service attack, Alice and Bob will
not 
detect the attack unless Bob authenticates his message.  Since
Bob's message to Alice is of length

\be
\label{E:c4}
c_4=g_{EC}~,
\ee

\noindent
and the authentication tag is of length $g_{auth}$, the authentication cost 
for the transmission is

\be
\label{E:w4}
w_4=4\left( g_{auth} + \log_2\log_2 g_{EC}\right)
      \log_2 g_{EC}~.
\ee

\noindent
If Alice determines that her equivalence tag matches the one Bob sent to her,
and if 
the authentication tag agrees as well, she indicates that the authentication
was 
successful be sending $\tilde g_{EC}$ secret bits to Bob.  The authentication
cost 
for this step is

\be
\label{E:w5}
w_5 = \tilde g_{EC}~.
\ee

\noindent
bits to signal her agreement to Bob.  Alice must also authenticate this
message 
to protect against a man-in-the-middle attack by Eve.  The length of the 
message to be authenticated is:

\be
\label{E:c6}
c_6 = \tilde g_{EC}
\ee

\noindent
and the authentication tag is of length $g_{auth}$ so that the authentication 
cost is 

\be
\label{E:w6}
w_6 = 4\left( g_{auth} + \log_2\log_2 \tilde g_{EC}\right)
      \log_2 \tilde g_{EC}~.
\ee

\noindent
Classical communications between Alice and Bob are required in each step 
of the error correction phase.  
During each iteration of the error detection and correction step, Bob sends 
to Alice a single transmission containing the parities computed for each of
the 
$J^{\left( i\right)}$ blocks.  Alice sends a similar transmission to Bob
containing 
her parities.  For each such iteration, the communication load in bits in 
each direction is approximately

\be
\left(1+{f_o \over m_p}\right) \chi_{EC}J^{\left(i\right)}~,
\ee

\noindent
where we have used the packetization approximation described in the appendix.  
Then, for each block containing a detected error, Alice and Bob bisect the
block and 
exchange the parities of the the bisected blocks.  This requires a
communication in 
each direction of 

\be
\left(1+{f_o \over m_p}\right) \chi_{EC}e_f^{\left(i\right)}
\ee

\noindent
bits.  The bisective search repeats for a total of 
$\lceil \log_2 k^{\left( i\right)}\rceil$ iterations.  The total
communications load 
in bits for this step is thus
\be
\Delta{\cal C}_{B\rightarrow A}^{EC,1} = \Delta{\cal C}_{A\rightarrow B}^{EC,1} = 
\sum_{i=1}^{N_1} \left[
   \left(1+{f_o \over m_p}\right) \chi_{EC}J^{\left(i\right)} +
   \lceil \log_2 k^{\left( i\right)}\rceil 
      \left(1+{f_o \over m_p}\right) \chi_{EC}e_f^{\left(i\right)}
\right]~.
\ee
For the next step, Alice and Bob each compute the parity for a random subset
of 
their strings and exchange the results.  One bit of parity information is sent
in each 
direction, so that the communication load is $\chi_{EC}+f_o$ bits in each
direction.  
If the parities are different, Alice and Bob carry out a bisective search for
the 
error, resulting in $\lceil 1+\log_2{n \over 2}\rceil$ transmissions of
$\chi_{EC}+f_o$ 
bits in each direction.  This is repeated until $N_2$ successive parities
match.  
The communications load for this step is then
\be
\Delta{\cal C}_{B\rightarrow A}^{EC,2} = \Delta{\cal C}_{A\rightarrow B}^{EC,2} = 
   N_2^{\left( n\right)} \left(\chi_{EC}+f_o\right) +
   N_2^{\left( f\right)} \Big\lceil 1+\log_2{n \over 2}\Big\rceil
\left(\chi_{EC}+f_o\right)~,
\ee
\noindent
where $N_2^{\left( n\right)}$ is the number of iterations which do not find an
error 
and $N_2^{\left( f\right)}$ is the number of iterations which do find an
error.  

In the third step, authentication, Bob sends to Alice an equivalence tag of
length $g_{EC}$ 
and an 
authentication tag of length $g_{auth}$ giving a communication load of
\be
\Delta{\cal C}_{B\rightarrow A}^{EC,3} = 
   \chi_{EC}\left( g_{EC}+g_{auth}\right) +f_o
\ee
\noindent
bits.  Alice sends a confirmation string of length $\tilde g_{EC}$ and an
authentication 
tag of length $g_{auth}$ for a communications load of 
\be
\Delta{\cal C}_{A\rightarrow B}^{EC,3} = 
   \chi_{EC}\left(\tilde g_{EC}+g_{auth}\right) +f_o
\ee
\noindent
bits.  Note that we have implicitly assumed that the tags are short enough to
send 
in a single packet.  Since the tags are typically less than 50 bits, this
amounts 
to a requirement that the packet size $m_p$ exceed 200 bits so as to
accomodate 
two tags plus error correction codes. This is a modest constraint on the
optical 
communications system.  

Collecting all of the contributions to the communications load, we obtain the
following 
expressions for the load during the error correction phase:
\bea
\label{E:cbaec1}
{\cal C}_{B\rightarrow A}^{EC}&=&
\sum_{i=1}^{N_1} \left[
        \left(1+{f_o \over m_p}\right) \chi_{EC}J^{\left(i\right)} +
        \Big\lceil \log_2 k^{\left( i\right)}\Big\rceil 
           \left(1+{f_o \over m_p}\right) \chi_{EC}e_f^{\left(i\right)}
        \right] \nonumber\\
   &&+N_2^{\left( n\right)} \left(\chi_{EC}+f_o\right) +
     N_2^{\left( f\right)} 
        \Big\lceil 1+\log_2{n \over 2}\Big\rceil \left(\chi_{EC}+f_o\right)
\nonumber\\
   &&+\chi_{EC}\left(g_{EC}+g_{auth}\right) + f_o~,
\eea
and
\bea
{\cal C}_{A\rightarrow B}^{EC}&=&
\sum_{i=1}^{N_1} \left[
        \left(1+{f_o \over m_p}\right) \chi_{EC}J^{\left(i\right)} +
        \Big\lceil \log_2 k^{\left( i\right)}\Big\rceil 
           \left(1+{f_o \over m_p}\right) \chi_{EC}e_f^{\left(i\right)}
        \right] \nonumber\\
   &&+N_2^{\left( n\right)} \left(\chi_{EC}+f_o\right) +
     N_2^{\left( f\right)} 
        \Big\lceil 1+\log_2{n \over 2}\Big\rceil \left(\chi_{EC}+f_o\right)
\nonumber\\
   &&+\chi_{EC}\left(\tilde g_{EC}+g_{auth}\right) + f_o~.
\eea

\noindent
The communications load can be expressed in terms of the fundamental
quantities 
$n, e_T^{\left( 0\right)}, {\rm and}~\varrho$ by using eqs. (\ref{E:blocks}), 
(\ref{E:bitsinblock}), (\ref{E:errsfound}), (\ref{E:n2n}), and (\ref{E:n2f}).  
The necessary 
summations are elementary:
\bea
\label{E:sumji}
\sum_{i=1}^{N_1} J^{\left( i\right)}
   &\simeq&{e_T^{\left( 0\right)} \over
\varrho}\sum_{i=1}^{N_1}\beta^{i-1}\nonumber\\
   &\simeq&{e_T^{\left( 0\right)} \over \varrho}
           {{1-\beta^{N_1}}\over {1-\beta}}\nonumber\\
   &\simeq&{e_T^{\left( 0\right)} \over \varrho}
           {{1-{{2\varrho} \over {e_T^{\left( 0\right)}}}  \over 
              {1-{{2\varrho - 1 + e^{-2\varrho}}\over{2\varrho}}}}}\nonumber\\
   &\simeq&2~{{e_T^{\left( 0\right)}-2\varrho} \over {1-e^{-2\varrho}}}~,
\eea
\noindent
where we have approximated the results by disregarding the rounding up of real
quantities to integers, and 
\bea
\label{E:sumlogkiefi}
\sum_{i=1}^{N_1} \log_2 k^{\left( i\right)} e_f^{\left( i\right)}
   &\simeq&\sum_{i=1}^{N_1}
      \log_2\left({\varrho n}\over {e_T^{\left( i-1\right)}}\right)
e_f^{\left( i\right)}
      \nonumber\\
   &\simeq&\sum_{i=1}^{N_1}
      \log_2\left({{\varrho n}\over {\beta^{i-1}e_T^{\left( 0\right)}}}\right)
      \left( 1-\beta\right) \beta^{\left( i-1\right)}e_T^{\left( 0\right)}
      \nonumber\\
   &\simeq&e_T^{\left( 0\right)} \left\{
      \log_2\left({{\varrho n}\over e_T^{\left( 0\right)}}\right) \left(
1-\beta^{N_1}\right) -
      {{\beta\log_2\beta} \over {1-\beta}}
         \left[ 1-N_1\beta^{N_1-1}+\left( N_1-1\right) \beta^{N_1}\right]
      \right\}~,
\nonumber\\
\eea
\noindent
where a similar approximation is made.  Inserting these results and
eqs.(\ref{E:n2n}) and (\ref{E:n2f}) into eq.(\ref{E:cbaec1})
gives the following expressions for the communications load:
\bea
{\cal C}_{B\rightarrow A}^{EC} &\simeq&2~\left(1+{f_o \over m_p}\right) \chi_{EC}
        {{e_T^{\left( 0\right)}-2\varrho} \over {1-e^{-2\varrho}}}
   \nonumber\\
   &&+\left(1+{f_o \over m_p}\right) \chi_{EC}~e_T^{\left( 0\right)}
   \nonumber\\
   &&\cdot\left\{
      \log_2\left({{\varrho n}\over e_T^{\left( 0\right)}}\right) \left(
1-\beta^{N_1}\right) -
      {{\beta\log_2\beta} \over {1-\beta}}
         {\Bigg [} 1-N_1\beta^{N_1-1}+\left( N_1-1\right) \beta^{N_1}{\Bigg ]}
     \right\}
   \nonumber\\
   &&+\left(N_2+2\varrho\right) \left(\chi_{EC}+f_o\right)
   \nonumber\\
   &&+2\varrho\Bigg\lceil 1+\log_2{n \over 2}\Bigg\rceil \left(\chi_{EC}+f_o\right)
\nonumber\\
   &&+\chi_{EC}\left(g_{EC}+g_{auth}\right) + f_o~,~
\eea
and
\bea
{\cal C}_{A\rightarrow B}^{EC} &\simeq&
2~\left(1+{f_o \over m_p}\right) \chi_{EC}
        {{e_T^{\left( 0\right)}-2\varrho} \over {1-e^{-2\varrho}}}
   \nonumber\\
   &&+\left(1+{f_o \over m_p}\right) \chi_{EC}~e_T^{\left( 0\right)}
   \nonumber\\
   &&\cdot\left\{
      \log_2\left({{\varrho n}\over e_T^{\left( 0\right)}}\right) \left(
1-\beta^{N_1}\right) -
      {{\beta\log_2\beta} \over {1-\beta}}
         {\Big [} 1-N_1\beta^{N_1-1}+\left( N_1-1\right) \beta^{N_1}{\Big ]}
     \right\}
   \nonumber\\
   &&+\left(N_2+2\varrho\right) \left(\chi_{EC}+f_o\right)
   \nonumber\\
   &&+2\varrho\Bigg\lceil 1+\log_2{n \over 2}\Bigg\rceil \left(\chi_{EC}+f_o\right)
\nonumber\\
   &&+\chi_{EC}\left(\tilde g_{EC}+g_{auth}\right) + f_o~.
\eea

Since parity information is exchanged over a classical channel, and since we 
assume that all classical communications are intercepted and correctly
interpreted by 
Eve, we must therefore assume that some information about the strings shared
by Alice and Bob 
is leaked to Eve during the error correction phase.  The degree to which 
this protocol leaks such information is an important characteristic of the
protocol.  
As was seen in Section 3.1.3, the theoretical lower bound on this leakage is given
by ({\it cf} eq. (\ref{42}))
\be
q_{\rm min} = n h\left({e_T^{\left( 0\right)} \over n}\right)~,
\ee
\noindent
where $h$ is the binary entropy function.  We estimate the leakage $q_p$
associated 
with our error correction protocol by counting the parity bits that were
exchanged 
during error correction.  During each iteration of the first step, 1 bit of
parity is 
leaked for each of the $J^{\left( i\right)}$ blocks, and an additional 
$\lceil\log_2 k^{\left( i\right)}\rceil$ bits is leaked for each block in
which 
an error was detected.  During the second step, one bit is leaked for each
iteration 
that does not reveal an error and $\lceil 1+\log_2{n\over 2} \rceil$ bits are
leaked 
for each iteration that does reveal an error.  The total is thus

\bea
q_p &=&\sum_{i=1}^{N_1} \left( J^{\left( i\right)} +
           \log_2 k^{\left( i\right)}~e_f^{\left( i\right)} \right)
      \nonumber\\
      && + N_2^{\left( n\right)} + N_2^{\left( f\right)}\Big\lceil 1+\log_2{n\over 2}
\Big\rceil~.
\eea

\noindent
Using eqs. (\ref{E:sumji}), (\ref{E:sumlogkiefi}, (\ref{E:n2n}), and (\ref{E:n2f}) , 
this becomes

\bea
q_p &\simeq& 
    2~{{e_T^{\left( 0\right)}-2\varrho} \over {1-e^{-2\varrho}}}
   \nonumber\\
   &&+e_T^{\left( 0\right)} \left\{
      \log_2\left({{\varrho n}\over e_T^{\left( 0\right)}}\right) \left(
1-\beta^{N_1}\right) -
      {{\beta\log_2\beta} \over {1-\beta}}
         \left[ 1-N_1\beta^{N_1-1}+\left( N_1-1\right) \beta^{N_1}\right]
      \right\}
   \nonumber\\
   &&+\left(N_2+2\varrho\right) + 2\varrho\Big\lceil 1+\log_2{n \over 2}\Big\rceil~.
\eea

\begin{figure}[htb]
\vbox{
\hfil
\scalebox{0.6}{\rotatebox{0}{\includegraphics{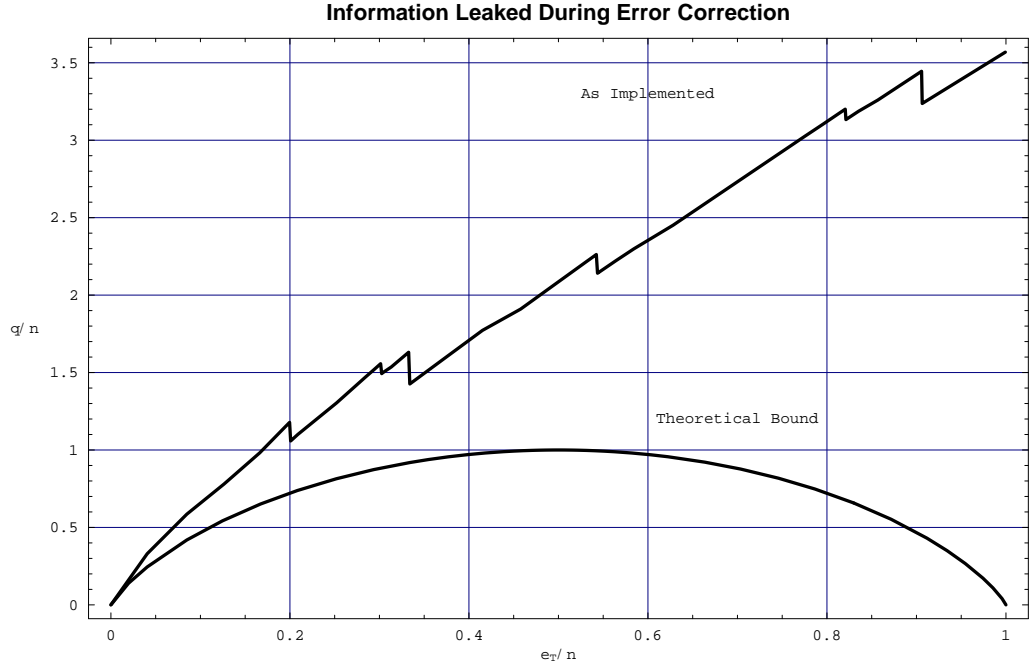}}}
\hfil
\hbox to -1.25in{\ } 
}
\bigskip
\caption{%
Information Leaked During Error Correction
}
\label{F:qandhvsrc}
\end{figure}

\begin{figure}[htb]
\vbox{
\hfil
\scalebox{0.6}{\rotatebox{0}{\includegraphics{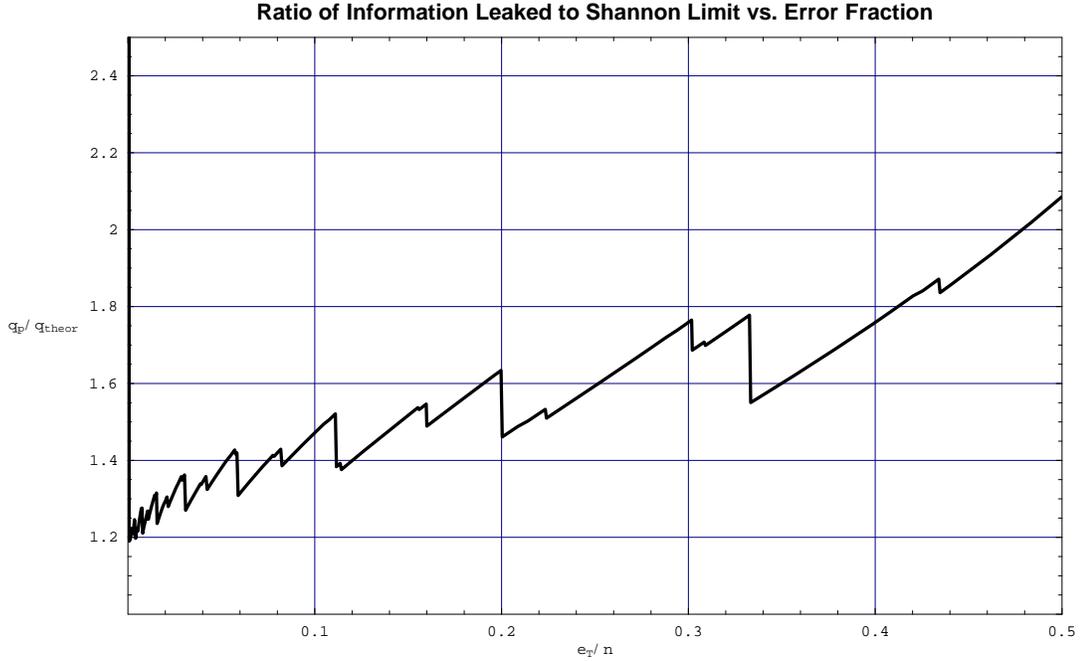}}}
\hfil
\hbox to -1.25in{\ } 
}
\bigskip
\caption{%
Ratio of Information Leaked to Shannon Limit {\it versus} Error Fraction
}
\label{F:qoverhvsrc}
\end{figure}

Figure \ref{F:qandhvsrc} is a comparison of the information leaked by the protocol to the
theoretical 
minimum for error correction parameters $\varrho = 0.5$ and 
$N_2=30$ and for a sifted string 
blocksize $n = 2 \times 10^5$ bits.  The results are plotted as a function of 
the error fraction 
$e_T^{\left( 0\right)}/n$.  As expected, the theoretical minimum represents a 
lower bound for the result predicted for the protocol.  The ratio of the
predicted leakage 
to the theoretical minimum, $q_p/q_{\rm min}$, is shown in Figure \ref{F:qoverhvsrc}.  
The ratio diverges 
at very low error rates due to the fact that some parity bits are exchanged
even if there 
are no errors in the string.  (This divergence is not apparent on the scale of the 
figure.)  The discontinuities in the curve occur at points where an increase in the 
error rate causes additional iterations 
at some point in the protocol.  For error fractions between
2\% and 10\%, 
the ratio fluctuates between about 1.2 and 1.5, indicating that the actual
protocol can 
be expected to leak up to 50\% more information than the theoretical minimum.  

\begin{figure}[htb]
\vbox{
\hfil
\scalebox{0.6}{\rotatebox{0}{\includegraphics{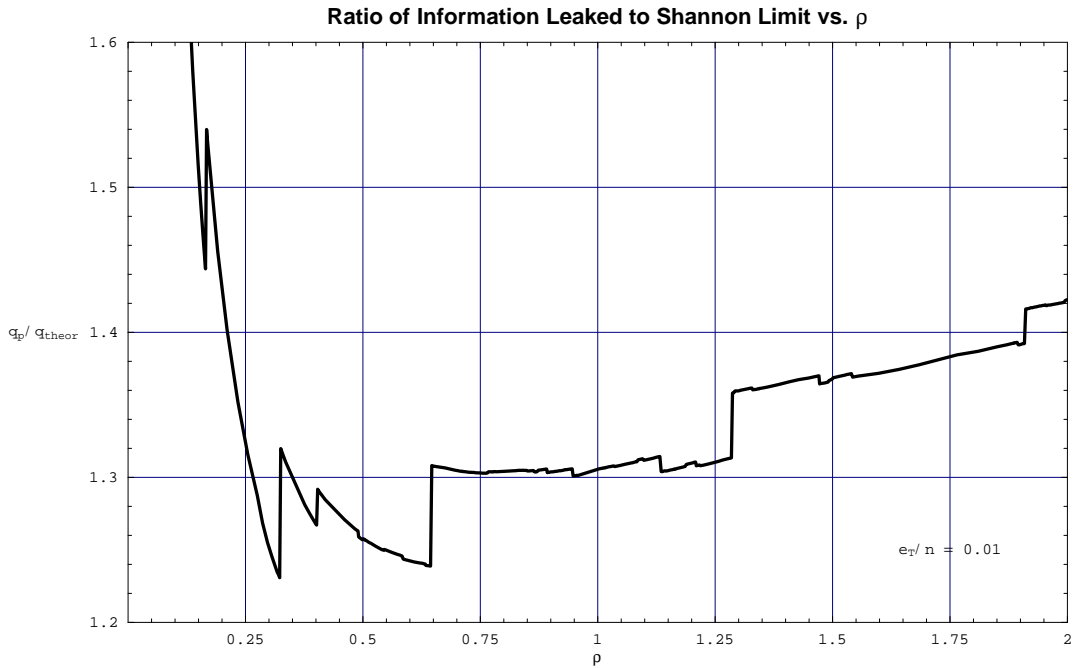}}}
\hfil
\hbox to -1.25in{\ } 
}
\bigskip
\caption{%
Ratio of Information Leaked to Shannon Limit {\it versus} $\varrho$
}
\label{F:qoverhvsrho}
\end{figure}

Figure \ref{F:qoverhvsrho} shows the ratio $q_p/q_{\rm min}$ as 
a function of $\varrho$ for an
error fraction 
$e_T^{\left( 0\right)}/n$ of 1\%.  This indicates that a choice of $\varrho$ in the
neighborhood 
of 0.5 results in minimum leakage of information relative to the theoretical
minimum 
for this choice of sifted string size and error fraction.

\vskip 10pt
\noindent {\it Privacy Amplification Phase}

The general scheme of privacy amplification is described in \cite{bbcm} and
\cite{cw} .  
The hash functions map a sifted, error corrected string of length $n$ to a
string of length {\it L}, 
where
\be
L \equiv n-e_T^{(0)}-q-t-\nu-a-g_{pa}~.
\ee
\noindent
The resulting string is shorter than the sifted string by the number of bits
that Eve 
may have obtained by listening to the classical discussion, plus an additional
parameter $g_{pa}$.  
In \cite{bbcm} it is shown that the expected information, $I$, that Eve can retain
about the 
hashed string is bounded by a quantity that can made very small by a suitable
choice 
of $g_{pa}$:

\be
\label{E:infopa}
I \leq {{2^{-g_{pa}}}\over{\ln 2}}~.
\ee

Hash functions appropriate for privacy amplification are described by Carter 
and Wegman 
\cite{cw}.  The class
of hash functions used for authentication and equivalence checking is not 
practical for privacy amplification due to 
the much larger size of the output string.  The authentication hash functions 
are designed to produce output strings that are no more than half as long as the input 
string.  Since we wish to retain as much information as possible, it is clearly 
advantageous to use hash functions that can produce an output string that 
is nearly as long as the input 
string.  Furthermore, recall that the length of the
index for choosing 
an authentication hash function is 

\be
w\left( g,c\right) = 4\left( g+\log_2 \log_2 c\right) \log_2 c
\ee

\noindent
where $c$ and $g$ are the lengths of the input and output strings,
respectively.  For purposes 
of authentication and error correction, an output string of length $g \leq 50$
is adequate, 
and the length of the index is relatively short even for long input strings
due to the 
logarithmic factors.  In privacy amplification, where the output
string is 
nearly as long as the input string, this index is roughly 4 
times as 
long as the string to be hashed.  In contrast, the 
hash functions suitable for privacy
amplification 
are described by
two parameters, each as long as the input string, so that the total size of
the index is only twice as long as the input string.  Since the index represents 
shared secret bits that must be sacrificed in order to achieve privacy amplification, 
it is desirable to use the class of hash functions that requires the shorter index.  
The Carter-Wegman functions described 
in \cite{cw} are a good choice for privacy amplification since they are 
capable of producing keys nearly as long as the input and since they require 
shorter indices for their definition given the large size of the output strings.    

The error correction phase guarantees that the strings Alice and Bob have
obtained 
are identical to a high probability.  Bob and Alice implement privacy
amplification 
by agreeing on an index and applying the hash functions separately to their
strings.  
The resulting strings are identical and secret in the sense of privacy 
amplification ({\it cf} eq.(\ref{E:infopa})).
Note that the 
sifting protocol itself supplies random strings of sufficient length to define
the required 
hash index.  Bob's choice of basis for the $2n$ pulses he receives is one such
source.  
Another alternative is to compute the parities of the indices Bob sends to
Alice by 
which he identifies which pulses were detected by his equipment.  

The privacy amplification protocol requires no communications between Alice
and Bob, as described in Section 3.1.6 above.  
The security parameter $g_{pa}$ is an additional secrecy cost incurred due to
privacy 
amplification, but 
privacy amplification entails no additional authentication cost.

\vskip 10pt
\noindent {\it Total Continuous Authentication Cost}

The total continuous authentication cost is the number of bits from each block of sifted
bits 
that need to be sacrificed during the processing of the subsequent block to
provide 
authentication and equivalence checking as described above.  Collecting the
contributions
from eqs.(\ref{E:c1}), (\ref{E:w1}), (\ref{E:c2}), (\ref{E:w2}), 
(\ref{E:c3}), (\ref{E:w3}), (\ref{E:c4}), (\ref{E:w4}), (\ref{E:w5}), 
(\ref{E:c6}), and (\ref{E:w6}) the result is the following sum of six terms:
\bea
a &=& a\left(n,m\right)
\nonumber\\
&=&4{\Big \{} g_{auth} + \log_2\log_2{\Big [} 2n\left(
1+\log_2m\right){\Big ]}{\Big \}}
      \log_2{\Big [} 2n\left( 1+\log_2m\right){\Big ]}
   \nonumber\\
   &&+ 4{\Big [} g_{auth}+\log_2\log_2\left( 2n\right){\Big ]} \log_2\left(
2n\right)
   \nonumber\\
   &&+ 4\left( g_{EC} + \log_2\log_2 n\right)\log_2 n
   \nonumber\\
   &&+ 4\left( g_{auth} + \log_2\log_2 g_{EC}\right)\log_2 g_{EC}
   \nonumber\\
   &&+ \tilde g_{EC}
   \nonumber\\
   &&+ 4\left( g_{auth} + \log_2\log_2 \tilde g_{EC}\right)
      \log_2 \tilde g_{EC}
\nonumber\\
&=&\tilde g_{EC}+\sum_{j=1}^5w_j\left(g_j,c_j\left(\mu\right)\right)~,
\eea
as in eq.(\ref{152}).

For example, if we take $m=2\times 10^8$ bits, and $n=2\times 10^5$ bits to be the
processing 
block lengths of the raw and sifted strings, and if we set all security parameters
$g_i$ to 30, we obtain  
a total authentication cost of $9.5 \times 10^3$ bits per processing block.  
For a laser pulse repetition rate
of 10 GHz, 
$\tau = 10^{-10}$ sec, and the rate at which secret bits are consumed is 

\be
{a \over {m\tau}} = 4.7 \times 10^5 ~{\rm bits/second}~.
\ee

\subsubsection{System Load: Total Communications Requirements}

The total communications load is the number of bits transmitted in either
direction 
over the classical communications channel to support the sifting and error
correction 
protocols for a single block of data.  Combining eqs.(\ref{E:cbasift})
and (\ref{E:cbaec1}), the result for the Bob-to-Alice link is
\bea
  {\cal C}_{B\rightarrow A} &\simeq&
      \left( 1+{f_o\over m_p}\right) 
          \left[\chi_{EC}2n\left( 1+\log_2m\right)\right]\nonumber\\
      &&+ \left(\chi_{EC} g_{auth}+f_o\right)\nonumber\\
   &&+ \sum_{i=1}^{N_1} \left[
        \left(1+{f_o \over m_p}\right) \chi_{EC}J^{\left(i\right)} +
        \lceil \log_2 k^{\left( i\right)}\rceil 
           \left(1+{f_o \over m_p}\right) \chi_{EC}e_f^{\left(i\right)}
        \right] \nonumber\\
   &&+ N_2^{\left( n\right)} \left(\chi_{EC}+f_o\right) +
     N_2^{\left( f\right)} 
        \lceil 1+\log_2{n \over 2}\rceil \left(\chi_{EC}+f_o\right)
\nonumber\\
   &&+ \chi_{EC}\left(g_{EC}+g_{auth}\right) + f_o~,
\label{270}
\eea
\noindent
and the result for the Alice-to-Bob link is
\bea
  {\cal C}_{A\rightarrow B} &\simeq&
      \left( 1+{f_o\over m_p}\right) 
          \left(\chi_{EC}2n\right)\nonumber\\
      &&+ \left(\chi_{EC} g_{auth}+f_o\right) \nonumber\\
   &&+ \sum_{i=1}^{N_1} \left[
        \left(1+{f_o \over m_p}\right) \chi_{EC}J^{\left(i\right)} +
        \lceil \log_2 k^{\left( i\right)}\rceil 
           \left(1+{f_o \over m_p}\right) \chi_{EC}e_f^{\left(i\right)}
        \right] \nonumber\\
   &&+ N_2^{\left( n\right)} \left(\chi_{EC}+f_o\right) +
     N_2^{\left( f\right)} 
        \lceil 1+\log_2{n \over 2}\rceil \left(\chi_{EC}+f_o\right)
\nonumber\\
   &&+ \chi_{EC}\left(\tilde g_{EC}+g_{auth}\right) + f_o~.
\label{271}
\eea
\noindent
For large $m$ and $n$, the sifting transmission from Bob to Alice is by far
the largest term.  Eqs. (\ref{E:sumji}), (\ref{E:sumlogkiefi}, 
(\ref{E:n2n}), and (\ref{E:n2f})  can be used to express these in terms of 
$N_2, e_T^{\left( 0\right)}, {\rm and}~\varrho$ if desired.  
The throughput requirement ${\cal R}^{comm}$
is found by dividing the total load by the
transmission time $m\tau$ for a single block on the quantum channel:
\be
{\cal R}^{comm} = {{\cal C} \over {m\tau}}~.
\ee
\noindent
Evaluated for reasonable, conservative, values of the parameters ($m=2\times
10^8$ bits, 
$n=2\times 10^5$ bits, $e_T^{\left( 0\right)} = 2 \times 10^3$ bits, $\tau =
10^{-10}$ sec, 
$m_p = 1000$ bits, $\chi_{EC} = 2$, 
$f_o = 400$ bits, $\varrho = 0.5$, $g_{EC} = \tilde g_{EC} = 
g_{auth} = N_2 = 30$), we obtain a throughput requirement of
\be
\label{273}
{\cal R}^{comm}\Big\vert_{Bob-to-Alice}=1600 ~{\rm Mbps}
\ee
for the Bob-to-Alice link, and 
\be
\label{274}
{\cal R}^{comm}\Big\vert_{Alice-to-Bob}=60 ~{\rm Mbps}
\ee
for the Alice-to-Bob link.\footnote{
These values of the required communications bandwidth do not exceed the
capabilities of
currently available optical
classical communications technology. As we will see in Section 5.3.2 below,
the specific numerical
parameter values chosen in this example correspond to the case of various
free space quantum cryptography systems set up between a LEO satellite node and a
ground-based (or airplane-based) node. Existing classical optical satcom links
operating at 1550 nanometers wavelength providing duplex communications at rates
ranging from 51 Megabits per second to 1.244 Gigabits per second between a ground
station and GEO satellite have been developed \cite{geolite}. Thus
it is clear that these
communications requirements, in particular the 1.6 Gigabits per second requirement,
can be satisfied for a LEO satellite link, for which there is much less attenuation
than for the GEO satellite link. As discussed
below, it turns out that there are in fact {\it smaller} classical
communications bandwidth requirements than those given in
eqs.(\ref{273}) and (\ref{274}) for a free space quantum cryptography system between
a ground station and a GEO satellite, which can easily be accomodated by existing
communications systems, and the same is true for fiber-optic cable quantum
cryptography systems.}

\begin{figure}[htb]
\vbox{
\hfil
\scalebox{0.6}{\rotatebox{0}{\includegraphics{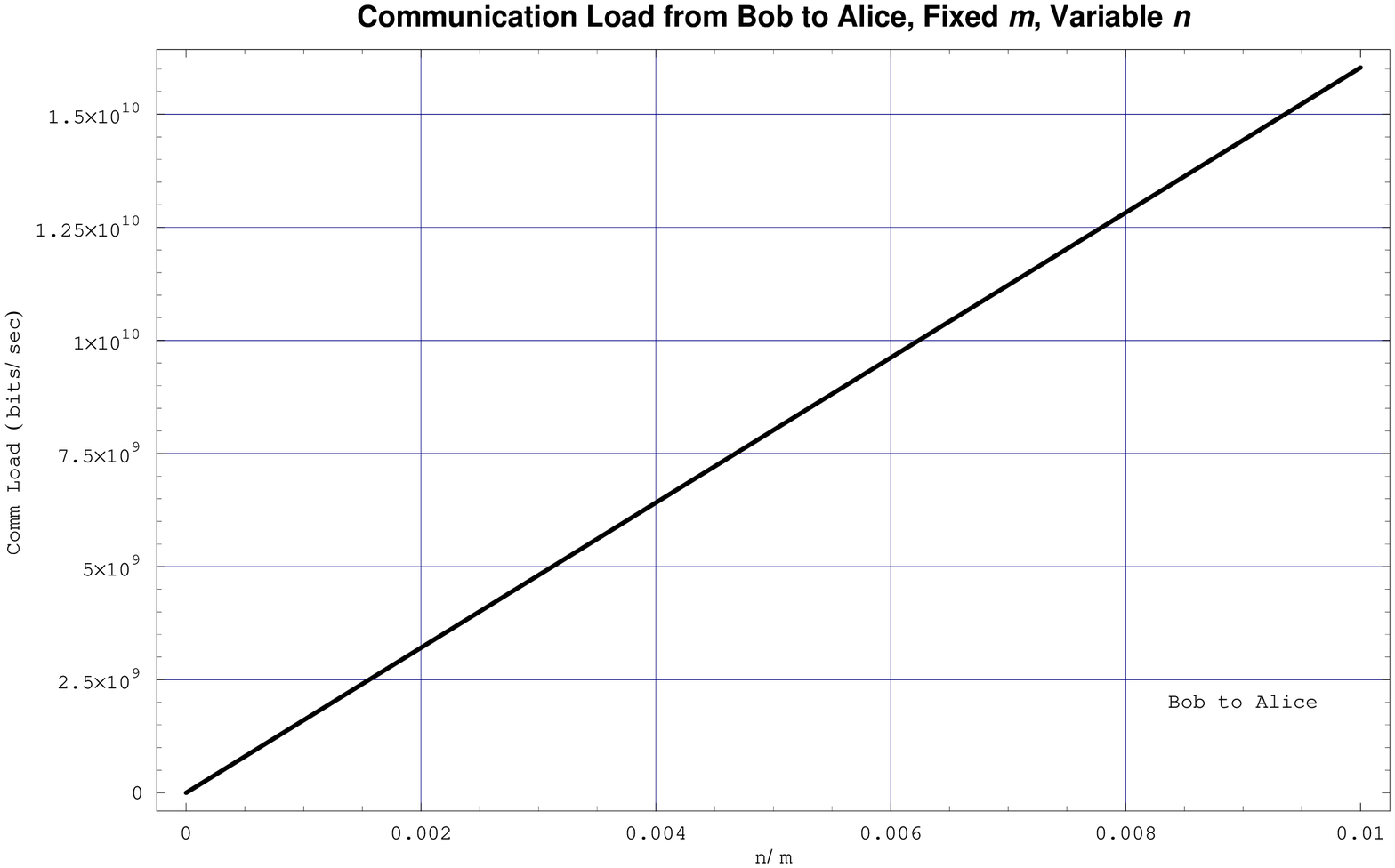}}}
\hfil
\hbox to -1.25in{\ } 
}
\bigskip
\caption{%
Communication Load from Bob to Alice, Fixed $m$, Variable $n$
}
\label{F:rbavsnoverm}
\end{figure}

\begin{figure}[htb]
\vbox{
\hfil
\scalebox{0.6}{\rotatebox{0}{\includegraphics{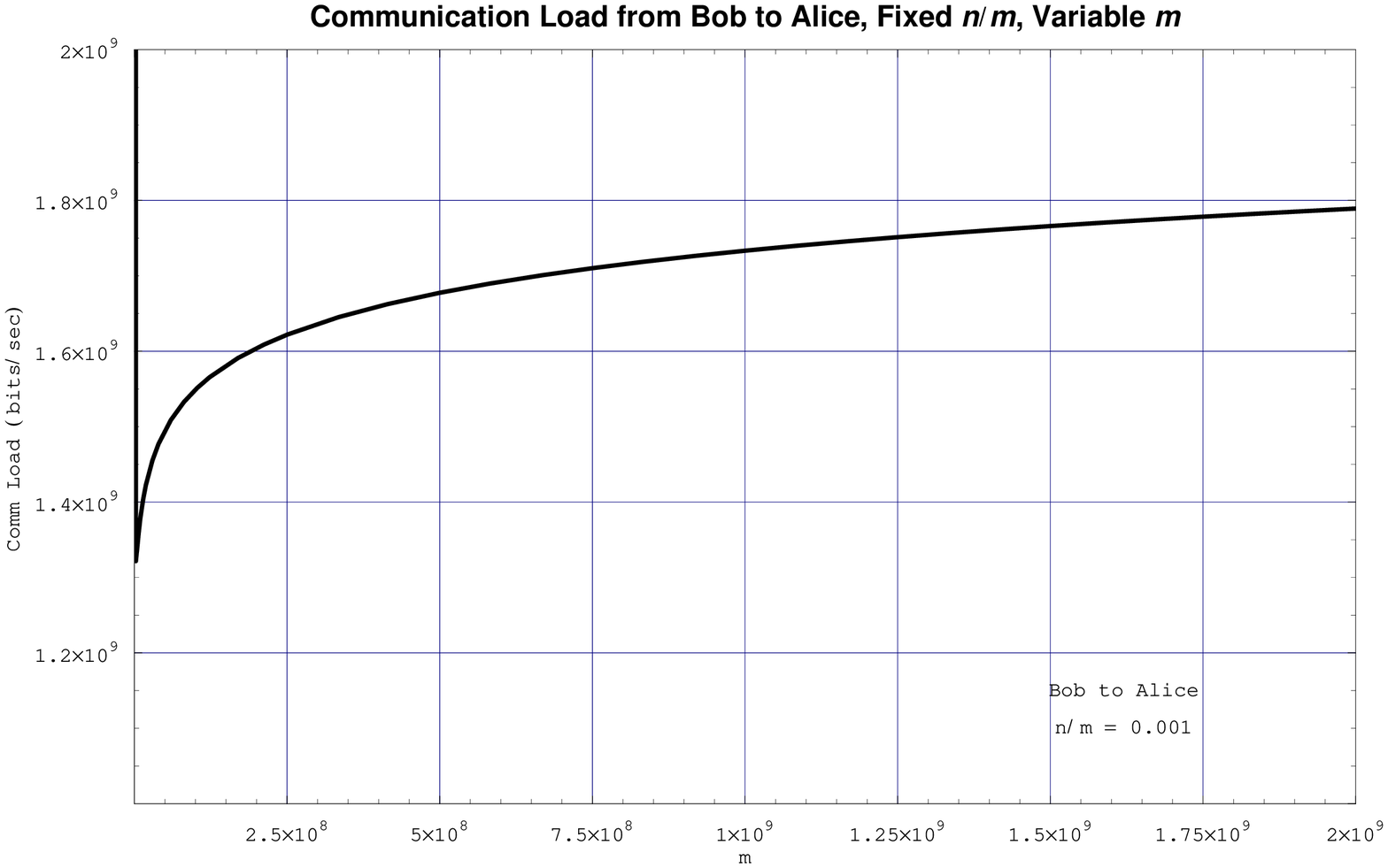}}}
\hfil
\hbox to -1.25in{\ } 
}
\bigskip
\caption{%
Communication Load from Bob to Alice, Fixed $n/m$, Variable $m$
}
\label{F:rbavsm}
\end{figure}

\begin{figure}[htb]
\vbox{
\hfil
\scalebox{0.6}{\rotatebox{0}{\includegraphics{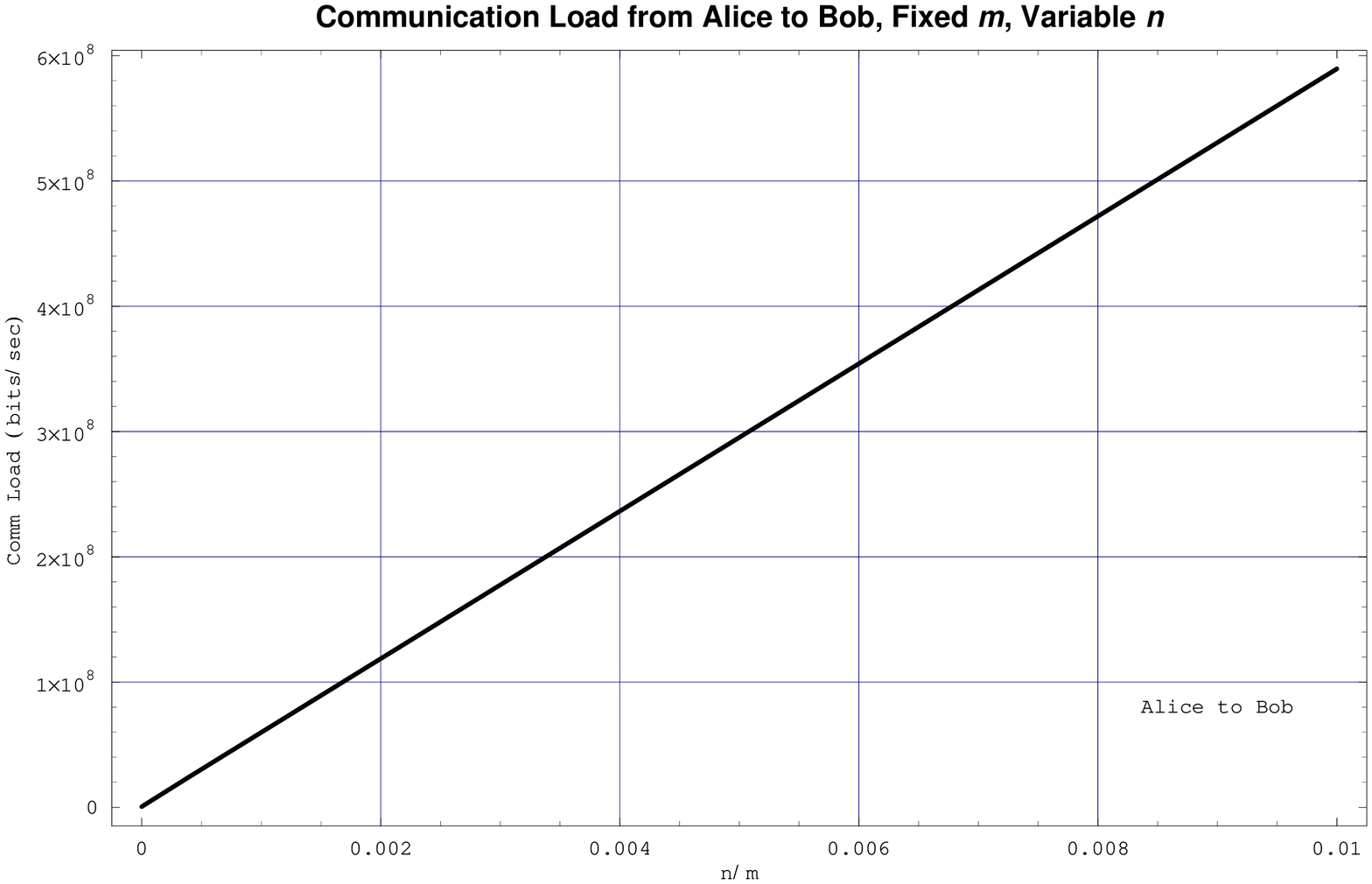}}}
\hfil
\hbox to -1.25in{\ } 
}
\bigskip
\caption{%
Communication Load from Alice to Bob, Fixed $m$, Variable $n$
}
\label{F:rabvsnoverm}
\end{figure}

\begin{figure}[htb]
\vbox{
\hfil
\scalebox{0.6}{\rotatebox{0}{\includegraphics{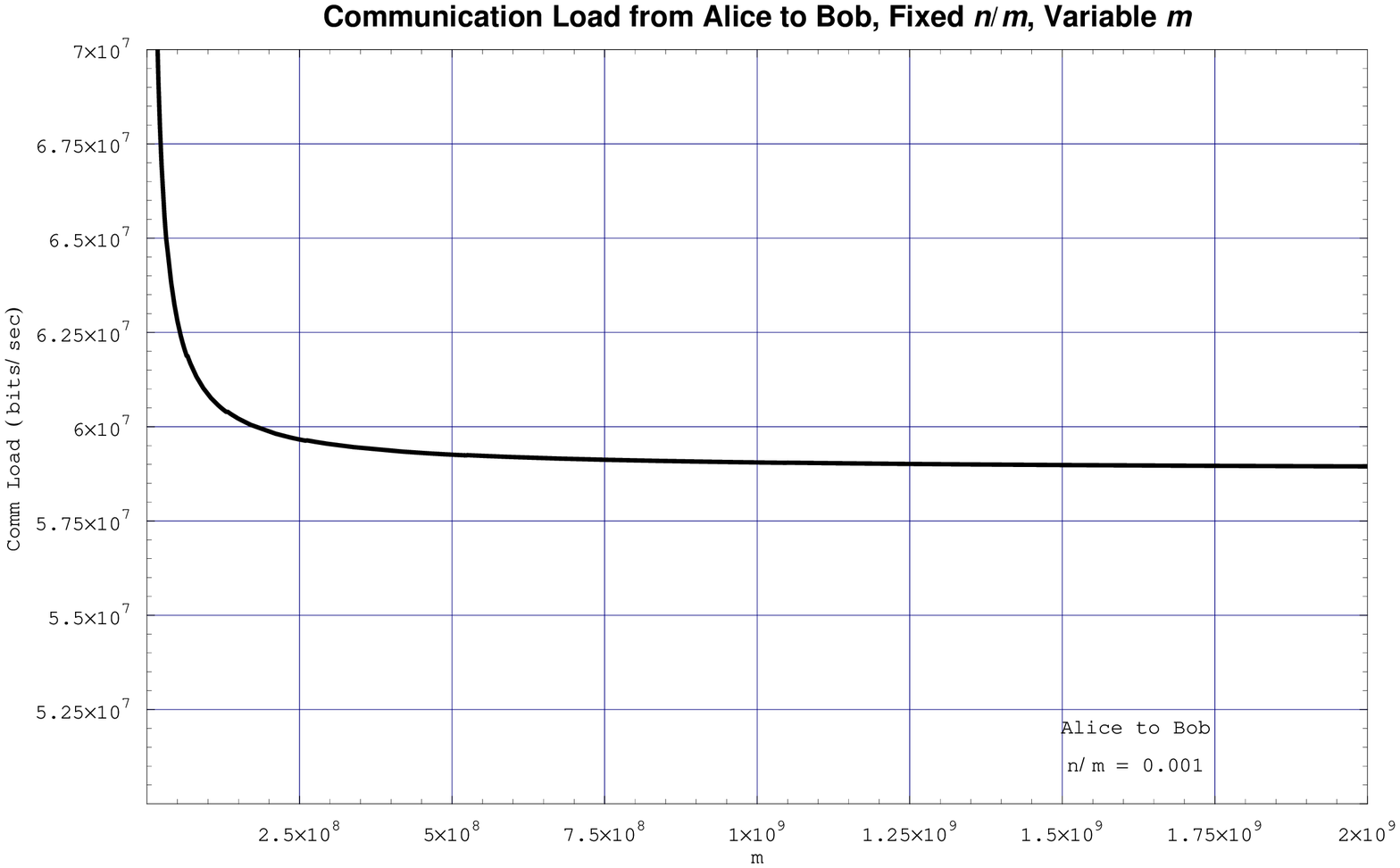}}}
\hfil
\hbox to -1.25in{\ } 
}
\bigskip
\caption{%
Communication Load from Alice to Bob, Fixed $n/m$, Variable $m$
}
\label{F:rabvsm}
\end{figure}

Figures \ref{F:rbavsnoverm}, \ref{F:rbavsm}, \ref{F:rabvsnoverm}, and \ref{F:rabvsm} 
show the communication load between Bob and Alice.  The dependence on $n$ is roughly 
linear for fixed $m$.  The dependence on $m$ is relatively weak for large $m$ and for a  
fixed ratio $n/m$.

\subsubsection{System Load: Total Computational Requirements}

In this 
section we investigate the computational resources that are required to implement the 
sifting, error correction, and privacy amplification algorithms we have been discussing.  
In principle, the processing could be done with special purpose hardware that is designed 
to perform the necessary operations as efficiently as possible.  In this analysis, however, 
we will assume that the operations are to be carried out using a general purpose computer.  
As the instruction set of a general purpose computer is not particularly well suited to 
operations such as finding the parities of long bit strings, the results of the analysis 
should represent a conservative upper bound for the processing requirements as compared to 
what is achievable with special purpose devices.  

We first analyze the algorithms for sifting, error correction, and privacy amplification 
into processing steps of a size suitable for implementation as assembly language 
subroutines.  The steps for Bob's computation that require iteration on the 
entire string are:

Pack received polarizations and indices \\
\hspace* {0.5 in}    ($2n(1+\log_2 m)$ bits) \\
Compute authentication tag \\
\hspace* {0.5 in}    ($2n(1+\log_2 m)$ bits in, $g_{auth}$ bits out) \\
Unpack Alice's response \\
\hspace* {0.5 in}    ($2n$ bits) \\
Check authentication tag from Alice \\
\hspace* {0.5 in}    ($2n$ bits in, $g_{auth}$ bits out) \\
Sift bit string \\
\hspace* {0.5 in}    ($2n$ bits in, $n$ bits out) \\
Perform random shuffle \\
\hspace* {0.5 in}    ($n$ bits) \\
Compute block parities for error detection and correction step \\
\hspace* {0.5 in}    ($N_1$ iterations, $n$ bits processed) \\
Perform bisective search \\
\hspace* {0.5 in}    ($N_1$ iterations, each performing $e_f^{\left( i \right)}$ searches, 
     each search involving $k^{\left( i \right)}$ bits) \\
Extract blocks for validation step \\
\hspace* {0.5 in}    ($N_2^{\left( n\right)}+N_2^{\left( f\right)}$ iterations, $n$ bits processed) \\
Compute block parities for validation \\
\hspace* {0.5 in}    ($N_2^{\left( n\right)}+N_2^{\left( f\right)}$ iterations, 
     ${n \over 2}$ bits processed) \\
Perform bisective search \\
\hspace* {0.5 in}    ($N_2^{\left( f\right)} = e_T^{\left( r\right)}$ blocks of size ${n \over 2}$) \\
Pack error corrected string \\
\hspace* {0.5 in}    ($n$ bits) \\
Compute equivalence checking tag \\
\hspace* {0.5 in}    ($n$ bits in, $g_{EC}$ bits out) \\
Get privacy amplification key \\
\hspace* {0.5 in}    ($2n$ bits) \\
Compute privacy amplification tag \\
\hspace* {0.5 in}    ($n$ bits in, $L \approx n$ bits out) \\

Bob's computational load represents an approximate upper bound for Alice's 
computational load, since 
Alice does not need to compute an equivalence checking tag, but otherwise has to 
perform computations of comparable complexity.  We may thus restrict our discussion 
to Bob's computations without loss of generality.  

We obtain a rough estimate of the total computational load by addressing the 
computational loads associated with each of the above steps.  
First, we assume that the operations of 
packing, unpacking, sifting, block extraction, bisective search, and
parity computation require 25 assembly language statements (or operations) 
per bit processed on 
each iteration.   Sections of assembly code developed to implement a subset of these 
operations indicate that this should consistently overestimate the required computations. 
For example, the code segment for computing parity included in the appendix requires only 5 
operations per iteration.  
The intent of this analysis is to use these conservative estimates for most 
of the steps in the computation, but to analyze more carefully those parts of the 
computation whose contribution to the rates is more sensitive to the block size.  (Note 
that computations with loads that are strictly linear in the block size lead to rates 
that are independent of the blocksize, since the amount of time available for computation 
for each block is proportional to the block size.)  

In applying this assumption to the above steps, we note that the initial block parity 
calculations for the error detection and correction step process every bit in the 
string.  The bisective searches process every bit in the substring under examination, 
except for 1 bit.  (For the purposes of this analysis, we simply include the extra bit.)  
The initial block parity calculations for the validation step process
half of the bits in the string.  

The estimate of the load for the bisective search in the error detection and correction 
step involves a summation over the iterations internal to the step:

\be
{\cal L}_B^{\left( EDC, BS \right)} = 25 \cdot 
   \sum_{i=1}^{N_1} e_f^{\left( i \right)} k^{\left( i \right)}~.
\ee

This may be expressed in terms of fundamental parameters by using results from 
the appendix ``Statistical Results for Error Correction."  We obtain:

\bea
{\cal L}_B^{\left( EDC, BS \right)}
  & =& 25 \cdot \sum_{i=1}^{N_1} \left( 1-\beta \right) \beta^{i-1} e_T^{\left(0\right)}
       { {\varrho n} \over {\beta^{i-1} e_T^{\left( 0 \right)}} } \nonumber\\
  & =& 25 \cdot \varrho n \left( 1-\beta \right) N_1 \nonumber\\
  & =& 25 \cdot \varrho 
       \left( 1 - {{2 \varrho - 1 + e^{-2\varrho}} \over {2 \varrho}} \right)
       {{\log_2 {{2\varrho} \over {e_T^{\left( 0 \right)}}}} \over {\log_2 \beta}}
       n \nonumber\\
  & =& 12.5 \left( 1 - e^{-2\varrho} \right) 
       {{\log_2 {{2\varrho} \over {e_T^{\left( 0 \right)}}}} \over {\log_2 \beta}} 
       n \nonumber\\
  & =& 12.5 \left( 1 - e^{-2\varrho} \right) N_1 n~.
\eea

The computations for the authentication and error correction tags are relatively complex, 
and require more detailed attention.  
As described by Wegman and Carter, \cite{wc}, the algorithm proceeds by partitioning 
the input string into substrings of length $2s$, where

\be
\label{E:shash}
s = g_{auth} + \log_2 \log_2 c~,
\ee

\noindent
and $c$ is the length of the input string.  
An auxiliary hash function is applied to each of the substrings resulting in a set of 
strings of length $s$.  The results are concatenated and repartitioned into 
substrings of length $2s$.  The process is repeated until the concatenated string 
is of length $s$.  The final tag is taken from the lower order bits of this string.  
The hash function is applied $\lceil{c \over {2 s}}\rceil$ times in the first iteration.  
The length of the input string is reduced by one half in each successive iteration, so 
that the hash function is applied $\lceil{c \over {2^i s}}\rceil$ times during the 
$i$th iteration.  The process continues until 

\be
{c \over {2^{i_{max}} s}} \leq 1~,
\ee

\noindent
or, equivalently, 

\be
i_{max} \geq \log_2 {c \over s}~.  
\ee

\noindent
The total number of times the hash function is applied is then

\be
n_{hash} \simeq \sum_{i=1}^{\log_2 {c \over s}} {c \over {2^i s}}~.
\ee

\noindent
We obtain a rough upper bound by extending the summation to infinity, yielding
the simple estimate

\be
n_{hash} \simeq {c \over s}~.
\ee

The hash function itself involves the multiplication and an addition 
of integers encoded as bit strings 
of length $2s$.  For $g \sim 30$ and $c \sim 10^{12}$ bits, we have, using 
eq. (\ref{E:shash}),

\be
2s = 2g + 2 \log_2 \log_2 c \sim 70~.
\ee

This is slightly longer than 64 bits, so we will assume that the integer 
operations operate on double words.  Each application of the hash function requires 
three operations: a substring of length $2s$ is extracted from the string, the 
hash operation is applied to the substring, and the substring is inserted into the 
output string.  Assembly code segments for extracting the substring and applying 
the hash operation are presented 
in the appendix.  These segments contain 26 and 43 instructions respectively.  
The code for inserting the result in the output string should be roughly as 
complex as the extraction and will thus require an additional 26 operations.  If 
we add 15 instructions to handle the loop control, we obtain a total of 110 
operations for each application of the hash operation.   
The resulting estimate of the 
computational load for the authentication and equivalence checking steps 
is then 

\be
{\cal L}_B^{auth} = {\cal L}_B^{EC} \simeq 110 \cdot {c \over s}~.
\ee  

Potentially the largest contribution to the computational load is the privacy amplification 
hash function.  This is due largely to the presence of nested loops in the code
that result in a quadratic 
dependence on the size of the sifted string, $n$.  The assembly code for this function 
is given in the appendix. The resulting load is given by\footnote{
The authors of \cite{cw} introduce an alternative class of hash functions
the computational complexity of which is linear in the key size. Use of this class
of hash function in privacy amplification could result in a moderate
reduction in the computational load (as computed in eq.(\ref{56billion}) below), and/or
allow for a significant increase in the allowed processing block
size \cite{linhashwork}. In this case the block size is limited by memory requirements
and the $n\left(1+\log_2m\right)$ term in eq.(\ref{precompload}).}

\be
{\cal L}_B^{PA} \approx 43\left({n\over w}\right) + 46\left({n\over w}\right)^2~.
\ee

\noindent
This represents the number of instructions required to perform the hash computation for 
a single block of sifted, error corrected bits of length $n$.  The parameter $w$ is the 
wordsize of the processor in bits.

We find the total load by summing the contributions from each of the individual steps:  

\bea
\label{precompload}
{\cal L}_B &\leq&
   {\cal L}_B^{\left( 0 \right)}  \nonumber\\
  &&+ 25 \cdot \left( 2n \left( 1+\log_2 m \right) \right)  \nonumber\\
  &&+ 110 \cdot 
        {{ 2n \left( 1+\log_2 m \right)} \over 
         {g_{auth} + \log_2 \log_2 \left[ 2n \left( 1+\log_2 m \right) \right]}} 
                                                                       \nonumber\\
  &&+ 25 \cdot 2n  \nonumber\\
  &&+ 110 \cdot {{2n} \over {g_{auth} + \log_2 \log_2 \left( 2n \right)}}  \nonumber\\
  &&+ 25 \cdot 2n  \nonumber\\
  &&+ 25 \cdot n  \nonumber\\
  &&+ N_1 \cdot 25 \cdot n  \nonumber\\
  &&+ 12.5 \left( 1 - e^{-2\varrho} \right) N_1 n  \nonumber\\
  &&+ \left( N_2^{\left( n \right)} + N_2^{\left( f \right)} \right) \cdot
     25 \cdot n  \nonumber\\
  &&+ \left( N_2^{\left( n \right)} + N_2^{\left( f \right)} \right) \cdot
     25 \cdot {n \over 2}  \nonumber\\
  &&+ e_T^{\left( r \right)} \cdot 25 \cdot {n \over 2}  \nonumber\\
  &&+ 25 \cdot n  \nonumber\\
  &&+ 110 \cdot {n \over {g_{EC} + \log_2 \log_2 n}}  \nonumber\\
  &&+ 25 \cdot \left( 2n \right)  \nonumber\\
  &&+ 43 \cdot \left({n\over w}\right) + 46 \cdot \left({n\over w}\right)^2~.  
\eea

Each term corresponds to one of the steps in the processing.  
${\cal L}_B^{\left( 0 \right)}$ is the ``non-iterative" portion of the load,
representing code that executes once for each block of data processed. (Note
that there may 
be iterative loops in this code as well.  The point is that these loops do not 
represent processing that iterates bit-by-bit through the string.)  
We simplify the 
result by collecting terms, noting from eq.(\ref{E:errsrem}) that the
residual error count after 
error correction and detection is given approximately by

\be
e_T^{\left( r \right)} \simeq 2 \varrho~.  
\ee

We also drop the double log terms in the denominators, thus 
replacing those terms by larger quantities.  The resulting expression is 

\bea
\label{E:compload}
{\cal L}_B &\leq&
   {\cal L}_B^{\left( 0 \right)} \nonumber\\
  &&+ \left( 50 + {220 \over g_{auth}} \right) n \left(1+ \log_2 m \right) \nonumber\\
  &&+ {\Bigg [} 200 + 25 N_1 + 
            12.5 \left( 1 - e^{-2\varrho} \right) N_1 + 
            25 \varrho + 
            37.5 \left( N_2^{\left( n \right)} + N_2^{\left( f \right)} \right)\nonumber\\ 
     &&\qquad+{43 \over w} + {220 \over g_{auth}} + {110 \over g_{EC}} {\Bigg ]} n\nonumber\\
  &&+ {46 \over {w^2}} n^2 ~.  
\eea

It is instructive to evaluate this expression for the same example used in 
finding the communications load.  We take the non-iterative contribution to the load to 
be substantial:

\be
{\cal L}_B^{\left( 0 \right)} = 10^6 {\rm ~operations~per~block}~.
\ee

We take the wordsize of the processor to be 64 bits.  The other parameters are 
as before ($m=2\times 10^8$ bits, $n=2\times 10^5$ bits, 
$e_T^{\left( 0\right)} = 2 \times 10^3$ bits, $\tau = 10^{-10}$ sec, $\varrho = 0.5$, 
$g_{EC} = g_{auth} = N_2 = 30$).  
The resulting estimate of the load is 1.1 billion operations per block.   The quadratic 
term contributes 450 million operations to the total.  Of the other terms, the dominant 
contributions are the term in in $N_2^{\left( n \right)} + N_2^{\left( f \right)}$, 
which is due to parity checks and random 
block extractions during the validation step of error correction, and the term in 
$\left( 1+\log_2 m\right)$, which is due to sifting.  
Note that the non-iterative overhead load is negligible in comparison with the other 
contributions.  This indicates that a substantial amount of ``bookkeeping" code can 
be included along with the core software that is essential to arriving at the final 
secret key without significantly affecting the processing requirements.  One of the uses 
of eq.(\ref{E:compload}) is to establish a load budget for such code 
during software design and implementation 
to ensure that the bulk of the 
processing resources are available for the core software functions.  

The computation rate ${\cal R}_B^{comp}$ required to support key distribution 
is found by dividing the load per 
block by $m\tau$, 
the time required to transmit one block over the quantum channel:  

\be
{\cal R}_B^{comp} = {{\cal L}_B\over m\tau}~.
\ee

Applying this to our preceding example yields an estimated processing rate requirement of 

\be
\label{56billion}
{\cal R}_B^{comp} = 56 {\rm ~billion~operations/sec}~.
\ee

This is rather high for a single general purpose 
processor, but should be achievable in a parallel architecture in which each block 
of the input data is allocated to a single processor as it becomes available.  
Recall also that general purpose computers are far from optimal for this type of operation.  
Most of the processing steps involving the packing and unpacking of the bits would not be 
necessary in a special purpose device, and many of the other processing steps, 
notably block parity 
computations and random selection of substrings, could be accomplished much more 
efficiently using special purpose hardware.

\clearpage

\section{High-Speed Quantum Cryptography}
\label{S:HSQKD}

In this section of the paper we analyze the possibility of achieving very high data
throughput rates
for a QKD system. We first discuss essential elements that such a system requires, and
then make use of the analysis in the preceding sections of effective secrecy capacity,
system losses and loads to deduce universal maximal rates achievable in QKD.

\subsection{Methods to Achieve High-Speed Quantum Cryptography}

There are three different techniques that may be applied to achieve
high data throughput values:

$\bullet$ Reducing the value of the bit cell period, $\tau$, in the effective
secrecy capacity

or

$\bullet$ Combining some number of quantum bit transmitting setups together,
{\it i.e.}, combining some number of Alice systems together,
by multiplexing the outputs into a common bit stream

or

$\bullet$ Applying both of the above techniques together.\footnote{
The throughput rate could also in principle be increased by employing a quantum
bit generating device at the Alice end that, through whatever
means, does not generate any multiple particle states. In this idealized
case the multi-photon privacy amplification function $\nu$ employed by
Alice and Bob could be set to $\nu=0$, which would result in a significant
increase in the throughput rate. If the qubit source produced multiple states,
but with a smaller likelihood per bit cell than for the Poisson distribution considered
in detail in Section 3 of this paper, the resulting required
value for $\nu$ would lie between 0 and the values of
the expressions found in eqs.(141) through
(151), resulting in an improved throughput rate.}

The optimal effective secrecy rate ${\cal R}_{{\rm opt}}$ is inversely
proportional to the bit cell period $\tau$. For the first technique identified
above, in decreasing the size of the bit
cell period to increase the effective rate we need to ensure that Bob's
detector apparatus can count incoming photons at a correspondingly higher speed as well.
A decrease in the bit cell period means that the source laser must operate at a higher
pulse repetition frequency (PRF) than before, so we must also make use of pulsed
lasers with the necessary stability to operate at the required PRF values. Furthermore, we
must ensure that the various opto-electronic components and switches can likewise
operate at the required high frequencies. The collection of various components, all
operating at very high repetition frequencies, must be properly synchronized together
for the protocol to be properly executed. In addition, real-time data recording of
all necessary quantities must be taken at the required high rate.
We discuss in the sections below various practical approaches to each of these
critical requirements on a high speed quantum cryptography system. All of them are
requirements that must be satisifed if we are to increase the rate by making the bit
cell period smaller.

The second technique identified above, the multiplexing together of a number of
output streams into a common transmission, leads us to different concerns. As
deduced in Section 4.4.3 above, the execution
of the QKD protocol imposes calculable requirements on the necessary computing resources,
and for a system operating at a high speed, this is already a significant burden for
a single Alice device. The required computing resources will be even larger for several
Alices operating in tandem. As discussed
briefly in Section 3.1.7 above, for the multiplexing of several data streams to
succeed we must ensure that we are not exceeding the computing capacity available to
the system. In Section 5.2.6 below
we work out how the system throughput can be increased
while taking into account these computing requirements constraints.

\subsection{System Components and Constraints}

\begin{figure}[htb]
\vbox{
\hfil
\scalebox{0.66}{\rotatebox{270}{\includegraphics{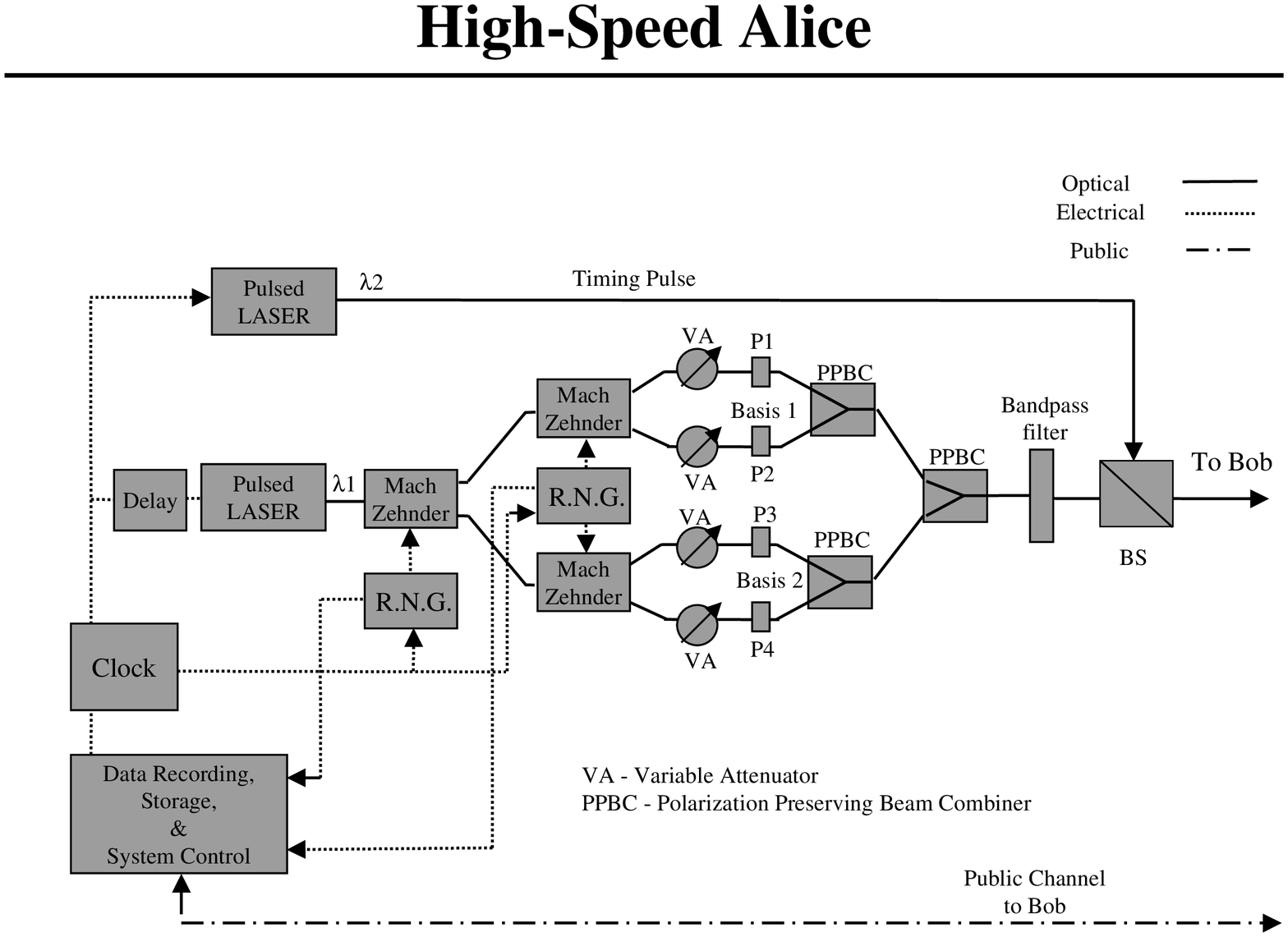}}}
\hfil
\hbox to -1.25in{\ } 
}
\bigskip
\caption{%
Block Diagram for Alice System
}
\label{F:u_alice}
\end{figure}

\begin{figure}[htb]
\vbox{
\hfil
\scalebox{0.66}{\rotatebox{270}{\includegraphics{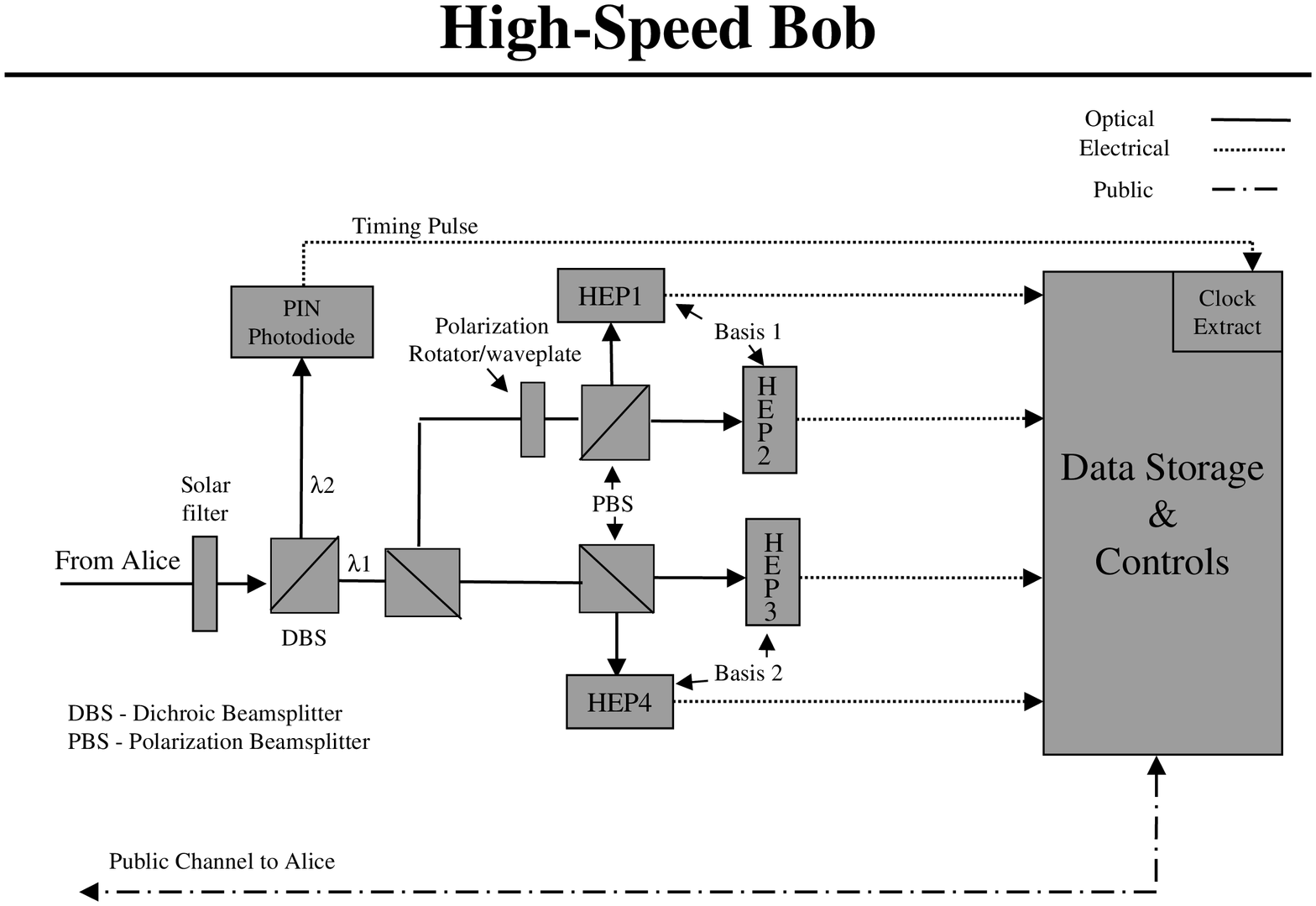}}}
\hfil
\hbox to -1.25in{\ } 
}
\bigskip
\caption{%
Block Diagram for Bob System
}
\label{F:u_bob}
\end{figure}

In this section we address the practical capabilities of a proposed realistic
high-speed quantum cryptography
system, taking into account engineering constraints on the various system
components. Our purpose here is to illustrate with specific examples that such
high-speed systems are in principle {\it feasible}. These examples of specific
components are chosen to support the argument that it is possible to design and
build such a system entirely out of currently available, commercially
produced equipment, with the exception of the required high-speed photon detectors.
For these, we have identified and analyzed the possibilities inherent in a promising
new approach to high-speed photon detection, as described in Section 5.2.1
below.\footnote{
We do not specifically advocate the use of the {\it particular} components described
herein as necessarily
comprising the ``best" approach to the problem of building a high-speed
quantum cryptography system.}
The overall design of the Alice and Bob systems, respectively, are
illustrated in Figures \ref{F:u_alice} and \ref{F:u_bob}.
The general structure
of both systems is based on previously
proposed design concepts \cite{qcdesigns,lutkenhaus3,demorefs1}. The
primary innovation here is the proposed very high speed
of operation of the system, which we briefly sketch now.
The basic source of the quantum bits is a high-speed pulsed
laser producing pulses at a PRF of 10 GHz, along
with a second synchronized laser producing a bright timing pulse. In the
Alice system the random choices of both
polarization bases and states are implemented with two suitable high speed random
number generators, which operate on a set of three high-speed Mach-Zender
interferometers, the first intended to select the polarization basis, and the second
and third to select the specific polarization states. The outputs are balanced by
being fed through variable attenuators, after which they are passed through appropriate
filters and put into states of definite polarization.
These polarized streams are then
combined into a common output stream to be transmitted through the
quantum channel to the Bob system. The necessary real-time records of all the random
selections for bases and states are obtained {\it via} suitable high-speed data
recording and processing devices, for which appropriate
de-multiplexing techniques are required, constrained by the speed of available
computer data bus rates.
The Bob system is purely passive, so that all
basis and state selection is accomplished, purely randomly,
solely {\it via} optical components. After being passed through a solar filter,
the received stream of pulses is separated into the
data and timing parts by a dichroic beamsplitter,
after which the cipher stream is passed through a passive beamsplitter.
The ``basis stream" outputs
are sent through two additional polarizing beamsplitters, after which
the four separated streams are then
focussed onto four high-speed photon detectors. As with the Alice system, accurate
real-time, high-speed recording of the random choices of basis and state are
obtained {\it via} a suitable de-multiplexing scheme.

\vskip 10pt
\noindent {\it The Significance of the 10 GHz System Clock Speed}

What is the significance of the 10 GHz system clock speed we have discussed above?
We are interested in exploring conditions and constraints that will allow quantum
cryptography to be implemented at high data throughput rates. The overriding
requirement for the success of such a scheme is to employ robust,
stable equipment built from high-speed components. As outlined above,
this can be achieved by exploiting
recent developments in high-speed optical communications technology, along with the
proposed use of a novel experimental technique for high-speed single photon counting
that employs cooled thin-film devices.
There are a variety of commercially available high-speed components
for optical classical communications
({\it i.e.}, conveying information in pulses containing very many
photons) with 10 GHz throughput.
This motivates the possibility of carrying out quantum cryptography
at the same basic clock rate.
We have chosen the clock speed of 10 GHz as representative of what can be
achieved in classical optical communications systems solely with commercially
available equipment. However, in classical optical communications it is
not necessary to count individual photons, and in particular not at a rate
of 10 GHz. Thus the only additional component required for quantum cryptography
at such a clock speed beyond that which is commercially available
is a suitable means of achieving high-speed single photon detection.
(Higher system clock speeds would require additional components that are net yet
commercially available.)
Of course, the actual {\it throughput} rate that can be
achieved will be considerably lower than the basic system clock rate, as
expected based on the analysis in Sections 3 and 4 and as discussed with specific
examples in Section 5.3 below.

In the sections below we describe in more detail
the various high-speed components that should allow such a system to be implemented
in practice.


\subsubsection{Fast Photon Detectors: Hot-Electron Photo-Effect}

It is an essential requirement in achieving high-speed quantum
cryptography that we make use of a fast data
source for the quantum bits transmitted
by Alice to Bob. It is necessary that the qubit detector
apparatus keep strict pace with the rate of qubit generation.
There are a number of different approaches to the detection of single
photons, including the use of single photon avalanche diodes, photomultiplier tubes,
single-electron transistors and superconducting tunnel junctions.
Unfortunately none of these approaches to photon detection 
are ideally suited for the high-speed quantum cryptography system that we propose.
We require a method to detect individual photons at a wavelength of 1550 nanometers
and at a sustained rate of 10 GHz, with a suitably high quantum efficiency of detection
and a very low intrinsic dark count rate.
With respect to these requirements
various drawbacks characterize the existing approaches listed above. These include
insufficient sensitivity in the required wavelength window, low photon counting
rates, requirements for cooling to millikelvin temperatures, {\it etc.}
However, recent advances \cite{sobopapers} in work on picosecond response time
single photon detectors based on the use
of superconducting thin films of Niobium Nitride (NbN) to exploit the so-called
``hot electron photo-effect" suggest that a new
approach, well suited to the requirements for high-speed quantum cryptrography,
can be developed. 

Here
we will sketch the main features of the proposed approach.\footnote{
All other elements of our proposed approach to high speed quantum cryptography
are based on existing, mature technology. High-speed (10 GHz) detection of photons
at 1550 nanometers wavelength, however, is not an existing, mature technology. We
have identified and here discuss
an approach to high-speed photon counting for quantum communications at
telecommunications wavelengths that is very promising, based on initial
experimental results.}
Hot-electron photodetectors (HEPs) based on ultrathin niobium nitride
films can operate as single-photon counters in the 
wavelength range from below 0.5 micrometers to at least 
2 micrometers. The NbN single-photon counter is 
characterized by a high (40\%) intrinsic quantum efficiency and 
practically negligible dark counts \cite{sobo1}.\footnote{
\label{darkcount}
The dark count probability can be estimated to be
no greater than approximately $e^{-40}\approx
4.25\times 10^{-18}$
based on an experimentally measured signal-to-noise ratio of
$40$ in a given HEP cycle period \cite{sobotalk}.}
The counting rate is intrinsically limited by the 
electron-phonon interaction time, measured to be 
10 picosecond. The response of practical devices is 
further limited by the phonon escape time from the film 
to the substrate and is equal to approximately 30 picosecond \cite{sobo2}.
The primary detector element
consists of an ultrathin (5 nanometers), very narrow (0.2 micrometers) NbN strip,
maintained at a temperature of 4.2 K. Although this low temperature requirement
may be problematic for standard ``long-haul" telecommunications applications,
its proposed use here is for the specialized area of secure quantum
communications for which it is entirely acceptable. In this connection we
emphasize that the systems we propose, in the case of free space implementations
of quantum cryptography, involve placing the Alice system on the orbiting
satellite, so that the Bob system, which is where the necessary cryogenic
appartus will have to be situated, is either on the ground or on an airplane.
In either case it is much easier to arrange for the operation of the cryogenic system
than would be the case if we envisaged placing the Bob system on the satellite.
In early experiments
such detectors have already been demonstrated to
be able to count individual photons, characterized thus far by a {\it measured}
response time of 100 picoseconds, which although lower than the theoretically
predicted maximum of 30 picoseconds is already fast enough to allow photon
counting at 10 GHz. The actual PRF of the source laser in completed experiments
was much slower, pulsing
in different experimental setups at a 76 MHz PRF producing 790 nanometers wavelength
light, or at 1 KHz PRF producing pulses at wavelengths of 500 nanometers, 1500
nanometers and 2100 nanometers.\footnote{
Work is in progress \cite{soboprivate}
to obtain results at higher PRFs, specifically at 1550 nanometers wavelength.}
In early experimental results the initial estimated
quantum efficiency was determined to be 20\%, which is not yet as high as the
theoretically achievable value.
An advantage of the HEP approach to single photon counting, apart
from the ability to achieve extremely high detection rates, is the lack of the
so-called ``afterpulsing" problem that plagues other approaches. 

The extremely narrow width of the NbN strip in the detecting element necessitates the
manufacturing and use of long-microbridge and meander structures to increase the
active area so as to mitigate the ``behind-the-telescope" loss discsussed in
Section 4.1.4 above.
The overall design would require the side-by-side placement of a small number
of detector ``chips," on each of which would be affixed a single meander structure
of NbN thin film. Each such meander structure would incorporate
a single input lead and single output lead, so that capacitance constraints on the
set of chips can be expected to be minimal.

A fuller discussion of this emerging technology and its application to
high speed quantum communication will be presented elsewhere \cite{sobomitre}.

\subsubsection{High Pulse-Repetition-Frequency Lasers}

The state-of-the-art in experimental research in high-speed pulsed lasers 
({\it e.g.}, \cite{fastlasers1}) makes the
use of 10 GHz PRF sources a completely
realistic possibility, assuming that we can also count
single photons at the same rate. Commercial fiber mode-locked lasers operating
at this PRF value are available, making this a very practical instrument to
incorporate into an actual QC system implementation.

Recent ``heroic" experiments \cite{fastlasers2} in
which pulsed lasers operating at a PRF of over
450 GHz have been carried out, demonstrating the possibility that at some point
in the future it {\it might} be possible to implement a practical QC system operating
at a bit cell period of 2.2 picoseconds if it should prove
possible to detect individual photons at the same rate. Of
course, as we discuss below, the critical difficulty here is to maintain real-time
date recording at this rate.

\subsubsection{High Speed Opto-electronics Components}

The use of Mach-Zender interferometers that operate at switching speeds of
10 GHz has been common in the laboratory for many years. Much recent
work has been done on achieving substantially higher switching speeds for Mach-Zender
interferometers \cite{MZrefs1,MZrefs2,MZrefs3,MZrefs4}, leading
to the current situation in which it is should be feasible as such
to incorporate 40 GHz devices in practical quantum cryptography systems, if
photon detection and
suitable real-time data recording can also be accomplished at the same rate.

\subsubsection{Synchronization Constraints}

As has been proposed and demonstrated elsewhere ({\it e.g.}, \cite{demorefs1}), a
bright timing pulse generated
by a suitable pulsed laser operating at the same PRF as the source laser can be
used to provide necessary system synchronization across the various components. In the
system that we envisage for high speed quantum cryptography, the bright timing
pulse laser will be connected to the 10 GHz
system master clock. In addition to this, the stringent synchronization
requirements that are dictated by a high-speed system
also require that the master clock must be connected {\it via}
4- and 10-way splitters to the data recording de-multiplexing system, as illustrated in
Figures \ref{F:u_alice}, \ref{F:u_bob} and \ref{F:demux}, which serves to synchronize
the ``internal" oscillators of the data recording computers to the ``external"
pulse rate of the Alice laser system.

\subsubsection{Data Recording De-multiplexing Constraints}
\label{S:PSD}

In considering a practical design for a quantum cryptography
system intended to achieve high-speed
throughput, an important requirement
is to isolate the potential ``clogging points" of the
overall scheme. The crucial question is: What is the limiting engineering design issue
that slows
down the system operation? It is clear from all the preceding analysis in this paper
that,
having properly accounted for the many system losses and loads, and assuming that very
fast
photon detection at a suitably high value of quantum efficiency is possible, the main
``engineering" issue to address is that of keeping proper, error-free,
{\it real-time} records, at
both the Alice {\it and} Bob ends of the system, of the continuous stream of
information,
such as polarization basis and state information, that must be carried out
in order to perform the
processing required in the protocol. Here we are limited mainly by the current state of
the
art in achieving {\it sufficiently high data bus speeds}.
The speeds, as such, of the central
processing units in the various control computers do {\it not} furnish
the limiting
constraint here: it is essential to keep a running tally of the polarization state of
each and every bit cell transmitted from Alice to Bob, and to do so without errors.

\begin{figure}[htb]
\vbox{
\hfil
\scalebox{0.66}{\rotatebox{270}{\includegraphics{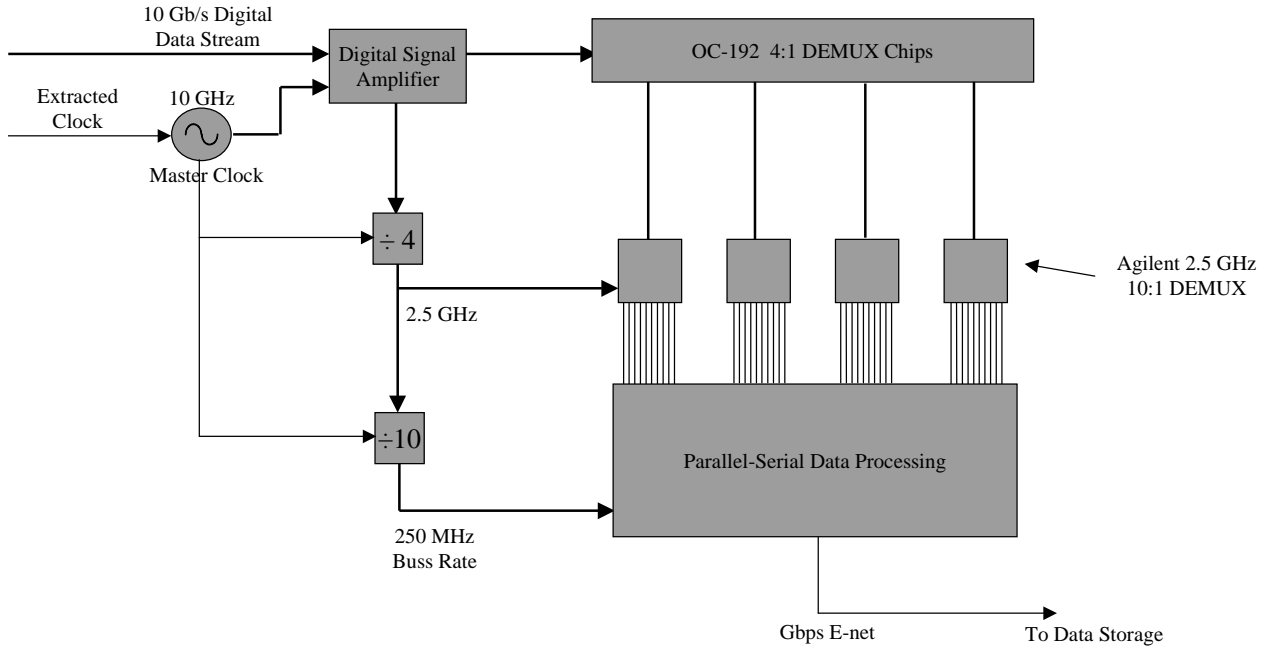}}}
\hfil
\hbox to -1.25in{\ } 
}
\bigskip
\caption{%
Block Diagram for Data Recording System
}
\label{F:demux}
\end{figure}

In our analysis we will consider a situation in which we are constrained to make use of
practical data bus devices which can accomodate incoming data at a rate of
250 Megabits per second, which is many times slower than the
10 GHz PRF of our proposed laser source.\footnote{
Although this rate is also somewhat faster than the bus {\it speeds} currently found in
typical, commercially available personal computers, this apparent
incompatibility in fact does not pose a problem. The
{\it effective} data recording rate is dictated by both the intrinsic bus rate {\it and}
the width of the bus. Taking both these factors into account it is clear that the
requirement of 250 Megabits per second is well within the capabilities of
commerically available high-end personal computers today.}
The solution to this problem is to design a
suitable de-multiplexing system that can connect these very different
rates, making sure to include appropriate error correction capability. One possible
(although expensive) approach to this problem is to employ existing OC-192
telecommunications equipment intended to operate at a 10 GHz repetition rate. As
illustrated in Figure \ref{F:demux}, 
one may envisage employing commercially available OC-192 4:1
DEMUX chips in a practical quantum cryptography implementation. These typically
include a framer to support channelized OC-192 ``synchronous optical network" (SONET)
and ``synchronous digital hierarchy" (SDH) traffic and 10 GHz
bit error rate testing in various telecommunications applications.\footnote{
The standard commerical OC-192 bit error rate test
units are typically designed to be compatible with the
SONET family of protocols, which would require that appropriate SONET pattern framing
be encoded in suitable segments of
the bright timing pulse in order to use such equipment ``as is" for
high speed quantum cryptography. Other framing schemes are obviously possible as well.}
In the same
system one may also employ
commercially available 10:1 DEMUX chips that operate at a repetition frequency
of 2.5 GHz. A parallelized set of four of these can be linked in sequence
to the OC-192 DEMUX unit to achieve a net demultiplexing ratio of 40:1, which then
produces an integrated output suitable for routing through a 250 Megabit per second
data bus. The data stream would be fed {\it via} Gigabit per second
ethernet link to the data storage
components of the system for real-time processing according to the QKD protocol.

The principal
objective here is to be able to carry out {\it real-time} quantum cryptography, where
at
any given moment during the transmission of qubits a previously transmitted batch
are being processed.
The data recording de-multiplexing solution described above,
and other approaches to the problem of real-time data recording furnish
practical
solutions that can be implemented entirely with currently existing technology. A
more thorough analysis \cite{demuxpaper} of
the actual requirements and system details,
including a comparison with related real-time data recording solutions from both
telecommunications and experimental particle physics applications can be carried out
to demonstrate that this problem can be completely (at some cost) solved for
practical quantum cryptography systems.

\subsubsection{Multiple Transmitter-Receiver Multiplexing Analysis}

Here we outline the methods whereby the throughput of a quantum cryptography system
can be increased by multiplexing together the outputs of some number of transmitters
into a common data stream.\footnote{
Note that the ``multiplexing" referred to in this section is different than the
``(de)multiplexing" referred to in the previous section.
In the previous section ``(de)multiplexing" pertained
to data {\it recording} associated with
a {\it single} Alice-and-Bob system, while the in the present section
``multiplexing" refers to data {\it transmission} associated with
{\it several} Alice-and-Bob systems.}
The notion of multibeam
transmission has been previously proposed for deep space, classical lasercomm
applications \cite{deepspace2}. Here we extend that concept to the
area of quantum communications.

To determine what are the throughput possibilities and
constraints for any scheme involving the pooling together of
the output streams of more than one Alice device we need two mathematical
relationships: (1) the explicit function that provides a relationship
between processing block size and the
associated computing resource requirements, and (2) the explicit relationship
between processing
block size and throughput rate. We have derived both of these in this paper.
Together, these two pieces of information
allow us to determine the relationship between
fiducial block size and throughput rate for
any individual Alice-and-Bob system, as well as
for any combination of Alice-and-Bob systems multiplexed together with different
chosen block sizes.

The general approach to this problem is as follows:

$\bullet$ ~First we establish, for a given single
Alice-and-Bob system operating at a specified inverse bit cell period $\tau^{-1}$,
what is the largest allowable processing block size based on the maximum computing
machine power available using the known relationship between
block size and computing resources
({\it cf} eq.(\ref{E:compload})). We denote the maximum available computing resources
(measured in terms of basic computer instructions per second)\footnote{
The ``maximum available computing resources" will be determined on
a case-by-case basis for the particular application that is envisaged. In the
case of a free space implementation for which Alice is placed on an orbiting satellite,
for example, there may be more stringent constraints (dictated at least
in part by how much computing machinery can be physically fitted on board the
satellite) than those that apply when both Alice and Bob are on the ground.}
by $C_{ceiling}$, and
we denote the associated processing block size, referred
to as the ``ceiling" block size, by the symbol $B_{ceiling}$.
Thus $B_{ceiling}$ is the largest possible processing block size on which a single
Alice-and-Bob system can operate as constrained by the available computing
resources.
The block size $B_{ceiling}$ is then used in the known relationship between
block size and rate, {\it i.e.}, the equation for the optimal effective secrecy
rate ({\it cf} eq.(\ref{173})), to determine what the highest rate is, constrained
by the available computing resources, for a single Alice and Bob setup.\footnote{
We now see explicitly why it is impossible to analyze any such multiplexing
scheme by
using expressions for the effective secrecy capacity and effective secrecy rate that
are only valid in the abstract limit of an infinitely long cipher. In the
infinitely long cipher limit, the transmit block size, which is simply some specified
number of raw bits $m_0$,
{\it completely drops out} of the expressions for $\cal S$ and $\cal R$.}

$\bullet$ ~Second, we examine the consequences of
reducing the ceiling block size to some proposed smaller size $B_{smaller}$. Using
again the known relationship between the block size and computing resources,
we determine how many {\it copies} of
Alice-and-Bob systems can be operated at the smaller
processing
block size without exceeding the overall computing resource constraint dictated
by the value $C_{ceiling}$.

$\bullet$ ~Third, we again
employ the known relationship between the system throughput and
processing block size to determine the rate that can be achieved by a single
Alice-and-Bob system operating on processing blocks of the smaller size
$B_{smaller}$. By comparing this with the previously determined rate that applies
for a single Alice-and-Bob system operating on the larger block size $B_{ceiling}$,
we may obtain the decrease in rate that arises upon replacement of $B_{ceiling}$
with $B_{smaller}$.

$\bullet$ ~Fourth, we combine the above results to obtain the final change in
total throughput rate that
occurs when some number of copies of single Alice-and-Bob systems, each
operating on processing blocks of size $B_{smaller}$, are multiplexed together.

Going through the above steps in detail, we write the expression that
relates the general block size $B$ to the associated required
computing resources $C$ as
\bea
\label{290}
C&=&C\left(B\right)
\nonumber\\
&=&a_1B^2+\cdots
\nonumber\\
&\simeq& a_1B^2~,
\eea
where the detailed form of this relation,
including the explicit value of the leading coefficient
$a_1$, was derived in Section 4.4.3 above and
is given in eq.(\ref{E:compload}). It
is sufficient for the present discussion to note that
the computing resources scale quadratically with the processing block size.

We now take as given some fixed amount of computing power that for whatever reason
may not be exceeded for {\it any}
implementation, and call this amount $C_{ceiling}$. Using the above equation we find
the associated block size, $B_{ceiling}$, determined by:
\bea
\label{291}
C_{ceiling}&\equiv& C\left(B_{ceiling}\right)
\nonumber\\
&\simeq& a_1B_{ceiling}^2~,
\eea
associated to which we find the largest possible rate for a single Alice-and-Bob
system, ${\cal R}_{max}^{(single)}$, as
\be
\label{292}
{\cal R}_{max}^{(single)}\equiv{\cal R}_{{\rm opt}}\left(B_{ceiling}\right)~,
\ee
where we have displayed only the processing block size dependence in the argument
of ${\cal R}_{{\rm opt}}$ and suppressed all other
dependences ({\it cf} eq.(\ref{168})).

We now propose a new, smaller processing block size $B_{smaller}$, that is related to
the ceiling block size by the reduction factor $b<1$:
\be
\label{293}
B_{smaller}=bB_{ceiling}~.
\ee
The computing resources consumed for a single Alice-and-Bob system that employs the
smaller block size $B_{smaller}$ are calculated to be
\bea
\label{294}
C_{smaller}&\equiv&C\left(B_{smaller}\right)
\nonumber\\
&=&C\left(bB_{ceiling}\right)
\nonumber\\
&=&a_1b^2B_{ceiling}^2+\cdots
\nonumber\\
&\simeq&a_1b^2B_{ceiling}^2
\nonumber\\
&\simeq&b^2C_{ceiling}~,
\eea
and thus we find
\be
\label{295}
C_{ceiling}\simeq b^{-2}\times C_{smaller}~.
\ee

Thus, we may simultaneously run altogether as many as $b^{-2}$ parallel
implementations of {\it single} Alice-and-Bob systems, each
employing a processing block of size $B_{smaller}$ (and thus each
consuming an amount $C_{smaller}$ of computing resources), and
{\it still} satisfy the overall
computing resource constraint dictated by the value $C_{ceiling}$. This can
be done simultaneously by interleaving the output streams of the individual
Alice-and-Bob systems into a common stream using any one of several multiplexing
schemes, including for instance simple time division multiplexing, space division
mutiplexing, {\it etc.} ({\it e.g.}, \cite{TDM}).\footnote{
Of course, there is a computing resource cost associated to the actual multiplexing
implementation, {\it per s\'e}, but this is very small for the simple time division
multiplexing which would be adequate to achieve the objective under discussion
here.}

To determine the total throughput rate that would be achieved in such a multiplexed
system, we first determine the rate that applies for a single Alice-and-Bob system
operating on the smaller processing block
\be
\label{296}
{\cal R}_{smaller}^{(single)}\equiv{\cal R}_{{\rm opt}}\left(B_{smaller}\right)~,
\ee
from which we deduce the relative rate decrease, $r$, that characterizes the
replacement of $B_{ceiling}$ with $B_{smaller}$:
\be
\label{297}
r\equiv{{\cal R}_{smaller}^{(single)}\over{\cal R}_{max}^{(single)}}~.
\ee

The main point is that the total throughput rate that can be achieved in the
multiplexed system will be better than that which can be achieved with a single
system operating on the ceiling processing block size by a factor of as much
as $b^{-2}\times r$, if as many as $b^{-2}$ Alice-and-Bob systems operating on
blocks of size $B_{smaller}$ are multiplexed together.
Depending on the competing values of $b$ and $r$, this can be a quite significant
increase in rate, and as long as the rate decrease $r$ is larger
than $O\left(b^2\right)$, there will at least be some increase in the
rate.\footnote{
One may of course choose to multiplex fewer than $b^{-2}$ Alice-and-Bob systems
together, and still achieve a (smaller) increase in rate, depending on the actual
value of $r$.}
We may now obtain the final throughput rate that will be achieved in the multiplexed
system, ${\cal R}_{max}^{(multiplexed)}$, as
\be
\label{298}
{\cal R}_{max}^{(multiplexed)}=\left(b^{-2}\times r\right)
\times{\cal R}_{max}^{(single)}~,
\ee
where this maximal result specifically applies to the case that a total of
$b^{-2}$ Alice-and-Bob systems are multiplexed together.

\subsubsection{High Speed Random Number Generation}

Although for the purely passive Bob setup that we advocate there is no need
to actively generate random numbers with which to associate the successive
choices of polarization basis, such an active choice may be required at the Alice
end of the system. This topic, which is of crucial importance to the successful
execution of quantum cryptography, cannot be discussed in an unrestricted
publication and will be treated elsewhere.

\subsection{Universal Maximal Rate Predictions}

In this section we make use of the different results obtained in this paper
to deduce universal bounds on the maximal throughput rates that can be achieved
with various quantum cryptography systems and scenarios. We consider examples of
ground-ground, ground-satellite, air-satellite and satellite-satellite links.

\subsubsection{Necessary Condition for Unconditional Secrecy}

Although the full derivation of the complete effective secrecy capacity is rather
complicated, the necessary condition that must be satisfied in order to ensure that
Alice and Bob
share at least {\it some} unconditionally secret bits (in the sense of privacy
amplification) is extremely simple. There will be at least some number of secret bits
shared between Alice and Bob if the optimal effective secrecy capacity is positive:
\be
\label{299}
{\cal S}_{{\rm opt}}>0~.
\ee
If this simple necessary condition is satisfied then we are guaranteed that there
will be some secret bits, as we have constructed ${\cal S}_{{\rm opt}}$ to account
for {\it all} system effects, so that there are by definition and construction no
losses suffered by the system other than those which are already incorporated
in ${\cal S}_{{\rm opt}}$. 

Given that the condition in eq.(\ref{299}) is satisfied, {\it any} quantum cryptography
system will produce some number of
secret bits shared between Alice and Bob. The
particular {\it rate} at which these secret bits are generated will then
be entirely
determined by the value of the bit cell period and by the number of multiple
beams, if any, that are multiplexed together.

\subsubsection{Systems with Single Transmitter-Receiver Arrangement}

In this section we consider a number of representative examples of
quantum cryptography systems to illustrate the
throughput rates that can be achieved for the exchange of unconditionally
secret Vernam cipher material between Alice and Bob. For all
of the examples below we assume for definiteness
that the laser at the Alice end of the system produces
pulses of light at a wavelength of 1550 nanometers. Although they are not
available today, we also {\it assume} for all but the last of these
examples that high-speed HEP
photon detectors, of the kind described in Section 5.2.1 above, will in the near future
be available to count
photons at a rate of 10 GHz corresponding to a bit cell period
$\tau=100~{\rm picoseconds}$, with a dark count per bit cell
of $4.25\times 10^{-18}$ (as discussed in footnote \ref{darkcount} above).
In the last example below we will calculate rates based on the use of generic,
currently available commercial photon detectors capable of detecting photons at a
rate of 1 MHz with an assumed dark count per bit cell rating
of $1\times 10^{-6}$.

\clearpage
\vskip 20pt
\leftline{({\it i}){\it ~Free Space Quantum Channel: Aircraft-to-Satellite (LEO) Link}}

For this example we
consider a quantum cryptography setup in which Alice is located on a LEO satellite
at an altitude of 300 kilometers above MSL and Bob is located on a platform
at an altitude that
is substantially above the bulk of the atmospheric turbulence, which we take to
be an aircraft flying at 35000 feet or
higher ({\it e.g.}, such as a suitably modified
Joint Surveillance Target Attack Radar System (Joint STARS) aircraft). As
with the various
examples given in Section 4.1.3 above, we take the diameter of the aperture
of Alice's transmitting instrument to be $D_A=30$ cm. Inspection of
Figure \ref{F:attenuation1} reveals that
in this situation the line attenuation will be given by $\alpha=-10~{\rm dB}$ if we
take a value of $D_B=58$ cm for the diameter of
the aperture of Bob's receiving instrument.\footnote{
This is a realistic value for the size of airborne optics. For example, it is
publicly known \cite{fasurl} that the U.S. Department of Defense U-2 and U-2R aircraft
have in the past
been equipped with the 30-inch (76.2 cm) Optical Bar Camera, in
addition to other sensors.}
\footnote{
We emphasize again, as discussed at the beginning of Section 4.1 above, that the
line attenuation $\alpha$
is not the ``total" attenuation suffered by the signal.}

\begin{figure}[htb]
\vbox{
\hfil
\scalebox{0.66}{\rotatebox{0}{\includegraphics{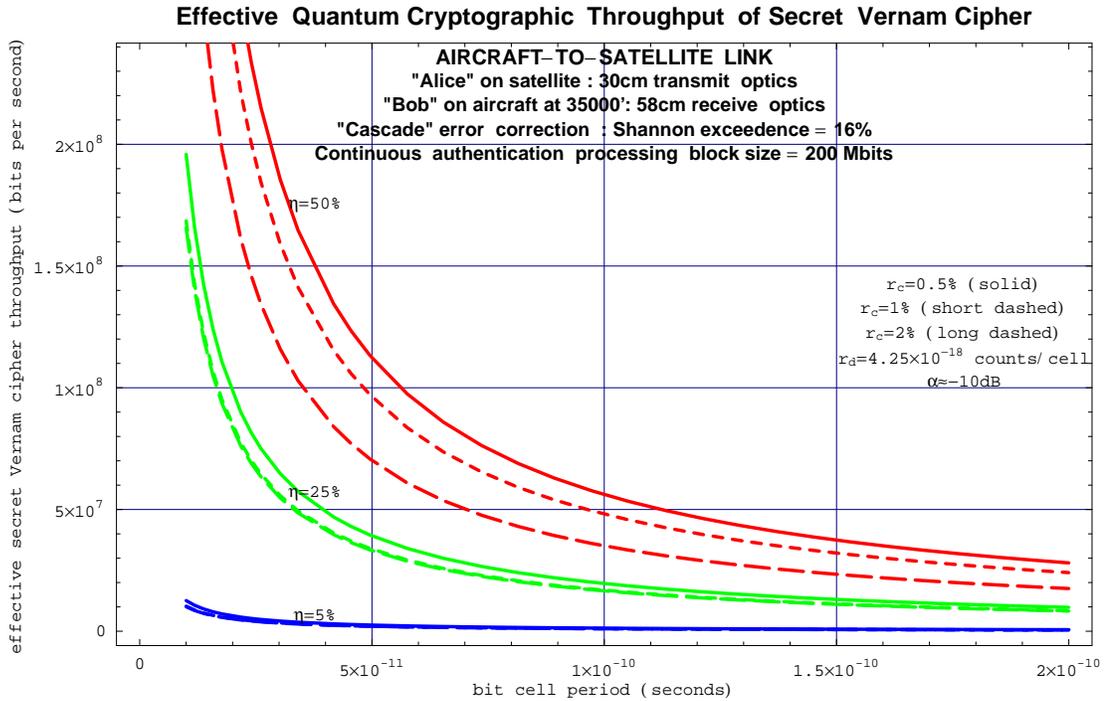}}}
\hfil
\hbox to -1.25in{\ } 
}
\bigskip
\caption{%
Effective Rate Graph for Aircraft-to-Satellite (LEO) Link: $\alpha=-10~{\rm dB}$
(``Bob" - 58 cm telescope at
35000' ; ``Alice" - 30 cm telescope on
LEO satellite)
}
\label{F:rate1}
\end{figure}

\begin{figure}[htb]
\vbox{
\hfil
\scalebox{0.66}{\rotatebox{0}{\includegraphics{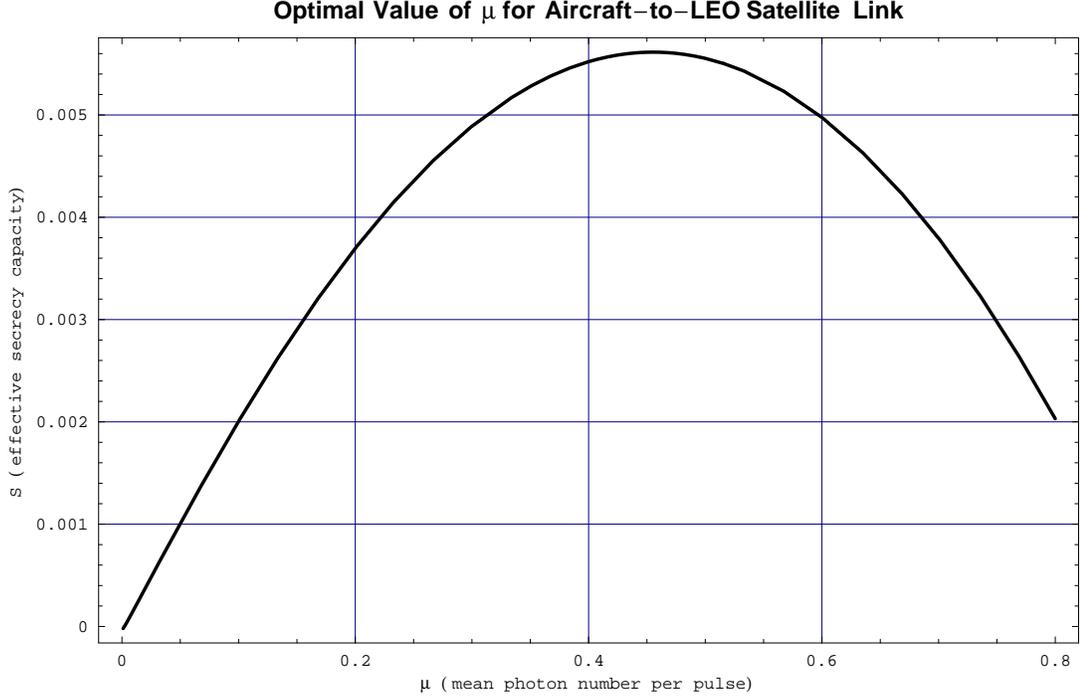}}}
\hfil
\hbox to -1.25in{\ } 
}
\bigskip
\caption{%
Effective Secrecy Capacity Graph for Aircraft-to-LEO Link:$\eta=50\%$; $r_c=0.005$
$\alpha=-10~{\rm dB}$ (58 cm telescope at 35000' - LEO satellite)
}
\label{F:muopt}
\end{figure}

In Figure \ref{F:rate1}
we plot the optimal effective secrecy
rate, ${\cal R}_{{\rm opt}}$, for the above QC system as a function of the bit
cell period. Inspection of the graph reveals that for a bit cell period of
100 picoseconds, corresponding to a laser with a PRF of 10 GHz, and a photon
detector device efficiency $\eta$ of 50\%,
the maximum rate
for the generation of Vernam cipher material is about 57 Megabits per second,\footnote{
Due to the assumption in this example that Bob is above most of the atmospheric
turbulence we note that the rates for this scenario are independent of the slant
angle between Bob and Alice.}
based on a calculated value for the optimal mean photon number per pulse of
$\mu_{{\rm opt}}=0.455$, which is obtained by numerically solving eq.(\ref{170}).
This is illustrated in Figure \ref{F:muopt}, where the (real) solution to
eq.(\ref{170}) is plotted.
For this example we have taken
an assumed fractional intrinsic channel error rate value of
$r_c=0.005$. From eq.(\ref{206}) we see that this value of $r_c$ requires active
control of the relative angular misalignment between the satellite and the
airborne platform so as to restrict overall relative motion within a cone of solid
angle
\bea
2\delta&\lsim& 2\arcsin\left(\sqrt {0.005}\right)
\nonumber\\
&=&0.142 ~{\rm radian}
\nonumber\\
&=& 8.11~{\rm degrees}~.
\eea
This throughput rate is higher than that provided by a standard
T3 telecommunications link, and in particular is faster than the 45 Megabits per second
data rate of the TACLANE encryptor system \cite{TACLANE}.

We see that for $r_c=0.01$, which (as described in the text below eq.(\ref{206}))
requires attitude control within a cone of solid angle 11.5 degrees, the optimal
effective secrecy rate decreases to ${\cal R}_{{\rm opt}}=49$ Megabits per
second, in this case for a lower optimal mean photon number per pulse value of
$\mu_{{\rm opt}}=0.426$.
If we allow instead
for a value of $r_c=0.02$, which only requires attitude control within
a cone of solid angle of 16.3 degrees, we find an optimal throughput rate of
${\cal R}_{{\rm opt}}=36$ Megabits per second, for $\mu_{{\rm opt}}=0.37$.

If we consider now a photon detector apparatus with a lower intrinsic device efficiency
of $\eta=25\%$, we find that for a value of $r_c=0.005$
the optimal throughput rate drops to about 21 Megabits per second (for a value
of $\mu_{{\rm opt}}=0.131$), with rate values
of about 18 Megabits per second for quantum channels characterized by fractional
intrinsic error values of either 1\% or 2\% (for $\mu_{{\rm opt}}=0.125$ and
$\mu_{{\rm opt}}=0.122$, respectively). Finally, we note that when the device
efficiency of the photon detector apparatus drops below about $\eta\approx 5\%$,
there can be no unconditionally
secret Vernam cipher material exchanged {\it at all} between Alice and Bob.

For all of the above scenarios we have taken a value for the Shannon deficit parameter
$x$ of $x=1.16$ ({\it cf}
eq.(\ref{43})), which means that we are assuming that an efficient method of error
correction has been employed that approaches the Shannon limit to within 16\%, and
we use a raw bit processing block size of $m=200$ Megabits. In addition, we have
also set all of the continuous authentication security parameter values, $g_i$,
({\it cf} eq.(\ref{152})), as well as the privacy amplification security parameter
$g_{pa}$, equal to 30, and we have employed a value of $\epsilon=10^{-9}$ for the
selectable infinitesimal quantity that determines the success likelihood for attacks
on single-photon pulses ({\it cf} eq.(\ref{49}) and the discussion in the text above
eq.(\ref{45})).

\vskip 20pt
\leftline{({\it ii}){\it ~Free Space Quantum Channel: Earth-to-Satellite (LEO) Link;
clear weather}}

For this example we consider a quantum cryptography setup in which Alice is
on a LEO satellite at an altitude of 300 kilometers and Bob is on the ground
at MSL. Unlike the previous example, in this scenario the full effects of
atmospheric turbulence are important.
As before we take the diameter of the aperture of the transmit optics to
be $D_A=30$ cm, and we see from Figure \ref{F:attenuation1} that
in order to achieve a value for the line
attenuation of $\alpha=-20~{\rm dB}$ we must take a value of $D_B=50$ cm for the
diameter of the aperture of Bob's reeciving instrument.
As before, we assume a value for the Shannon deficit parameter of
$x=1.16$ and employ a raw bit processing block size of $m=200$ Megabits.

\begin{figure}[htb]
\vbox{
\hfil
\scalebox{0.66}{\rotatebox{0}{\includegraphics{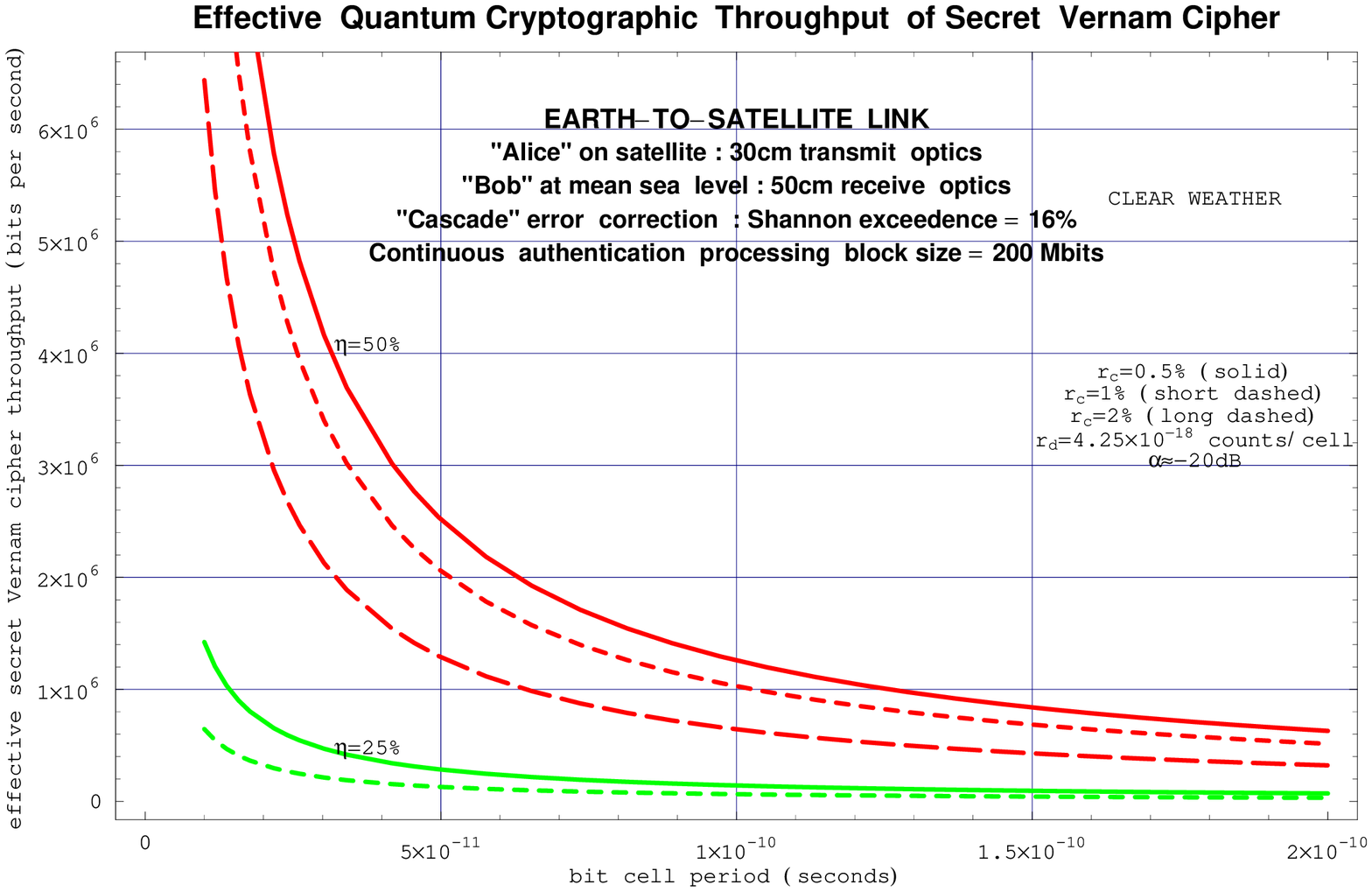}}}
\hfil
\hbox to -1.25in{\ } 
}
\bigskip
\caption{%
Effective Rate Graph for Earth-to-Satellite (LEO) Link: $\alpha=-20~{\rm dB}$
(``Bob" - 50 cm telescope at
mean sea level ; ``Alice" - 30 cm telescope on
LEO satellite; clear weather)
}
\label{F:rate2}
\end{figure}

In Figure \ref{F:rate2} we plot the optimal effective secrecy rate for this
QC system as a function of the bit cell period. Inspection of the graph reveals
that, for a pulsed laser source with a PRF of 10 GHz and a photon detector with
an intrinsic device efficiency of $\eta=50\%$, the maximum rate for the generation
of Vernam cipher material is now about 1.3 Megabits per second, for a calculated
value of the optimal mean photon number per pulse of $\mu_{{\rm opt}}=0.131$,
where we have assumed that $r_c=0.005$. If we instead consider
values of $r_c=0.01$ and $r_c=0.02$, we find corresponding
throughput rates of 1.05 Megabits per
second and 665 Kilobits per second for values of $\mu_{{\rm opt}}=0.125$ and
$\mu_{{\rm opt}}=0.111$, respectively.

We note that the effective throughput
rates drop precipitously if we consider photon detector apparatuses with a smaller
intrinsic device efficiency of $\eta=25\%$. In this case we find that, for a value
of $r_c=0.005$ there is a maximum throughput rate of unconditionally secret Vernam
cipher material of about 165 Kilobits per second (for a value of
$\mu_{{\rm opt}}=0.0898$), and for a value of $r_c=0.01$ the maximum rate is
about 105 Kilobits per second (for a value of $\mu_{{\rm opt}}=0.0848$). Poorer values
for {\it either} the detector efficiency $\eta$
or the fractional intrinsic quantum channel
error $r_c$ produce essentially no viable throughput of Vernam cipher material at all.

For the above examples we have assumed ``clear" weather conditions for the input to
the FASCODE runs (as described in Section 4.1.2 above, this is defined as yielding 23
kilometers visibility), and we have taken the LEO satellite to be located at
{\it zenith} above Bob. We have also taken values of $g_i=g_{pa}=30$ and
$\epsilon=10^{-9}$.

In contrast to the previous example in which maximal rates in excess of those
characterized by T3 telecommunications lines are possible, we see that the
effects of atmospheric turbulence reduce the maximal possible
rate for a ground-to-LEO satellite link in clear weather conditions
to slightly less than that provided by a standard T1 telecommunications
line, if we make use of a small (50 cm) receiving telescope.\footnote{
Note that the publicly acknowledged data rate for the radiation-hardened
U.S. Department of Defense MILSTAR satellite is that of a T1 link \cite{MILSTAR}.}

\begin{figure}[htb]
\vbox{
\hfil
\scalebox{0.66}{\rotatebox{0}{\includegraphics{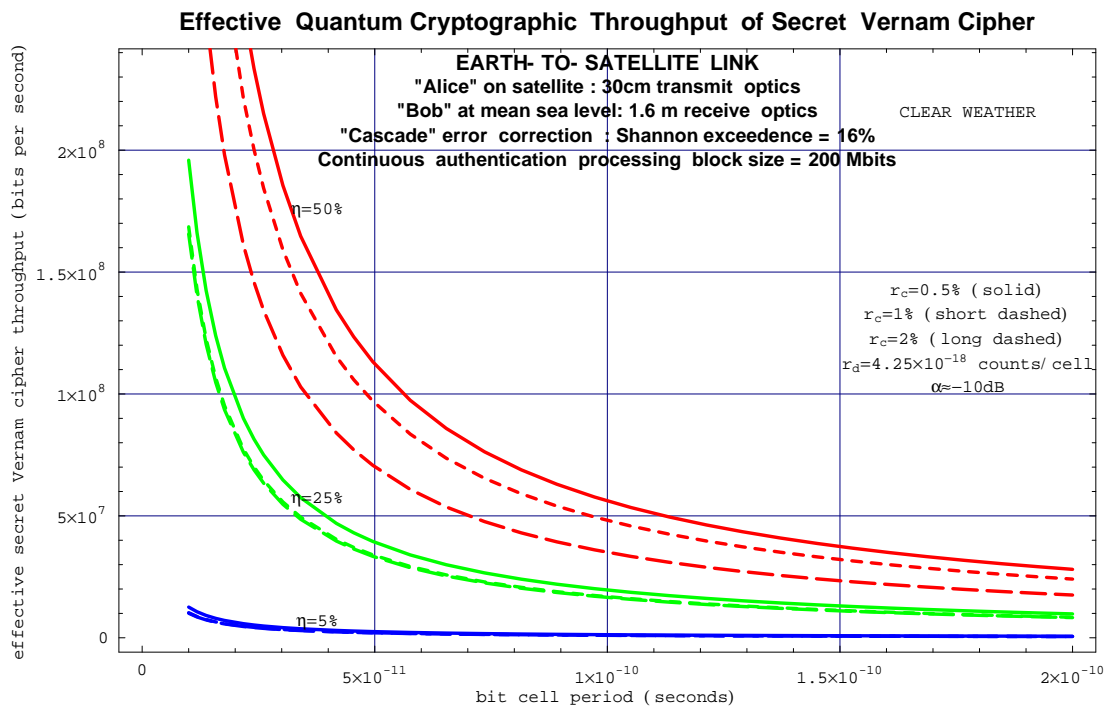}}}
\hfil
\hbox to -1.25in{\ } 
}
\bigskip
\caption{%
Effective Rate Graph for Earth-to-Satellite (LEO) Link: $\alpha=-10~{\rm dB}$
(``Bob" - 1.6 m telescope at
mean sea level ; ``Alice" - 30 cm telescope on
LEO satellite; clear weather)
}
\label{F:rate2a}
\end{figure}

However, further inspection of Figure \ref{F:attenuation1} reveals that we may
achieve the much better line attenuation value of $\alpha=-10~{\rm dB}$ by employing
instead a receiving
telescope with an optical aperture of $D_B=1.6$ m. (Since the Bob apparatus is on the
ground in this example, the larger size of the receiving telescope optics is
acceptable compared to the previously considered
scenario in which Bob is located on an airborne platform.) In this case,
the effective
throughput rates will be identical to those obtained for the aircraft-to-saltellite
link example considered above, as we illustrate in Figure \ref{F:rate2a}.\footnote{
In this example we keep all parameters at the same values used for the calculation
of the ground-to-LEO
satellite link rate with the 50 cm receive optics, except that we use the values
for $\mu_{{\rm opt}}$ that we found in Scenario ({\it i})
above for the aircraft-to-LEO satellite case (because the line attenuation has the
same value).}
Thus it
should be possible to establish
between a ground station and a LEO satellite a clear weather quantum cryptography
link that operates at a rate somewhat faster
than that provided by a T3 line when the satellite is at the zenith location. (When the
satellite drops to the 45 degree declination position the value of the
line attenuation will change to about $\alpha=-11.8~{\rm dB}$, causing a drop in the
throughput rate.)

\vskip 20pt
\leftline{({\it iii}){\it ~Free Space Quantum
Channel: Earth-to-Satellite (LEO) Link; light
rain}}

\begin{figure}[htb]
\vbox{
\hfil
\scalebox{0.66}{\rotatebox{0}{\includegraphics{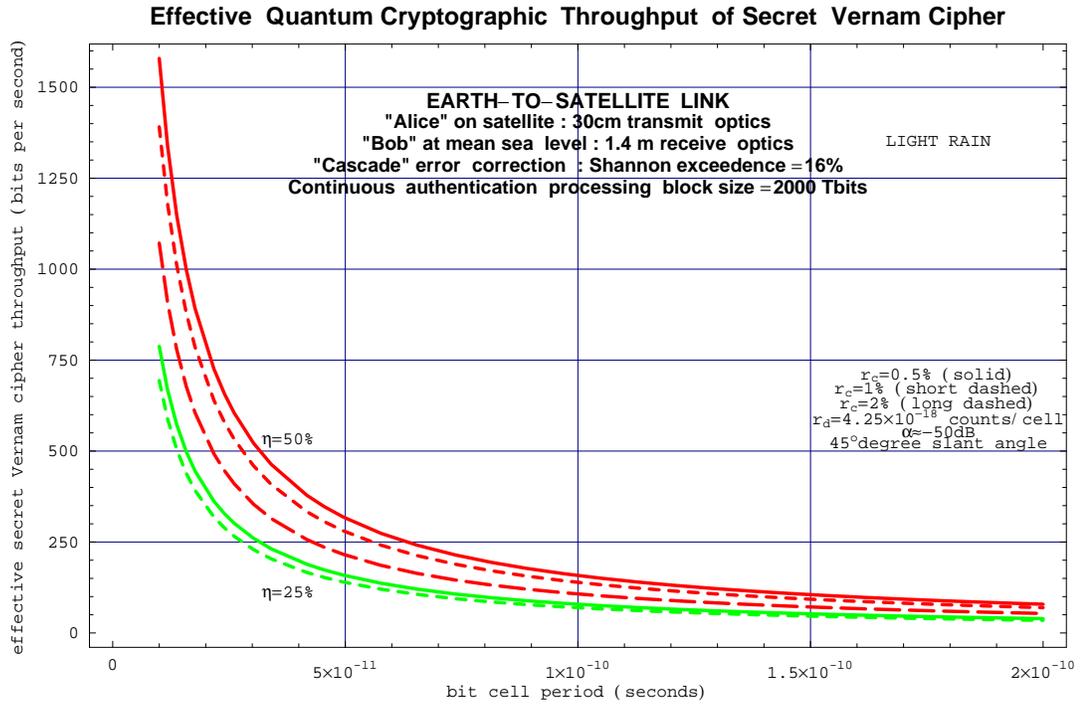}}}
\hfil
\hbox to -1.25in{\ } 
}
\bigskip
\caption{%
Effective Rate Graph for Earth-to-Satellite (LEO) Link: $\alpha=-50~{\rm dB}$
(``Bob" - 1.4 m telescope at
mean sea level ; ``Alice" - 30 cm telescope on
LEO satellite; light rain)
}
\label{F:rate3a}
\end{figure}

To illustrate the severe link degradation that can spoil a quantum cryptography
system in the presence of adverse weather conditions, we again consider an example in
which, as before, Alice and Bob are located, respectively,
on a LEO satellite and at a ground
station at MSL. In this case, we replace the assumption of clear weather conditions
with the assumption of ``light rain" conditions. This is quantitatively
incorporated in the problem by running the FASCODE computer program with the
appropriate
corresponding input parameters, for which ``light rain" is defined (as in standard
meteorological analysis) as comprising 5 mm per hour of precipitation. Numerical results
analogous to those displayed in the graph in Figure \ref{F:attenuation1}
\begin{figure}[htb]
\vbox{
\hfil
\scalebox{0.66}{\rotatebox{0}{\includegraphics{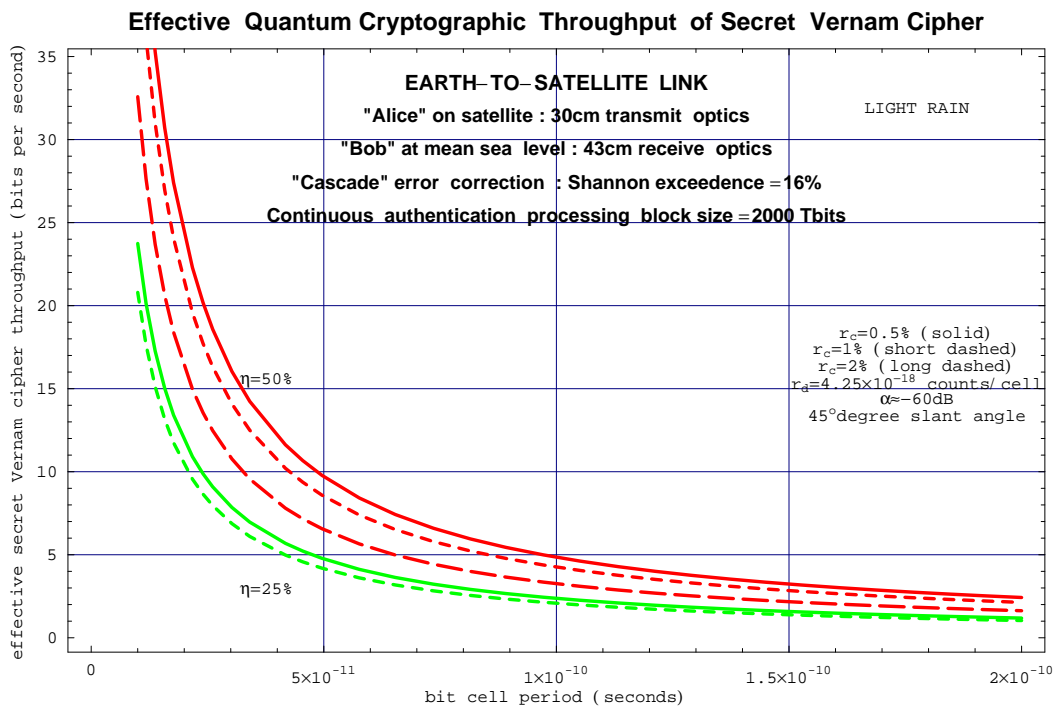}}}
\hfil
\hbox to -1.25in{\ } 
}
\bigskip
\caption{%
Effective Rate Graph for Earth-to-Satellite (LEO) Link: $\alpha=-60~{\rm dB}$
(``Bob" - 43 cm telescope at
mean sea level ; ``Alice" - 30 cm telescope on
LEO satellite; light rain)
}
\label{F:rate3}
\end{figure}
indicate
that if we take
the diameter of the aperture of the transmit optics to be $D_A=30$ cm and the
diameter of the aperture of the receive optics to be $D_B=43$ cm (slightly smaller than,
but
roughly comparable to, the smaller of the two receive apertures used
in Scenario ({\it ii}) above in the clear weather example), the
line attenuation will be given by $\alpha=-60~{\rm dB}$ for a slant angle of
45 degrees. For this scenario we observe
a very severe decrease in the throughput rate of the system, as can be seen
from the results plotted in Figure \ref{F:rate3}. With the parameters
$g_i$, $g_{pa}$, $\epsilon$ and $x$ set
identically to the values taken in the clear weather example above, we find that
the maximum throughput rate for a ground-to-LEO satellite link
in the presence of light rain is given by about 5 bits per second for a bit cell
period of 100 picoseconds and a detector efficiency of 50\%, assuming a quantum
channel fractional intrinsic error of $r_c=0.005$ and employing a calculated
optimal value of $\mu_{{\rm opt}}=0.00328$. This drops to a rate of about 2.25 bits
per second for a system with a photon detector efficiency of 25\% and a quantum
channel fractional error of $r_c=0.02$, corresponding to a value
of $\mu_{{\rm opt}}=0.0029$.
Note that, in order to achieve even these low throughput rates it is necessary to
take a value for the raw bit processing block of $m=2000~{\rm Terabits}$ (as opposed
to the value of $m=200~{\rm Megabits}$ employed in the previous examples). Smaller
values for the raw bit processing block do not yield any shared secret Vernam cipher
material at all, due to the very severe degradation to the quantum channel caused
by the rain. Although the use of such a large processing block
can in principle
be arranged so as {\it not} to introduce a larger classical communications
bandwidth requirement between Alice and Bob,\footnote{
The communications load in bits per second is actually smaller for a larger raw
block size, $m_0$, provided the sifted block size, $n_0$, does not
change.  This may be understood on the basis of two observations: the
number of transmitted bits per block is roughly linear in the size of
the sifted block and depends only weakly on the size of the raw block,
and the amount of time available to carry out the transmission increases
linearly with the raw block size.  The communications load then varies
roughly as $n_0/m_0$, which decreases with increasing $m_0$ at constant
$n_0$.}
a processing block of this large size is nevertheless
not realistic for satcom quantum cryptography applications.
This is because the {\it physical
size} requirements on the memory that Alice must utilize to accept such a large
raw bit processing block are incompatible with typical satellite space (and power
and cooling) constraints.\footnote{
A rough estimate based on the characteristics of
currently available memory modules indicates that such
a processing block would require enough space on the
satellite - for the memory alone - to accomodate a large vehicle such as a truck.}
Furthermore, without
monitoring of click statistics by
Bob it is impossible to achieve any viable throughput in this scenario, so that
we have made use of the appropriately modified form of the privacy amplification
function in computing these rates ({\it cf} eq.(\ref{81})), wherein the the leading
term is down by a factor of $\eta$ compared to the version without monitoring
of click statistics.
In addition to the preceding, we have {\it
also} had to employ a slightly modified value of ${\hat z}_E\left(\mu\right)$
({\it cf} eq.(\ref{57})) to achieve even these rates: following the discussion
in the text between
eqs.(\ref{57}) and (\ref{58}), we have (in this example only) assumed
that the enemy is unable to achieve the full strength direct attack implied by
eq.(\ref{57}), which varies as ${\hat z}_E\left(\mu\right)={1\over 12}\mu^3+O\left(
\mu^4\right)$, and instead taken a value of one-third of this, so that in the
corresponding privacy amplification function we have
${\hat z}_E\left(\mu\right)={1\over 36}\mu^3+O\left(\mu^4\right)$.
Of course, it would be possible to achieve higher
throughput in the presence of adverse weather conditions by employing a larger
aperture for the receive optics at the Bob site (modulo the above-mentioned
concerns regarding physical constraints on memory that can be placed on a realistic
satellite).
If we instead use a receiving telescope with an optical aperture
of $D_B=1.4$ m we obtain the better line attenuation value
of $\alpha=-50~{\rm dB}$. The results for this case are illustrated in
Figure \ref{F:rate3a},
inspection of which reveals
a maximum throughput rate of 164 bits per second for a detector efficiency
of $\eta=50\%$, a bit cell period of 100 picoseconds, a quantum channel fractional
error of $r_c=0.005$ and at a calculated optimal
value of $\mu_{{\rm opt}}=0.0106$ with a slant angle of 45 degrees.

\vskip 20pt
\leftline{({\it iv}){\it ~Free Space Quantum Channel: Earth-to-Satellite (LEO) Link;
moderate rain}}

In the presence of ``moderate rain," which is defined for FASCODE runs as
comprising 12.5 mm per hour of precipitation, the line attenuation $\alpha$ becomes
much worse, never any better than -76 dB even for a LEO satellite located at zenith
above the ground station with a receiving optics aperture of $D_B=1.6$ m diameter.
In this case the only means of producing {\it any} shared,
unconditionally secret Vernam cipher at all is to increase the processing block
size to a value which is not practical for currently available computing machinery that
can be fitted on a satellite.

\vskip 20pt
\leftline{({\it v}){\it ~Free Space Quantum Channel: Earth-to-Satellite
(GEO) Link; clear
weather}}

We now consider the example of a ground-to-GEO satellite link. Thus, we assume that
Alice is located on a geosynchronous satellite above the Earth at an altitude
of 35783 kilometers (22236 miles) above mean sea level.
In order to achieve a viable effective throughput
rate of Vernam cipher material we envisage for this scenario that Bob is located at
a sufficiently high altitude so as to mitigate somewhat the effects of atmospheric
turbulence. Unlike the example of the aircraft-to-satellite link considered above, we
here want to describe a situation in which there would be a more-or-less
permanent link between the Bob and Alice, so that Bob should be located at a ground
receiving station. An important point here is that the GEO link should be available
for a much more extended period of time that is the case for the LEO link, which would
provide access for a roughly nine minute period before the satellite drops below
the horizon visible to Bob. Thus, we may hope to achieve effective, {\it integrated}
throughput values that are higher in the GEO link case than in the LEO link case.

For this computation we have
taken the parameters $g_i$, $g_{pa}$, $\epsilon$ and $x$ as in the previous
examples, but we now choose a value of $m=2 ~{\rm Gigabits}$ for the raw
bit processing block size. Unlike the example of the Earth-to-LEO satellite link in
the presence of light rain, for which it was necessary to take a much larger value
for the raw bit processing block size, {\it this} value imposes no practical
difficulties regarding either communications bandwidth or physical space
requirements for memory size on the satellite. We do adopt, however,
as in the example above
of the Earth-to-LEO satellite scenario, the assumption that Bob actively monitors
click statistics, and we use the corresponding privacy amplification function for
the direct attack as a result. Without the use of the form of the privacy amplification
function associated to active click statistics monitoring by Bob, no viable
throughput can be established for the Earth-to-GEO satellite link.\footnote{
We do assume that the enemy can mount the strongest possible direct attack, as
measured by ${\hat z}_E\left(\mu\right)$ given in eq.(\ref{57}).}
An important requirement to achieve viable throughput
is to employ a sufficiently large receiveing telescope aperture, since the line
attenuation due to the spreading of the beam is otherwise completely prohibitive.

\begin{figure}[htb]
\vbox{
\hfil
\scalebox{0.66}{\rotatebox{0}{\includegraphics{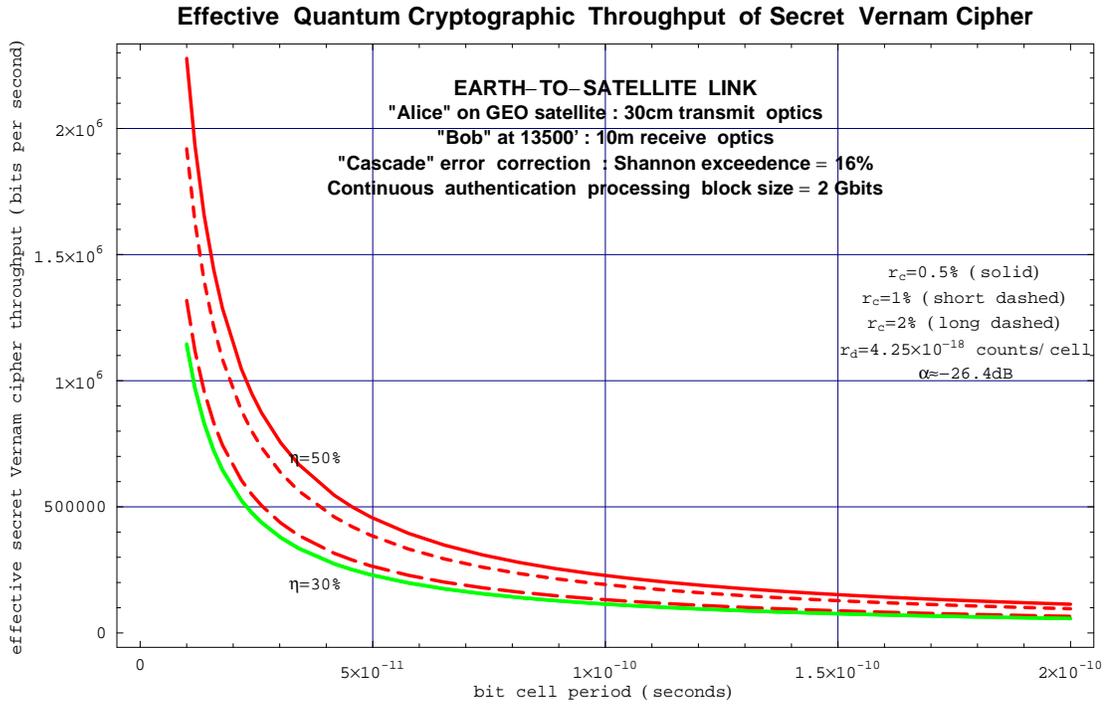}}}
\hfil
\hbox to -1.25in{\ } 
}
\bigskip
\caption{%
Effective Rate Graph for Earth-to-Satellite (GEO) Link: $\alpha=-26.4~{\rm dB}$
(``Bob" - 10 m telescope at
13500' ; ``Alice" - 30 cm telescope on
GEO satellite)
}
\label{F:rate5}
\end{figure}

For this example we envisage a receiving telescope with an effective optical
aperture of 10 meters. Although this is a very large optical instrument, this
size of receive aperture is characteristic of what has been proposed in the
literature \cite{deepspace2} for use in optical
communications for deep space missions (coincidentally, such
proposals for classical optical deep space lasercomm
have also incorporated 30 cm transmit optics), and this is also available at the Keck
Telescope Facility on Mauna Kea mountain in Hawaii at an altitude of 13500 feet. For
the purposes of this example we will imagine that Bob is located at such a site and
has access to such an instrument. In
this case the line attenuation $\alpha$ can be calculated using eq.(\ref{204}) and
is found to
be given by
$\alpha=-26.4~{\rm dB}$, and we use this value in the numerical evaluation
of ${\cal R}_{{\rm opt}}$. In Figure \ref{F:rate5} we plot the optimal
effective secrecy rate that characterizes such a
ground-to-GEO satellite quantum cryptography system.
Inspection of the curves in the graph reveals that,
with a laser PRF of 100 picoseconds, a
photon detector device efficiency of $\eta=50\%$ and a value for the
quantum channel intrinsic error of $r_c=0.005$, such a system should achieve a
throughput rate of about 240 Kilobits per second, for a value of
$\mu_{{\rm opt}}=0.0891$. This throughput rate, which should be essentially
continuously available since Alice is on a GEO satellite, is roughly one-sixth the
rate of a standard T1 telecommunications link.\footnote{
Note that this rate is approximately equal to the throughput
rates currently available for the ``Mobile Subscriber
Equipment" (MSE) and ``Tri-Service Tactical Communications" (TRITAC) systems,
employed by the U.S. Army and U.S. Air Force/U.S. Marine Corps, respectively.}
The important point is that, as mentioned
above, such a link would be available for more than the approximately nine-minute
period provided by a LEO link prior to the disappearance of the latter type of
satellite below the horizon, thus potentially providing a comparable (or higher)
effective integrated throughput
value compared to the latter. If we now consider a photon detector with a lower
device efficiency given by $\eta=30\%$, we find (again for an intrinsic
channel error value of $r_c=0.005$) an effective throughput rate of about 118
Kilobits per second (for a value of $\mu_{{\rm opt}}=0.0884$).

\vskip 20pt
\leftline{({\it vi}){\it ~Free Space Quantum Channel: Satellite-to-Satellite
(GEO-to-GEO) Link}}

It is interesting to consider the possible use of quantum cryptography for
satellite-to-satellite communications. There are a wide range of scenarios that might
be considered: to illustrate the general problem we will only briefly discuss a
GEO-to-GEO satellite link. We will take a very simple model of a three-satellite
constellation set up to provide a combined
footprint covering most of the Earth (with the
exception of the polar regions), here for simplicity assumed to be situated at
120 degree angles with respect to each other. Assuming a GEO altitude of 35783
kilometers and noting that the radius of the Earth is 6378 kilometers, we consider
a satellite-to-satellite crosslink distance of about 48683 kilometers (30250 miles). It
is clear from the previous example that the line attenuation due to beam spreading
will be even larger here than for the Earth-to-GEO scenario.
In order to achieve useful throughput values
this large amount of line attenuation would presumably
require that the Bob satellite be equipped
with a receiving instrument that has an effective optical aperture of at least 10
meters in size, which is considerably larger than the 2.6 meter
aperture of the Hubble Space Telescope. It might be possible in the future
to obtain such an effective aperture for a spaceborne light-collecting instrument.
In this connection the
U.S. National Aeronautics and Space Administration (NASA) has recently announced the
``Gossamer Spacecraft Initiative" \cite{Gossamer1,Gossamer2} which
is intended to result in a spaceborne telescope with
an effective aperture of 50 meters or more in size. Although this initiative is still
in the earliest planning stages, such an instrument could be used for quantum
cryptography as well as astronomical research activities.

\clearpage
\vskip 20pt
\leftline{({\it vii}){\it ~Fiber Optic Quantum Channel}}

In this example we consider a fiber-optic cable implementation of quantum cryptography.
We envisage for this example the use of high-quality, polarization-preserving
fiber characterized by an intrisic attenuation characteristic of 0.2 dB per kilometer,
and for purposes of illustration compare the associated throughput values
with results for lower quality fiber with an attenuation characteristic
of 0.3 dB per kilometer. We take the photon detector device efficiency to be
$\eta=50\%$, and we assume that appropriate splicing and insertion of suitable
dispersion-compensating fiber segments, as discussed
in Section 4.2 above, has been carried out
so as to mitigate the dispersion losses described and
analyzed there. To account for the
associated splicing loss and other system imperfections we assume that the quantum
channel is characterized by a total bulk loss of -5 dB, in addition to the
losses associated with the attenuation per unit length. 

\begin{figure}[htb]
\vbox{
\hfil
\scalebox{0.66}{\rotatebox{0}{\includegraphics{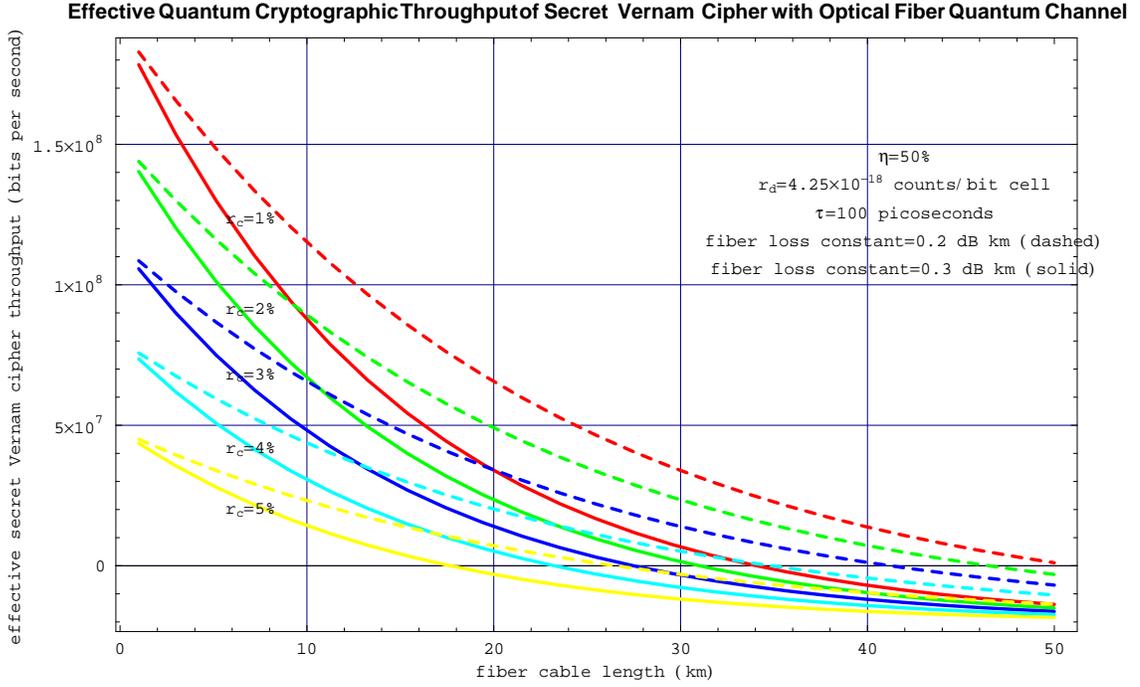}}}
\hfil
\hbox to -1.25in{\ } 
}
\bigskip
\caption{%
Effective Rate Graph for Fiber-Optic Cable Link without Surreptitious Cable Replacement
}
\label{F:rate6}
\end{figure}

In Figure \ref{F:rate6} we plot several
effective secrecy rate curves for this system. In this example, we compute
throughput rates in the case that the cable remains untouched by the enemy. We
see that, for a good quality cable with an attenuation characteristic of
0.2 dB per kilometer and an intrinsic channel error value
of $r_c=0.01$, the rate to a distance of 10 kilometers along the cable should
be at least as high as about 115 Megabits per second, whereas for a cable with
an attenuation characteristic of 0.3 dB per kilometer (and the same value for
$r_c$) the corresponding rate should be at least as high as about 88 Megabits per
second (in each case using a value of $\mu_{{\rm opt}}=0.4$). We note that, for
{\it any} of the illustrated parameter values,\footnote{
We display curves with values of $r_c=\{0.01,0.02,0.03,0.04,0.05\}$.}
for the cable with the attenuation characteristic
of 0.3 dB per kilometer {\it any} exchange
of unconditionally secret Vernam cipher material
beyond about 33 kilometers is impossible. For the higher-quality cable with
a loss characteristic of 0.2 dB per kilometer, throughput of at least {\it some}
number of secret key bits is just barely possible to the 50 kilometer point in the
case of a low error value of $r_c=0.01$.

\begin{figure}[htb]
\vbox{
\hfil
\scalebox{0.66}{\rotatebox{0}{\includegraphics{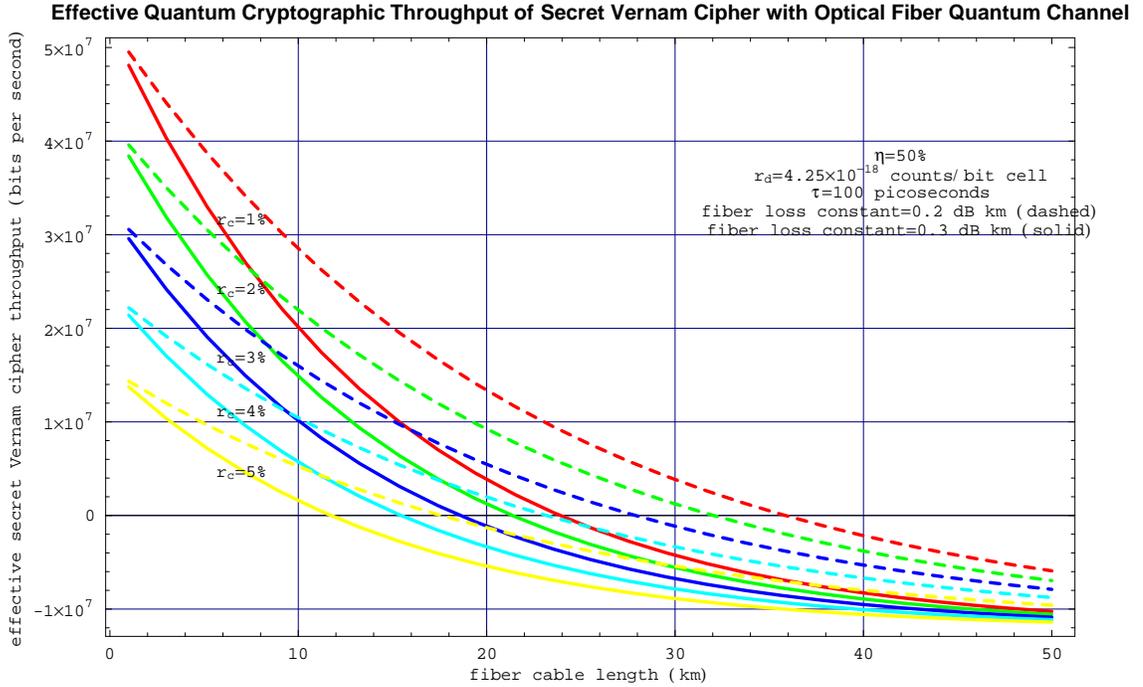}}}
\hfil
\hbox to -1.25in{\ } 
}
\bigskip
\caption{%
Effective Rate Graph for Fiber-Optic Cable Link with Surreptitious Cable Replacement
}
\label{F:rate6a}
\end{figure}

We now consider the case that the enemy has somehow
been able to surreptitiously replace
the cable with one which is effectively lossless. In Figure \ref{F:rate6a} we
plot several
effective secrecy rate curves for this system. We
see that, for a good quality cable as above with an attenuation characteristic of
0.2 dB per kilometer and an intrinsic channel error value
of $r_c=0.01$, the rate to a distance of 10 kilometers along the cable should
drop to a value at least as high as about 29 Megabits per second, whereas for
a cable with
an attenuation characteristic of 0.3 dB per kilometer (and the same value for
$r_c$) the corresponding rate should drop to a value at
least as high as about 20 Megabits per
second (in each case in this example we use a value of $\mu_{{\rm opt}}=0.1$). We
note that, for any of the illustrated parameter values,
for the cable with the attenuation characteristic
of 0.3 dB per kilometer any exchange
of unconditionally secret Vernam cipher material
beyond about 24 kilometers is impossible. For the higher-quality cable with
a loss characteristic of 0.2 dB per kilometer, throughput of at least {\it some}
number of secret key bits is just barely possible to the 36 kilometer point in the
case of a low error value of $r_c=0.01$.

\vskip 20pt
\leftline{({\it viii}){\it ~Free Space Quantum Channel: Aircraft-to-Satellite (LEO) Link;
1 MHz Photon Detector}}

In the above examples we analyzed the throughput rates that will be available
with the potential development in the future of high-speed photon detectors capable of
counting photons at a rate of 10 GHz. Here, we compare the rates predicted
for the aircraft-to-LEO
satellite link considered in Figure \ref{F:rate1} with the effective throughput rates
of unconditionally secret Vernam cipher that are possible with the use of a
``generic" commercially available
photon detector capable of counting at a rate of 1 MHz, with an assumed
value of $r_d=1\times 10^{-6}$.

\begin{figure}[htb]
\vbox{
\hfil
\scalebox{0.66}{\rotatebox{0}{\includegraphics{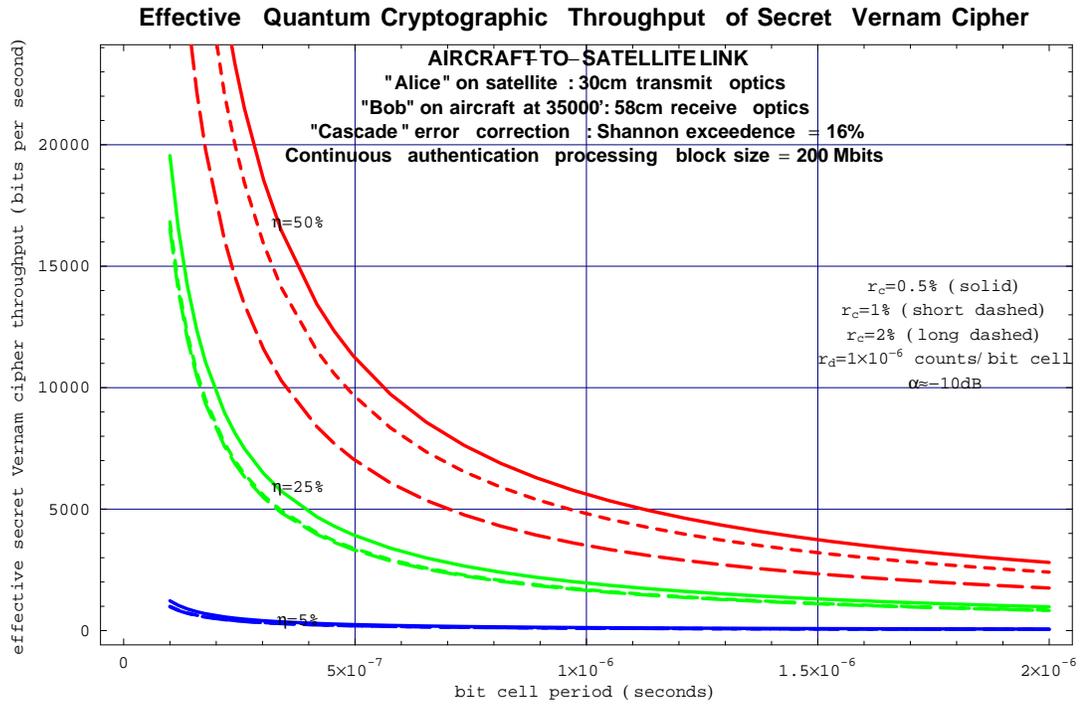}}}
\hfil
\hbox to -1.25in{\ } 
}
\bigskip
\caption{%
Effective Rate Graph for Aircraft-to-Satellite (LEO) Link: $\alpha=-10~{\rm dB}$
(``Bob" - 58 cm telescope at
35000' ; ``Alice" - 30 cm telescope on
LEO satellite; commercial (1 MHz) photon detector)
}
\label{F:rate8}
\end{figure}

In Figure \ref{F:rate8} we plot curves corresponding to those shown in Figure
\ref{F:rate1}. The various environmental conditions are taken to be identical here
as for the previous case, so that the various values for $\mu_{{\rm opt}}$ are identical
for each curve (the replacement of $r_d=4.25\times 10^{-18}$ with
$r_d=1\times 10^{-6}$ makes a negligible change in the solution to the optimization
equation, eq.(\ref{170})). We find that the new rates reflect
the simple, inverse dependence of ${\cal R}_{{\rm opt}}$ on the bit cell period
$\tau$: the highest rate possible in this scenario is about 5700 bits per second for
a good photon detector with a quantum efficiency of 50\% and a high quality
quantum channel with $r_c=0.005$, with various lower rates for the other combinations,
as expected. The case of $\eta=25\%$ and $r_c=0.01$, which is more realistic, yields
a highest possible rate of about 1760 bits per second.

\subsubsection{Systems with Multiple Transmitter-Receiver Arrangement}

The rates presented in the preceding section were calculated based on the assumption
that the Alice-and-Bob systems comprise a single transmitter and receiver combination.
In
considering how the data throughput rate may be increased by multiplexing together a
number of Alice-and-Bob systems, we may make use of the analysis presented in
Section 5.2.6. It is possible to show that in various situations the
effective system throughput rate can be increased by making use of a suitably
multiplexed multi-beam transmission, just as
is done in classical lasercomm systems \cite{laserrefs1,laserrefs2}. An analysis
of the
various rate improvements that are possible through the use of the multiplexing
technique described in Section 5.2.6 above will be presented in a future paper
\cite{demuxpaper}.

\subsubsection{Rate Improvement with Additional Emerging and
Possible Future Technology}

A number of other emerging commercial technology developments, as well as
research efforts that are currently underway, may provide additional means
to achieve and improve the overall performance characteristics
of high speed quantum cryptography systems in the future.

An emerging commercial technology area that may find useful application
in high speed quantum cryptography systems in the future would be the
incorporation of tiny, fast mirror switches, such as the mirror
components in the recently introduced
Lucent WaveStar ${\rm LambdaRouter}^{\rm TM}$ \cite{LambdaRouter}. The incorporation
of these small optical components in networked
systems, extending the application of quantum cryptography to allow for integration
into existing, multi-node communications architectures, is an area that has not been
thoroughly studied.

A promising research activity currently supported by the
U.S. Defense Advanced Research Projects Agency (DARPA) is
a project called ``Steered Agile Beams" (STAB) conducted by the U.S.
Air Force Research Laboratory, which is intended to develop
chip-scale laser beam control components for a number of applications \cite{STABref}.
If successful,
this work should in particular benefit applications that depend on adaptive optics
to correct for the types of atmospheric-induced losses that were considered and
calculated in detail in Section 4.1 above.

\clearpage

\section{Discussion}

We have carried out a detailed
analysis of the various processes and constraints that
determine the operating characteristics of a practical quantum key distribution system.
This analysis applies generally to quantum cryptography systems built entirely
from currently available commercial technology.
Our results also allow us
to establish the requirements that need to be satisfied in order
to execute practical, unconditionally secret
high-speed quantum cryptography in actual physical environments.
An important objective has been to determine the extent to which it is possible
to increase system throughput rates to high values for actual
quantum cryptography systems. Insofar as is possible, we
have been guided by the desire to achieve this
solely with mature, currently available commercial technology.
Based on the results of our analysis,
we have proposed a general quantum cryptography
design that meets this criterion, with the only
system element that is not available today as mature technology identified
as the necessary fast photon detection apparatus.
We have mathematically
shown that it should be possible to achieve high-speed transmission of secret Vernam
cipher throughput in various scenarios
with sufficiently fast photon detectors that can operate at a speed of 10
Gigabits-per-second, and we have identified HEP photon detection as a
promising approach to solving this
technological problem. Although our calculations show that, due
to the extremely large losses and loads that
characterize and constrain any practical QC system
even this very high speed of photon detection will not suffice to achieve
Gigabit-per-second throughput values for a single beam arrangement, a properly
multiplexed multiple
beam architecture should be able to achieve rates approaching this.

{\bf {Added Note:}}
In the earlier versions of this paper it was asserted without proof
that the attacks considered in our analysis, namely, direct ({\it {i.e.}},
unambiguous state discrimination or ``USD"),
indirect ({\it {i.e.}},photon number splitting or ``PNS") and combined
(hybrid PNS and USD),
exhaust all possible individual attacks. In our paper we defined
individual attacks as
``those attacks that do not require that the enemy apply unitary transformations to
the intercepted state with a quantum computing device" ({\it {cf}} Footnote 33).
Adhering to this notion of individual attack, in this revision of our
paper we retract our unproved assertion that the three aforementioned
attacks exhaust all possible individual attacks. We thank Professor Jeffrey Shapiro
for having first pointed this out to us (Private communication, 21 November, 2000.)
What we do assert is that
our attack analysis (contained in Chapter 3) furnishes a general comparison
of the relative strengths of direct, indirect and combined attacks, subject to the
specific conditions prescribed in the text (such as the condition that the enemy cannot
manipulate the detector efficiency). As shown in Chapter 3, we demonstrate that the
combined attack is never the best choice, and our analysis determines the circumstances
under which
indirect (PNS) or direct (USD) constitute the stronger attack.

\section{Acknowledgements}

The members of the MITRE Quantum Information Processing Group, including
A. Donadio, M. Drake, R. Ewing, J. Guttman, P. Henry and J. Thayer are thanked
for many useful
discussions and comments, with special thanks to A. Donadio for carrying
out the FASCODE computer simulations and to R. Ewing and J. Guttman for
specific helpful
contributions to the research reported in this paper. J. Babcock, T. Elkins and R. Fante
are thanked for reading the draft version of this document and for providing
various helpful suggestions. M. Visser (Washington
University in St. Louis) is thanked for
several discussions and for carrying out the calculation in Appendix A.
The authors wish to thank R. Sobolewski (University of Rochester) and
I. Duling, III (U.S. Naval Research Laboratory) for useful discussions and
comments. The authors also wish to thank particular employees of the
U.S. National Security Agency for helpful comments and questions. GG in
addition wishes to thank Ja. F. Providakes and Ji. F. Providakes for encouragement,
and especially D. Lehman and the MITRE Technology Program for supporting this
work and helping him to establish the MITRE Quantum Information Processing Group.

\clearpage

\section{Appendices}
\appendix

\section{Derivation of the Relation between Intrinsic Channel Error and Polarizer Misalignment}

\def\N{{\mathcal{N}}}
\def\ab{{\alpha\beta}}
\def\d{{\mathrm{d}}}
\def\ew{\leftrightarrow}
\def\ns{\updownarrow}
\def\ne{\nearrow\hspace{-12pt}\swarrow}
\def\nw{\searrow\hspace{-12pt}\nwarrow}
\def\Prob{\mathrm{Prob}}
\def\Crossover{\mathrm{Crossover}}
\def\d{{\mathrm{d}}}
\def\half{{1\over2}}
\def\etal{{\em et al}}
\def\ie{{\em i.e.}}
\def\implies{\Rightarrow}
\def\average{{\mathrm{average}}}

In this section we shall develop for the case of a free-space implementation
of quantum cryptography an explicit estimate for $r_c$, the
intrinsic channel error rate. The most obvious of the problems that
could lead to intrinsic channel error are due to the misalignment of
the polarizers: if they are not quite at 45 degrees or 90 degrees the
analysis of the protocol must be altered; we now proceed to
perform such an analysis in terms of the angular mismatch of the
polarizers.

Suppose we have two polarizers, one at the emitting end and one at the
receiving end, that are rotated with respect to each other by an angle
$\Delta \theta$. Then the probability that a photon will get through
the second, given that it gets through the first (and so be detected
if it is emitted in the first place) is given by Malus's law:
\begin{equation}
\Prob(\Delta\theta) = \cos^2(\Delta\theta).
\end{equation}

In the situation we are interested in, the photon is emitted in one of
four polarization states, conventionally at roughly 0, 45, 90, or 135
degrees, ($\theta\in\{0,\pi/4,\pi/2,3\pi/4\}$: we shall denote these
nominal states by $\ew,\ne,\ns,\nw$ respectively) with roughly equal
(classical) probabilities of $\approx1/4$. Call the actual
probabilities $p_e^i$ (here $e$ stands for emission; $i$ stands for one of the four
orientations), and call the actual angles the emission polarizers are
set at $\theta_e^i$.

Now call the actual angles the reception polarizers are set at
$\theta_r^i$; and the actual probability that the reception polarizer is
set to this position $p_r^i$.

The probability that the detector actually sees a photon if it is at a
particular setting $\theta_r^i$ is
\begin{equation}
\Prob\left({\rm detect~at}~i\right) = \sum_j \; p_e^j \; \cos^2(\theta_e^j - \theta_r^i).
\end{equation}
(We note that, if all values are nominal so that $p_e\to1/4$,
$\Delta\theta\in\{0,\pi/4,\pi/2,3\pi/4\}$, this reduces to
$\Prob({\rm detect~at}~i)\to\half$, independent of $i$, as it should.)

If we now average over all detector settings, the probability of
detecting a photon is
\begin{equation}
\Prob\left({\rm detect}\right) =
\sum_i \sum_j \; p_e^j \; p_r^i \; \cos^2(\theta_e^j - \theta_r^i)~.
\end{equation}

This is rather complicated in general. We now consider a useful special
case by assuming that the probabilities are nominal, so that
$p_e^i=p_r^i=1/4$, and that the reception and emission polarizers are
each perfectly aligned internally, with at most a relative mismatch of
$\delta$. That is:
\begin{equation}
\theta_r^i\in\{0,\pi/4,\pi/2,3\pi/4\}
\end{equation}
and
\begin{equation}
\theta_e^i\in\{0+\delta,\pi/4+\delta,\pi/2+\delta,3\pi/4+\delta\}~.
\end{equation}
Then we find
\begin{eqnarray}
\Prob({\rm detect}) &=& (1/4) \cdot (1/4) \cdot 4 \cdot
\nonumber\\
&&
\quad
\bigg[
\cos^2(\delta) + \cos^2(\pi/4+\delta)
\nonumber\\
&&
\quad
+\cos^2(\pi/2+\delta) + \cos^2(3\pi/4+\delta)
\bigg]~,
\end{eqnarray}
so that
\begin{eqnarray}
\Prob({\rm detect}) &=& (1/4)
\left[
\cos^2(\delta) + \cos^2(\pi/4+\delta)
+\sin^2(\delta) + \sin^2(\pi/4+\delta)
\right]
\nonumber\\
&=& \half~.
\end{eqnarray}
This is exactly the same as if there were no angular mismatch, so we
do not lose here, but this is only the ``raw'' rate before any
processing. The next step proceeds as follows.
If Bob detects a photon and knows that the receiving
polarizer is at angle $\theta_r^i$, he may query Alice as to
whether her polarizer was nominally set at
$P_+=\{0,\pi/2\}=\{\ew,\ns\}$ or $P_\times=\{\pi/4,3\pi/4\}=
\{\ne,\nw\}$.  When $\theta_r^i$ as used by Bob is in the same nominal
polarization class as ($P_+$ or $P_\times$) as the bit sent by Alice,
Bob keeps it, and otherwise discards it, attributing it to a case of
crossed polarizers. Doing so nominally throws away half the received
bits.

If we allow $p_e^i$ to deviate from 1/4, the probability that Bob keeps
the received bit (received at $\theta_r^i$) is then given by summing the
probabilities that this bit was sent when Alice had her polarizers at
one of the nominally compatible positions:
\begin{equation}
\Prob\left({\rm keep~at}~i~{\Big\vert}~{\rm detect~at}~i\right) = \sum_j \; p_e^j \;
\Theta(j~{\rm compatible~with}~i)~.
\end{equation}
(Here $\Theta$ is 1 or 0 depending on whether the polarizer orientations
are compatible or incompatible.)

To be more explicit, we may write
\begin{equation}
\Theta(\ns,\ns) = \Theta(\ns,\ew) = \Theta(\ew,\ns) = \Theta(\ew,\ew) = 1~,
\end{equation}
\begin{equation}
\Theta(\ne,\ne) = \Theta(\ne,\nw) = \Theta(\nw,\ne) = \Theta(\nw,\nw) = 1~,
\end{equation}
%
%
\begin{equation}
\Theta(\ns,\nw) = \Theta(\ns,\nw) = \Theta(\ew,\nw) = \Theta(\ew,\ne) = 0
\end{equation}
and
\begin{equation}
\Theta(\nw,\ns) = \Theta(\nw,\ew) = \Theta(\ne,\ns) = \Theta(\ne,\ew) = 0~.
\end{equation}
This can be made more compact by defining the simple function:
\begin{equation}
\wp: \{\ew,\ne,\ns,\nw\} \to \{+,\times\}
\end{equation}
so that in terms of the ordinary Kronecker delta one has
\begin{equation}
\Theta(i~{\rm compatible~with}~j) = \delta_{\wp(i)\wp(j)}~,
\end{equation}
and thus
\begin{equation}
\Prob\left({\rm keep~at}~i~{\Big\vert}~{\rm detect~at}~i
\right) = \sum_j \; p_e^j \;  \delta
_{\wp(i)\wp(j)}~.
\end{equation}
Note that this is independent of the actual mismatch angles, and
depends only on {\it a priori} conventional decisions of what constitutes a
nominal match or mismatch of the polarization orientations. (We note that if
all $p_e^j=1/4$ we
have $\Prob({\rm keep~at}~i~{\vert}~{\rm detect~at}~i)\to (1/4) \cdot
(1+0+1+0) = (1/4) \cdot 2 = \half$, which is independent of $i$ as it should be.)

Now, the key step addresses the question:
What is the probability that an error is made in this last
negotiation?

An error occurs if, despite the fact that Bob and Alice agree on the
nominal polarization class ($P_+$ or $P_\times$) they disagree on the
value of the bit that was transmitted. This happens if $\theta_e
\neq\theta_r$ but $\theta_e$ and $\theta_r$ are in the same
polarization class. (That is, $\theta_e\approx \theta_r\pm\pi/2$;
we will write this as $\theta_e = \Crossover(\theta_r)$, meaning pick
the other supposedly orthogonal element of the polarization basis.) This
is explicitly given by:
\begin{equation}
\Crossover(\theta_r[\ew]) = \theta_e[\ns]~,
\end{equation}
\begin{equation}
\Crossover(\theta_r[\ne]) = \theta_e[\nw]~,
\end{equation}
\begin{equation}
\Crossover(\theta_r[\ns]) = \theta_e[\ew]
\end{equation}
and
\begin{equation}
\Crossover(\theta_r[\nw]) = \theta_e[\ne]~.
\end{equation}
The probability of such an error occurring is then:
\begin{equation}
\Prob\left({\rm error~at}~i~{\Big\vert}~{\rm keep~at}~i\right) = \cos^
2(\theta_r^i-\Crossover(\theta_r^i))~.
\end{equation}
The easiest case to deal with is when the emitting and receiving
polarizers are each internally properly aligned, and the only mismatch
is due to an overall rotation between the two. (We have already seen
that in this case $\Prob({\rm detect})=\half$ and $\Prob({\rm keep})=\half$ are
unaffected.) Recall that now we have as {\em exact} statements that:
\begin{equation}
\theta_r\in\{0,\pi/4,\pi/2,3\pi/4\}
\end{equation}
and
\begin{equation}
\theta_e\in\{0+\delta,\pi/4+\delta,\pi/2+\delta,3\pi/4+\delta\}~.
\end{equation}
In this situation
\begin{equation}
\Crossover(\theta_r) = \theta_r+\pi/2+\delta~,
\end{equation}
and the probability of error is independent of $i$ and equals
\begin{equation}
\label{misalign}
\Prob\left({\rm error}~{\Big\vert}~{\rm keep}
\right) = \cos^2(\pi/2+\delta) = \sin^2(\delta)~.
\end{equation}
In particular, even if $\delta$ is as big as 1/10 radian (5.7
degrees), the probability of error is less than $1\%$, which is a very useful
result, as this amount of airborne platform attitude control is easily achievable
in practice.

We now consider the completely general case with arbitrary values
for the system angles. We already have
\begin{equation}
\Prob\left({\rm detect}\right) =
\sum_i \sum_j \; p_e^j \; p_r^i \; \cos^2(\theta_e^j - \theta_r^i)~.
\end{equation}
In the same way we find
\begin{eqnarray}
\Prob\left({\rm keep}\right)
&=& \sum_i p_r^i \; \Prob\left({\rm keep~at}~i\right)
\\
&=& \sum_i p_r^i \; \Prob\left({\rm keep~at}~i~{\Big\vert}~{\rm detect~at}~i
\right) \; \Prob\left({\rm detect~at}~i\right)
\\
&=& \sum_i p_r^i \;
\left( \sum_j \; p_e^j \;  \delta_{\wp(i)\wp(j)} \right)
\left( \sum_j \; p_e^j \; \cos^2(\theta_e^j - \theta_r^i) \right).
\end{eqnarray}
We also have
\begin{eqnarray}
\Prob\left({\rm error}\right)
&=& \sum_i p_r^i \; \Prob\left({\rm error~at}~i\right)
\\
&=& \sum_i p_r^i \; \Prob\left({\rm error~at}~i~{\Big\vert}~{\rm keep~at}~i\right)
\nonumber\\
&&
\times \Prob\left({\rm keep~at}~i~{\Big\vert}~{\rm detect~at}~i
\right) \; \Prob\left({\rm detect~at}~i\right)
\\
&=& \sum_i p_r^i \;
\left( \cos^2(\theta_r^i-\Crossover(\theta_r^i)) \right)
\left( \sum_j \; p_e^j \;  \delta_{\wp(i)\wp(j)} \right)
\nonumber\\
&& \times
\left( \sum_j \; p_e^j \; \cos^2(\theta_e^j - \theta_r^i) \right).
\end{eqnarray}
Although these expressions are rather complicated
and too cumbersome to be analytically useful, they
do have all the correct limits. The important point is that with the application
of suitably high quality design control the possibility of
intrinsic mismatch should not constitute a high priority issue.

Thus, the final, practical result for actual systems is given
by eq.(\ref{misalign}):
The fractional error rate due to net polarizer misalignment
is given by
\begin{equation}
\label{332}
r_c = \sin^2(\delta)
\end{equation}
as presented in the text in eq.(\ref{206}).

\section{Packetization Approximation}

Consider a message of length $M$ bits.  Applying an error correction code to the message 
results in a string of length $\chi_{EC}$ bits.  If the communications protocol supports 
data frames of length $m_p$, then the number of packets transmitted is

\be
{\cal N} = \Big\lceil {{\chi_{EC} M} \over m_p} \Big\rceil  ~.  
\ee

All packets but the last are of length $m_p + f_o$, where $f_o$ is the frame 
overhead of the communications protocol.  The last packet is shorter, due to the 
fact that its data frame is not full, but contains only $\left( \chi_{EC} M \right) 
{\rm mod}~m_p$ bits.  The total number of bits in the transmission is then

\be
\label{E:pack1}
{\cal C} = \left( m_p + f_o \right) \Big\lfloor {{\chi_{EC} M} \over m_p} \Big\rfloor +
   \left( \chi_{EC} M \right) {\rm mod}~m_p + 
   f_o ~{\cal D} \left( \chi_{EC} M,~m_p \right)~,  
\ee

where

\be
{\cal D} \left( a,~b \right) \equiv
   \Bigg\{ {{1{\rm ~if~} a {\rm ~mod~} b \neq 0} \atop 
            {0{\rm ~if~} a {\rm ~mod~} b = 0}} ~.
\ee

The idea of the packetization approximation is to simplify this expression for large 
numbers of packets so as to avoid the mathematical complication introduced by 
treating the last packet as a special case.  We begin by writing the identity

\be
\chi_{EC} M = \Big\lfloor {{\chi_{EC} M} \over m_p} \Big\rfloor m_p + 
   \left( \chi_{EC} M \right) {\rm mod}~m_p ~,
\ee

from which it follows that 

\be
\Big\lfloor {{\chi_{EC} M} \over m_p} \Big\rfloor = 
   {{\chi_{EC} M} \over m_p} - {{\left( \chi_{EC} M \right) {\rm mod}~m_p} \over m_p}~.  
\ee

Substituting this in eq.(\ref{E:pack1}) gives

\bea
\label{E:pack2}
{\cal C}  
 &=& \left( m_p + f_o \right) {{\chi_{EC} M} \over m_p} - 
     \left( m_p + f_o \right) {{\left( \chi_{EC} M \right) {\rm mod}~m_p} \over m_p} 
     \nonumber\\
 &&+  \left( \chi_{EC} M \right) {\rm mod}~m_p + 
      f_o ~{\cal D} \left( \chi_{EC} M,~m_p \right) \nonumber\\
 &=& \left( 1 + {f_o \over m_p} \right) \chi_{EC} M + 
     f_o \left[ {\cal D} \left( \chi_{EC} M,~m_p \right) - 
               {{\left( \chi_{EC} M \right) {\rm mod}~m_p} \over m_p} \right]~.
\eea

The quantity in square brackets is always in the range $[0,1)$, so that 

\be
\label{E:packineq1}
f_o \left[ {\cal D} \left( \chi_{EC} M,~m_p \right) - 
          {{\left( \chi_{EC} M \right) {\rm mod}~m_p} \over m_p} \right] < f_o~.
\ee

If the message is long enough to require several packets, that is, if 

\be
\label{E:packcond}
{{\chi_{EC} M} \over m_p} >> 1 ~,
\ee

then

\be
f_o << f_o {{\chi_{EC} M} \over m_p} < 
  \left( 1 + {f_o \over m_p} \right) \chi_{EC} M ~,
\ee

which, with eq.(\ref{E:packineq1}), gives

\be
f_o \left[ {\cal D} \left( \chi_{EC} M,~m_p \right) - 
           {{\left( \chi_{EC} M \right) {\rm mod}~m_p} \over m_p} \right] <<
   \left( 1 + {f_o \over m_p} \right) \chi_{EC} M ~.
\ee

We may therefore neglect the term in square brackets in eq.(\ref{E:pack2}) to obtain

\be
\label{E:packapprox}
{\cal C} \simeq \left( 1 + {f_o \over m_p} \right) \chi_{EC} M ~,
\ee

which is valid as long as eq.(\ref{E:packcond}) is satisfied.  

\section{Statistical Results for Error Correction}

The length of the entire sifted string is $n$.  The number of errors in the string after the 
$i$th iteration of the error detection and correction step is denoted by 
$e_T^{\left( i \right)}$, and the number of errors before error correction begins 
is 

\be
e_T^{\left( 0 \right)} \equiv e_T~.
\ee

Each iteration of the error detection and correction step begins by breaking the 
string into blocks such that the expected number 
of errors in each block is less than or equal to the parameter $\varrho$.  The number of 
blocks in the string is then
\be
J^{\left( i\right)} = {\Big\lceil}{e_T^{\left( i-1\right)}\over\varrho}{\Big\rceil}~.
\ee
\noindent
and the average number of bits per block is

\be
k^{\left( i\right)} = {n\over J^{\left( i\right)}}~.
\ee

We wish to find an expression for the number of errors remaining after each iteration, 
from which the other parameters of interest can be obtained.  At the beginning of the 
$i$th iteration, there are $e_T^{\left( i-1 \right)}$ errors distributed among 
$J^{\left( i \right)}$ blocks, so that the probability of a given error being in a 
particular block is

\be
{\cal P} \left( {\rm one~specific~error~in~a~given~block} \right) =
{1 \over J^{\left( i \right)}} \simeq {\varrho \over e_T^{\left( i-1 \right)}} ~.  
\ee

The probability of $l$ errors occurring in a given block is given by a binomial 
distribution:

\be
{\cal P} \left( l~{\rm errors~in~a~given~block} \right) \simeq
      {\Bigg (} {e_T^{\left( i-1 \right)} \atop l} {\Bigg )} 
      {\Bigg (} {\varrho \over e_T^{\left( i-1 \right)}} {\Bigg )}^l 
      {\Bigg [} 1 - {\Bigg (} {\varrho \over e_T^{\left( i-1 \right)}} {\Bigg )}
                  {\Bigg ]}^{e_T^{\left( i-1 \right)} - l}~.
\ee

If the number of errors, and thus the number of blocks, is large, this can be 
approximated by a Poisson distribution \cite{papoulis} as follows:

\be
{\cal P} \left( l~{\rm errors~in~a~given~block} \right) \simeq
   e^{-\varrho} {\varrho^l \over l!} ~.  
\ee

The parity check will reveal an error in the block if and only if there is an 
odd number of errors in the block.  The probability of detecting an error is 
thus

\be
\label{E:parityfinderr1}
{\cal P} \left( {\rm {finding~an~error~in~the~block}} \right) \simeq
   e^{-\varrho} \sum_{l~{\rm odd}}^\infty {\varrho^l \over l!} ~.  
\ee

To find the sum over odd $l$, note first that

\be
e^\varrho = \sum_{l=0}^\infty {\varrho^l \over l!} ~,
\ee

and

\be
e^{-\varrho} = \sum_{l=0}^\infty \left( -1 \right)^l {\varrho^l \over l!} ~,
\ee

so that

\be
e^\varrho - e^{-\varrho} = 2 \sum_{l~{\rm odd}}^\infty {\varrho^l \over l!} ~,
\ee

or

\be
\sum_{l~{\rm odd}}^\infty {\varrho^l \over l!} = {1 \over 2} 
   \left( e^\varrho - e^{-\varrho} \right) ~,
\ee

and eq.(\ref{E:parityfinderr1}) becomes

\be
\label{E:parityfinderr2}
{\cal P} \left( {\rm {finding~an~error~in~the~block}} \right) \simeq
   {{1-e^{-2\varrho}} \over 2}~.  
\ee

The expected number of errors found in the $i$th iteration is then

\bea
e_f^{\left( i \right)}
&\simeq& J^{\left( i \right)} {{1-e^{-2\varrho}} \over 2} \nonumber\\
   &\simeq& e_T^{\left( i-1 \right)} {{1-e^{-2\varrho}} \over {2\varrho}}~.
\eea

The number of errors remaining after the $i$th iteration is 

\bea
\label{E:errsleft1}
e_T^{\left( i \right)}
&\equiv& e_T^{\left( i-1 \right)} - e_f^{\left( i \right)} \nonumber\\
   &\simeq& e_T^{\left( i \right)} 
         {\Bigg [} 1 - {{1-e^{-2\varrho}} \over {2\varrho}} {\Bigg ]} \nonumber\\
   &\simeq& e_T^{\left( i \right)} 
         {\Bigg (} {{2\varrho - 1 + e^{-2\varrho}} \over {2\varrho}} {\Bigg )}~.
\eea

The quantity in parentheses is a function of $\varrho$ only.  We introduce the notation 

\be
\beta \equiv {{2\varrho - 1 + e^{-2\varrho}} \over {2\varrho}}~,
\ee

in terms of which eq.(\ref{E:errsleft1}) becomes

\be
e_T^{\left( i \right)} \simeq \beta e_T^{\left( i-1 \right)}~.  
\ee

We proceed by induction to obtain the number of errors remaining after the $i$th 
iteration in 
terms of the initial number of errors and $\beta$, which is a function of the 
parameter $\varrho$ introduced to establish the block size for error detection:  

\be
\label{E:errsleft}
e_T^{\left( i \right)} \simeq \beta^i e_T^{\left( 0 \right)}~.  
\ee

The number of errors found during the $i$th iteration is then 

\bea
\label{E:errsfound}
e_f^{\left( i \right)}
&\simeq& e_T^{\left( i-1 \right)} - e_T^{\left( i \right)} \nonumber\\
   &\simeq& \left( 1-\beta \right) \beta^{i-1} e_T^{\left( 0 \right)}~. 
\eea

The number of blocks for the $i$th pass is approximately

\bea
\label{E:blocks}
J^{\left( i \right)}
&\simeq& {e_T^{\left( i-1 \right)} \over \varrho} \nonumber\\
   &\simeq& \beta^{i-1} {e_T^{\left( 0 \right)} \over \varrho}~, 
\eea

and the number of bits in a block is 

\bea
\label{E:bitsinblock}
k^{\left( i \right)}
&=& {n \over J^{\left( i \right)}}  \nonumber\\
   &\simeq& {{\varrho n} \over {\beta^{i-1} e_T^{\left( 0 \right)}}}~.
\eea

The error detection and correction step ends when there would be two or fewer 
blocks for the 
subsequent iteration.  The number of iterations completed, $N_1$, thus satisfies

\be
J^{\left( N_1+1 \right) } \leq 2~,
\ee

or

\be
\beta^{N_1} {{e_T^{\left( 0 \right)}} \over \varrho} \leq 2~,
\ee

so that

\be
\beta^{N_1} \leq {{2 \varrho} \over {e_T^{\left( 0 \right)}}}~,
\ee

and, since $\beta < 1$,

\be
\label{E:itern1}
N_1 \simeq {\Bigg\lceil} {{\log_2 {{2 \varrho} \over {e_T^{\left( 0 \right)}}}}
       \over {\log_2 \beta}} {\Bigg\rceil}~.  
\ee

The expected number of errors remaining after this step is then 

\bea
\label{E:errsrem}
e_T^{\left( r \right)} 
   &\simeq& \beta^{N_1} e_T^{\left( 0 \right)} \nonumber\\
   &\simeq& 2 \varrho ~.  
\eea

We next estimate the number of iterations in the validation step.  Each iteration detects 
either a single error or no errors.  Let $N_2^{\left( n \right)}$ denote the number 
of iterations in which no error is detected, and let $N_2^{\left( f \right)}$ denote 
the number of iterations in which an error is detected.  The iterations continue 
until there are $N_2$ successive iterations in which no error is detected.  

First we find $N_2^{\left( n \right)}$, which can be written

\be
\label{E:n2n0}
N_2^{\left( n \right)} \equiv 
   \sum_{l=0}^\infty {\hat N}^{\left( n \right)} \left( l \right)
      {\cal P} \left( l~{\rm residual~errors} \right) ~,
\ee

where ${\hat N}^{\left( n \right)} \left( l \right)$ is the expected number of 
error-free iterations when there are initially $l$ errors.  We find this by induction.  
If there are no errors, no iterations will detect an error, and the process will 
end after $N_2$ iterations:

\be
{\hat N}^{\left( n \right)} \left( 0 \right) = N_2~.  
\ee

If there are $l+1$ errors, then one of two things can occur.  There may be 
$N_2$ successive iterations without errors, or an error may be detected on the 
iteration following $N$ error-free iterations, where $0 \leq N \leq N_2-1$.  
If an error is found, the process repeats from a starting point of $l$ residual 
errors.  This gives the following recurrence relation:  

\bea
\label{E:n2n1}
{\hat N}^{\left( n \right)} \left( l+1 \right) &=&
N_2 {\cal P} \left( N_2~{\rm successive~error-free~iterations}~{\Big\vert}~
      l+1~{\rm residual~errors} \right)  \nonumber\\
   &&+
   \sum_{N=0}^{N_2-1} \left[ N + {\hat N}^{\left( n \right)} \left( l \right) \right] 
      {\cal P} \left( N~{\rm successive~error-free~iterations} ~{\Big\vert}~
         l+1~{\rm residual~errors} \right) \nonumber\\
   &&\cdot {\cal P} \left( 1 ~{\rm error-detected~iteration} ~{\Big\vert}~
         l+1~{\rm residual~errors} \right)~.
\eea

We need to find several probabilities to make use of this formula.  We begin by 
investigating the probability of an error-detected iteration given $l$ residual errors.  
The error detection algorithm first builds a block of bits by selecting 
bits from the original string at random.  The parity of the block is computed, and 
an error will be detected if there is an odd number of errors in the randomly selected 
block.  This gives the following expression for the desired probability:  

\be
{\cal P} \left( 1 ~{\rm error-detected~iteration} ~{\Big\vert}~ l~{\rm residual~errors}
\right)
   = \sum_{l'~{\rm odd}}^l \left({l \atop l'}\right) 
      \left({1 \over 2}\right)^{l'} \left({1 \over 2}\right)^{l-l'}
\ee

We use the identities

\be
\sum_{l'=0}^l \left( {l \atop l'} \right) \left({1 \over 2}\right)^{\lp}\left({1
\over 2}\right)^{l-\lp} -
\sum_{l'=0}^l \left( {l \atop l'} \right) \left({1 \over 2}^{\lp}\right)  
   \left( {-{1 \over 2}} \right)^{l-l'}   =
2 \sum_{l'~{\rm odd}}^l \left( {l \atop l'} \right) \left({1 \over 2}
\right)^{\lp}\left({1 \over 2}\right)^
{l-\lp}~,
\ee

and

\bea
\sum_{l'=0}^l \left( {l \atop l'} \right)\left({1 \over 2}\right)^{\lp}\left(
{1 \over 2}\right)^{l-\lp} -
\sum_{l'=0}^l \left( {l \atop l'} \right)\left({1 \over 2}\right)^{\lp} 
   \left( {- {1 \over 2}} \right)^{l-l'}
&=& \left( {1 \over 2} + {1 \over 2} \right)^l - 
         \left[ {1 \over 2} + \left( {-{1 \over 2}} \right) \right]^l \nonumber\\
&=& 1 ~,
\eea

which holds for $l \geq 1$.  This gives 

\be
\sum_{l'~{\rm odd}}^l \left({l \atop l'}\right)\left({1 \over 2}\right)^{\lp}
\left({1
\over 2}\right)^{l-\lp} = 
{1 \over 2} ~,
\ee

from which we obtain the result

\be
{\cal P} \left( 1 ~{\rm error-detected~iteration} ~{\Big\vert}~ l~{
\rm residual~errors} \right)
   = {1 \over 2}~,
\ee

provided $l {\not =} 0$.  If $l = 0$, there is no error to detect, and the probability 
is $0$, so that our final result is

\be
\label{E:probdet}
{\cal P} \left( 1 ~{\rm error-detected~iteration} ~{\Big\vert}~ l~{
\rm residual~errors} \right)
   = {1 \over 2} \left( 1 - \delta_{l,0} \right)~.  
\ee

This is a sensible result.  Since the block of bits is selected at random, there 
should be no bias towards an even or an odd number of errors in the block, unless 
there are no errors to start with.  

The probability of an error-free iteration is found by subtracting this result from 1:  

\be
\label{E:probnodet}
{\cal P} \left( 1 ~{\rm error-free~iteration} ~{\Big\vert}~ l~{
\rm residual~errors} \right)
   = {1 \over 2} \left( 1 + \delta_{l,0} \right)~.  
\ee

The probability of $N$ successive error-free iterations is thus 

\be
\label{E:probsuccnodet}
{\cal P} \left( N~{\rm successive~error-free~iterations} ~{\Big\vert}~ l~{
\rm residual~errors} \right) 
   = \delta_{l,0} + \left( 1 - \delta_{l,0} \right) \left({1 \over 2}\right)^N~.  
\ee

With these results, eq.(\ref{E:n2n1}) becomes

\bea
{\hat N}^{\left( n \right)} \left( l+1 \right) &=&  
   N_2 \left({1 \over 2}\right)^{N_2} \nonumber\\
   &&+
   \sum_{N=0}^{N_2-1} \left[ N + {\hat N}^{\left( n \right)} \left( l \right) \right] 
      \left({1 \over 2}\right)^N \cdot \left({1 \over 2}\right)  \nonumber\\
   &=&
   \left({1 \over 2}\right)^{N_2} N_2 \nonumber\\
   &&+
   \sum_{N=0}^{N_2-1} \left({1 \over 2}\right)^{N+1}
   \left[ N + {\hat N}^{\left( n \right)} \left( l \right) \right] \nonumber\\
   &=&
   \left[ 1 - \left( {1 \over 2 } \right)^{N_2} \right] + 
   \left[ 1 - \left( {1 \over 2 } \right)^{N_2} \right] 
   {\hat N}^{\left( n \right)} \left( l \right)~.  
\eea

Introducing 

\be
A \equiv \left[ 1 - \left( {1 \over 2 } \right)^{N_2} \right]~,
\ee

this becomes

\bea
{\hat N}^{\left( n \right)} \left( l+1 \right) 
   &=&  
   A + A {\hat N}^{\left( n \right)} \left( l \right) \nonumber\\
   &=& 
   A^{l+1} {\hat N}^{\left( 0 \right)} \left( l \right) + \sum_{s=1}^l A^s \nonumber\\
   &=&
   A^{l+1} N_2 + {{A-A^{l+2}} \over {1-A}}
\eea

Renaming the argument of the function to $l$ gives, for $l > 0$, 

\bea
\label{E:n2n2}
{\hat N}^{\left( n \right)} \left( l \right) 
   &=&
   A^{l} N_2 + {{A-A^{l+1}} \over {1-A}} \nonumber\\
   &=&
   A^{l} N_2 + {{1-A^l} \over {1-A}} A \nonumber\\
   &=&
   \left[ 1 - \left( {1 \over 2 } \right)^{N_2} \right]^l N_2 +
   2^{N_2} \left\{ 1 - \left[ 1 - \left( {1 \over 2 } \right)^{N_2} \right]^l \right\}
   \nonumber\\
   &=&
   \left[ 1 - \left( {1 \over 2 } \right)^{N_2} \right]^l 
   \left( N_2 - 2^{N_2} + 1 \right) + \left( 2^{N_2} -1 \right)
\eea

Practical values of $N_2 \sim 30$ imply that 

\be
\left( {1 \over 2 } \right)^{N_2} << 1~.  
\ee

In this limit, eq.(\ref{E:n2n2}) becomes

\bea
\label{E:n2n3}
{\hat N}^{\left( n \right)} \left( l \right) 
   &\simeq&
   \left[ 1 - l \left( {1 \over 2 } \right)^{N_2} \right]
   \left( N_2 - 2^{N_2} + 1 \right) + \left( 2^{N_2} -1 \right) \nonumber\\
   &\simeq&
   N_2 \left[ 1 - l \left( {1 \over 2 } \right)^{N_2} \right] +
   l \left( {1 \over 2 } \right)^{N_2} \left( 2^{N_2} - 1 \right) \nonumber\\
   &\simeq&
   N_2 + l~.  
\eea

Substituting this in eq.(\ref{E:n2n0}) yields

\bea
\label {E:n2n}
N_2^{\left( n \right)} 
   &\simeq& N_2 + <l> \nonumber\\
   &\simeq& N_2 + e_T^{\left( r \right)} \nonumber\\
   &\simeq& N_2 + 2 \varrho~,
\eea

where we have used eq.(\ref{E:errsrem}) to estimate the
number of errors at the beginning of the process.  

The recurrence relation for $N_2^{\left( f \right)}$, the number of iterations during 
which an error is detected is

\bea
\label{E:n2f1}
{\hat N}^{\left( f \right)} \left( l+1 \right) &=&  
   0 \cdot {\cal P} \left( N_2~{\rm successive~error-free~iterations} ~{\Big\vert}~ 
      l+1~{\rm residual~errors} \right) \nonumber\\
   &&+
   \sum_{N=0}^{N_2-1} \left[ 1 + {\hat N}^{\left( f \right)} \left( l \right) \right] 
      {\cal P} \left( N~{\rm successive~error-free~iterations} ~{\Big\vert}~
         l+1~{\rm residual~errors} \right) \nonumber\\
   &&\cdot ~{\cal P} \left( 1 ~{\rm error-detected~iteration} ~{\Big\vert}~
         l+1~{\rm residual~errors} \right)~.
\eea

The value at $l=0$ is

\be
{\hat N}^{\left( f \right)} \left( 0 \right) = 0~,
\ee

since there are no errors to be detected in this case.  Use of eq.(\ref{E:probdet}) and 
eq.(\ref{E:probsuccnodet}) yields the result

\bea
{\hat N}^{\left( f \right)} \left( l \right) 
   &\simeq& 
   \left( 2^{N_2} - 1 \right) 
   \left\{ 1 - \left[1 - \left( {1 \over 2} \right)^{N_2} \right]^l \right\} \nonumber\\
   &\simeq&
   l~,
\eea

from which we find

\bea
\label{E:n2f}
N_2^{\left( f \right)} 
   &\simeq& <l> \nonumber\\
   &\simeq& e_T^{\left( r \right)} \nonumber\\
   &\simeq& 2 \varrho~.  
\eea

Finally, we estimate an upper bound for the probability that an error remains after 
the validation step. The probability of one or more residual errors is 

\bea
\label{E:presid1}
p_{resid} 
   &=& {\cal P} \left( {\rm residual~errors} ~{\Big\vert}~ N_2~{
\rm error~free~iterations} \right)
   \nonumber\\
   &=& {{{\cal P} \left( {\rm residual~errors~and}~N_2~{\rm error~free~iterations}
\right)}
        \over
        {{\cal P} \left( N_2~{\rm error~free~iterations} \right)}}
   \nonumber\\
   &=& {{\sum_{l=1}^\infty 
           {\cal P} \left( N_2~{\rm error~free~iterations} ~{\Big\vert}~ l~{
\rm residual~errors}
\right)
           {\cal P} \left( l~{\rm residual~errors} \right)}
        \over
        {\sum_{l=0}^\infty 
           {\cal P} \left( N_2~{\rm error~free~iterations} ~{\Big\vert}~ l~{
\rm residual~errors}
\right)
           {\cal P} \left( l~{\rm residual~errors} \right)}}
   \nonumber\\
   &\leq& 
       {{\sum_{l=1}^\infty 
           {\cal P} \left( N_2~{\rm error~free~iterations} ~{\Big\vert}~ l~{
\rm residual~errors}
\right)
           {\cal P} \left( l~{\rm residual~errors} \right)}
        \over
        {{\cal P} \left( N_2~{\rm error~free~iterations} ~{\Big\vert}~ 0~{
\rm residual~errors}
\right)
         {\cal P} \left( 0~{\rm residual~errors} \right)}}~.\nonumber\\
\eea

By the argument that led to eq.(\ref{E:probsuccnodet}), if
there are residual errors in the 
string, the probability of $N_2$ successive error-free iterations is

\be
{\cal P} \left( N_2~{\rm error~free~iterations} ~{\Big\vert}~ l~{
\rm residual~errors} \right) = 
\left( {1 \over 2} \right)^{N_2}~,
\ee

and the probability of any number of error free iterations in the absence of residual 
errors is 1, so that eq.(\ref{E:presid1}) becomes

\bea
\label{E:presid2}
p_{resid} 
   &\leq&
   {{\sum_{l=1}^\infty \left( {1 \over 2} \right)^{N_2} 
      {\cal P} \left( l~{\rm residual~errors} \right)}
   \over
   {{\cal P} \left( 0~{\rm residual~errors} \right)}} \nonumber\\
   &\leq&
   {{\left( {1 \over 2} \right)^{N_2} {\cal P} \left( {\rm residual~errors} \right)}
   \over 
   {{\cal P} \left( 0~{\rm residual~errors} \right)}} \nonumber\\
   &\leq&
   {{\left( {1 \over 2} \right)^{N_2}}
   \over 
   {{\cal P} \left( 0~{\rm residual~errors} \right)}}~.
\eea

The probability of 0 residual errors after the validation step is no less than 
the probability of 0 residual errors after the error detection and correction step, since 
the validation step introduces no errors.  We expect the errors remaining after the 
error detection step to be Poisson distributed with mean $\sim 2 \varrho $, so
that

\be
{\cal P} \left( 0~{\rm residual~errors} \right) \geq e^{-2 \varrho}
\ee

and

\bea
\label{E:presid}
p_{resid} 
   &\leq&
   {{\left( {1 \over 2} \right)^{N_2}} \over {e^{-2 \varrho}}} \nonumber\\
   &\leq&
   e^{2 \varrho} \left( {1 \over 2} \right)^{N_2}~.
\eea

For a reasonable choice of $\varrho \sim 0.5$, the exponential factor is less than 
an order of magnitude, so that we may write

\be
\label{E:presid3}
p_{resid} \leq O\!\left(10\right)\cdot\left( {1 \over 2} \right)^{N_2}
\ee

for an upper bound on the probability that residual errors remain in the string 
after the validation step.  This is a crucial result, since even a single error 
in the string will render the entire key useless after the privacy amplification 
hash function is applied.  

\section{Assembly Code Segments}

These segments of assembly code were developed to support an estimate of the 
number of operations required to carry out the computations in sifting, 
error correction, and privacy amplification.  The emphasis is on the code 
that executes within loops that iterate through the bits of the key material 
being processed. In each case, a description of the context required for 
the code segment precedes the code itself.  

The assembly language used is a variant of the IBM 370 assembly 
language.  The principal extension is the use of register increment and decrement 
instructions.  We have also ignored the distinction between incrementing a counter 
by one and incrementing an index register by the size of an array element, since 
either operation requires exactly one instruction.  Note also the use of unsigned 
arithmetic operations when operating on multi-word integers.  

\subsection{Code to compute block parity}

The bits have been unpacked so that each bit occupies one word

Register B contains the address of the beginning of the block

Register END contains the address of the first word beyond the end of the block

When done, register PARITY will contain the parity of the block

\vspace{22 pt}

\halign {
# & \hskip 0.5 in # \hfil & \hskip 0.5 in # \hfil & \hskip 0.5 in # \hfil \cr
& SR & PARITY, PARITY & Clear parity register \cr
STARTLOOP & CR & B, END & Check if end of block \cr
&&&                       is reached \cr
& BGE & DONE \cr
& XOR & PARITY, 0(B) & Update parity register \cr
& INC & B & Increment to get address of \cr
&&&         next bit \cr
& B & STARTLOOP \cr
DONE & EQU & * & Exit from loop \cr
}

Instruction count: 5 instructions in loop

Iteration count: 1 iteration per input bit

\vspace{22 pt}

\subsection{Code to extract substring for hash function}

BUF contains the full bit string packed into memory

I0 contains the word index for the start of the substring

I1 contains the bit offset for the start of the substring

J0 contains the word index for the start of the next substring

J1 contains the bit offset for the start of the next substring

MASKA and MASKB are arrays containing bit masks for right and left justified 
substrings within a word

SHIFTA = $W - \log_2 W$ and SHIFTB = $\log_2 W$ are shift counts used 
to update the start of string indices, I0, I1, J0, and J1, for the next iteration, 
where $W$ 
is the wordsize of the machine

LEN = $2s$ is the size of the substring to which the hash function is applied

Register names of form R0, R1 indicate linked even-odd register pairs

When done, register pair A0, A1 will contain the substring, right justified

\vspace{22 pt}

\halign {
# & \hskip 0.5 in # \hfil & \hskip 0.5 in # \hfil & \hskip 0.5 in # \hfil \cr
& L & A1,BUF(I0) \cr
& N & A1,MASKA(I1) \cr
& SR & A0, A0 & First part of substring, right \cr
&&&             justified in A0+A1\cr
& LR & DELTA, J0 \cr
& SR & DELTA, I0 \cr
& C & DELTA, ONE & Substring spans 2 words or 3? \cr
& BLE & TWOWORDS \cr
& C & J1, ZERO \cr
& BE & TWOWORDS \cr
& LR & A0, A1 & Move first part of substring into \cr
&&&             higher register \cr
& LR & K, I0 \cr
& INC & K \cr
& L & A1, BUF(K) & Move second part of substring into \cr
&&&                lower register \cr
TWOWORDS & L & B1, BUF(J0) \cr
& N & B1, MASKB(J1) \cr
& SR & B0, B0 \cr
& SLDLR & B0, J1 & Final part of substring, right \cr
&&&                justified in B0 \cr
& SLDLR & A0, J1 \cr
& OR & A1, B0 & Substring right justified in A0+A1 \cr
& LR & I0, J0 & Increment substring indices for \cr
&&&                  next iteration \cr
& LR & I1, J1 \cr
& SLL & J1, SHIFTA \cr
& SLDL & J0, SHIFTB \cr
& A & J0, LEN \cr
& SRDL & J0, SHIFTB \cr
& SRL & J1, SHIFTA \cr
}

Instruction count: 26 instructions

\vspace{22 pt}

\subsection{Code to compute Carter-Wegman affine hash function for double-word integers}

Input in linked pair of registers A0, A1

Multiplier in M0, M1

Additive parameter in P0, P1

When done, result is in registers E0, E1, E2, and E3

\vspace{22 pt}

\halign {
# & \hskip 0.5 in # \hfil & \hskip 0.5 in # \hfil & \hskip 0.5 in # \hfil \cr
& LR & B1, A0 & Upper word \cr
& LR & C1, A1 & Lower word \cr
& LR & D1, A0 & Upper word \cr
& SR & A0, A0 & Clear upper words for \cr
&&&             multiplication \cr
& SR & B0, B0 \cr
& SR & C0, C0 \cr
& SR & D0, D0 \cr
& MU & A0, M1 & Multiply pairwise \cr
& MU & B0, M1 \cr
& MU & C0, M0 \cr
& MU & D0, M0 \cr
& SR & CARRY, CARRY & Clear carry for additions \cr
& LR & E3, A1 & Lowest order word \cr
& AUR & E3, P1 \cr
& BCZ & NOCARRY1 & Branch if carry is zero \cr
& INC & CARRY \cr
NOCARRY1 & LR & E2, CARRY & Next higher order word \cr
& SR & CARRY, CARRY \cr
& AUR & E2, A0 \cr
& BCZ & NOCARRY2 \cr
& INC & CARRY \cr
NOCARRY2 & AUR & E2, B1 \cr
& BCZ & NOCARRY3 \cr
& INC & CARRY \cr
NOCARRY3 & AUR & E2, C1 \cr
& BCZ & NOCARRY4 \cr
& INC & CARRY \cr
NOCARRY4 & AUR & E2, P0 \cr
& BCZ & NOCARRY5 \cr
& INC & CARRY \cr
NOCARRY5 & LR & E1, CARRY & Next higher order word \cr
& SR & CARRY, CARRY \cr
& AUR & E1, B0 \cr
& BCZ & NOCARRY6 \cr
& INC & CARRY \cr
NOCARRY6 & AUR & E1, C0 \cr
& BCZ & NOCARRY7 \cr
& INC & CARRY \cr
NOCARRY7 & AUR & E1, D1 \cr
& BCZ & NOCARRY8 \cr
& INC & CARRY \cr
NOCARRY8 & LR & E0, CARRY & Highest order word \cr
& AUR & E0, D0 \cr
}

Instruction count: 43 instructions

\vspace{22 pt}

\subsection{Code to compute multi-word hash function for privacy amplification}

Input string is the multiplicand ($N$ words)

Multiplier is first parameter of hash function ($N$ words)

Output array initially contains the second, additive parameter of hash function in 
the lower order $N$ words, the higher order $N+1$ words are clear ($2N+1$ words)

ASTART and AEND are addresses of highest and lowest order words of multiplier

BSTART and BEND are addresses of highest and lowest order words of multiplicand

CSTART is address of array of partial products

CROW is the size of a row of partial products ($2N + 1$ words)

CSIZE is the full size of the array of partial products ($N(2N+1)$ words)

CEND is address of first word beyond partial products array

DSTART and DEND are addresses of the highest and lowest order words in output array

When done, the output array contains the full result of the affine transformation.  
The selection 
of the hashed substring is accomplished by saving only the portions of the output 
array that constitute the hashed substring.  

\vspace{22 pt}

\halign {
# & \hskip 0.25 in # \hfil & \hskip 0.25 in # \hfil & \hskip 0.25 in # \hfil \cr
& SR & Z, Z & Zero \cr
& L & I, CSTART & Start of output array \cr
& LR & J, I \cr
& A & J, CSIZE & One word beyond end of array \cr
CLEAR & CR & I, J \cr
& BGE & MULTIPLY \cr
& ST & Z, 0(I) & Clear output array entry \cr
& INC & I \cr
& B & CLEAR \cr
MULTIPLY & L & I, AEND & Multiplier index \cr
& L & K, CSTART \cr
& A & K, CROW \cr
& DEC & K & Partial products end of row index \cr
MPLIER & C & I, ASTART \cr
& BL & ADD & Branch when done multiplication \cr
& L & MP, 0(I) & Get multiplier word \cr
& L & J, BEND & Multiplicand index \cr
& LR & L, K & Partial products entry index \cr
& LR & LP, L \cr
& DEC & LP & Next higher order entry index \cr
& LR & LPP, LP \cr
& DEC & LPP & Next higher order entry index \cr
MCAND & C & J, BSTART \cr
& BL & MCANDEND & Branch when this row done \cr
& L & MC1,0(J) & Get multiplicand word \cr
& SR & HCARRY, HCARRY & Clear high order carry \cr
& SR & MC0, MC0 \cr
& MUR & MC0, MP & Multiply unsigned \cr
& AU & MC1, 0(L) & Add carry from previous multiply \cr
& BCZ & NOCARRY1 & Carry? \cr
& INC & MC0 & Add in the carry \cr
& BCZ & NOCARRY1 &  Carry? \cr
& INC & HCARRY & Increment high order carry \cr
NOCARRY1 & A & MC0, 0(LP) & Add previous high order carry \cr
& BCZ & NOCARRY2 & Carry? \cr
& INC & HCARRY & Increment high order carry \cr
NOCARRY2 & ST & MC1, 0(L) & Store results \cr
& ST & MC0, 0(LP) \cr
& ST & HCARRY, 0(LPP) \cr
& DEC & J & Decrement indices for next word \cr
&&&         of multiplicand \cr
& DEC & L \cr
& DEC & LP \cr
& DEC & LPP \cr
& B & MCAND \cr
MCANDEND & DEC & I & Update indices for next word \cr
&&&                  of multiplier \cr
& A & K, CROW \cr
& DEC & K \cr
& B & MPLIER \cr
ADD & L & L, CSTART \cr
& A & L, CROW \cr
& DEC & L & Low order word in row \cr
& LR & M, DEND & Result index \cr
& SR & CARRY, CARRY & Clear carry \cr
OUTER & C & M, DSTART \cr
& BL & OUTEREND \cr
& LR & SUM, CARRY & Start with carry from previous sum \cr
& SR & CARRY, CARRY & Clear carry \cr
& LR & K, L & Entry index \cr
INNER & C & K, CEND \cr
& BGE & INNEREND \cr
& AU & SUM, 0(K) & Add unsigned \cr
& BCZ & NOCARRY3 \cr
& INC & CARRY \cr
NOCARRY3 & A & K, CROW & Next word to add \cr
& B & INNER \cr
INNEREND & ST & SUM, 0(M) & Store result in output array \cr
& DEC & L & Start for next column \cr
& DEC & M & Index for next entry in \cr
&&&         output array \cr
& B & OUTER \cr
OUTEREND & EQU & * & DONE \cr
}

\vspace{22 pt}

We introduce the notation

\be
N_w \equiv {n \over w}
\ee

for the number of words in the string to be hashed, where $n$ is the number of bits in 
the string and $w$ is the wordsize.  

Instruction and iteration counts:

CLEAR loop - 5 instructions, $(2N_w+1)N_w$ iterations

MPLIER loop - 13 instructions, $N_w$ iterations

MCAND loop - 22 instructions, $N_w^2$ iterations

OUTER loop - 9 instructions, $(2N_w+1)$ iterations

INNER loop - 7 instructions, $(2N_w+1)N_w$ iterations

Total: $9 + 43 N_w + 46 N_w^2$ instructions

\newpage


\end{document}